\documentclass{pas}

\usepackage{upgreek}
\usepackage{multirow}
\usepackage{bm}

\usepackage{amssymb}

\usepackage{makecell}
\usepackage{tikz-cd}
\usetikzlibrary{arrows.meta,positioning,shapes.geometric,calc,backgrounds}

\newcommand{\CL}{0}      
\newcommand{\CA}{4.6}
\newcommand{\CB}{9.2}
\newcommand{\CC}{13.2}

\newcommand{\RH}{2.6}    
\newcommand{\RA}{0}
\newcommand{\RB}{-3.5}
\newcommand{\RC}{-7}
\newcommand{\RD}{-10.5}
\newcommand{\RE}{-14}

\begin{document}

\lefttitle{Publications of the Astronomical Society of Australia}
\righttitle{Fortune et al.}

\jnlPage{1}{41}
\jnlDoiYr{2026}
\doival{10.1017/pasa.xxxx.xx}

\articletitt{Research Paper}

\title{New optimisations for exact, scalable 2D Gaussian process inference, with applications to astronomical time-series}
    
\author{Mark Fortune$^{1, 2}$\orcid{0000-0002-8938-9715}, Neale P. Gibson$^{2}$\orcid{0000-0002-9308-2353}, and Benjamin Pope$^{1}$\orcid{0000-0003-2595-9114}}

\affil{$^1$School of Mathematical \& Physical Sciences, Macquarie University, 12 Wally’s Walk, Macquarie Park, NSW 2113, Australia\\$^2$School of Physics, Trinity College Dublin, University of Dublin, Dublin 2, Ireland}

\corresp{Mark Fortune, Email: mark.fortune@mq.edu.au}
\history{(Received xx xx xxxx; revised xx xx xxxx; accepted xx xx xxxx)}

\begin{abstract}
Astronomical datasets often consist of a grid of datapoints in two or more dimensions, such as images, spectroscopic time-series, and imaging time-series. Gaussian processes (GPs) are a powerful tool for dealing with correlated noise in these data, arising from both astrophysical and instrumental systematics. Despite this, popular GP optimisations such as \textsc{celerite} are largely restricted to 1D datasets. Extensions to 2D data typically come with limited flexibility or poor scaling. As a motivating example, this limitation is significant for transiting exoplanet spectroscopy with JWST, as current analyses typically assume each spectroscopic light curve is statistically independent, despite many potential sources of wavelength-correlation, including detector readout noise and stellar granulation. This introduces a potential bias in atmospheric retrievals which is typically ignored.

To solve this problem, we present new, exact GP optimisations implemented in the \textsc{JAX}-based package \textsc{luas}, which extend 1D GP optimisations to multidimensional data using Kronecker product algebra. These GP optimisations are highly flexible, permit efficient analysis of 2D datasets containing millions of datapoints, and in many cases have a cheaper cost than transit light curve modelling. These optimisations are key to pushing the limits of exoplanet spectroscopy to smaller targets around bright host stars, as we increasingly hit instrumental and astrophysical noise floors. We suggest further applications to the analysis of ground-based high resolution spectroscopic time-series, extreme precision radial velocities, and high-contrast imaging, drawing on the connection that certain 2D GP models represent a probabilistic generalisation of PCA. Applications may even extend to machine learning and geostatistics, as we present novel optimisations for the Linear Mixing Model, and a special case of the Linear Model of Coregionalisation.
\end{abstract}

\begin{keywords}
methods: statistical, methods: data analysis, planets and satellites: atmospheres, planets and satellites: detection
\end{keywords}

\maketitle

\section{Introduction}

Gaussian processes (GPs) are a non-parametric fitting tool, which makes them very useful at modelling signals we cannot describe explicitly. They have therefore seen widespread use across many areas of astronomy, such as modelling stellar activity \citep{aigrain2012, Rajpaul2015}, asteroseismology and stellar granulation \citep{Brewer2009, Foreman_Mackey_2017}, brown dwarf variability \citep{Vos2022}, gravitational lensing \citep{Press1992}, pulsar timing delays \citep{Haasteren2014, Susobhanan2025}, atmospheric retrievals \citep{Rotman2025}, supernova light curves \citep{Lochner2016}, and many more (see \citealt{Aigrain2022} for a review). Outside of astronomy, they have also seen very diverse applications, including their original application for geostatistics \citep{krige1951}, and later uses within epidemiology \citep{Diggle2013}, ecology \citep{Dovers2024}, genetics \citep{Rakitsch2013, Velten2022}, and neuroimaging \citep{Cai2020}.

This work has mainly been motivated by their applications for modelling correlated noise in exoplanet transit time-series, originally introduced in \citet{Gibson2012}. Noise which is correlated in time, often appearing as `bumps' in the light curves, can be referred to as either correlated noise, red noise, or systematics, while noise which is independent in time (e.g. due to photon noise) is referred to as `white noise'. Correlated noise may arise from a variety of sources, typically split between instrumental sources (e.g. pointing instabilities, thermal breathing), and astrophysical sources (e.g. stellar granulation, asteroseismology and stellar activity). The key reason why we may want to joint-fit transit observations with both a deterministic transit model and a GP for modelling systematics, is that the GP can marginalise over an infinite space of possible systematics models and weight towards the best-fitting ones, which propagates the uncertainty introduced by complex noise sources into our measured transit depth uncertainties. \citet{Gibson2014} demonstrated using simulated data that GPs can produce robust results despite contamination from a wide variety of systematics models, and are often more robust than common alternatives, such as marginalising over a set of detrending vectors.

While Gaussian processes are typically applied to one-dimensional datasets (e.g. a time-series of flux values), there are many cases where it is useful to apply them to 2D datasets, modelling correlations across two input dimensions, such as wavelength and time. Despite being a standard approach within exoplanet atmospheres, fitting transit light curves one wavelength band at a time ignores the presence of wavelength-correlated noise, which can arise from effects such as detector readout noise or stellar granulation (e.g. \citealt{Schlawin2020, Gordon2020}). \citet{Fortune2024} demonstrated using simulated data that simultaneously fitting full spectroscopic time-series with a 2D Gaussian process can produce more robust transmission spectra in the presence of both wavelength- and time-correlated systematics. This could enable us to reliably analyse spectroscopic time-series at higher spectroscopic-resolution, while accounting for wavelength correlation due to instrumental effects such as interpixel capacitance, and 1/f readout noise (e.g. \citealt{Schlawin2020, Ahrer2022}). In the case of exoplanet detection, multi-band transit fitting \citep{Gordon2020}, as well as joint-fitting radial velocities with multiple stellar activity indicators \citep{Rajpaul2015}, have been shown to more reliably recover exoplanet signals in the presence of stellar variability. Additional uses of GPs applied to multi-dimensional datasets include joint-fitting pixel light curves for eclipse photometry \citep{Fortune2025}, modelling pulsar profile variations \citep{pulsar2DGPs}, and fitting multi-band supernova light curves \citep{Villar2020}.

Unfortunately, GPs can be expensive to scale to large datasets, especially data with multiple dimensions. In the most general case, GP log-likelihood evaluations scale as $\mathcal{O}(N^3)$ in runtime and $\mathcal{O}(N^2)$ in memory for $N$ datapoints. Fortunately, plenty of optimisations exist with significantly improved scaling in both runtime and memory, which either approximate log-likelihood calculations, or use a restricted set of kernel functions which have special mathematical structure. A popular example of the latter is \textsc{celerite}, a GP optimisation developed to fit one-dimensional inputs \citep{Foreman_Mackey_2017}, which has been used extensively for modelling systematics in transit light curves (e.g. \citealt{Espinoza2019, Barros2020, Alderson2023}), as well as for modelling stellar activity, granulation and oscillations (e.g. \citealt{Angus2018, Morris2020}). In fact, \textsc{celerite} is a special case of a broader class of GP optimisations which scale linearly on 1D (sortable) inputs. Quasiseparable kernels, as implemented in \textsc{tinygp} \citep{Foreman-Mackey2024}, \textsc{s+leaf} matrices from the package \textsc{spleaf} \citep{Delisle2020, sleaf}, and state-space models \citep{Hartikainen2010, Jordan2021} implemented in \textsc{smolgp} \citep{Rubenzahl2026}, can all be viewed as generalisations of \textsc{celerite} which can implement all \textsc{celerite} kernels, plus exact implementations of Matérn half-integer kernels, GP derivatives, and banded covariance matrices.

Various GP optimisations have been introduced for 2D `rectangular' datasets, such as a spectroscopic time-series, images, parallel time-series of stellar activity indicators, etc. The various 2D GP optimisations introduced include extensions of \textsc{celerite} \citep{Gordon2020, pulsar2DGPs}, extensions of state-space models \citep{sarkka2013}, and various approximations including inducing point formulations (e.g. \citealt{Snelson2005, Wilson2015}) and HODLR \citep{Ambikasaran2015}, not all of which require a complete rectangular dataset. Some very flexible approaches work by exploiting Kronecker product algebra, which use clever linear transformations to split up covariance matrix decomposition into separate covariance matrices in each dimension, such as in work by \citet{Saatchi2011}, \citet{Rakitsch2013}, and \citet{Bruinsma2019}. These algorithms do tend to be restricted to complete rectangular data without missing values, but non-uniform spacing in each dimension is allowed. In this work, we expand on such optimisations to include a much broader range of kernel functions, as well as identifying which algorithms can be combined with common 1D GP optimisations, including \textsc{celerite}, to achieve excellent scaling in one or multiple dimensions, while log-likelihood calculations remain exact.

These optimisations could be relevant to a wide range of exoplanet applications, including transit spectroscopy at both low- and high-resolution, as well as extreme precision radial velocities, and high-contrast imaging. In the discussion section, we demonstrate the close connections between these methods and popular data-driven algorithms like PCA and SysRem, but implemented as a likelihood-based approach which marginalises over components, can include smoothing priors on components, and features a principled treatment of noise. Of relevance outside of astronomy, where 2D GPs may instead be referred to as spatiotemporal GPs, multi-output GPs, or multi-task GPs, we provide a new, exact optimisation for the Linear Mixing Model, which combined with \citet{Gordon2020} and quasiseparable kernels, extends the Orthogonal Instantaneous Linear Mixing Model (OILMM) from \citet{Bruinsma2019} to drop any orthogonality requirement for the mixing matrix, while maintaining a linear runtime scaling in the number of datapoints.

Each optimisation identified has been implemented and validated in a new update to our open-source Python package \textsc{luas}, with tutorials and extensive documentation available\footnote{https://github.com/markfortune/luas}.

\section{Methods}
\subsection{Introduction to Gaussian processes}

Gaussian processes provide a very versatile method for calculating the mean and uncertainty of deterministic model parameters, given a dataset containing complex noise sources. As an example, suppose that we have a single transit light curve with time-correlated noise, where the data can be described as:
\begin{equation}
\mathbf{y} = \boldsymbol{\mu}(\boldsymbol{\theta}, \mathbf{t}) + \mathbf{f}_\text{sys} + \boldsymbol{\epsilon},
\end{equation}
 where $\boldsymbol{\mu}(\boldsymbol{\theta}, \mathbf{t})$ is an analytical model of the transit light curve observed at times $\mathbf{t}$, and parameterised by model parameters $\boldsymbol{\theta}$, $\mathbf{f}_\text{sys}$ is an unknown flux time-series due to some form of systematics (e.g. stellar variability), and $\boldsymbol{\epsilon}$ is the observed white noise arising from Poisson processes like photon counting statistics, with $\boldsymbol{\epsilon} \sim \mathcal{N}(\mathbf{0}, \bm{\Sigma})$, where the covariance matrix describing the white noise $\bm{\Sigma}$ is diagonal. $\boldsymbol{\mu}(\boldsymbol{\theta}, \mathbf{t})$ is typically referred to as the `mean function', and for transits is often modelled following \citet{mandelagol}. For any fixed mean function and systematics model, the likelihood function (the probability of the data given the model) is simply the probability of generating $\boldsymbol{\epsilon}$ from $\bm{\Sigma}$, which is given by the Gaussian probability density function:
  \begin{equation}
p(D|\boldsymbol{\theta}, \mathbf{t}, \mathbf{f}_\text{sys}) = \frac{1}{\sqrt{(2 \pi)^N|\bm{\Sigma}|}}\exp\left({-\frac{1}{2}\boldsymbol{\epsilon}^T\bm{\Sigma}^{-1} \boldsymbol{\epsilon}}\right)
\label{eq:likelihood_wn}
\end{equation}
which after taking the logarithm (useful for numerical stability) yields
 \begin{equation}
\log p(D|\boldsymbol{\theta}, \mathbf{t}, \mathbf{f}_\text{sys})= -\frac{1}{2}\boldsymbol{\epsilon}^T\bm{\Sigma}^{-1}\boldsymbol{\epsilon}-\frac{1}{2} \log |\bm{\Sigma}| - \frac{N}{2}\log 2 \pi.
\label{eq:logL_wn}
\end{equation}
 The systematics model may in reality describe the host star flux varying in time due to chaotic processes such as stellar granulation, it could describe variations in the throughput of the telescope, or it could be a combination of multiple processes varying over time which we may have no way to analytically model. To span this full space of models, we could fit for every element of $\mathbf{f}_\text{sys}$, introducing $N$ free parameters for $N$ datapoints, but doing this without any constraints would completely overfit our data. So we add a prior constraint on $\mathbf{f}_\text{sys}$ which penalises complexity, normally we impose that it should vary with some amplitude $h$, and it should have a correlation timescale it changes over, with nearby points in time having similar values, and widely separated points in time being almost fully independent. We can implement this by choosing a multivariate Gaussian prior $\mathbf{f}_\text{sys} \sim \mathcal{N}(\mathbf{0}, \mathbf{K}_\text{sys})$, where the covariance matrix $\mathbf{K}_\text{sys}$ is calculated using a kernel function which describes the covariance between each pair of points. For example, a common choice of kernel function is the exponential kernel
\begin{equation}
    k_t(t, t') = h^2 \exp\left(-\frac{|t - t'|}{l_{t}}\right),
    \label{eq:exp_kernel}
\end{equation}
where the covariance between two points decays exponentially with their separation in time, parameterised by the timescale $l_{t}$. The kernel function can be thought of as a `smoothness prior' on our systematics model $\mathbf{f}_\text{sys}$, i.e. whatever process is the cause of these systematics, we assume that it is continuous, varying over some characteristic time scale $l_{t}$, and with an average amplitude $h$.


Now that we have this smoothing prior on $\mathbf{f}_\text{sys}$, we can freely fit $N$ parameters to $N$ datapoints. However, if we optimised $\mathbf{f}_\text{sys}$ to its best-fit values and then kept it fixed, we would essentially be assuming that we have perfectly modelled out the systematics without any error, but the presence of white noise introduces uncertainty in the exact systematics model. Instead, we marginalise over all of the parameter space $\mathbf{f}_\text{sys} \in \mathbb{R}^N$, weighting towards the best-fitting models. This could be performed by marginalising all $N$ parameters of $\mathbf{f}_\text{sys}$ in an MCMC, while simultaneously inferring our model parameters $\boldsymbol{\theta}$. Thankfully, GPs offer a much more efficient approach, since $\mathbf{f}_\text{sys}$ consists of linear parameters with the multivariate Gaussian prior $p(\mathbf{f}_\text{sys}) = \mathcal{N}(\mathbf{0}, \mathbf{K}_\text{sys})$, we can exactly marginalise over it analytically (e.g. \citealt{Luger2017}). The result of integrating Equation~\ref{eq:logL_wn} over all values of $\mathbf{f}_\text{sys}$ results in the marginal log-likelihood
\begin{equation}
    \log p(D|\boldsymbol{\theta}, \mathbf{t}) = -\frac{1}{2}\mathbf{r}^T \mathbf{K}^{-1}\mathbf{r} - \frac{1}{2}\log|\mathbf{K}| - \frac{N}{2}\log(2\pi),
    \label{eq:loglikelihood}
\end{equation}
where we have written our GP covariance matrix as $\mathbf{K} = \mathbf{K}_\text{sys} + \bm{\Sigma}$, and our data residuals as $\mathbf{r} = \mathbf{y} - \boldsymbol{\mu}(\boldsymbol{\theta}, \mathbf{t})$. Combining this marginalised log-likelihood with Bayesian inference methods (e.g. MCMC, Nested Sampling) allows us to propagate the uncertainty introduced by the presence of unknown systematics into our transit model parameters $\boldsymbol{\theta}$, within a statistically rigorous and interpretable framework, and without sampling a huge number of free parameters.

If the constraint on the systematics model $\mathbf{f}_\text{sys}$ is desired, it is described by the GP predictive mean and predictive covariance $\mathbf{f}_\text{sys} \sim \mathcal{N}(\boldsymbol{\mu}_{f_\text{sys}}, \mathbf{K}_{f_\text{sys}})$, as given by the equations:
\begin{align}
\boldsymbol{\mu}_{f_\text{sys}} &= \mathbf{K}_\text{sys}\mathbf{K}^{-1}\mathbf{r} \\
\mathbf{K}_{f_\text{sys}} &= \mathbf{K}_\text{sys} - \mathbf{K}_\text{sys}^T\mathbf{K}^{-1}\mathbf{K}_\text{sys}.
\end{align}
This can be used to visualise the most strongly weighted systematics models the GP is fitting, and in fact the distribution of systematics can even be predicted for unseen observations using similar formulae (see Appendix~\ref{sec:GP_predictions}). The reason why we can get a constraint on unseen data is because the correlation described by the kernel function restricts how much the systematics can deviate for predicted locations close to the observed data, which should give a robust predictive distribution if our kernel functions are an appropriate model of the systematics. The predictive mean and covariance on unseen data can be used to perform cross-validation tests, which can help to validate the accuracy of a GP model.

\subsection{Implementing GP Inference}
\label{subsec:1D_GP_opt}

Performing inference with GPs typically involves calculating the log-likelihood function in Equation~\ref{eq:loglikelihood} many times ($\sim 10^5 - 10^6$ times for MCMC inference), which requires calculating both $\mathbf{r}^T \mathbf{K}^{-1} \mathbf{r}$, and the log-determinant $\log|\mathbf{K}|$ each time. There are various approaches to do this depending on the problem, and methods like Cholesky decomposition can avoid explicitly calculating $\mathbf{K}^{-1}$.

\subsubsection{Using Cholesky Decomposition}

A common approach is to calculate the Cholesky decomposition $\mathbf{K} = \mathbf{L} \mathbf{L}^T$, where $\mathbf{L}$ is a lower triangular matrix. Since $\mathbf{K}^{-1} = \mathbf{L}^{-T} \mathbf{L}^{-1}$, we can calculate:
\begin{align}
    \mathbf{r}^T \mathbf{K}^{-1} \mathbf{r} &= \mathbf{r}^T( \mathbf{L}^{-1})^T \mathbf{L}^{-1} \mathbf{r},\\
    &= \tilde{\mathbf{r}}^T \tilde{\mathbf{r}},
\end{align}
where $\tilde{\mathbf{r}} = \mathbf{L}^{-1} \mathbf{r}$. We can use the fact that $\mathbf{L}$ is lower triangular to efficiently solve the linear system $\mathbf{L}\tilde{\mathbf{r}} = \mathbf{r}$ using an algorithm known as forward-substitution, which scales as $\mathcal{O}(N^2)$ for a dense triangular solve. The log-determinant is simply calculated as $\log|\mathbf{K}| = \sum_i 2\log(L_{ii})$.
\subsubsection{Using Eigendecomposition}

\begin{figure}
    \centering
    \includegraphics[width=\columnwidth]{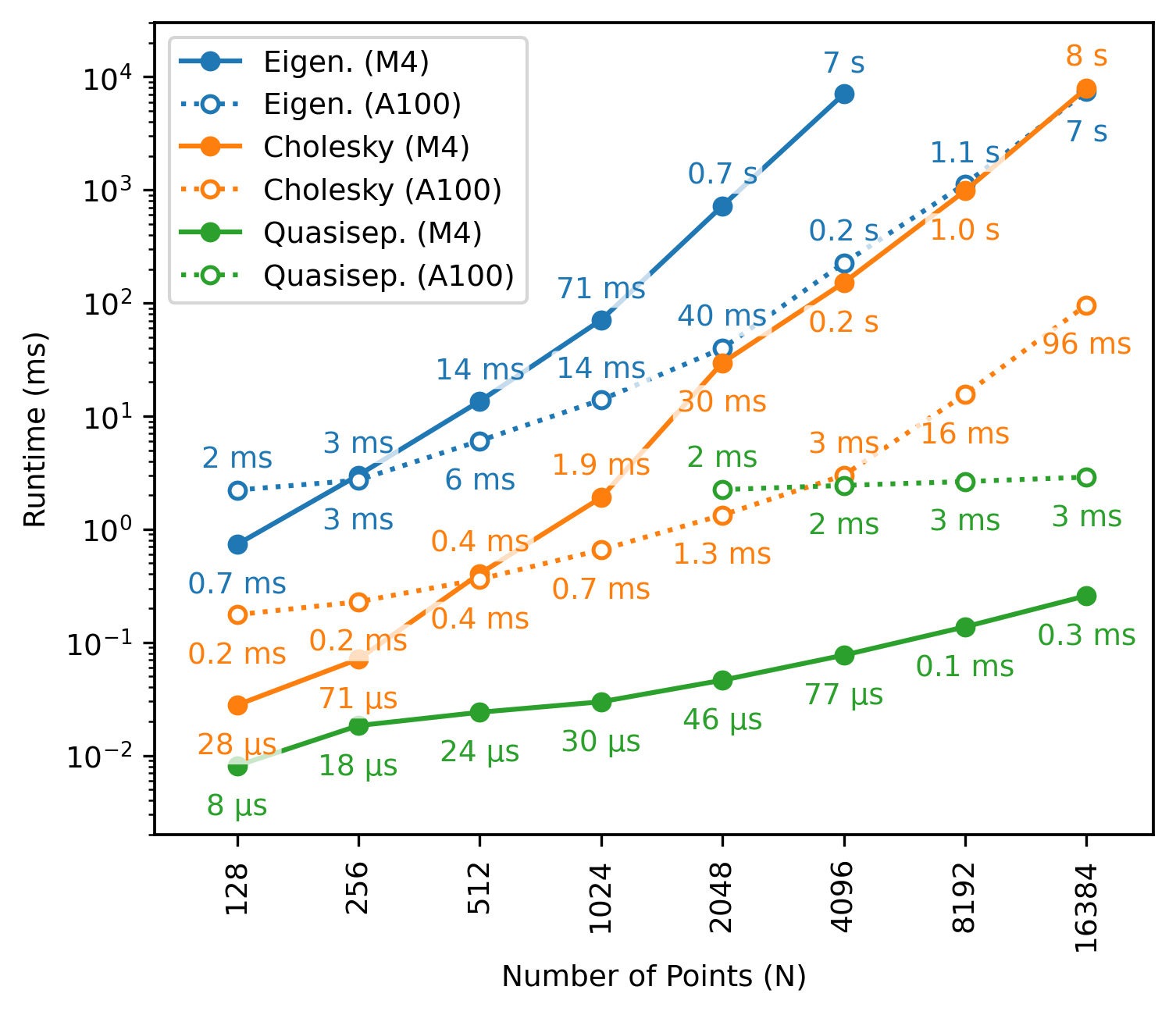}
    \caption{Runtime to decompose matrices using an M4 MacBook Air and an NVIDIA A100 GPU for eigendecomposition (using \textsc{jax.numpy.linalg.eigh}), general Cholesky decomposition (using \textsc{jax.scipy.linalg.cholesky}), and quasiseparable Cholesky decomposition of an exponential kernel (using \textsc{tinygp}). The GPU is typically slower for small matrices, but outperforms the CPU for larger matrices by exploiting more parallel workers.}
    \label{fig:matrix_decomp_cost}
\end{figure}

Cholesky decomposition is not the only matrix decomposition that can be used for GP log-likelihood calculations. We could instead calculate the eigendecomposition $\mathbf{K} = \mathbf{Q} \bm{\Lambda} \mathbf{Q}^T$, where $\mathbf{Q}$ is a matrix of eigenvectors of $\mathbf{K}$, and $\bm{\Lambda}$ is a diagonal matrix formed using the corresponding eigenvalues of those eigenvectors. While typically more expensive than Cholesky decomposition (see benchmarking in Figure~\ref{fig:matrix_decomp_cost}), eigendecomposition can be used to calculate the log-likelihood in a similar way
\begin{align}
    \mathbf{r}^T \mathbf{K}^{-1} \mathbf{r} &= \mathbf{r}^T (\bm{\Lambda}^{-\frac{1}{2}} \mathbf{Q}^T)^T (\bm{\Lambda}^{-\frac{1}{2}} \mathbf{Q}^T)\mathbf{r},\\
    &= \tilde{\mathbf{r}}^T \tilde{\mathbf{r}},
\end{align}
now with $\tilde{\mathbf{r}} = \bm{\Lambda}^{-\frac{1}{2}} \mathbf{Q}^T\mathbf{r}$, and the log-determinant is simply $\log |\mathbf{K}| = \sum_i \log \Lambda_{ii}$.

\subsubsection{Using Scalable 1D Approaches}
\label{sec:celerite}

Despite their performance differences, Cholesky decomposition and eigendecomposition both scale as $\mathcal{O}(N^3)$ in runtime and $\mathcal{O}(N^2)$ in memory for a general covariance matrix. These approaches can always be used for any valid covariance matrix, regardless of kernel function choice or the dimensions of the dataset. However, if we build the covariance matrix using \textsc{celerite} terms, the Cholesky decomposition can be computed in $\mathcal{O}(NJ^2)$ time, forward substitution can be applied in $\mathcal{O}(NJ)$ time, and the memory cost also scales as $\mathcal{O}(NJ)$ \citep{Foreman_Mackey_2017}, where $J$ refers to the number of \textsc{celerite} terms in the kernel function, which can be expressed as
\begin{align}
    k(t, t') = \sum_{j=1}^{J} e^{-c_j |t - t'|}\left[a_j\cos(d_j |t - t'|) + b_j \sin(d_j |t - t'|)\right].
    \label{eq:celerite_terms}
\end{align}
\textsc{celerite} can implement the exponential kernel and the simple harmonic oscillator kernel, which can be very useful for modelling stellar activity \citep{Foreman_Mackey_2017, Angus2018, Morris2020}.

For our work, we use the package \textsc{tinygp}, which implements a generalisation of \textsc{celerite} \citep{Foreman-Mackey2024}, using the algebra of quasiseparable matrices \citep{Eidelman1999}, and state-space models (e.g. \citealt{Hartikainen2010, Jordan2021}). It can implement all \textsc{celerite} kernels with the same runtime and memory scaling, but it also features numerically stable, exact implementations of the Matérn-3/2 and Matérn-5/2 kernels, and banded covariance matrices \citep{Foreman-Mackey2024}. It can also calculate GP derivatives following the framework of state-space models, similar to \citet{Rubenzahl2026}. While similar extensions have also been implemented in the package \textsc{s+leaf} \citep{Delisle2020, sleaf}, the quasiseparable kernels in \textsc{tinygp} are implemented entirely within \textsc{JAX}, which combined with our \textsc{JAX}-based implementation of \textsc{luas}, allows for efficient auto-differentiation and GPU hardware acceleration \citep{JAX}. This is useful for scaling to problems involving large datasets, and for performing high-dimensional parameter inference using efficient gradient-based methods. For this work, we will use $J$ to refer to the quasiseparable rank, where the exponential kernel is quasiseparable rank $1$, the Matern-3/2 kernel is rank $2$, and the kernel in Equation~\ref{eq:celerite_terms} actually has rank $2J$. Summing kernels will sum their quasiseparable ranks, while multiplying kernels multiplies their ranks. We note that some quasiseparable kernels theoretically follow a $\mathcal{O}(NJ^3)$ runtime scaling and $\mathcal{O}(NJ^2)$ memory scaling, such as high-order Matérn kernels and wide banded matrices, but these are not typically used in practice, and sums of arbitrary quasiseparable kernels will still tend towards \textsc{celerite} scalings by exploiting block diagonal transition matrices \citep{Foreman-Mackey2024}, so we use the scalings associated with \textsc{celerite} from \citet{Foreman_Mackey_2017} for calculations in this work. 

Unfortunately, while all these methods have excellent scaling, they depend on the sort order of their inputs, and cannot add or multiply kernels whose inputs have a different sort order. In practice, this makes them challenging to extend to 2D datasets, for example, if wavelength and time are kernel inputs on a spectroscopic time-series, it is not possible to sort all datapoints into chronological order, while never decreasing in wavelength. Creative approaches do exist to avoid this problem (e.g. \citealt{sarkka2013, Gordon2020, pulsar2DGPs}), but they can come at significant computational cost for large datasets.

\subsubsection{Using Any Matrix Square Root}
\label{sec:matrix_sqrt}

Both the Cholesky factor $\mathbf{L}$, and the term $\mathbf{Q}\bm{\Lambda}^{\frac{1}{2}}$, are distinct examples of matrix square roots $\mathbf{K}^{\frac{1}{2}}$ of $\mathbf{K}$, where we will use the notation that a matrix square root satisfies
\begin{equation}
    \mathbf{K} = \mathbf{K}^{\frac{1}{2}}(\mathbf{K}^{\frac{1}{2}})^T,
\end{equation}
and a matrix inverse square root satisfies
\begin{equation}
    \mathbf{K}^{-1} = (\mathbf{K}^{-\frac{1}{2}})^T\mathbf{K}^{-\frac{1}{2}},
\end{equation}
where for our purposes, we never require the matrix square root to be symmetric, but this notation does vary across the literature. By multiplying these two equations together, we see that if $\mathbf{K}^{\frac{1}{2}}$ is invertible, then its inverse must be a matrix inverse square root $\mathbf{K}^{-\frac{1}{2}}$.

We introduce this concept to notate a general framework to calculate $\mathbf{r}^T \mathbf{K}^{-1} \mathbf{r}$ in the GP log-likelihood, where we first multiply our data residuals by any matrix inverse square root $\tilde{\mathbf{r}} = \mathbf{K}^{-\frac{1}{2}}\mathbf{r}$, followed by squaring and summing the results to calculate $\tilde{\mathbf{r}}^T \tilde{\mathbf{r}}$. The log-determinant can often be calculated easily once a matrix square root is calculated, as noted for both the Cholesky decomposition and the eigendecomposition.

In addition, if we want to take a draw from the GP covariance matrix $\mathbf{K}$, which is equivalent to sampling from the prior the GP places on the space of systematics models $\mathbf{f}_\text{sys}$, we simply generate a random normal vector $\mathbf{z} \sim \mathcal{N}(\mathbf{0}, \mathbf{I}_N)$ and multiply by any matrix square root $\mathbf{r} = \mathbf{K}^{\frac{1}{2}}\mathbf{z}$, which will result in $\mathbf{r}$ having covariance $\mathbf{K}$, which can be shown from the definition of covariance $\mathbf{K} = \mathbb{E}((\mathbf{r} - \bar{\mathbf{r}})(\mathbf{r} - \bar{\mathbf{r}})^T)$. This can even be applied to sampling from the GP predictive covariance matrix (see Appendix~\ref{sec:GP_predictions}). Matrix square roots can therefore perform all of the typical calculations GPs are used for. We use this generalisation of factorising matrices into arbitrary matrix square roots to introduce modularity in our GP optimisations, scalable approaches using \textsc{celerite} and \textsc{tinygp} can be used for calculations with a linear runtime scaling when appropriate, but if an arbitrary covariance model is needed, general Cholesky decomposition or eigendecomposition can alternatively be used.

\subsection{Introduction to 2D Gaussian processes}
\label{sec:types_of_2D_kernels}

Consider the example of a spectroscopic time-series of a transiting exoplanet, containing $N_\lambda$ wavelengths and $N_t$ time points. We would like to recover the transit depth at each wavelength, accounting for some unknown, two-dimensional systematics model $\mathbf{F}_\text{sys}$, i.e.
\begin{equation}
    \mathbf{Y} = \mathbf{M}(\boldsymbol{\theta}, \boldsymbol{\lambda}, \mathbf{t}) + \mathbf{F}_\text{sys} + \mathbf{E}_{\lambda, t},
    \label{eq:2D_systematics}
\end{equation}
where our mean function $\mathbf{M}(\boldsymbol{\theta}, \boldsymbol{\lambda}, \mathbf{t})$ now produces a $(N_\lambda, N_t)$ 2D array of transit light curves parameterised by $\boldsymbol{\theta}$ (e.g. including transit depths at every wavelength), our systematics model $\mathbf{F}_\text{sys}$ is an unknown 2D array, and $\mathbf{E}_{\lambda, t}$ is a 2D array of independent white noise, which may be homoscedastic or heteroscedastic. Similar to the 1D case, we want to marginalise over $\mathbf{F}_\text{sys}$ by computing a marginal log-likelihood that avoids us needing to freely fit for every parameter in $\mathbf{F}_\text{sys}$ (which could require millions of free parameters!).

Note that typical transit spectroscopy analyses fit each spectroscopic light curve separately, and assume that the systematics in each light curve are independent (i.e. each row of $\mathbf{F}_\text{sys}$ is statistically independent). However, effects such as readout noise and various sources of stellar activity typically do not occur in isolation for each wavelength (e.g. \citealt{Schlawin2020, Morris2020}). Uncertainties in our mean function parameters $\boldsymbol{\theta}$ will not be correctly propagated if $\mathbf{F}_\text{sys}$ is not marginalised over accurately, typically if the process which produces $\mathbf{F}_\text{sys}$ is correlated in wavelength, then the uncertainty introduced to the transit depths will have a similar correlation in wavelength \citep{Holmberg2023, Fortune2024}. By joint-fitting the full dataset with a 2D GP, we can recover our transit depths reliably despite correlations in both wavelength and time. To help distinguish different noise sources, we divide them into a few useful scenarios.

\subsubsection{White noise}
\label{sec:white_noise}

We can expect some level of white noise in any observation, such as from photon noise. White noise has no correlation in time or wavelength, and is included by adding to the diagonal of the covariance matrix. Noise in real observations is almost always heteroscedastic, but an important distinction for this work is whether the amplitude of the white noise can be expressed in a separable way in each dimension, or if the amplitude of white noise must be specified explicitly for each datapoint. Mathematically we can separate these cases as
\begin{align}
    k_\text{separable}(\lambda, \lambda', t, t') &=  [\sigma^2(\lambda)\delta_{\lambda\lambda'}][\sigma^2(t)\delta_{tt'}],\\
    k_\text{non-separable}(\lambda, \lambda', t, t') &=  \sigma^2(\lambda, t)\delta_{\lambda\lambda'}\delta_{tt'},
    \label{eq:readout_noise}
\end{align}
where each Kronecker delta $\delta_{ij} = 1$ if $i=j$ and zero otherwise. The separable form produces structure in the covariance matrix on rectangular datasets which is important for our Kronecker product approach. Alternative 2D GP optimisations exist which can model non-separable noise amplitudes \citep{sarkka2013, Gordon2020}, but we will exclusively focus on optimisations for the separable case.

\subsubsection{Measurement Covariance}
\label{subsubsec:measurement_cov}

\begin{figure*}
        \centering
        \includegraphics[width=\textwidth]{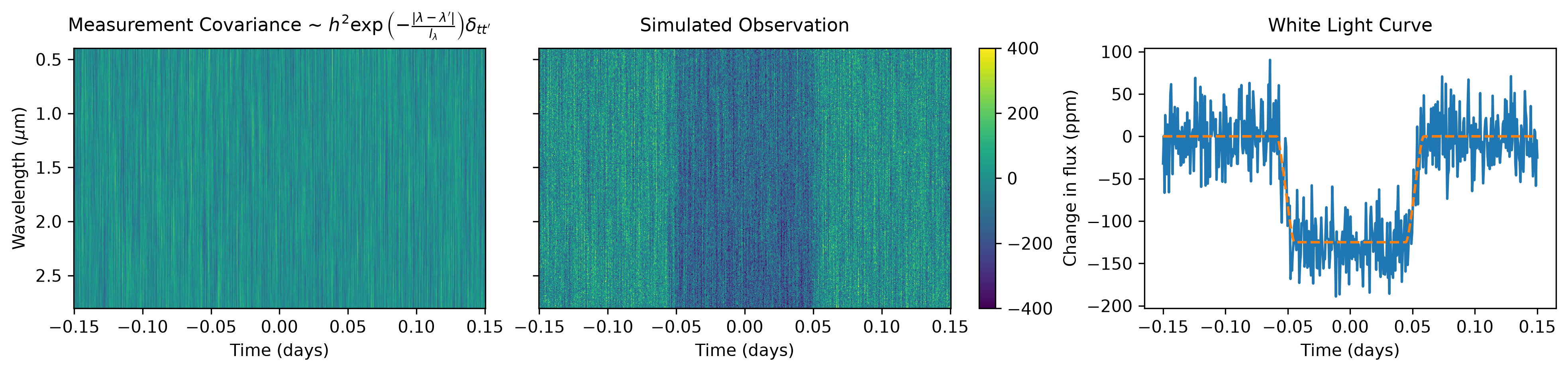}
        \caption{Simulated spectroscopic time-series of an exoplanet eclipse with measurement covariance added following the kernel function given in Equation~\ref{eq:measurement_cov}. Note the vertical streaks in the residuals which resemble real JWST/NIRCam data reductions, such as in \citet{Ahrer2022}. Despite simulating with very low white noise, the white light curve appears particularly noisy (compare with Figures~\ref{fig:low_rank_in_wv}, \ref{fig:low_rank_in_time}, \ref{fig:kernel_2D} which have the same amplitude of white noise), a consequence of how wavelength correlations bin down gradually compared to white noise.}
        \label{fig:readout_noise}
    \end{figure*}

In addition to white noise, we may also have noise which is uncorrelated in one of the dimensions, but correlated in the other. The case of time-correlated noise which is uncorrelated in wavelength is already an existing model, when time-correlated 1D GPs are fit to each spectroscopic light curve independently (which we return to in Section~\ref{subsubsec:2D_corr_terms}). For this section, we will focus on wavelength-correlated noise which is independent for every time point. For example, each spectrum we measure may include some wavelength correlation when being read out, such as due to residual 1/f readout noise in JWST data (e.g. \citealt{Schlawin2020, Ahrer2022}). See Figure~\ref{fig:readout_noise} for a simulated example of measurement covariance. In this case, our kernel function may be of the form
\begin{equation}
    k(\lambda, \lambda', t, t') =  [s_\lambda(\lambda, \lambda')] [\sigma^2(t)\delta_{tt'}],
    \label{eq:readout_noise}
\end{equation}
where the wavelength-correlation given by the kernel function $s_\lambda(\lambda, \lambda')$ models the \textit{measurement covariance}, but this wavelength-correlation is independent for each time point (i.e. each integration has an independent draw of readout noise). The square brackets highlight that this kernel function is also fully separable as a product of kernel functions in wavelength and time, important for the optimisations in this work. Within the notation of Equation~\ref{eq:2D_systematics}, this noise can be thought of as adding independent wavelength-correlated noise draws to each column of $\mathbf{F}_\text{sys}$. Equation~\ref{eq:readout_noise} includes a function $\sigma(t)$, which can be used to vary the amplitude of the measurement covariance in time, useful if the exposure time varies across the time-series.

Since both measurement covariance and white noise are independent in time, it is natural to include both terms within the kernel function $s_\lambda(\lambda, \lambda')$. For example, if readout noise can be described by an exponential kernel in wavelength, then we may have
\begin{equation}
   s_\lambda(\lambda, \lambda') = \sigma^2(\lambda)\delta_{\lambda\lambda'} + h^2_\textrm{readout}\exp\left(-\frac{|\lambda - \lambda'|}{l_\textrm{readout}}\right).
    \label{eq:measurement_cov}
\end{equation}
Alternatively, rather than parameterising using kernel functions, we could instead build the measurement covariance matrix $\mathbf{S}_\lambda$ directly using data from darks or from the autocovariance of the data residuals (e.g. \citealt{Schlawin2020, Holmberg2023}). However, parameterising measurement covariance in terms of kernel functions allows us to quantify the strength of each term, which could be important for comparing reduction choices, and it avoids conditioning on the best-fit residuals.

For the optimisations in this work, we will assume that the covariance matrices produced by $s_\lambda(\lambda, \lambda')$ and $s_t(t, t')$ are both positive definite. This will always be true if non-zero white noise is included in these terms plus any valid kernel terms, since the diagonal covariance matrix describing the white noise is guaranteed to be positive definite, and adding valid kernel terms to a positive definite matrix will always maintain positive definiteness. We label kernels as $s_\lambda$ and $s_t$ to highlight that these must produce positive definite terms, while any kernels labelled as $k_\lambda$ or $k_t$ can be either positive definite or positive semi-definite, which all kernels must satisfy by definition.

\subsubsection{Low-Rank Kernel Terms}
\label{subsubsec:low_rank_terms}

\begin{figure*}
        \centering
        \includegraphics[width=\textwidth]{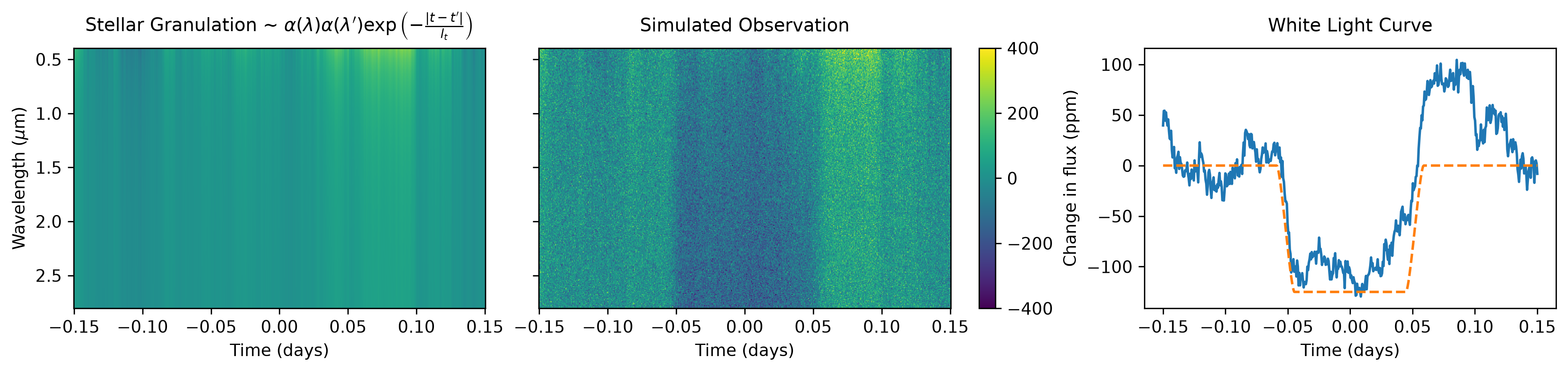}
        \caption{Same as Figure~\ref{fig:readout_noise} but injecting a stellar granulation signal following the kernel in Equation~\ref{eq:gordon_single}. The wavelength-dependence $\alpha(\lambda)$ follows the model in \citet{Morris2020} with an arbitrary scaling, while the time-dependence is stochastic and follows an exponential kernel in time. The systematics have the same shape in all light curves, but their amplitude is largest at optical wavelengths and gradually weakens into near-infrared wavelengths.}
        \label{fig:low_rank_in_wv}
    \end{figure*}

Some systematics sources may follow a Gaussian process in time, with only their amplitude varying in wavelength, i.e. they have the same shape in all light curves. Common-mode systematics are an example, where systematics have both the same shape and the same amplitude across all light curves. Stellar granulation has been suggested to approximately follow this model with a wavelength-dependent amplitude (e.g. \citealt{Gordon2020, Morris2020}). Figure~\ref{fig:low_rank_in_wv} simulates a simple stellar granulation signal following an exponential kernel function in time (see \citealt{Kallinger2014} for relevant time kernels), and a wavelength-dependence from \citet{Morris2020} for a $T_\mathrm{eff} = 6000$ K star, scaled to $80$ ppm at $0.4 \upmu$m. We can describe such a model with the kernel
\begin{equation}
    k(\lambda, \lambda', t, t') = [\alpha(\lambda)\alpha(\lambda')] [k_t(t, t')],
    \label{eq:gordon_single}
\end{equation}
where $\alpha(\lambda)$ is a deterministic function we can use to describe the amplitude in wavelength. Alternatively, we could freely fit for the amplitude at each wavelength in an amplitude vector $\boldsymbol{\alpha}_\lambda$, rather than expecting it to follow a physically-motivated model. While current data analyses may attempt to remove these systematics before spectroscopic light curve fitting (e.g. performing a common-mode correction), we can use such a kernel to joint-fit these systematics and propagate the uncertainty induced into the transit spectrum \citep{Fortune2024}. 

We will refer to these kernel terms as \textit{low-rank kernel terms} (or just \textit{low-rank terms} for short), since the wavelength covariance matrix built using $k_\lambda(\lambda, \lambda') = \alpha(\lambda)\alpha(\lambda')$ only has a single non-zero eigenvalue. It is also equivalent to a low-rank systematics model, of the form $\mathbf{F}_\text{sys} = \boldsymbol{\alpha}_\lambda \mathbf{v}_t^T$, where $\mathbf{v}_t \sim \mathcal{N}(\mathbf{0}, \mathbf{K}_t)$ is a 1D GP in time. In fact, we will show in Section~\ref{subsec:GPPCA} that this connects to a probabilistic interpretation of a single PCA component, with a GP smoothing prior placed on its time component.

\begin{figure*}
        \centering
        \includegraphics[width=\textwidth]{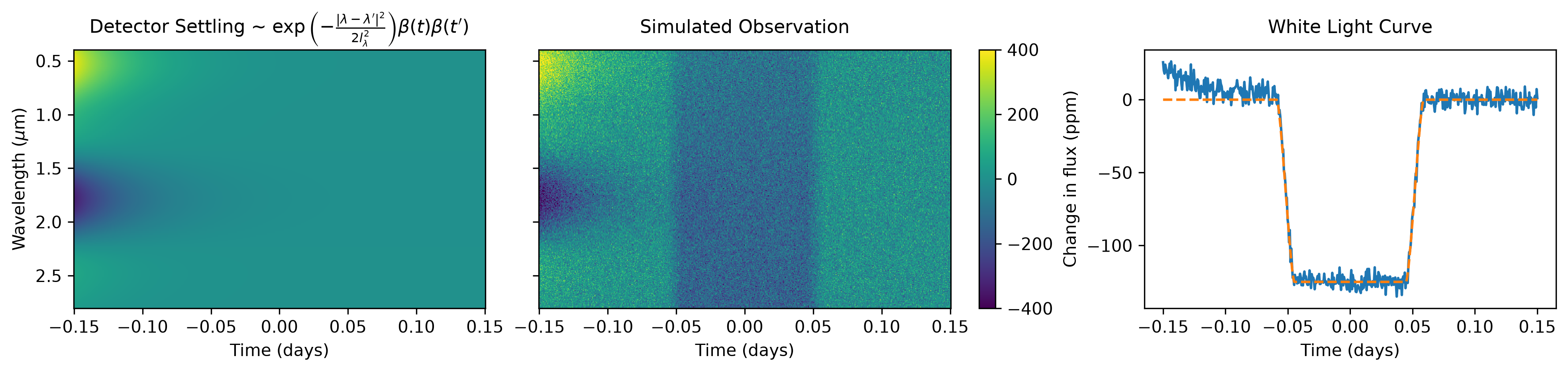}
        \caption{Analogous plot to Figure~\ref{fig:low_rank_in_wv} but with the kernel function flipped, injecting a low-rank term whose amplitude follows a deterministic exponential decay in time, but treating the amplitude in wavelength as a GP following a squared exponential kernel. An example use case is fitting spectroscopic light curves with wavelength-dependent settling, imposing that the settling amplitude varies gradually between wavelengths as a GP, but still explicitly defining how the settling varies in time within the function $\beta(t)$. Adding this term to the GP covariance matrix allows us to analytically marginalise out the wavelength-dependent settling amplitude.}
        \label{fig:low_rank_in_time}
    \end{figure*}

We might also consider a low-rank kernel term with the dimensions flipped, i.e.
\begin{equation}
    k(\lambda, \lambda', t, t') = [k_\lambda(\lambda, \lambda')][\beta(t)\beta(t')].
    \label{eq:low_rank_in_time}
\end{equation}
This type of kernel could be useful if fitting a systematic with a fixed shape in time, but with an unknown shape in wavelength. This could be most useful for analytically marginalising out baseline or detrending parameters that are fit for each light curve. As an example, modelling initial detector settling in spectroscopic time-series could be a use case, where the time-dependence could be explicitly modelled as a decaying exponential with some settling time length scale $l_\text{ramp}$, i.e.
\begin{equation}
    \beta(t) = \exp\left(-\frac{t - t_0}{l_\text{ramp}}\right)
\end{equation}
which is a deterministic function, not to be confused with the exponential kernel function. The wavelength-dependence could be treated as stochastic and independent for each wavelength, i.e.
\begin{equation}
    k_\lambda(\lambda, \lambda') = \sigma^2_\text{ramp}\delta_{\lambda\lambda'}.
\end{equation}
Such a model would be exactly equivalent to freely marginalising over an additive settling ramp model to every spectroscopic light curve independently, with a Gaussian prior on its amplitude $\mathcal{N}(0, \sigma^2_\text{ramp})$, and a shared decay length scale $l_\text{ramp}$. By adding this to the GP covariance matrix, we save ourselves fitting $N_\lambda$ free parameters, instead marginalising over their values analytically. Alternatively, we could use a kernel function in wavelength, if a smoothing prior is desired (see Figure~\ref{fig:low_rank_in_time} as a simulated example). Imposing a constraint that the settling amplitude varies smoothly in wavelength may help to improve constraints if pushing to analyse data at higher spectroscopic resolution, as freely fitting a ramp to each pixel channel could be poorly constrained. Some caveats of this approach are that anomalous ramps in individual pixels may be poorly modelled, and that this ramp model would have to be applied additively rather than being multiplicative (although if the transit model is close to unity at the beginning of the observation then this may be a negligible distinction). Other possible parameters which could be marginalised out include linear baseline terms, such as an offset or slope at each wavelength, or fitting regularised detrending vectors to each light curve, such as the $y$-offset of the spectrum, or the FWHM of the spectral trace.

    
\begin{figure*}
        \centering
        \includegraphics[width=\textwidth]{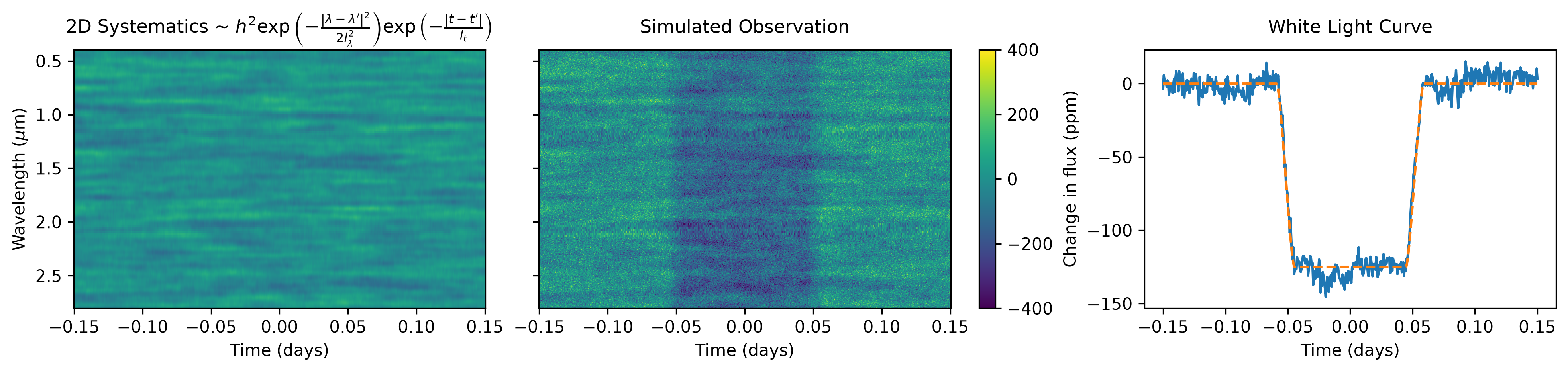}
        \caption{Similar to Figures~\ref{fig:readout_noise}, \ref{fig:low_rank_in_wv}, and \ref{fig:low_rank_in_time}, but with systematics simulated using a 2D kernel term (a squared exponential in wavelength times an exponential kernel in time), with a short wavelength length scale to simulate systematics that are close to independent for each light curve. This is similar to what is currently modelled when fitting a GP to each spectroscopic light curve independently. The height scale could also be varied in wavelength, in case there are specific wavelengths with strong systematics.}
        \label{fig:kernel_2D}
    \end{figure*}
    
\subsubsection{2D Kernel Terms}
\label{subsubsec:2D_corr_terms}

The final scenario we consider is systematics which change shape as a function of both time-separation and wavelength-separation. We focus on the case when the kernel function is formed from multiplying together separate kernel functions in wavelength and time:
\begin{equation}
    k(\lambda, \lambda', t, t') = k_\lambda(\lambda, \lambda') k_t(t, t'),
    \label{eq:2D_correlation}
\end{equation}
which we will refer to as a \textit{2D kernel term}. For example, we could have a squared exponential kernel with both wavelength and time as inputs
\begin{align}
    k(\lambda, \lambda'&, t, t') = h^2 \exp\left(-\frac{1}{2}\left[\frac{|\lambda - \lambda'|^2}{l_{\lambda}^2} +\frac{|t - t'|^2}{l_{t}^2}\right]\right)\\
    &= \left[h^2 \exp\left(-\frac{|\lambda - \lambda'|^2}{2 l_{\lambda}^2}\right)\right] \left[\exp\left(-\frac{|t - t'|^2}{2l_{t}^2}\right)\right],
    \label{eq:kernel2D}
\end{align}
which can be viewed as the product of separate wavelength and time squared exponential kernels. We can also just pick arbitrary kernels in wavelength and time to multiply together, e.g. a squared exponential in wavelength times a Matérn-5/2 kernel in time. Note that 2D kernel terms can produce a full-rank $\mathbf{F}_\text{sys}$ systematics model, they are not to be confused with the model $\mathbf{F}_\text{sys} = \mathbf{u}_\lambda \mathbf{v}_t^T$, where both $\mathbf{u}_\lambda$ and $\mathbf{v}_t$ represent independent GPs in wavelength and time.

In the limit as the wavelength length scale $l_{\lambda} \to 0$, these 2D kernels become equivalent to time-correlated systematics which are statistically independent for each wavelength. Figure~\ref{fig:kernel_2D} visualises this by simulating systematics with the 2D kernel using a very short wavelength length scale. This model is therefore similar to fitting 1D GPs to each spectroscopic light curve separately, just with a shared time length scale (these approaches are compared directly in \citealt{Fortune2024}). We could even have a wavelength-dependent height scale $h^2 \to h(\lambda)h(\lambda')$, if we want to account for specific wavelengths having larger systematics amplitudes.

\subsubsection{Combining All Terms: The \textsc{ExtendedLuasKernel}}

The primary focus of this work is a family of seven GP optimisations, which are all fully spanned by the \textsc{ExtendedLuasKernel}, which we define as
\begin{align}
    k(\lambda, \lambda', t, t') = s_\lambda(\lambda, \lambda') s_t(t, t') + k_{\lambda; \text{2D}}&(\lambda, \lambda')k_{t; \text{2D}}(t, t')\nonumber\\
    +\sum_{i=1}^{N_\alpha}\alpha_i(\lambda) \alpha_i(\lambda') k_{t_i}(t, t') + 
    \sum_{j=1}^{N_\beta}k_{\lambda_j}&(\lambda, \lambda')\beta_j(t) \beta_j(t'),
    \label{eq:full_luas}
\end{align}
where $s_\lambda$, $s_t$, $k_{\lambda; \text{2D}}$, $k_{t; \text{2D}}$, $k_{\lambda_j}$, $k_{t_i}$ can be arbitrary kernel functions in wavelength and time, and $\alpha_i(\lambda)$ and $\beta_j(t)$ are explicit user-defined functions (they may just build vectors of free parameters). The scaling in runtime and memory depends on which (if any) of these terms are non-zero, and which kernel terms can utilise GP optimisations, e.g. if they are quasiseparable. \textsc{luas} will automatically select from these optimisations based on the kernel function input by the user, although the user can also select the desired optimisation for cases where multiple are applicable.

\subsection{Implementing 2D GP Inference}

General approaches like Cholesky decomposition or eigendecomposition quickly become intractable beyond $N \sim 10^3$ datapoints, which is severely limiting for most 2D datasets. In fact, simply storing the covariance matrix $\mathbf{K}$ in memory can be prohibitive, e.g. a 2D dataset of shape $(10^3, 10^3)$ has a $(10^6, 10^6)$ covariance matrix, requiring 8 terabytes of memory at double precision. Methods which are scalable in both runtime and memory are needed to implement 2D GP inference for most datasets. Thankfully, there are already many existing approaches for this which we build from, with the most relevant for our work being \textsc{celerite}-based approaches, and Kronecker product algebra approaches.

\subsubsection{Existing \textsc{celerite}-based Approaches}

Simply multiplying together \textsc{celerite} kernels in wavelength and time is generally invalid, since the kernels must share the same input sort order\footnote{although there are some special cases, such as for periodic kernels \citep{pulsar2DGPs}}. However, low-rank kernel terms avoid this problem by only having the kernel applied to inputs along a single dimension. This was used in \citet{Gordon2020} to implement kernels of the form:
\begin{equation}
    k(\lambda, \lambda', t, t') = \sigma^2(\lambda, t)\delta_{\lambda\lambda'}\delta_{tt'} + \sum_{i=1}^{N_\alpha}\alpha_i(\lambda)\alpha_i(\lambda') k_{t_i}(t, t').
    \label{eq:gordon_kernel}
\end{equation}
It is equivalent to treating the 2D dataset as a single long time-series which can be fit using \textsc{celerite} kernels, first sorting all datapoints in chronological order, and then applying \textsc{celerite} while letting the white noise amplitude and kernel height scales vary for each point. Note that the white noise amplitude does not need to be separable in wavelength and time (distinguished in Section~\ref{sec:white_noise}). This approach naturally extends to quasiseparable kernels in \textsc{tinygp} with the same scaling as in Section~\ref{sec:celerite}, but for $N = N_\lambda N_t$ datapoints. For $N_\alpha$ quasiseparable kernels, we get a total quasiseparable rank of $J = \sum_i^{N_\alpha} J_{k_{t_i}} = N_\alpha \bar{J}_{k_t}$. This gives a total scaling of $\mathcal{O}(N_\alpha^2 N_\lambda N_t \bar{J}_{k_t}^2)$ in runtime, and $\mathcal{O}(N_\alpha N_\lambda N_t \bar{J}_{k_t})$ in memory. We will demonstrate how this can be further optimised with Kronecker product algebra using our \textsc{MultiLowRankKernel}.

If we instead have the kernel function
\begin{equation}
    k(\lambda, \lambda', t, t') = \sigma^2(\lambda, t)\delta_{\lambda\lambda'}\delta_{tt'} + k_\lambda(\lambda, \lambda') k_{t}(t, t'),
\end{equation}
we can express the wavelength covariance matrix built by $k_\lambda(\lambda, \lambda')$ as a sum of low-rank terms (e.g. using its eigendecomposition $\mathbf{K}_{\lambda} = \sum_i^{N_\lambda} \lambda_i \mathbf{v}_i \mathbf{v}_i^T$), and then use Equation~\ref{eq:gordon_kernel} with $N_\alpha = N_\lambda$, to decompose the matrix with a runtime scaling of $\mathcal{O}(N_\lambda^3 N_t J_{k_t}^2)$. Accounting for measurement covariance is not explored in \citet{Gordon2020}, but adding a \textsc{leaf} matrix as introduced in \citet{Delisle2020}, or adding a banded covariance matrix as implemented within \textsc{tinygp}, could add arbitrary measurement covariance with a similar $\mathcal{O}(N_\lambda^3 N_t)$ runtime scaling. While these optimisations are significantly faster than general matrix decomposition, they can still be prohibitive when $N_\lambda$ is large, which is a common case for many JWST spectroscopic time-series.

\subsubsection{Kronecker Product Algebra}

Our work combines various Kronecker product algebra methods, particularly inspired by the work of \citet{Saatchi2011} and \citet{Rakitsch2013}, to extend scalable 1D GP optimisations to 2D datasets for rectangular datasets. These are exact optimisations which can make the analysis of 2D datasets containing millions of datapoints tractable on a laptop. So far we have expressed our GP models in terms of kernel functions, but these optimisations are best derived by describing the covariance matrix of the GP in terms of its linear algebra structure. When we have a kernel term which is separable in each dimension, such as
\begin{equation}
    k(\lambda, \lambda', t, t') = k_{\lambda}(\lambda, \lambda') k_{t}(t, t'),
\end{equation}
and if our dataset lies on a complete 2D grid of input locations, e.g. we have flux values observed at wavelengths $\boldsymbol{\lambda}$ and times $\mathbf{t}$ without any missing values, the covariance matrix this builds will be a Kronecker product of separate covariance matrices in each dimension:
\begin{align}
(\mathbf{K}_\lambda)_{ij} &= k_{\lambda}(\lambda_i, \lambda_j)\\
(\mathbf{K}_t)_{ij} &= k_{t}(t_i, t_j)\\
\mathbf{K} &= \mathbf{K}_\lambda \otimes \mathbf{K}_t.
\end{align}
The Kronecker product operation multiplies each element in the left matrix times the full matrix on the right, for example:
\begin{equation}
    \mathbf{A} \otimes \mathbf{B} =
    \begin{bmatrix}
    a_{11} & a_{12} & a_{13}\\
    a_{21} & a_{22} & a_{23}\\
    a_{31} & a_{32} & a_{33}
    \end{bmatrix} \otimes \mathbf{B} = \begin{bmatrix}
    a_{11}\mathbf{B} & a_{12}\mathbf{B} & a_{13}\mathbf{B}\\
    a_{21}\mathbf{B} & a_{22}\mathbf{B} & a_{23}\mathbf{B}\\
    a_{31}\mathbf{B} & a_{32}\mathbf{B} & a_{33}\mathbf{B}
    \end{bmatrix}.
\end{equation}
For a spectroscopic time-series with $N_\lambda$ wavelengths and $N_t$ time points, this Kronecker product would form a $(N_\lambda N_t, N_\lambda N_t)$ matrix. Since this full covariance matrix could be prohibitively expensive to store in memory, methods that only require storing individual covariance matrices in each dimension can have significantly better scaling.

For the low-rank kernel terms in Section~\ref{subsubsec:low_rank_terms}, we can represent the covariance matrix they build using vector outer products. For example, our kernel term
\begin{equation}
    k(\lambda, \lambda', t, t') = [\alpha(\lambda)\alpha(\lambda')] [k_t(t, t')],
\end{equation}
builds a covariance matrix of the form
\begin{equation}
\mathbf{K} = \boldsymbol{\alpha}\boldsymbol{\alpha}^T \otimes \mathbf{K}_t,
\end{equation}
and the kernel term
\begin{equation}
    k(\lambda, \lambda', t, t') = [k_\lambda(\lambda, \lambda')][\beta(t)\beta(t')],
    \label{eq:low_rank_in_time}
\end{equation}
can be written as
\begin{equation}
\mathbf{K} = \mathbf{K}_\lambda \otimes \boldsymbol{\beta}\boldsymbol{\beta}^T,
\end{equation}
where the elements of these vectors are $\alpha_i = \alpha(\lambda_i)$, and ${\beta}_i = \beta(t_i)$.

Many useful properties of Kronecker product algebra are laid out in \citet{Saatchi2011}, we list some of these here
\begin{align}
    &(\mathbf{A} \otimes \mathbf{B})(\mathbf{C} \otimes \mathbf{D}) = \mathbf{A}\mathbf{C} \otimes \mathbf{B}\mathbf{D}, \label{eq:kronecker_distributive1}\\
    &\mathbf{A} \otimes (\mathbf{B} + \mathbf{D}) = \mathbf{A} \otimes \mathbf{B} + \mathbf{A} \otimes \mathbf{D},\label{eq:kronecker_distributive2}\\
    &(\mathbf{A} \otimes \mathbf{B})^{-1} = \mathbf{A}^{-1} \otimes \mathbf{B}^{-1},\label{eq:kronecker_inverse}\\
    &(\mathbf{A} \otimes \mathbf{B})^{T} = \mathbf{A}^{T} \otimes \mathbf{B}^{T},\label{eq:kronecker_transpose}\\
    &|\mathbf{A} \otimes \mathbf{B}| = |\mathbf{A}|^{N_B}|\mathbf{B}|^{N_A}
    \label{eq:kronecker_determinant}
\end{align}
where the last equation assumes $\mathbf{A}$ is of shape $(N_A, N_A)$ and similarly for $\mathbf{B}$. These properties demonstrate that Kronecker product algebra can dramatically simplify calculating the inverse and determinant of a large covariance matrix of the form $\mathbf{A} \otimes \mathbf{B}$, by breaking up the calculation into simpler decompositions of the individual matrices $\mathbf{A}$ and $\mathbf{B}$.

Finally, \citet{Saatchi2011} also present an efficient algorithm for performing matrix multiplication of a Kronecker product matrix times a vector, for an arbitrary $N_A N_B$ long vector $\mathbf{c}$:
    \begin{equation}\label{eq:Algo14}
        [\mathbf{A} \otimes \mathbf{B}] \mathbf{c} = \text{vec}(\mathbf{A} \mathbf{C} \mathbf{B}^T),
    \end{equation}
    where we reshape $\mathbf{c}$ into a $(N_A, N_B)$ matrix with $\mathbf{C}_{ij} = c_{(i-1)N_B + j}$ and the $\text{vec}()$ operator reverses this, such that $\text{vec}(\mathbf{C}) = \mathbf{c}$. This trick reduces what would be a runtime scaling of $\mathcal{O}(N_A^2 N_B^2)$, down to $\mathcal{O}(N_A^2 N_B + N_A N_B^2)$ in the worst case, and even faster if we can exploit special structure in the matrices $\mathbf{A}$ and/or $\mathbf{B}$ for fast matrix multiplications (e.g. for covariance matrices which are diagonal, low-rank, or built using \textsc{celerite} kernels). We make frequent use of this algorithm, and in our implementation we typically use the $(N_\lambda, N_t)$ residuals array $\mathbf{R}$, rather than a $N_\lambda N_t$-long vector $\mathbf{r}$.

\section{Results}
\label{sec:3}

We will build up to the \textsc{ExtendedLuasKernel} in Equation~\ref{eq:full_luas} by adding terms one-at-a-time. In practice, these more restricted optimisations are likely to be more useful for typical analyses, since they scale more efficiently, and most datasets likely won't require the full flexibility of the \textsc{ExtendedLuasKernel}.

\subsection{\textsc{SeparableKernel}: Accounting for Measurement Covariance}

Suppose we have any kernel which is fully separable in each dimension
\begin{equation}
    k(\lambda, \lambda', t, t') =  s_\lambda(\lambda, \lambda') s_t(t, t'),
    \label{eq:readout_noise2}
\end{equation}
which could be useful to account for both white noise and measurement covariance (e.g. following Equations~\ref{eq:readout_noise} and \ref{eq:measurement_cov}). When evaluated on a 2D grid of input locations, this can be written as a single Kronecker product of positive definite covariance matrices in each dimension:
\begin{equation}
    \mathbf{K} = \bm{\Sigma}_\lambda \otimes \bm{\Sigma}_t.
\end{equation}
\citet{Saatchi2011} introduces an efficient approach to decompose this matrix, which we reformulate in terms of arbitrary matrix square roots. We factor our covariance matrix into a product of terms using Kronecker product algebra
\begin{align}
\mathbf{K} &= 
   \bm{\Sigma}_\lambda^{\frac{1}{2}}(\bm{\Sigma}_\lambda^{\frac{1}{2}})^T \otimes \bm{\Sigma}_t^{\frac{1}{2}}(\bm{\Sigma}_t^{\frac{1}{2}})^T,\\
    &= 
    [\bm{\Sigma}_\lambda^{\frac{1}{2}} \otimes \bm{\Sigma}_t^{\frac{1}{2}}]
    [(\bm{\Sigma}_\lambda^{\frac{1}{2}})^T \otimes (\bm{\Sigma}_t^{\frac{1}{2}})^T],\\
   &= 
    \underbrace{[\bm{\Sigma}_\lambda^{\frac{1}{2}} \otimes \bm{\Sigma}_t^{\frac{1}{2}}]}_{\displaystyle\mathbf{K}_\mathrm{transf}}
    \underbrace{[\mathbf{I}_{N_\lambda} \otimes \mathbf{I}_{N_t}]}_{\displaystyle\tilde{\mathbf{K}}}
    \underbrace{[(\bm{\Sigma}_\lambda^{\frac{1}{2}})^T \otimes (\bm{\Sigma}_t^{\frac{1}{2}})^T]}_{\displaystyle\mathbf{K}_\mathrm{transf}^T},
\end{align}
where this factorisation includes a redundant identity matrix term labelled $\tilde{\mathbf{K}}$, which will be a useful concept for later optimisations.
We can invert the covariance matrix by inverting the individual terms:
\begin{equation}
    \mathbf{K}^{-1} = 
    \underbrace{[(\bm{\Sigma}_\lambda^{-\frac{1}{2}})^T \otimes (\bm{\Sigma}_t^{-\frac{1}{2}})^T]}_{\displaystyle\mathbf{K}_\mathrm{transf}^{-T}}
    \underbrace{[\mathbf{I}_{N_\lambda} \otimes \mathbf{I}_{N_t}]}_{\displaystyle\tilde{\mathbf{K}}^{-1}}
    \underbrace{[\bm{\Sigma}_\lambda^{-\frac{1}{2}} \otimes \bm{\Sigma}_t^{-\frac{1}{2}}]}_{\displaystyle\mathbf{K}_\mathrm{transf}^{-1}},
\end{equation}

\subsubsection{The Transformed Log-Likelihood}

We make use of an idea from \citet{Rakitsch2013}, where we interpret this log-likelihood calculation as taking the data residuals $\mathbf{r} = 
\mathbf{y} - \boldsymbol{\mu}$, performing the linear transformation $\tilde{\mathbf{r}} = \mathbf{K}_\mathrm{transf}^{-1}\mathbf{r}$ using Equation~\ref{eq:Algo14}, and then calculating the standard GP log-likelihood with the transformed residuals $\tilde{\mathbf{r}}$ and transformed covariance matrix $\tilde{\mathbf{K}}$
\begin{align}
\log p(D|\mathbf{\theta}) = &-\frac{1}{2}\mathbf{r}^T \mathbf{K}_\mathrm{transf}^{-T}\tilde{\mathbf{K}}^{-1}\mathbf{K}_\mathrm{transf}^{-1}\mathbf{r} - \frac{1}{2}\log|\mathbf{K}_\mathrm{transf}\tilde{\mathbf{K}}\mathbf{K}_\mathrm{transf}^T| \nonumber\\
&- \frac{N}{2}\log(2\pi),\\
\log p(D|\mathbf{\theta}) = &- \log|\mathbf{K}_\mathrm{transf}| +\nonumber\\
&\underbrace{\left[-\frac{1}{2}\tilde{\mathbf{r}}^T \bm{\tilde{K}}^{-1} \tilde{\mathbf{r}} - \frac{1}{2}\log|\tilde{\mathbf{K}}| - \frac{N}{2}\log(2\pi)\right]}_{\text{transformed log-likelihood}}
,
\label{eq:loglikelihood_transform}
\end{align}
where we introduce the \textit{transformed log-likelihood}, which differs from the normal GP log-likelihood by $\log |\mathbf{K}_\mathrm{transf}|$. For the \textsc{SeparableKernel}, we have
\begin{equation}
    \log |\mathbf{K}_\text{transf}| = \frac{N_t}{2}\log |\bm{\Sigma}_\lambda| + \frac{N_\lambda}{2}\log |\bm{\Sigma}_t|
    \label{eq:K_transf_det}
\end{equation}
and in fact this will be similar for many of the optimisations introduced, since additional orthogonal transformations in $\mathbf{K}_\mathrm{transf}$ will not affect its determinant.

For the \textsc{SeparableKernel}, $\tilde{\mathbf{K}}$ is just the identity matrix, so our transformed log-likelihood reduces to a simple least-squares likelihood on the transformed residuals. We can think of the transformation as `whitening' the data residuals; if measurement covariance is present and is modelled accurately, the effect would be to transform the residuals to remove any correlations, resulting in white noise with unit variance.

\subsubsection{Computational Scaling}

We note that since the transformation of the residuals can be performed with any matrix inverse square roots, \textsc{celerite} or quasiseparable covariance matrices can be combined with the \citet{Saatchi2011} approach to achieve a linear runtime scaling. If both $\bm{\Sigma}_\lambda$ and $\bm{\Sigma}_t$ are quasiseparable, it costs $\mathcal{O}(N_\lambda J_{\Sigma_\lambda}^2 + N_t J_{\Sigma_t}^2)$ to decompose both matrices, and $\mathcal{O}(N_\lambda N_t (J_{\Sigma_\lambda} + J_{\Sigma_t}))$ to transform the residuals. As an example, this could be applied for measurement covariance described by Equations~\ref{eq:readout_noise} and \ref{eq:measurement_cov}, an approach which is benchmarked in Figure~\ref{fig:SingleKronTermKernel_quasireadout_cost}. Even for a large spectroscopic time-series of $N_\lambda = 1024$ wavelengths and $N_t = 16384$ time points, we recover a log-likelihood runtime of just $27$ ms on CPU. For reference, using \textsc{jaxoplanet} to model transit light curves for the same dataset size took 0.2 seconds on CPU (see Figure~\ref{fig:jaxoplanet_transit_cost}). Our memory scaling is also linear, scaling as $\mathcal{O}(N_\lambda J_{\Sigma_\lambda} + N_t J_{\Sigma_t} + N_\lambda N_t)$.

If we instead have arbitrary matrices $\bm{\Sigma}_\lambda$ and $\bm{\Sigma}_t$, then we can simply use the existing \citet{Saatchi2011} approach with general Cholesky decomposition applied to both matrices. If we are just accounting for white noise and measurement covariance, then $\bm{\Sigma}_t$ must be diagonal and is therefore trivial to decompose. In this case, we get a runtime scaling of $\mathcal{O}(N_\lambda^3 + N_\lambda^2 N_t)$ and memory scaling of $\mathcal{O}(N_\lambda^2 + N_\lambda N_t)$ for any arbitrary measurement covariance $\bm{\Sigma}_\lambda$. The log-likelihood calculation for a $(N_\lambda, N_t) = (1024, 16384)$ dataset took 0.1 seconds with this model, see Figure~\ref{fig:SingleKronTermKernel_densereadout_cost} for more benchmarking.

\subsection{\textsc{LuasKernel}: Measurement Covariance Plus a 2D Kernel Term}

\citet{Rakitsch2013} identified an approach used within computational biology, which can fit any kernel function of the form
\begin{equation}
k(\lambda, \lambda', t, t') = s_{\lambda}(\lambda, \lambda')s_{t}(t, t') + k_{\lambda}(\lambda, \lambda')k_{t}(t, t')
\end{equation}
on a rectangular dataset, which corresponds to a covariance matrix which is a sum of two Kronecker products
\begin{equation}
    \mathbf{K} = \bm{\Sigma}_\lambda \otimes \bm{\Sigma}_t + \mathbf{K}_\lambda \otimes \mathbf{K}_t.
\label{eq:LuasKernel}
\end{equation}
This increase in complexity in allowing two separate time covariance matrices is extremely useful for simultaneously accounting for effects which are independent in time, such as white noise and measurement covariance, in addition to various sources of time-correlated systematics. This approach was adapted and applied to transit spectroscopy in \citet{Fortune2024}, motivating the creation of the original \textsc{luas} package, which is why we label it as the \textsc{LuasKernel}. The optimisation works by performing simultaneous diagonalisation on pairs of matrices, which we first describe in general terms.

\subsubsection{Simultaneous Diagonalisation}
\label{subsec:simult_diag}

Suppose we have two symmetric matrices $\bm{\Sigma}$ and $\mathbf{K}$, where at least $\bm{\Sigma}$ is positive definite. Can we perform the following transformation
\begin{equation}
\bm{\Sigma} + \mathbf{K} = \mathbf{S}[\mathbf{D}_\Sigma + \mathbf{D}_K]\mathbf{S}^T
\end{equation}
 using some invertible matrix $\mathbf{S}$, such that both $\mathbf{D}_\Sigma$ and $\mathbf{D}_K$ are diagonal?
 
In fact, we can always construct $\mathbf{S}$. First, we use a matrix inverse square root of $\bm{\Sigma}$ to transform $\mathbf{K}$ as follows:
\begin{equation}
    \mathbf{K}' = \bm{\Sigma}^{-\frac{1}{2}}\mathbf{K}(\bm{\Sigma}^{-\frac{1}{2}})^T
\end{equation}
where $\bm{\Sigma}^{-\frac{1}{2}}$ must exist because $\bm{\Sigma}$ is positive definite. This is followed by computing the eigendecomposition $\mathbf{K}' = \mathbf{Q}_{K'}^{}\bm{\Lambda}_{K'}^{}\mathbf{Q}_{K'}^T$, which doesn't require $\mathbf{K}'$ to be positive definite. $\mathbf{S}$ is then calculated as $\mathbf{S} = \bm{\Sigma}^{\frac{1}{2}}\mathbf{Q}_{K'}$, which we can verify solves the problem for $\mathbf{D}_\Sigma = \mathbf{I}_N$ and $\mathbf{D}_K = \bm{\Lambda}_{K'}$
\begin{align}
\label{eq:simult1_verify_start}
\bm{\Sigma} + \mathbf{K} &= \bm{\Sigma}^{\frac{1}{2}}\mathbf{Q}_{K'}[\mathbf{I}_N + \bm{\Lambda}_{K'}](\bm{\Sigma}^{\frac{1}{2}}\mathbf{Q}_{K'})^T\\
&= \bm{\Sigma}^{\frac{1}{2}}\mathbf{Q}_{K'}\mathbf{Q}_{K'}^T(\bm{\Sigma}^{\frac{1}{2}})^T + \bm{\Sigma}^{\frac{1}{2}}\mathbf{Q}_{K'}\bm{\Lambda}_{K'}\mathbf{Q}_{K'}^T(\bm{\Sigma}^{\frac{1}{2}})^T\\
&= \bm{\Sigma}^{\frac{1}{2}}(\bm{\Sigma}^{\frac{1}{2}})^T + \bm{\Sigma}^{\frac{1}{2}}\mathbf{K}'(\bm{\Sigma}^{\frac{1}{2}})^T\\
&= \bm{\Sigma} + \mathbf{K}
\label{eq:simult1_verify_end}
\end{align}
We see that simultaneous diagonalisation can always be performed on two symmetric matrices, provided that at least one of them is positive definite \citep{simult_diag_2mat}. In the case that $\mathbf{K}$ is positive semi-definite, then $\mathbf{K}'$ will also be positive semi-definite, which will simply result in some of the eigenvalues in $\bm{\Lambda}_{K'}$ equalling zero.


\subsubsection{The \textsc{LuasKernel} Factorisation}

The method in \citet{Rakitsch2013} performs a simultaneous diagonalisation in both dimensions. We use the inverse matrix square roots of $\bm{\Sigma}_\lambda$ and $\bm{\Sigma}_t$ to perform the transformations
\begin{align}
    \mathbf{K}'_\lambda &= \bm{\Sigma}_\lambda^{-\frac{1}{2}}\mathbf{K}_\lambda (\bm{\Sigma}_\lambda^{-\frac{1}{2}})^T,\\
    \mathbf{K}'_t &= \bm{\Sigma}_t^{-\frac{1}{2}}\mathbf{K}_t (\bm{\Sigma}_t^{-\frac{1}{2}})^T,
\end{align}
where we've assumed $\bm{\Sigma}_\lambda$ and $\bm{\Sigma}_t$ are both positive definite to do this, while $\mathbf{K}_\lambda$ and $\mathbf{K}_t$ can either be positive semi-definite or positive definite.

Next, we calculate the eigendecompositions of these transformed covariance matrices:
\begin{align}
 \mathbf{K}'_\lambda &= \mathbf{Q}_{\mathbf{K}'_\lambda}^{}\bm{\Lambda}_{\mathbf{K}'_\lambda}^{}\mathbf{Q}_{\mathbf{K}'_\lambda}^T\\
 \mathbf{K}'_t &= \mathbf{Q}_{\mathbf{K}'_t}^{}\bm{\Lambda}_{\mathbf{K}'_t}^{}\mathbf{Q}_{\mathbf{K}'_t}^T
 \end{align}
and we can then factorise the overall covariance matrix as
 \begin{align}
  \mathbf{K} &= \bm{\Sigma}_\lambda \otimes \bm{\Sigma}_t + \mathbf{K}_\lambda \otimes \mathbf{K}_t\nonumber\\
     &= 
    \underbrace{[\bm{\Sigma}_\lambda^{\frac{1}{2}}\mathbf{Q}_{\mathbf{K}'_\lambda} \otimes \bm{\Sigma}_t^{\frac{1}{2}}\mathbf{Q}_{\mathbf{K}'_t}]}_{\displaystyle\mathbf{K}_\mathrm{transf}}
    \underbrace{[\mathbf{I}_{N_\lambda} \otimes \mathbf{I}_{N_t} + \bm{\Lambda}_{K'_\lambda} \otimes \bm{\Lambda}_{K'_t}]}_{\displaystyle\tilde{\mathbf{K}}}\nonumber\\
    &\phantom{=} * \underbrace{[\mathbf{Q}_{\mathbf{K}'_\lambda}^T (\bm{\Sigma}_\lambda^{\frac{1}{2}})^T \otimes \mathbf{Q}_{\mathbf{K}'_t}^T (\bm{\Sigma}_t^{\frac{1}{2}})^T]}_{\displaystyle\mathbf{K}_\mathrm{transf}^T}.
    \label{eq:LuasKernelfactor}
\end{align}
Similar to the \textsc{SeparableKernel}, we can calculate our log-likelihood by first transforming our residuals vector as
\begin{equation}
\tilde{\mathbf{r}} = \mathbf{K}_\mathrm{transf}^{-1} \mathbf{r} = [\mathbf{Q}_{\mathbf{K}'_\lambda}^T\bm{\Sigma}_\lambda^{-\frac{1}{2}} \otimes \mathbf{Q}_{\mathbf{K}'_t}^T\bm{\Sigma}_t^{-\frac{1}{2}}]\mathbf{r},
\end{equation}
making use of Equation~\ref{eq:Algo14} to optimise this calculation. Our transformed log-likelihood in Equation~\ref{eq:loglikelihood_transform} will use the transformed covariance matrix $\tilde{\mathbf{K}}$, which is diagonal in this case and therefore trivial to invert and calculate its log-determinant. The log-determinant of the transformation $\log |\mathbf{K}_\text{transf}|$ will be the same as for the \textsc{SeparableKernel} (Equation~\ref{eq:K_transf_det}).

We can verify that this factorisation is correct by multiplying out the terms in Equation~\ref{eq:LuasKernelfactor}, similar to Equations~\ref{eq:simult1_verify_start}-\ref{eq:simult1_verify_end} for the one dimensional case
 \begin{align}
    \mathbf{K} &= \bm{\Sigma}_\lambda^{\frac{1}{2}}\mathbf{Q}_{\mathbf{K}'_\lambda} \mathbf{I}_{N_\lambda}\mathbf{Q}_{\mathbf{K}'_\lambda}^T (\bm{\Sigma}_\lambda^{\frac{1}{2}})^T \otimes  \bm{\Sigma}_t^{\frac{1}{2}}\mathbf{Q}_{\mathbf{K}'_t} \mathbf{I}_{N_t} \mathbf{Q}_{\mathbf{K}'_t}^T (\bm{\Sigma}_t^{\frac{1}{2}})^T\nonumber\\
    &+ \bm{\Sigma}_\lambda^{\frac{1}{2}}\mathbf{Q}_{\mathbf{K}'_\lambda} \bm{\Lambda}_{K'_\lambda} \mathbf{Q}_{\mathbf{K}'_\lambda}^T (\bm{\Sigma}_\lambda^{\frac{1}{2}})^T \otimes \bm{\Sigma}_t^{\frac{1}{2}}\mathbf{Q}_{\mathbf{K}'_t}\bm{\Lambda}_{K'_t} \mathbf{Q}_{\mathbf{K}'_t}^T (\bm{\Sigma}_t^{\frac{1}{2}})^T, \\
    &= \bm{\Sigma}_\lambda^{\frac{1}{2}}(\bm{\Sigma}_\lambda^{\frac{1}{2}})^T \otimes \bm{\Sigma}_t^{\frac{1}{2}}(\bm{\Sigma}_t^{\frac{1}{2}})^T + \bm{\Sigma}_\lambda^{\frac{1}{2}} \mathbf{K}'_\lambda (\bm{\Sigma}_\lambda^{\frac{1}{2}})^T \otimes \bm{\Sigma}_t^{\frac{1}{2}} \mathbf{K}'_t (\bm{\Sigma}_t^{\frac{1}{2}})^T,\\
    &= \bm{\Sigma}_\lambda \otimes \bm{\Sigma}_t + \mathbf{K}_\lambda \otimes \mathbf{K}_t.
    \label{eq:LuasKernelmultout}
\end{align}

\subsubsection{Computational Scaling}

The \textsc{LuasKernel} has a runtime scaling of $\mathcal{O}(N_\lambda^3 + N_t^3 + N_\lambda N_t(N_\lambda + N_t))$, and a memory scaling of $\mathcal{O}(N_\lambda^2 + N_t^2 + N_\lambda N_t)$. While it has a nice level of flexibility, capable of being used with any valid kernel functions, the need to eigendecompose two matrices of shape $(N_\lambda, N_\lambda)$ and $(N_t, N_t)$ can be a major limitation, restricting us from easily scaling it beyond $\sim 1000-4000$ points in either dimension. This can be seen from the benchmarking in Figures~\ref{fig:LuasKernelND_cost} and \ref{fig:LuasKernelND_A100_cost} for CPU and GPU respectively. As we highlight in Figure~\ref{fig:matrix_decomp_cost}, eigendecomposition can be significantly slower than general Cholesky decomposition - despite both scaling as $\mathcal{O}(N^3)$ - while Cholesky decompositions using \textsc{celerite} and quasiseparable matrices are dramatically faster. Thus, the \textsc{LuasKernel} is typically bottlenecked by the eigendecomposition of the larger of the two covariance matrices.

Compared with \citet{Rakitsch2013}, the only further optimisation we introduce with the \textsc{LuasKernel} is that we generalise to any matrix square root to decompose $\bm{\Sigma}_\lambda$ and $\bm{\Sigma}_t$ (e.g. Cholesky decomposition, \textsc{celerite}), rather than specifically using eigendecomposition, which can roughly halve the runtime cost.

\subsection{\textsc{LuasLasrachKernel}: Extending 1D GP Optimisations to 2D}

To avoid the cost of eigendecomposition in both dimensions, an alternative approach is to perform the simultaneous diagonalisation in just one of the dimensions:
\begin{align}
    \mathbf{K} &= \bm{\Sigma}_\lambda \otimes \bm{\Sigma}_t + \mathbf{K}_\lambda \otimes \mathbf{K}_t,\\
    \mathbf{K} &= \underbrace{[\bm{\Sigma}_\lambda^{\frac{1}{2}}\mathbf{Q}_{\mathbf{K}'_\lambda} \otimes \mathbf{I}_{N_t}]}_{\displaystyle\mathbf{K}_\mathrm{transf}}
    \underbrace{[\mathbf{I}_{N_\lambda} \otimes \bm{\Sigma}_{t} + \bm{\Lambda}_{K'_\lambda} \otimes \mathbf{K}_t]}_{\displaystyle\tilde{\mathbf{K}}}
    \underbrace{[\mathbf{Q}_{\mathbf{K}'_\lambda}^T(\bm{\Sigma}_\lambda^{\frac{1}{2}})^T \otimes \mathbf{I}_{N_t}]}_{\displaystyle\mathbf{K}_\mathrm{transf}^T},
\end{align}
The transformed covariance matrix $\tilde{\mathbf{K}}$ is now block-diagonal
\begin{equation}
\tilde{\mathbf{K}} = 
\begin{bmatrix}
\bm{\Sigma}_t + \lambda_{K'_\lambda; 1} \mathbf{K}_t & 0 & \cdots & 0 \\
0 & \bm{\Sigma}_t + \lambda_{K'_\lambda; 2} \mathbf{K}_t & \cdots & 0 \\
\vdots & \vdots & \ddots & \vdots \\
0 & 0 & \cdots & \bm{\Sigma}_t + \lambda_{K'_\lambda; N_\lambda} \mathbf{K}_t
\end{bmatrix},
\label{eq:luaslasrachKtilde}
\end{equation}
\pagebreak
which means our transformed log-likelihood is equivalent to
separately decomposing each light curve in parallel
\begin{align}
\log p(\mathbf{r}|\mathbf{K}) = &- \log|\mathbf{K}_\mathrm{transf}| + \sum_{i=1}^{N_\lambda} \left[-\frac{1}{2}\tilde{\mathbf{r}}_i^T (\bm{\Sigma}_t + \lambda_{K'_\lambda; i} \mathbf{K}_t)^{-1} \tilde{\mathbf{r}}_i \right. \nonumber\\
&\left.- \frac{1}{2}\log|\bm{\Sigma}_t + \lambda_{K'_\lambda; i} \mathbf{K}_t| - \frac{N_t}{2}\log(2\pi)\right], \\
\log p(\mathbf{r}|\mathbf{K}) = &-N_t\log|\bm{\Sigma}_\lambda^{\frac{1}{2}}|  + \sum_{i=1}^{N_\lambda}\log p(\tilde{\mathbf{r}}_i|\bm{\Sigma}_t + \lambda_{K'_\lambda; i} \mathbf{K}_t),
\label{eq:blackbox_logL}
\end{align}
where we've labelled the $i^\mathrm{th}$ transformed light curve as $\tilde{\mathbf{r}}_i$. Transforming the residuals with $\mathbf{K}_\mathrm{transf}$ therefore decouples a dense 2D GP into a sum of independent 1D GPs. The covariance matrices $\mathbf{K}_t$ and $\bm{\Sigma}_t$ are completely unmodified by this transformation, with the exception that $\mathbf{K}_t$ gets multiplied by a scalar and is then added to $\bm{\Sigma}_t$ (which is often diagonal). Therefore, a wide range of 1D GP optimisations will be valid to decompose each block, including \textsc{celerite}, quasiseparable kernels, and state-space models, as well as approximate methods like HODLR and inducing point methods.

\subsubsection{Computational Scaling}

Our transformation cost with $\mathbf{K}_\text{transf}^{-1}$ is now $\mathcal{O}(N_\lambda^3 + N_\lambda^2 N_t)$ including decomposition, avoiding the $\mathcal{O}(N_t^3)$ eigendecomposition of the time covariance matrix. Instead, we may need to perform as many as $N_\lambda$ separate $(N_t, N_t)$ matrix decompositions, one for every non-zero eigenvalue $\lambda_{K'_\lambda; i}$, which naively should scale even worse. However, we are not limited to eigendecomposition, so we can decompose using faster methods, and each of these decompositions can be performed in parallel.

Suppose that $\mathbf{K}_\lambda$ has $R_\lambda$ non-zero eigenvalues, then the transformed matrix $\mathbf{K}'_\lambda$ will have the same number (since they are congruent). If the runtime cost of each decomposition is $\mathcal{O}(C_t)$, then the decomposition of all blocks should scale linearly as $\mathcal{O}(R_\lambda C_t)$. For example, if we have a 2D kernel term $\mathbf{K}_\lambda \otimes \mathbf{K}_t$ with a full-rank wavelength covariance matrix $\mathbf{K}_\lambda$ (i.e. $R_\lambda = N_\lambda$), and a quasiseparable rank of $J_{K_t}$ for the time covariance matrix, then we get a scaling of $\mathcal{O}(N_\lambda N_t J_{K_t}^2)$ for the decomposition of all blocks, linear in the number of datapoints. Including the cost of transforming the residuals with any arbitrary measurement covariance matrix $\bm{\Sigma}_\lambda$, we get a total runtime scaling of $\mathcal{O}(N_\lambda^3 + N_\lambda^2 N_t + N_\lambda N_t J_{K_t}^2)$. The memory cost is only $\sim\mathcal{O}(N_\lambda^2 + N_\lambda N_t J_{K_t})$.

Note that we can also parallelise the transformation of the residuals and the decomposition of each block, so for $T \le \min(N_\lambda, N_t)$ parallel workers, the runtime scaling can reduce to $\mathcal{O}(N_\lambda J_{\Sigma_\lambda}^2 + N_\lambda N_t (J_{\Sigma_\lambda} + J_{K_t}^2)/T)$, which suggests that GPUs should be particularly fast for this optimisation. See Figures~\ref{fig:LuasLasrachKernel_cost} and \ref{fig:LuasLasrachKernel_A100_cost} for log-likelihood runtimes comparing CPU and GPU runtimes (for arbitrary $\bm{\Sigma}_\lambda$), on CPU the performance can be comparable to transit light curve modelling (Figure~\ref{fig:jaxoplanet_transit_cost}), and was always faster when $N_\lambda \le 64$ and $N_t \ge 1024$. The GPU is faster than the CPU for larger datasets, and achieves a 61 ms log-likelihood runtime for $(N_\lambda, N_t) = (1024, 16384)$.

Note that this is not just limited to \textsc{celerite} and other quasiseparable kernels, even approximations like HODLR \citep{Ambikasaran2015} could be used to extend to a more general range of time kernels, later discussed in Section~\ref{sec:other_GP_opt}. In fact, due to the difference in runtime cost between eigendecomposition and general Cholesky decomposition (Figure~\ref{fig:matrix_decomp_cost}), the \textsc{LuasLasrachKernel} with general Cholesky decompositions of $R_\lambda$ time covariance matrices can outperform a single eigendecomposition using the \textsc{LuasKernel} when $R_\lambda$ is small. Due to the similar flexibility of this optimisation to the \textsc{LuasKernel}, named after the Irish word for speed `luas', we refer to this optimisation as the \textsc{LuasLasrachKernel}, from the Irish `luas lasrach'\footnote{pronounced like `Lewis loss-rock'}, meaning `lightning speed'.

\subsection{\textsc{MultiLowRankKernel}: Summing Low-Rank Kernel Terms}
\label{subsec:multilowrankkernel}

Suppose we have a combination of multiple low-rank kernel terms, potentially in the presence of white noise or measurement covariance
\begin{align}
    k(\lambda, \lambda', t, t') &= s_\lambda(\lambda, \lambda')\sigma^2(t)\delta_{tt'} +\sum_{i=1}^{N_\alpha}\alpha_i(\lambda) \alpha_i(\lambda') k_{t_i}(t, t')
    \label{eq:low_rank_kernel}
\end{align}
corresponding to the covariance matrix
\begin{equation}
\mathbf{K} = \bm{\Sigma}_{\lambda} \otimes \mathbf{D}_{t} + \sum_{i=1}^{N_\alpha} \boldsymbol{\alpha}_{\lambda; i}^{} \boldsymbol{\alpha}_{\lambda; i}^T \otimes \mathbf{K}_{t; i}
\label{eq:sum_alpha}
\end{equation}
for a rectangular dataset.

As described in Section~\ref{subsubsec:low_rank_terms}, \citet{Gordon2020} already introduced an optimisation for this kernel in the case where $\bm{\Sigma}_{\lambda}$ is diagonal and $k_{t_i}$ are \textsc{celerite} kernels. This is a particularly important kernel to optimise, which we focus on extensively in Section~\ref{sec:4}, as it can be viewed as a generalisation of PCA/SysRem, it is a physically-motivated model of stellar granulation \citep{Gordon2020, Morris2020}, can model common-mode systematics, and this model is also used within machine learning, where it may be referred to as the linear mixing model. To understand how we implement the \textsc{MultiLowRankKernel}, we begin with the simplest case of $N_\alpha = 1$. 

\subsubsection{The Case of $N_\alpha = 1$}

We can already decompose $N_\alpha = 1$, as it is a special case of the \textsc{LuasLasrachKernel} with $\mathbf{K}_\lambda = \boldsymbol{\alpha}_\lambda\boldsymbol{\alpha}_\lambda^T$. First we factor out $\bm{\Sigma}_\lambda$
\begin{align}
\label{eq:gordon_singlevec}
\mathbf{K} &= \bm{\Sigma}_{\lambda}  \otimes \mathbf{D}_{t} + \boldsymbol{\alpha}_\lambda^{} \boldsymbol{\alpha}_\lambda^T \otimes \mathbf{K}_t\\
\mathbf{K} &= [\bm{\Sigma}_\lambda^{\frac{1}{2}} \otimes \mathbf{I}_{N_t}][\mathbf{I}_{N_\lambda}  \otimes \mathbf{D}_{t} + \boldsymbol{\alpha}_\lambda' \boldsymbol{\alpha}_\lambda'^T \otimes \mathbf{K}_t] [(\bm{\Sigma}_\lambda^{\frac{1}{2}})^T \otimes \mathbf{I}_{N_t}]
\label{eq:gordon_singlevec2}
\end{align}
where we define the whitened wavelength component $\boldsymbol{\alpha}_\lambda' = \bm{\Sigma}_\lambda^{-\frac{1}{2}}\boldsymbol{\alpha}_\lambda$, which is applying the same initial simultaneous diagonalisation step as before $\mathbf{K}_\lambda' = \bm{\Sigma}_\lambda^{-\frac{1}{2}} \mathbf{K}_\lambda(\bm{\Sigma}_\lambda^{-\frac{1}{2}})^T$.

However, instead of performing a general $\mathcal{O}(N_\lambda^3)$ eigendecomposition on $\mathbf{K}_\lambda' = \boldsymbol{\alpha}_\lambda' \boldsymbol{\alpha}_\lambda'^T$, a valid eigendecomposition can be calculated using the Householder matrix
\begin{equation}
    \mathbf{H}_{\alpha_\lambda'} = \mathbf{I}_{N_\lambda} - 2\hat{\mathbf{u}}\hat{\mathbf{u}}^T
\end{equation}
where $\hat{\mathbf{u}}$ is calculated as
\begin{align}
    \mathbf{u} &= \frac{\boldsymbol{\alpha}_\lambda'}{|\boldsymbol{\alpha}_\lambda'|} \pm \hat{\mathbf{e}}_i\\
    \hat{\mathbf{u}} &= \frac{\mathbf{u}}{|\mathbf{u}|}
\end{align}
where we are free to choose the sign of $\pm$ to maximise numerical stability (e.g. if $\boldsymbol{\alpha}_\lambda' = \hat{\mathbf{e}}_i$ we use this sign choice to avoid $\mathbf{u} = \mathbf{0}$). This Householder matrix has a number of useful properties, which include:
\begin{align}
    \mathbf{H}_{\alpha_\lambda'} \hat{\mathbf{e}}_i &= \mp \frac{\boldsymbol{\alpha}_\lambda'}{|\boldsymbol{\alpha}_\lambda'|}\\
    \mathbf{H}_{\alpha_\lambda'}\mathbf{H}_{\alpha_\lambda'}^T &= \mathbf{I}_{N_\lambda}\\
     \boldsymbol{\alpha}_\lambda' \boldsymbol{\alpha}_\lambda'^T &= \mathbf{H}_{\alpha_\lambda'}\left[|\boldsymbol{\alpha}_\lambda'|^2\hat{\mathbf{e}}_i^{} \hat{\mathbf{e}}_i^T\right]\mathbf{H}_{\alpha_\lambda'}^T
\end{align}
So $\mathbf{H}_{\alpha_\lambda'}$ maps the basis vector $\hat{\mathbf{e}}_i$ to our desired transformed vector $\boldsymbol{\alpha}_\lambda'$ (potentially with a sign flip), $\mathbf{H}_{\alpha_\lambda'}$ is unitary, and it diagonalises $\boldsymbol{\alpha}_\lambda' \boldsymbol{\alpha}_\lambda'^T$. Since $\mathbf{H}_{\alpha_\lambda'}$ is just a rank-1 update to an identity matrix, we can perform the matrix multiplication $\mathbf{H}_{\alpha_\lambda'}\mathbf{R}$ in linear time $\mathcal{O}(N_\lambda N_t)$, and because $\mathbf{H}_{\alpha_\lambda'}$ is completely defined by the $N_\lambda$-long vector $\hat{\mathbf{u}}$, each Householder matrix has a linear $\mathcal{O}(N_\lambda)$ memory scaling.

Factorising Equation~\ref{eq:gordon_singlevec2} further using the Householder matrix $\mathbf{H}_{\alpha_\lambda'}$ with $\hat{\mathbf{e}}_i = \hat{\mathbf{e}}_1$, we get
\begin{align}
\mathbf{K} &= \underbrace{[\bm{\Sigma}_\lambda^{\frac{1}{2}}\mathbf{H}_{\alpha_\lambda'} \otimes \mathbf{I}_{N_t}]}_{\displaystyle\mathbf{K}_\mathrm{transf}}\underbrace{[\mathbf{I}_{N_\lambda} \otimes \mathbf{D}_{t} + |\boldsymbol{\alpha}_\lambda'|^2\hat{\mathbf{e}}_1^{} \hat{\mathbf{e}}_1^T \otimes \mathbf{K}_{t}]}_{\displaystyle\tilde{\mathbf{K}}}\nonumber\\
&\phantom{=} \underbrace{[\mathbf{H}_{\alpha_\lambda'}^T(\bm{\Sigma}_\lambda^{\frac{1}{2}})^T \otimes \mathbf{I}_{N_t}]}_{\displaystyle\mathbf{K}_\mathrm{transf}^T}
\label{eq:gordon_singlevec3}
\end{align}
where $\tilde{\mathbf{K}}$ can be expressed as the block matrix
\begin{align}
\tilde{\mathbf{K}} &=
\begin{bmatrix}
\mathbf{A} & \mathbf{B}\\
\mathbf{C} & \mathbf{D}
\end{bmatrix} = \begin{bmatrix}
\mathbf{D}_t + |\boldsymbol{\alpha}_\lambda'|^2 \mathbf{K}_t & \mathbf{0}\\
\mathbf{0} & \mathbf{I}_{N_\lambda - 1} \otimes \mathbf{D}_t
\end{bmatrix},
\end{align}
which is diagonal apart from the $(N_t, N_t)$ $\mathbf{A}$ block. In the highly optimised scenario where both $\mathbf{K}_t$ and $\bm{\Sigma}_\lambda$ are quasiseparable, we achieve a log-likelihood runtime scaling of $\mathcal{O}(N_t J_{K_t}^2 + N_\lambda J_{\Sigma_\lambda}^2 + N_\lambda N_t (1 + J_{\Sigma_\lambda}))$.  

\subsubsection{Extending to $N_\alpha > 1$ Using Householder QR}

Now let's consider the general case of $N_\alpha \ge 1$ low-rank kernel terms. We start off with a similar trick as before, where we whiten all $\boldsymbol{\alpha}_{\lambda; i}$ components to factor out measurement covariance:
\begin{equation}
\mathbf{K} = [\bm{\Sigma}_{\lambda}^{\frac{1}{2}} \otimes \mathbf{I}_{N_t}][\mathbf{I}_{N_\lambda} \otimes \mathbf{D}_{t} + \sum_{i=1}^{N_\alpha} \boldsymbol{\alpha}_{\lambda; i}' \boldsymbol{\alpha}_{\lambda; i}'^T \otimes \mathbf{K}_{t; i}][(\bm{\Sigma}_{\lambda}^{\frac{1}{2}})^T \otimes \mathbf{I}_{N_t}]
\label{eq:sum_alpha_whitened}
\end{equation}
with $\boldsymbol{\alpha}_{\lambda; i}' = \bm{\Sigma}_{\lambda}^{-\frac{1}{2}}\boldsymbol{\alpha}_{\lambda; i}$. Our transformed covariance matrix $\tilde{\mathbf{K}}$ is now in a form that can be handled by the optimisation in \citet{Gordon2020} for \textsc{celerite}/quasiseparable kernels in $\mathcal{O}(N_\alpha^2 N_\lambda N_t \bar{J}_{K_t}^2)$ runtime, since we've diagonalised the measurement covariance. However, in the case that $N_\alpha < N_\lambda$, we introduce an additional transformation which can further reduce runtime costs.


We can make use of successive Householder transformations to make the problem more sparse and faster to compute. In fact, this process is equivalent to calculating the QR decomposition of the $(N_\lambda, N_\alpha)$ matrix $\mathbf{A}_\lambda'$, which we form using $\boldsymbol{\alpha}_{\lambda; i}'$ as column vectors
\begin{equation}
    \mathbf{A}_{\lambda}' = \begin{bmatrix}
\boldsymbol{\alpha}_{\lambda; 1}' & \cdots & \boldsymbol{\alpha}_{\lambda; N_\alpha}'\\
\end{bmatrix} = \mathbf{Q}_{\mathbf{A}_{\lambda}'}\begin{bmatrix}
\mathbf{\mathbf{R}_{A_{\lambda}'}}\\
\mathbf{0}
\end{bmatrix}\\[6pt]
\label{eq:QR_decomp}
\end{equation}
where $\mathbf{Q}_{A_{\lambda}'}$ is an $(N_\lambda, N_\lambda)$ orthonormal matrix and $\mathbf{R}_{A_{\lambda}'}$ is $(N_\alpha, N_\alpha)$ and upper-triangular. By using the Householder approach to calculating the QR decomposition, $\mathbf{Q}_{A'}$ will be equal to a product of $N_\alpha$ Householder matrices, and its inverse is simply the Householder matrices multiplied in reverse order
\begin{align}
    \mathbf{Q}_{A_{\lambda}'} &= \mathbf{H}_1\mathbf{H}_2\dots\mathbf{H}_{N_\alpha},\\
    \mathbf{Q}_{A_{\lambda}'}^T &= \mathbf{H}_{N_\alpha}\mathbf{H}_{N_\alpha-1}\dots\mathbf{H}_1,
    \label{eq:qr_inverse}
\end{align}
Representing $\mathbf{Q}_{A_{\lambda}'}$ as a stack of Householder transformations has a number of advantages when $N_\alpha < N_\lambda$, it allows us to transform our data residuals in linear time $\mathcal{O}(N_\alpha N_\lambda N_t)$ by performing $N_\alpha$ successive Householder transformations, and it comes with a cheap memory cost of $\mathcal{O}(N_\alpha N_\lambda)$ to represent $\mathbf{Q}_{A_{\lambda}'}$. The QR decomposition is cheap to calculate, with a runtime scaling of $\sim \mathcal{O}(N_\alpha^2 N_\lambda)$ for a $(N_\lambda, N_\alpha)$ matrix \citep{blockinverseformula}.

\subsubsection{The \textsc{MultiLowRankKernel} Factorisation}

Using our QR decomposition of $\mathbf{A}_{\lambda}'$ in Equation~\ref{eq:QR_decomp}, we can now factorise our \textsc{MultiLowRankKernel} covariance matrix:
\begin{align}
\mathbf{K} &= \bm{\Sigma}_{\lambda} \otimes \mathbf{D}_{t} + \sum_{i=1}^{N_\alpha} \boldsymbol{\alpha}_{\lambda; i}^{} \boldsymbol{\alpha}_{\lambda; i}^T \otimes \mathbf{K}_{t; i}\\
&= [\bm{\Sigma}_{\lambda}^{\frac{1}{2}} \otimes \mathbf{I}_{N_t}][\mathbf{I}_{N_\lambda} \otimes \mathbf{D}_{t} + \sum_{i=1}^{N_\alpha} (\mathbf{A}_{\lambda}'\hat{\mathbf{e}}_i)(\mathbf{A}_{\lambda}'\hat{\mathbf{e}}_i)^T \otimes \mathbf{K}_{t; i}][(\bm{\Sigma}_{\lambda}^{\frac{1}{2}})^T \otimes \mathbf{I}_{N_t}]\nonumber\\
    &= 
    \underbrace{[\bm{\Sigma}_{\lambda}^{\frac{1}{2}} \mathbf{Q}_{A_{\lambda}'} \otimes \mathbf{I}_{N_t}]}_{\displaystyle\mathbf{K}_\mathrm{transf}}\nonumber\\
    &* \underbrace{\left[\mathbf{I}_{N_\lambda} \otimes \mathbf{D}_{t} + \sum_{i=1}^{N_\alpha} \begin{bmatrix}
\mathbf{\mathbf{R}_{A_{\lambda}'}}\hat{\mathbf{e}}_i\\
\mathbf{0}
\end{bmatrix}\begin{bmatrix}
\mathbf{\mathbf{R}_{A_{\lambda}'}}\hat{\mathbf{e}}_i\\
\mathbf{0}
\end{bmatrix}^T \otimes \mathbf{K}_{t; i}\right]}_{\displaystyle\tilde{\mathbf{K}}}\nonumber\\
    &* \underbrace{[\mathbf{Q}_{A_{\lambda}'}^T(\bm{\Sigma}_{\lambda}^{\frac{1}{2}})^T \otimes \mathbf{I}_{N_t}]}_{\displaystyle\mathbf{K}_\mathrm{transf}^T}.
\label{eq:QR_rwrite}
\end{align}
Re-labelling the $N_\alpha$-long vectors $\mathbf{a}_{i} = \mathbf{R}_{A_{\lambda}'}\hat{\mathbf{e}}_i$, we write $\tilde{\mathbf{K}}$ as the block matrix
\begin{equation}
\tilde{\mathbf{K}} =
    \begin{bmatrix}
    \mathbf{I}_{N_\alpha} \otimes \mathbf{D}_{t} + \sum_{i=1}^{N_\alpha}\mathbf{a}_{i}^{}\mathbf{a}_{i}^T \otimes \mathbf{K}_{t; i} & \mathbf{0}\\
    \mathbf{0} & \mathbf{I}_{N_\lambda - N_\alpha} \otimes \mathbf{D}_{t}
    \end{bmatrix}
    \label{eq:multilowrank_ktilde}
\end{equation}
where our $\mathbf{B}$ and $\mathbf{C}$ blocks consist fully of zeros, and our $\mathbf{D}$ block is fully diagonal. The $\mathbf{A}$ block is still in the same form as $\tilde{\mathbf{K}}$ in Equation~\ref{eq:sum_alpha_whitened}, before transforming with the QR decomposition. However, we have exploited sparsity to reduce the problem from decomposing a $(N_\lambda N_t, N_\lambda N_t)$ matrix, down to a $(N_\alpha N_t, N_\alpha N_t)$ matrix. This QR decomposition transformation has an interesting interpretation, as it is closely related to PCA, where we are projecting the data residuals onto separate systematics components which can have different kernel functions. This is explained in more detail in Section~\ref{sec:PCA}, and visualised in Figure~\ref{fig:transformation_visual}. 

\subsubsection{Computational Scaling}
\label{sec:multilowrankscaling}

If our time kernels are valid quasiseparable kernels, then the method in \citet{Gordon2020} can decompose the $\mathbf{A}$ block of $\tilde{\mathbf{K}}$, rather than be applied to the full $\mathbf{K}$ matrix. The log-likelihood runtime scaling works out to $\mathcal{O}(N_\alpha^3 N_t \bar{J}_{K_t}^2 + N_\alpha N_\lambda N_t)$ when $\bm{\Sigma}_\lambda$ is diagonal, with an extra $\mathcal{O}(N_\lambda J_{\Sigma_\lambda}^2 + N_\lambda N_t J_{\Sigma_\lambda})$ runtime cost if $\bm{\Sigma}_\lambda$ is quasiseparable with rank $J_{\Sigma_\lambda}$, or an extra $\mathcal{O}(N_\lambda^3 + N_\lambda^2 N_t)$ cost for arbitrary $\bm{\Sigma}_\lambda$. The memory scaling will be at most $\sim\mathcal{O}(N_\lambda^2 + N_\alpha^2 N_t \bar{J}_{K_t}+ N_\lambda N_t)$ for arbitrary $\bm{\Sigma}_\lambda$, but this can be reduced to $\sim\mathcal{O}(N_\lambda J_{\Sigma_\lambda} + N_\alpha^2 N_t \bar{J}_{K_t} + N_\lambda N_t)$ with quasiseparable $\bm{\Sigma}_\lambda$.

We make an improvement to the existing runtime scaling of $\mathcal{O}(N_\alpha^2 N_\lambda N_t \bar{J}_{K_t}^2)$ from applying the \citet{Gordon2020} method without the QR decomposition, and we have also generalised it to include measurement covariance, although we cannot deal with non-separable noise (see Section~\ref{sec:white_noise}). Not only do we achieve a better runtime scaling, but even in the case of $N_\alpha = 1$ and $\bar{J}_{K_t} = 1$, benchmarking both approaches in Figures~\ref{fig:GordonKernel_Nalpha1_cost} and \ref{fig:MultiLowRankKernel_Nalpha1_cost} found that the \textsc{MultiLowRankKernel} can reduce runtimes by over an order of magnitude compared to \citet{Gordon2020}. For example, with $(N_\lambda, N_t) = (1024, 16384)$ we achieve a log-likelihood runtime of just 29 ms on CPU, compared to 0.6 s with the \citet{Gordon2020} approach implemented using \textsc{tinygp}. Increasing the \textsc{MultiLowRankKernel} to $N_\alpha = 4, 8, 16$ is benchmarked in Figures~\ref{fig:MultiLowRankKernel_Nalpha4_cost}, \ref{fig:MultiLowRankKernel_Nalpha8_cost} and \ref{fig:MultiLowRankKernel_Nalpha16_cost}, demonstrating highly scalable runtimes even for multiple components with different time kernels.

Unlike \citet{Gordon2020}, our approach even optimises the case where $\mathbf{K}_{t; i}$ are arbitrary covariance matrices with cubic runtime decompositions, as we reduce the dense covariance matrix decomposition from a runtime scaling of $\mathcal{O}(N_\lambda^3 N_t^3)$, down to $\sim\mathcal{O}(N_\alpha^3 N_t^3)$ by using a general Cholesky decomposition of the $\mathbf{A}$ block in Equation~\ref{eq:multilowrank_ktilde}. We have benchmarked this case for $N_\alpha = 1$ on both CPU and GPU in Figures~\ref{fig:MultiLowRankKernel_Nalpha1_poweredexp_cost} and \ref{fig:MultiLowRankKernel_Nalpha1_poweredexp_densereadout_A100_cost} respectively. Note that a similar optimisation has already been independently derived in \citet{Bruinsma2019} for machine learning applications, but they do not handle measurement covariance, and to achieve a linear scaling in $N_t$ they require orthogonal $\boldsymbol{\alpha}_{\lambda; i}$ components, which we avoid by combining with the \citet{Gordon2020} method and Householder QR. Restricting to orthogonal components can improve runtimes further however, as we discuss in the next subsection.

\subsubsection{Improved Scaling with Orthogonality}

If the \textit{whitened} vectors $\boldsymbol{\alpha}_{\lambda; i}' = \bm{\Sigma}_{\lambda}^{-\frac{1}{2}}\boldsymbol{\alpha}_{\lambda; i}$ are assumed to be orthogonal, then the matrix $\mathbf{R}_{A_\lambda'}$ from the QR decomposition will be diagonal, which results in $\tilde{\mathbf{K}}$ becoming fully block-diagonal. This is equivalent to the \textsc{LuasLasrachKernel}, but where we vary each time covariance matrix $\mathbf{K}_t$ within the diagonal blocks of $\tilde{\mathbf{K}}$ from Equation~\ref{eq:luaslasrachKtilde}. We derive a similar black-box log-likelihood as the \textsc{LuasLasrachKernel} for the decoupled terms:
\begin{align}
\log p(\mathbf{r}|\mathbf{K}) = &-N_t\log|\bm{\Sigma}_\lambda^{\frac{1}{2}}|  + \sum_{i=1}^{N_\alpha}\log p(\tilde{\mathbf{r}}_i|\mathbf{D}_t + |\boldsymbol{\alpha}_{\lambda; i}'|^2\mathbf{K}_{t; i})\nonumber\\
&+ \sum_{i=N_\alpha + 1}^{N_\lambda}\log p(\tilde{\mathbf{r}}_i|\mathbf{D}_t),
\label{eq:blackbox_logL_multilowrank}
\end{align}
where each of the 1D log-likelihood terms can use any desired GP optimisation on the transformed residuals. This orthogonal case essentially reduces to the GP optimisation introduced in \citet{Bruinsma2019}, although we have further generalised it to include measurement covariance.

Orthogonal components produce a more efficient runtime scaling of $\mathcal{O}(N_\alpha N_t \bar{J}_{K_t}^2 + N_\alpha N_\lambda N_t)$ for diagonal $\bm{\Sigma}_\lambda$, plus the same additional costs as Section~\ref{sec:multilowrankscaling} if $\bm{\Sigma}_\lambda$ is instead quasiseparable or arbitrary. Note that orthogonality allows for a linear rather than cubic scaling in $N_\alpha$, which can make a difference with many components, but if $N_\lambda N_t$ is large then they can both be bottlenecked by the same $\mathcal{O}(N_\alpha N_\lambda N_t)$ cost to transform the residuals, or from additional costs due to measurement covariance. This can be seen by examining the benchmarking plots in Figures~\ref{fig:MultiLowRankKernel_Nalpha16_cost} and ~\ref{fig:MultiLowRankKernel_Nalpha16_ortho_cost}, which compare running $N_\alpha = 16$ components without and with the orthogonality constraint respectively, and show that the greatest benefit of orthogonality is when $N_t >> N_\lambda$. However, we note that the relative cost to calculate gradients of the log-likelihood was also cheaper with orthogonality (as discussed in Section~\ref{sec:gradients}). One other difference worth noting, specifically for \textsc{celerite} and quasiseparable kernels, is that this decoupling of each time kernel with orthogonality means that each kernel $k_{t; i}$ has no requirement to share the same input sort order, e.g. allowing us to joint-fit low-rank \textsc{celerite} kernels with different input regressors along the time axis (e.g. FWHM, centroiding position).


\subsection{\textsc{DoubleMultiLowRankKernel}: Low-Rank Terms in Both Dimensions}
\label{subsec:double_low_rank_terms}

We can extend these tricks to deal with low-rank terms which have kernel functions in different dimensions, corresponding to kernel functions of the form
\begin{align}
    k(\lambda, \lambda', t, t') &= \sigma^2(\lambda)\sigma^2(t)\delta_{\lambda\lambda'}\delta_{tt'} +\sum_{i=1}^{N_\alpha}\alpha_i(\lambda) \alpha_i(\lambda') k_{t_i}(t, t') \nonumber\\
    &+ \sum_{j=1}^{N_\beta}k_{\lambda_j}(\lambda, \lambda')\beta_j(t) \beta_j(t'),
\end{align}
In this case, the factorisation we use is similar to the \textsc{MultiLowRankKernel}, but with a QR decomposition in both dimensions:
\begin{align}
\label{eq:double_sum_alpha}
\mathbf{K} &= \mathbf{D}_{\lambda}  \otimes \mathbf{D}_{t} + \sum_{i=1}^{N_\alpha} \boldsymbol{\alpha}_{\lambda; i}^{} \boldsymbol{\alpha}_{\lambda; i}^T \otimes \mathbf{K}_{t; i} + \sum_{j=1}^{N_\beta} \mathbf{K}_{\lambda; j} \otimes \boldsymbol{\beta}_{t; j} \boldsymbol{\beta}_{t; j}^T\\
&= \underbrace{[\mathbf{D}_\lambda^{\frac{1}{2}}\mathbf{Q}_{\mathbf{A}'} \otimes \mathbf{D}_t^{\frac{1}{2}}\mathbf{Q}_{\mathbf{B}'}]}_{\displaystyle\mathbf{K}_\mathrm{transf}}\nonumber\\
&* \underbrace{\left[\mathbf{I}_{N_\lambda N_t} + \sum_{i=1}^{N_\alpha} \left[\begin{smallmatrix}
\mathbf{a}_i\\
\mathbf{0}
\end{smallmatrix}\right]\left[\begin{smallmatrix}
\mathbf{a}_i\\
\mathbf{0}
\end{smallmatrix}\right]^T \otimes \mathbf{K}'_{t; i}+\sum_{j=1}^{N_\beta} \mathbf{K}'_{\lambda; j} \otimes \left[\begin{smallmatrix}
\mathbf{b}_j\\
\mathbf{0}
\end{smallmatrix}\right]\left[\begin{smallmatrix}
\mathbf{b}_j\\
\mathbf{0}
\end{smallmatrix}\right]^T\right]}_{\displaystyle\tilde{\mathbf{K}}}\nonumber\\
&* \underbrace{[\mathbf{Q}_{\mathbf{A}'}^T\mathbf{D}_\lambda^{\frac{1}{2}} \otimes \mathbf{Q}_{\mathbf{B}'}^T\mathbf{D}_t^{\frac{1}{2}}]}_{\displaystyle\mathbf{K}_\mathrm{transf}^T}
\label{eq:doublemultiterm}
\end{align}
where we've transformed all of the $\mathbf{K}_{\lambda; j}$ and $\mathbf{K}_{t; i}$ matrices following
\begin{align}
\mathbf{K}'_{\lambda; j} &= \mathbf{Q}_{\mathbf{A}'}^T\mathbf{D}_\lambda^{-\frac{1}{2}}\mathbf{K}_{\lambda; j}\mathbf{D}_\lambda^{-\frac{1}{2}}\mathbf{Q}_{\mathbf{A}'}^{}\\
\mathbf{K}'_{t; i} &= \mathbf{Q}_{\mathbf{B}'}^T\mathbf{D}_t^{-\frac{1}{2}}\mathbf{K}_{t; i}\mathbf{D}_t^{-\frac{1}{2}}\mathbf{Q}_{\mathbf{B}'}^{}
\end{align}
which can be performed while maintaining quasiseparability if we exclude measurement covariance ($\mathbf{D}_\lambda$ and $\mathbf{D}_t$ must be diagonal), which is demonstrated in Appendix~\ref{app:householder_quasisep}. Just note that each of the $N_\alpha$ Householder transformations in $\mathbf{Q}_{\mathbf{A}'}$ (and $N_\beta$ transformations in $\mathbf{Q}_{\mathbf{B}'}$) increases the quasiseparable rank of the transformed covariance matrix by 2. Therefore, all quasiseparable ranks get shifted from $J_{K_{\lambda; j}} \to J_{K_{\lambda; j}} + 2N_\alpha$ and $J_{K_{t; i}} \to J_{K_{t; i}} + 2N_\beta$.

Similar to the \textsc{MultiLowRankKernel}, the $\tilde{\mathbf{K}}$ term can be written as the block matrix:
\begin{align}
\label{eq:K_tilde_double}
\begin{bmatrix}
\left(\begin{aligned}
\mathbf{I}_{N_\alpha} &\otimes \mathbf{I}_{N_t} +\\
\sum_{i=1}^{N_\alpha} \mathbf{a}_i\mathbf{a}_i^T &\otimes \mathbf{K}'_{t;i} +\\
\sum_{j=1}^{N_\beta} [\mathbf{K}'_{\lambda;j}]_{A} &\otimes \left[\begin{smallmatrix}
\mathbf{b}_j\\
\mathbf{0}
\end{smallmatrix}\right]\left[\begin{smallmatrix}
\mathbf{b}_j\\
\mathbf{0}
\end{smallmatrix}\right]^T
\end{aligned}
\right)
& \sum_{j=1}^{N_\beta} [\mathbf{K}'_{\lambda;j}]_{B} \otimes \left[\begin{smallmatrix}
\mathbf{b}_j\\
\mathbf{0}
\end{smallmatrix}\right]\left[\begin{smallmatrix}
\mathbf{b}_j\\
\mathbf{0}
\end{smallmatrix}\right]^T\\
\rule{0pt}{4.5ex} \sum_{j=1}^{N_\beta} [\mathbf{K}'_{\lambda;j}]_{C} \otimes \left[\begin{smallmatrix}
\mathbf{b}_j\\
\mathbf{0}
\end{smallmatrix}\right]\left[\begin{smallmatrix}
\mathbf{b}_j\\
\mathbf{0}
\end{smallmatrix}\right]^T 
& \left(\begin{aligned}\mathbf{I}_{N_\lambda - N_\alpha} &\otimes \mathbf{I}_{N_t} +\\
\sum_{j=1}^{N_\beta} [\mathbf{K}'_{\lambda;j}]_{D} &\otimes \left[\begin{smallmatrix}
\mathbf{b}_j\\
\mathbf{0}
\end{smallmatrix}\right]\left[\begin{smallmatrix}
\mathbf{b}_j\\
\mathbf{0}
\end{smallmatrix}\right]^T\end{aligned}
\right)
\end{bmatrix}
\end{align}
where we can split up the decomposition of $\tilde{\mathbf{K}}$ using formulae for 2x2 block matrices. This works by first decomposing the $\mathbf{A}$ block, calculating the term $\mathbf{C}\mathbf{A}^{-1}\mathbf{B}$, and then decomposing $\mathbf{D} - \mathbf{C}\mathbf{A}^{-1}\mathbf{B}$, all described in detail in Appendix~\ref{app:K_tilde_doublemultilowrankkernel}.

\subsubsection{Computational Scaling}
\label{sec:doublemultilowrank_compscaling}

In Appendix~\ref{app:doublemultilowrankdecomp}, we derive a runtime scaling of $\mathcal{O}(N_\alpha^3 N_t (\bar{J}_{K_t} + 3 N_\beta)^2 + N_\beta^3 N_\lambda (\bar{J}_{K_\lambda} + 3N_\alpha)^2 + N_\lambda N_t (N_\alpha + N_\beta))$ assuming all $\mathbf{K}_{\lambda;j}$ and $\mathbf{K}_{t; i}$ are quasiseparable, with mean quasiseparable ranks of $\bar{J}_{K_\lambda}$ and $\bar{J}_{K_t}$ respectively. The memory scaling for this case is $\sim\mathcal{O}(N_\alpha^2 N_t (\bar{J}_{K_t} + 3 N_\beta) + N_\beta^2 N_\lambda (\bar{J}_{K_\lambda} + 3N_\alpha) + N_\lambda N_t)$. See Figures~\ref{fig:DoubleMultiLowRankKernel_Nalpha1_Nbeta1_cost}, \ref{fig:DoubleMultiLowRankKernel_Nalpha1_Nbeta4_cost}, \ref{fig:DoubleMultiLowRankKernel_Nalpha4_Nbeta1_cost} and \ref{fig:DoubleMultiLowRankKernel_Nalpha4_Nbeta4_cost} for benchmarking in the cases of $(N_\alpha, N_\beta) = (1, 1), (1, 4), (4, 1), (4, 4)$, respectively. For $(N_\alpha, N_\beta) = (1, 1)$ and $(N_\lambda, N_t) = (1024, 16384)$, we get a log-likelihood runtime of 80 ms on CPU, so this kernel remains viable even for large datasets. However, this runtime increases to 0.2-0.3 seconds when $N_\alpha = 4$ or $N_\beta = 4$, and if both $N_\alpha = N_\beta = 4$, then it increases to 2 seconds to calculate, as the runtime can grow significantly if the product $N_\alpha N_\beta$ increases.


\subsection{\textsc{LuasPlusMultiLowRankKernel}: Combining 2D Kernels with Low-Rank Kernel Terms}
\label{sec:luasplusmultilowrank}

We can keep extending to more complex kernels, with $\tilde{\mathbf{K}}$ becoming an increasingly complex 2x2 block matrix. Consider the case where we may have measurement covariance, a full-rank 2D kernel term, and a sum of low-rank terms, such as:
\begin{align}
    k(\lambda, \lambda', t, t') &= s_\lambda(\lambda, \lambda')s_t(t, t') + k_{\lambda; \text{2D}}(\lambda, \lambda') k_{t; \text{2D}}(t, t') \nonumber\\
    &+ \sum_{i=1}^{N_\alpha}\alpha_i(\lambda) \alpha_i(\lambda')k_{t_i}(t, t').
    \label{eq:half_luas_alpha}
\end{align}
which corresponds to covariance matrices of the form
\begin{equation}
\mathbf{K} = \bm{\Sigma}_{\lambda}  \otimes \bm{\Sigma}_{t} + \mathbf{K}_{\lambda; \text{2D}} \otimes \mathbf{K}_{t; \text{2D}} + \sum_{i=1}^{N_\alpha} \boldsymbol{\alpha}_{\lambda; i}^{} \boldsymbol{\alpha}_{\lambda; i} ^T \otimes \mathbf{K}_{t; i}.
\label{eq:luasplusmultimat}
\end{equation}
In this scenario, our \textsc{LuasPlusMultiLowRankKernel} optimisation can still exploit quasiseparability, but just in the wavelength dimension (i.e. $\bm{\Sigma}_{\lambda}$ and $\mathbf{K}_{\lambda; \text{2D}}$ must be quasiseparable/diagonal), in which case the runtime scales linearly with $N_\lambda$. However, if the dimensions of the low-rank terms are flipped, i.e.
\begin{align}
    k(\lambda, \lambda', t, t') &= s_\lambda(\lambda, \lambda')s_t(t, t') + k_{\lambda; \text{2D}}(\lambda, \lambda') k_{t; \text{2D}}(t, t') \nonumber\\
    &+
    \sum_{j=1}^{N_\beta}k_{\lambda_j}(\lambda, \lambda')\beta_j(t) \beta_j(t')
    \label{eq:half_luas}
\end{align}
then we can instead scale linearly with $N_t$, provided $\bm{\Sigma}_t$ and $\mathbf{K}_{t; \text{2D}}$ are quasiseparable (or diagonal). The algorithm is the same in either case, we can simply transpose our residuals array $\mathbf{R}$ before calculating to switch between them.

For the case in Equation~\ref{eq:luasplusmultimat}, our transformation performs a simultaneous diagonalisation in the time dimension of $\bm{\Sigma}_{t}$ and $\mathbf{K}_{t; \text{2D}}$, similar to the \textsc{LuasKernel}. In the wavelength dimension, we perform a similar transformation to the \textsc{MultiLowRankKernel}, without any whitening transformation. We form the matrix $\mathbf{A}_\lambda$ using $\boldsymbol{\alpha}_{\lambda; i}$ as column vectors, and instead of a QR decomposition, we perform Householder QL decomposition, where we instead project to a lower-triangular matrix $\mathbf{L}_{A}$, i.e.
\begin{align}
    &\mathbf{A}_{\lambda} = \begin{bmatrix}
\boldsymbol{\alpha}_{\lambda; 1} & \cdots & \boldsymbol{\alpha}_{\lambda; N_\alpha}\end{bmatrix} = \mathbf{Q}_{\mathbf{A}_\lambda}\begin{bmatrix}
\mathbf{0}\\
\mathbf{\mathbf{L}_{A_\lambda}}
\end{bmatrix}
\end{align}
which creates sparsity at the top of the vectors, as it happens to simplify the calculation of $\mathbf{C}\mathbf{A}^{-1}\mathbf{B}$. We now label $\mathbf{a}_{i} = \mathbf{L}_{A_{\lambda}}\hat{\mathbf{e}}_i$, and factorise $\mathbf{K}$ as follows
\begin{align}
    \mathbf{K} = 
    &\underbrace{[\mathbf{Q}_{\mathbf{A}_\lambda} \otimes \bm{\Sigma}_t^{\frac{1}{2}}\mathbf{Q}_{K'_{t; \text{2D}}}]}_{\displaystyle\mathbf{K}_\mathrm{transf}}\nonumber\\
    * &\underbrace{\left[\bm{\Sigma}'_{\lambda} \otimes \mathbf{I}_{N_t} + \mathbf{K}'_{\lambda; \text{2D}} \otimes \bm{\Lambda}_{K'_{t; \text{2D}}} + \sum_{i=1}^{N_\alpha} \left[\begin{smallmatrix}
\mathbf{0}\\
\mathbf{a}_i
\end{smallmatrix}\right]\left[\begin{smallmatrix}
\mathbf{0}\\
\mathbf{a}_i
\end{smallmatrix}\right]^T \otimes\mathbf{K}'_{t; i}\right]}_{\displaystyle\tilde{\mathbf{K}}}\nonumber\\
    * &\underbrace{[\mathbf{Q}_{\mathbf{A}_\lambda}^T \otimes \mathbf{Q}_{K'_{t; \text{2D}}}^T(\bm{\Sigma}_t^{\frac{1}{2}})^T]}_{\displaystyle\mathbf{K}_\mathrm{transf}^T},
\label{eq:half_luas_factorisation}
\end{align}
where the QL transformation has transformed the wavelength covariance matrices as
\begin{align}
    \bm{\Sigma}'_{\lambda} &= \mathbf{Q}_{\mathbf{A}_\lambda}^T\bm{\Sigma}_{\lambda}\mathbf{Q}_{\mathbf{A}_\lambda}^{}\\
    \mathbf{K}'_{\lambda; \mathrm{2D}} &= \mathbf{Q}_{\mathbf{A}_\lambda}^T\mathbf{K}_{\lambda; \mathrm{2D}}\mathbf{Q}_{\mathbf{A}_\lambda}^{}
\end{align}
which will increase each of their quasiseparable ranks by $2N_\alpha$ if they are quasiseparable (see Appendix~\ref{app:householder_quasisep}). Each of the time covariance matrices $\mathbf{K}_{t; i}$ are transformed as
\begin{equation}
    \mathbf{K}'_{t; i} = \mathbf{Q}_{K'_{t; \text{2D}}}^T\bm{\Sigma}_t^{-\frac{1}{2}} \mathbf{K}_{t; i}(\bm{\Sigma}_t^{-\frac{1}{2}})^T\mathbf{Q}_{K'_{t; \text{2D}}}
\end{equation}
by the simultaneous diagonalisation of the other two time covariance matrices $\bm{\Sigma}_t$ and $\mathbf{K}_{t; \text{2D}}$.

Similar to the \textsc{DoubleMultiLowRankKernel}, we can break the decomposition of $\tilde{\mathbf{K}}$ up as a 2x2 block matrix, where we can already decompose the $\mathbf{A}$ block using either the \textsc{LuasKernel} or \textsc{LuasLasrachKernel}, depending on whether quasiseparable kernels are used. Once we can calculate $\mathbf{C}\mathbf{A}^{-1}\mathbf{B}$, general Cholesky decomposition is used to decompose $\mathbf{D} - \mathbf{C}\mathbf{A}^{-1}\mathbf{B}$, which introduces a scaling of $\mathcal{O}(N_\alpha^3 N_t^3)$ which tends to bottleneck this optimisation. See Appendix~\ref{app:luasplusmultilowrankkernel} for more details on the decomposition, including the calculation of $\mathbf{C}\mathbf{A}^{-1}\mathbf{B}$.

\subsubsection{Computational Scaling}

In Appendix~\ref{app:luasplusmultilowrankkernel}, we derive an overall runtime scaling of $\mathcal{O}((1 + N_\alpha^3) N_t^3 + N_\lambda N_t^2 + N_\lambda N_t (J_{\Sigma_\lambda} + J_{K_\lambda} + 4 N_\alpha)^2)$ for the kernel function in Equation~\ref{eq:half_luas_alpha} assuming all wavelength kernels are quasiseparable, which is linear in $N_\lambda$. If we instead use the \textsc{LuasKernel} in our decomposition of $\tilde{\mathbf{K}}$, then we do not require any quasiseparable kernels, but our decomposition scales as $\mathcal{O}(N_\lambda^3 + (1 + N_\alpha^3) N_t^3 + N_\lambda N_t (N_\lambda + N_t + N_\alpha^2))$, cubic in both dimensions. However, if $N_t >> N_\lambda$ then this difference is negligible, and we present benchmarking with this \textsc{LuasKernel} approach in Figures~\ref{fig:LuasPlusMultiLowRankKernel_Nalpha1_cost} and \ref{fig:LuasPlusMultiLowRankKernel_Nalpha1_A100_cost} for $N_\alpha = 1$ on CPU and GPU respectively.

Alternatively, for the kernel function in Equation~\ref{eq:half_luas} we recover a runtime scaling of $\mathcal{O}((1 + N_\beta^3) N_\lambda^3 + N_\lambda^2 N_t  + N_\lambda N_t (J_{K_t} + 2 N_\beta)^2)$ with quasiseparable $k_{t; 2D}$ and where we've assumed $s_t$ is diagonal. This has a linear scaling in $N_t$, which can be seen from benchmarking in Figures~\ref{fig:LuasPlusMultiLowRankKernel_Nbeta1_cost} and ~\ref{fig:LuasPlusMultiLowRankKernel_Nbeta4_cost} for $N_\beta = 1$ and $N_\beta = 4$ respectively. For $N_\beta = 1$, fitting a large $(N_\lambda, N_t) = (1024, 16384)$ dataset had a log-likelihood runtime of 1.1 seconds on CPU.

\subsection{\textsc{ExtendedLuasKernel}: Putting it All Together}

We finally arrive at the most flexible of kernel functions in this work:
\begin{align}
    k(\lambda, \lambda'&, t, t') = s_\lambda(\lambda, \lambda') s_t(t, t') + k_{\lambda; \text{2D}}(\lambda, \lambda')k_{t; \text{2D}}(t, t')\nonumber\\
    +\sum_{i=1}^{N_\alpha}&\alpha_i(\lambda) \alpha_i(\lambda') k_{t_i}(t, t') + 
    \sum_{j=1}^{N_\beta}k_{\lambda_j}(\lambda, \lambda')\beta_j(t) \beta_j(t'),
    \label{eq:full_luas2}
\end{align}
which builds covariance matrices of the form
\begin{align}
\mathbf{K} &= \bm{\Sigma}_\lambda \otimes \bm{\Sigma}_t
+\mathbf{K}_{\lambda; \text{2D}} \otimes \mathbf{K}_{t; \text{2D}} \nonumber\\
&+ \sum_i^{N_\alpha} \boldsymbol{\alpha}_{\lambda; i}^{} \boldsymbol{\alpha}_{\lambda; i}^T \otimes \mathbf{K}_{t; i} + \sum_j^{N_\beta}\mathbf{K}_{\lambda; j} \otimes \boldsymbol{\beta}_{t; j}\boldsymbol{\beta}_{t; j}^T,
\label{eq:full_luas_covmat_2}
\end{align}
At this point, we abandon any attempt to exploit quasiseparability, and our scaling is dominated by $\mathcal{O}(N_\alpha^3 N_t^3 + N_\beta^3 N_\lambda^3)$ matrix decompositions. We perform a whitening transformation followed by a QL decomposition in both dimensions, factorising as follows:
\begin{align}
\mathbf{K} = &\underbrace{[\bm{\Sigma}_\lambda^{\frac{1}{2}}\mathbf{Q}_{\mathbf{A}_\lambda'} \otimes \bm{\Sigma}_t^{\frac{1}{2}}\mathbf{Q}_{\mathbf{B}_t'}]}_{\displaystyle\mathbf{K}_\text{transf}}\nonumber\\
* &\left[ \mathbf{I}_{N_\lambda} \otimes \mathbf{I}_{N_t} + \mathbf{K}'_{\lambda; \text{2D}} \otimes \mathbf{K}'_{t; \text{2D}} \right. \nonumber\\
&\underbrace{\left.+ \sum_{i=1}^{N_\alpha} \left[\begin{smallmatrix}
\mathbf{0}\\
\mathbf{a}_{i}
\end{smallmatrix}\right]\left[\begin{smallmatrix}
\mathbf{0}\\
\mathbf{a}_{i}
\end{smallmatrix}\right]^T \otimes \mathbf{K}'_{t; i}+\sum_{j=1}^{N_\beta} \mathbf{K}'_{\lambda; j} \otimes \left[\begin{smallmatrix}
\mathbf{0}\\
\mathbf{b}_{j}\\
\end{smallmatrix}\right]\left[\begin{smallmatrix}
\mathbf{0}\\
\mathbf{b}_{j}
\end{smallmatrix}\right]^T\right]}_{\displaystyle\tilde{\mathbf{K}}}\nonumber\\
* &\underbrace{[\mathbf{Q}_{\mathbf{A}_\lambda'}^T(\bm{\Sigma}_\lambda^{\frac{1}{2}})^T \otimes \mathbf{Q}_{\mathbf{B}_t'}^T(\bm{\Sigma}_t^{\frac{1}{2}})^T]}_{\displaystyle\mathbf{K}_\text{transf}^T}
\label{eq:extendedluaskernelfactorisation}
\end{align}
with $\mathbf{K}'_{\lambda; \text{2D}}$ and all $\mathbf{K}'_{\lambda; j}$ transformed as:
\begin{align}
    \mathbf{K}'_{\lambda; \mathrm{2D}} &= \mathbf{Q}_{\mathbf{A}_\lambda'}^T\bm{\Sigma}_\lambda^{-\frac{1}{2}}\mathbf{K}_{\lambda; \mathrm{2D}}(\bm{\Sigma}_\lambda^{-\frac{1}{2}})^T\mathbf{Q}_{\mathbf{A}_\lambda'}^{}\\
    \mathbf{K}'_{\lambda; j} &= \mathbf{Q}_{\mathbf{A}_\lambda'}^T\bm{\Sigma}_\lambda^{-\frac{1}{2}}\mathbf{K}_{\lambda; j}(\bm{\Sigma}_\lambda^{-\frac{1}{2}})^T\mathbf{Q}_{\mathbf{A}_\lambda'}^{}
\end{align}
and similarly for $\mathbf{K}'_{t; \text{2D}}$ and all $\mathbf{K}'_{t; i}$.

The decomposition of $\tilde{\mathbf{K}}$ makes use of the \textsc{LuasPlusMultiLowRankKernel} to decompose the $\mathbf{A}$ block of $\tilde{\mathbf{K}}$, while calculation of $\mathbf{C}\mathbf{A}^{-1}\mathbf{B}$ is performed by splitting the $\mathbf{A}$, $\mathbf{B}$ and $\mathbf{C}$ blocks into their own 2x2 block matrices and exploiting sparsity. The $\mathbf{D} - \mathbf{C}\mathbf{A}^{-1}\mathbf{B}$ term is decomposed as a general $(N_\alpha N_t, N_\alpha N_t)$ matrix. See Appendix~\ref{app:extendedluaskernel} for the full details of the decomposition. 

\subsubsection{Computational Scaling}

In Appendix~\ref{app:extendedluaskernel}, the runtime scaling is worked out to $\sim\mathcal{O}((1 + N_\alpha^3)N_t^3 + (1 + N_\beta^3)N_\lambda^3 + N_\alpha N_\beta^2 N_\lambda^2 N_t + N_\lambda N_t^2)$, with a memory scaling of $\sim\mathcal{O}(N_\alpha^2 N_t^2 + N_\beta^2 N_\lambda^2 + N_\alpha N_\beta N_\lambda N_t)$. It has a similar runtime to the \textsc{LuasKernel} when $N_\alpha$ and $N_\beta$ are both small (compare Figures~\ref{fig:LuasKernelND_cost} vs \ref{fig:ExtendedLuasKernel_Nalpha1_Nbeta1_cost}), which means it could certainly help to add flexibility to analyses where the \textsc{LuasKernel} has already been applied \citep{Fortune2024, Fortune2025}. Similar to the \textsc{LuasKernel}, this optimisation is significantly faster on GPUs, as seen by comparing Figures~\ref{fig:ExtendedLuasKernel_Nalpha1_Nbeta1_cost} and \ref{fig:ExtendedLuasKernel_Nalpha1_Nbeta1_A100_cost}. We also demonstrate GPU runtimes with $(N_\alpha, N_\beta) = (1, 4), (4, 1)$ in Figures~\ref{fig:ExtendedLuasKernel_Nalpha1_Nbeta4_A100_cost} and \ref{fig:ExtendedLuasKernel_Nalpha4_Nbeta1_A100_cost} respectively. 

\subsection{Summary and Further Calculations}

We also include calculation details for the GP predictive mean and covariance in Appendix~\ref{sec:GP_predictions}, which can exploit any of these optimisations, provided not only our input observed points lie on a grid, but also the combined observed and predicted points lie on a grid. This allows us to visualise draws from the 2D GP model conditioned on the data, and also to perform cross-validation using any of these models, e.g. leaving out particular wavelengths and/or time points. We also extend the \textsc{SeparableKernel}, \textsc{LuasKernel} and \textsc{LuasLasrachKernel} to n-dimensional datasets in Appendix~\ref{app:higher_dim}, and benchmark log-likelihood calculations on datacubes for both CPU and GPU in Figures~\ref{fig:LuasKernelND_3D_m64_cost} - \ref{fig:LuasLasrachKernel_3D_m64_A100_cost}. We demonstrate the high numerical stability of our algorithms in Appendix~\ref{app:stability}. Finally, Appendix~\ref{app:benchmarking} describes how benchmarking was performed, Figure~\ref{fig:scaling_flowchart} summarises the runtime scaling for the different optimisations introduced, and the cost of gradient calculations is explained in Appendix~\ref{sec:gradients}.


\section{Discussion}
\label{sec:4}

\subsection{Comparison to Existing Methods}
\label{sec:existing_methods}
\subsubsection{Principal Component Analysis (PCA)}
\label{sec:PCA}

PCA is a classic approach to fit, understand, compress, and clean data, which can be calculated using the Singular Value Decomposition (SVD) of a 2D data array. A PCA model of a spectroscopic time-series can be written as
\begin{equation}
\mathbf{M}_\mathrm{PCA} = \mathbf{U}_\lambda \bm{\Lambda}_\mathrm{SVD}\mathbf{V}_t^T,
\label{eq:SVD}
\end{equation}
where assuming we use $N_\alpha$ PCA components, $\mathbf{U}_\lambda$ is $(N_\lambda, N_\alpha)$, $\bm{\Lambda}_\mathrm{SVD}$ is $(N_\alpha, N_\alpha)$ and diagonal, and $\mathbf{V}_t$ is $(N_t, N_\alpha)$. Both $\mathbf{U}_\lambda$ and $\mathbf{V}_t$ consist of orthonormal vectors, so they both satisfy $\mathbf{U}_\lambda^T \mathbf{U}_\lambda = \mathbf{V}_t^T \mathbf{V}_t = \mathbf{I}_{N_\alpha}$. We can alternatively express our PCA model in Equation~\ref{eq:SVD} as a sum of outer products, i.e.
\begin{equation}
\mathbf{M}_\mathrm{PCA} = \sum_{i=1}^{N_\alpha} \lambda_i \mathbf{u}_{\lambda; i}^{} \mathbf{v}_{t; i}^T.
\label{eq:SVD_sum}
\end{equation}
where a classic result of PCA is that it produces the best low-rank model to minimise the sum of squared residuals from the data, i.e. it minimises the loss function
\begin{equation}
S^2 = \sum_{i=1}^{N_\lambda}\sum_{j=1}^{N_t}\frac{(\mathbf{R}_{ij} - \sum_{k=1}^{N_\alpha}\lambda_k u_{ik}v_{jk})^2}{\sigma^2},
\label{eq:PCA_loss_function}
\end{equation}
for uniform white noise.
\citet{PPCA} demonstrate that PCA can be converted to a likelihood-based method, which they refer to as Probabilistic Principal Component Analysis (PPCA). Consider fitting a 2D dataset with a GP log-likelihood using the covariance matrix:
\begin{equation}
    \mathbf{K} = \sigma^2 \mathbf{I}_{N_\lambda} \otimes \mathbf{I}_{N_t} + \sum_{i=1}^{N_\alpha} \boldsymbol{\alpha}_{\lambda;i}^{}\boldsymbol{\alpha}_{\lambda;i}^T \otimes \mathbf{I}_{N_t},
    \label{eq:PCA_logL}
\end{equation}
where we assume uniform white noise, a mean function of zero, and we freely fit for all of the elements within each vector $\boldsymbol{\alpha}_{\lambda; i}$. This model is equivalent to the \textsc{MultiLowRankKernel} with uncorrelated time kernels $k_{t; i} = \delta_{tt'}$. Alternatively, it is equivalent to measurement covariance, as it reduces to a single Kronecker term in the form of the \textsc{SeparableKernel}
\begin{equation}
    \mathbf{K} = {\underbrace{\left[\sigma^2 \mathbf{I}_{N_\lambda} + \sum_{i=1}^{N_\alpha} \boldsymbol{\alpha}_{\lambda;i}^{}\boldsymbol{\alpha}_{\lambda;i}^T\right]}_{\bm{\Sigma}_\lambda}} \otimes \mathbf{I}_{N_t}.
    \label{eq:PCA_logL2}
\end{equation}
\citet{PPCA} derive the best-fit values of $\boldsymbol{\alpha}_{\lambda; i}$ analytically for exactly this covariance model. The maximum GP log-likelihood occurs at
\begin{equation}
    \boldsymbol{\alpha}_{\lambda; i} = \mathbf{U}_\lambda  [\bm{\Lambda} _\lambda - \sigma^2 \mathbf{I}_{N_\alpha}]^{\frac{1}{2}}\mathbf{W}\hat{\mathbf{e}}_i,
    \label{eq:PCA_vecs}
\end{equation}
where $\mathbf{U}_\lambda$ and $\bm{\Lambda}_\lambda$ come from the eigendecomposition of the sample wavelength covariance of the data. Equivalently, $\mathbf{U}_\lambda$ match the SVD components in Equation~\ref{eq:SVD}, while the eigenvalues are related by $\bm{\Lambda}_{\lambda} = \bm{\Lambda}_\mathrm{SVD}^2/N_t$. $\mathbf{W}$ is an arbitrary $(N_\alpha, N_\alpha)$ unitary matrix satisfying $\mathbf{W} \mathbf{W}^T = \mathbf{I}_{N_\alpha}$, which appears due to a degeneracy in our model, which can be seen if Equation~\ref{eq:PCA_vecs} is plugged into Equation~\ref{eq:PCA_logL2}. Since $\mathbf{W}$ has no impact on our model, $\boldsymbol{\alpha}_{\lambda; i}$ matching the principal PCA components in $\mathbf{U}_\lambda$ maximises the log-likelihood when we weight each component by $(\lambda_i - \sigma^2)$.

The GP mean for this model is
\begin{equation}
\boldsymbol{\mu}_* = \mathbf{U}_{\lambda}[\mathbf{I}_{N_
\alpha} - \sigma^2 \bm{\Lambda}_\lambda^{-1}]\mathbf{U}_{\lambda}^T\mathbf{R},
\end{equation}
regardless of $\mathbf{W}$, while the PCA model in Equation~\ref{eq:SVD} can equivalently be calculated as
\begin{equation}
    \mathbf{M}_\mathrm{PCA} = \mathbf{U}_{\lambda}^{}\mathbf{U}_{\lambda}^T\mathbf{R}.
\end{equation}
We see that our GP predictive mean approaches the PCA model in the limit as $\sigma \to 0$. Also in this limit, the predictive covariance goes to zero, which means the GP probability distribution becomes fixed at the GP mean. Therefore, PCA is equivalent to the GP model in Equation~\ref{eq:PCA_logL}, using its best-fit hyperparameters, in the limit of negligible white noise. When we include white noise, this GP model can be interpreted as a likelihood-based version of PCA which downweights components in the presence of white noise; analytically marginalises over the components in precisely one of the two dimensions; and can be applied to the data residuals after the subtraction of a mean function.

\subsubsection{Gaussian Process PCA (GPPCA)}
\label{subsec:GPPCA}

Note that Equation~\ref{eq:PCA_logL} can be generalised to 
\begin{equation}
    \mathbf{K} = \sigma^2 \mathbf{I}_{N_\lambda} \otimes \mathbf{I}_{N_t} + \sum_{i=1}^{N_\alpha} \boldsymbol{\alpha}_{\lambda;i}^{}\boldsymbol{\alpha}_{\lambda;i}^T \otimes \mathbf{K}_{t; i}
\label{eq:GP_PCA_logL}
\end{equation}
which is simply the \textsc{MultiLowRankKernel}, in this case with uniform white noise. This is equivalent to PPCA, but with the option to place smoothness priors $\mathbf{v}_{t, i} \sim \mathcal{N}(\mathbf{0}, \mathbf{K}_{t; i})$ on the time-components. It can be referred to as Gaussian Process PCA, or Generalised Probabilistic PCA (GPPCA) \citep{Gu2018}, and this link between PCA and GPs is also made in \citet{Bruinsma2019} for their Orthogonal Instantaneous Linear Mixing Model (OILMM). Both of these works provide GP optimisations for this model, which become equivalent to the \textsc{MultiLowRankKernel} in the case of uniform white noise and orthogonal $\boldsymbol{\alpha}_{\lambda;i}$ vectors. The optimisations in our work extend further, able to handle heterogeneous noise, measurement covariance, and non-orthogonal $\boldsymbol{\alpha}_{\lambda;i}$ vectors, which make it more equivalent to factor analysis models, described in the following sections.

We argue for three potential advantages of GPPCA over ordinary PCA:
\begin{enumerate}
  \item GPPCA is a likelihood-based approach that can be joint-fit with a mean function describing physical signals in the data
  \item It marginalises over a space of models analytically
  \item We can specify `smoothness priors' on components, which may be physically-motivated
\end{enumerate}
The first advantage could be particularly useful compared to current approaches which apply PCA as a pre-cleaning step, such as ground-based high-resolution transit spectroscopy, where pre-cleaning using PCA or SysRem has the ability to remove some of the exoplanet signal before it is ever fit (e.g. \citealt{Birkby2013, Gibson2022, Maguire2024b}). Our optimisations could permit efficient joint-inference of a modelled exoplanet signal combined with a GPPCA systematics model applied to the residuals of the data, potentially leading to less signal loss.

The second advantage could also help avoid signal loss when data cleaning, since marginalising over a space of models may make it less likely for the model to subtract out the exact exoplanet signal, since the best-fit model may be overfitting the data. It also allows us to marginalise over the uncertainty in the systematics correction, propagating this uncertainty into the atmospheric constraints in a Bayesian framework.

The third option could be useful for more closely modelling real physical signals. For example, PCA is sometimes used in astronomy to separate spectra which vary as continuous, time-correlated signals (e.g. modelling brown dwarf variability; \citealt{Wang2026, Schrader2026}), despite PCA not being weighted towards fitting smooth signals at all. This gives us an option to weight towards physically-motivated time-correlations, including with periodic kernels, or damped-harmonic oscillator kernels. It could even help disentangle two signals with the same amplitude but different time-correlations, which PCA cannot do (see Section~\ref{sec:BSS}).

\subsubsection{Factor Analysis}
\label{subsec:FA}

Factor analysis, as described by \citet{Bartholomew1987}, generalises Probabilistic PCA to heterogeneous white noise\footnote{although PPCA was actually derived as a special case of factor analysis \citep{PPCA}}
\begin{equation}
    \mathbf{K} = \bm{\Sigma} + \sum_{i=1}^{N_\alpha} \boldsymbol{\alpha}_{\lambda;i}^{}\boldsymbol{\alpha}_{\lambda;i}^T \otimes \mathbf{I}_{N_t},
\end{equation}
where $\bm{\Sigma}$ is diagonal, but not necessarily uniform or separable. Note the similarity with the popular data cleaning algorithm SysRem \citep{Tamuz2005}, which also generalises PCA to non-uniform white noise. However, factor analysis is analogous to the PPCA generalisation of PCA, as it also marginalises over components, makes use of mean functions, and downweights components in the presence of noise.

If the noise $\bm{\Sigma}$ is not separable in each dimension, there is no analytical solution that immediately solves for the best principal components, like SVD does for PCA, with iterative methods normally used to find the optimal $\boldsymbol{\alpha}_{\lambda;i}$. We also do not expect to recover orthogonal vectors $\boldsymbol{\alpha}_{\lambda;i}$ for our best-fit solution. If $\bm{\Sigma} = \bm{\Sigma}_\lambda \otimes \mathbf{I}_{t}$ is separable in each dimension on a grid of inputs, we can factor in the low-rank terms similar to Equation~\ref{eq:PCA_logL}, and model the full covariance matrix $\mathbf{K}$ as a single Kronecker product using the \textsc{SeparableKernel}.

\subsubsection{Gaussian Process Factor Analysis (GPFA)}
\label{subsec:GPFA}

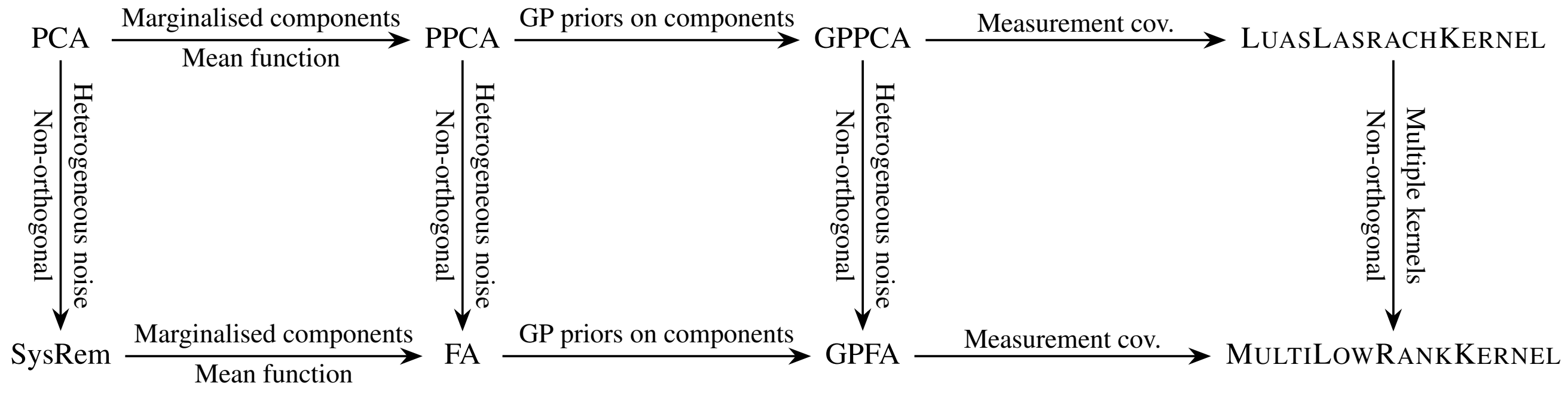
\begin{figure*}
\centering
\begin{tikzcd}[
    column sep = 9.5em,
    row sep    = 9em,
    arrows     = {-{Stealth[length=2.5mm]}, line width=0.7pt},
    every label/.append style = {font=\small}
]
    \text{ PCA }   \arrow[r, "\text{Marginalised components}", "\text{Mean function}"'] \arrow[d, "\text{Heterogeneous noise}"{sloped}, "\text{Non-orthogonal}"'{sloped}]
  & \text{PPCA}  \arrow[r, "\text{GP priors on components}"]\arrow[d, "\text{Heterogeneous noise}"{sloped}, "\text{Non-orthogonal}"'{sloped}]
  & \text{GPPCA} \arrow[r, "\text{Measurement cov.}"]\arrow[d, "\text{Heterogeneous noise}"{sloped}, "\text{Non-orthogonal}"'{sloped}]
  & \textsc{LuasLasrachKernel}\arrow[d, "\text{Multiple kernels}"{sloped}, "\text{Non-orthogonal}"'{sloped}] \\
    \text{SysRem}\arrow[r, "\text{Marginalised components}", "\text{Mean function}"']
  & \text{ FA }\arrow[r, "\text{GP priors on components}"]
  & \text{GPFA}\arrow[r, "\text{Measurement cov.}"]
  & \textsc{MultiLowRankKernel}
\end{tikzcd}
\caption{Comparison between different data-driven algorithms, with methods gaining in generality moving right and down. Therefore the \textsc{LuasLasrachKernel} and \textsc{MultiLowRankKernel} can be applied to any model to the left of them, provided the noise is separable, although the \textsc{SeparableKernel} can more efficiently handle Probabilistic PCA (PPCA) and Factor Analysis (FA). Note that both the \textsc{LuasLasrachKernel} and \textsc{MultiLowRankKernel} can account for heterogeneous noise, the difference is the \textsc{LuasLasrachKernel} is limited to a shared kernel function between all components, and those components come from eigendecomposition and therefore are all orthogonal, while the \textsc{MultiLowRankKernel} can have non-orthogonal components each with different kernel functions. Plot inspired by \citet{Bruinsma2019}, who present a model similar to GPPCA \citep{Gu2018}.}
\label{fig:model_relationships}
\end{figure*}

Similar to how GPPCA extends PCA to include a smoothness prior on time components, GP factor analysis makes the same generalisation for factor analysis, this time with heterogeneous noise present. It is equivalent to modelling using the covariance matrix:
\begin{equation}
    \mathbf{K} = \bm{\Sigma} + \sum_{i=1}^{N_\alpha} \boldsymbol{\alpha}_{\lambda;i}^{}\boldsymbol{\alpha}_{\lambda;i}^T \otimes \mathbf{K}_{t; i}
\end{equation}
Introduced in \citet{GPFA} to model time-series of multiple neurons in macaques, the \textsc{MultiLowRankKernel} generalises it further to include measurement covariance, provided $\bm{\Sigma}$ is separable in each dimension $\bm{\Sigma} = \bm{\Sigma}_\lambda \otimes \mathbf{D}_t$. While previous optimisations (e.g. \citealt{Gu2018, Bruinsma2019}) have been restricted to orthogonal components and/or homogeneous noise, we can handle non-orthogonal components and measurement covariance while still being exact and scalable. We summarise the links between the various data-driven methods introduced and two of our optimisations in Figure~\ref{fig:model_relationships}.

In practice, heterogeneous noise and measurement covariance are very common in astronomy, therefore GP factor analysis may be even more relevant than GPPCA. Astronomical examples equivalent to (or similar to) GP factor analysis include various approaches which attempt to disentangle different components of spectra, such as the use of `multidimensional GPs' within radial velocities to identify stellar activity components (e.g. \citealt{Rajpaul2015, Barragan2022}), distinguishing spectroscopic binaries \citep{Czekala2017}, or Heteroscedastic Matrix Factorisation (HMF), introduced to distinguish galactic spectra at different redshifts \citep{Tsalmantza2012}, although for HMF the smoothing priors are merely inspired by GP priors. Not all of these examples necessarily have complete, rectangular datasets and separable kernels in each dimension, required for the \textsc{MultiLowRankKernel} optimisation to be used, but we discuss applications to radial velocities in Section~\ref{sec:radial_velocity}.


\subsubsection{Analytical Marginalisation over Separable Linear Parameters}
\label{subsec:linearisation}

Consider a similar model to PCA or factor analysis, but where we have a model for building the wavelength components $\alpha(\lambda_i) = \boldsymbol{\alpha}_{\lambda; i}$, rather than freely fitting for their elements. In this case, our GP covariance matrix becomes equivalent to analytically marginalising out time components with Gaussian priors, convenient for reducing parameters to fit. For example, consider the GP model of a 2D dataset:
\begin{align}
    \mathbf{Y} &\sim \mathcal{N}(\mathbf{M}' + \boldsymbol{\alpha}_\lambda \boldsymbol{\omega}_t^T, \bm{\Sigma}),
\end{align}
where $\mathbf{M} = \mathbf{M}' + \boldsymbol{\alpha}_\lambda \boldsymbol{\omega}_t^T$ is our mean function model, equal to a vector outer product $\boldsymbol{\alpha}_\lambda \boldsymbol{\omega}_t^T$ plus some remaining model $\mathbf{M}'$. Suppose we want a Gaussian prior on $\boldsymbol{\omega}_t \sim \mathcal{N}(\mathbf{0}, \mathbf{K}_t)$, where $\mathbf{K}_t$ may or may not be diagonal, and our noise covariance is modelled by $\bm{\Sigma}$, which may be a sum of Kronecker products. Rather than fitting for the values of $\boldsymbol{\omega}_t$, we can analytically marginalise them out. Following \citet{Luger2017}, we can switch to the (marginal) GP log-likelihood:
\begin{equation}
        \mathbf{y} \sim \mathcal{N}(\mathbf{M}', \bm{\Sigma} + \boldsymbol{\alpha}_\lambda\boldsymbol{\alpha}_\lambda^T \otimes \mathbf{K}_t),
\end{equation}
where we've essentially moved some of our mean function into the GP covariance matrix. This new covariance matrix is equivalent to the \textsc{MultiLowRankKernel} with $N_\alpha = 1$ when the covariance matrix $\bm{\Sigma}$ can be expressed as a single Kronecker product, or the \textsc{LuasPlusMultiLowRankKernel} when $\bm{\Sigma}$ is a sum of two Kronecker products. The reverse of this process is also valid, any low-rank terms in our GP kernel function $\boldsymbol{\alpha}_\lambda\boldsymbol{\alpha}_\lambda^T \otimes \mathbf{K}_t$ could instead be marginalised over via MCMC sampling by adding a low-rank term $\boldsymbol{\alpha}_\lambda \boldsymbol{\omega}_t^T$ to the mean function and setting the prior $\boldsymbol{\omega}_t \sim \mathcal{N}(\mathbf{0}, \mathbf{K}_t)$. Naturally, the analytical marginalisation approach can be more computationally favourable if $N_t$ is very large, as it avoids freely-sampling thousands of parameters. That said, gradient-based inference can be highly scalable, so sampling low-rank kernel terms by moving them to the mean function may be more efficient if it permits a faster GP optimisation, particularly if the kernel function can also be approximated using a small number of low-rank components (see Section~\ref{sec:low_rank_approx}).

If we want distinct Gaussian priors on multiple linear parameters, and also to marginalise out parameters in both dimensions, e.g. 
\begin{equation}
        \mathbf{Y} = \boldsymbol{\mu}(\mathbf{\theta}) + \sum_i^{N_\alpha}\boldsymbol{\alpha}_{\lambda; i} \boldsymbol{\omega}_{t; i}^T + \sum_j^{N_\beta}\boldsymbol{\omega}_{\lambda; j}\boldsymbol{\beta}_{t; j}^T + \mathbf{E}_{\lambda, t},
\end{equation}
with $\boldsymbol{\omega}_{t; i} \sim \mathcal{N}(\mathbf{0}, \mathbf{K}_{t_i})$, $\boldsymbol{\omega}_{\lambda; j} \sim \mathcal{N}(\mathbf{0}, \mathbf{K}_{\lambda_j})$, and some 2D noise draw $\mathbf{E}_{\lambda, t} \sim \mathcal{N}(\mathbf{0}, \bm{\Sigma})$, then marginalising over all $\boldsymbol{\omega}$ is equivalent to the GP model
\begin{equation}
        \mathbf{y} \sim \mathcal{N}\left(\boldsymbol{\mu}, \bm{\Sigma} + \sum_i^{N_\alpha} \boldsymbol{\alpha}_{\lambda; i}^{} \boldsymbol{\alpha}_{\lambda; i}^T \otimes \mathbf{K}_{t; i} + \sum_j^{N_\beta} \mathbf{K}_{\lambda; j} \otimes \boldsymbol{\beta}_{t; j} \boldsymbol{\beta}_{t; j}^T\right),
\end{equation}
which is the \textsc{DoubleMultiLowRankKernel} when $\bm{\Sigma}$ is diagonal and separable. If $\bm{\Sigma}$ is the sum of two Kronecker products, i.e. if there are also wavelength- and time-correlated systematics we are marginalising over, then we can instead use the \textsc{ExtendedLuasKernel}.

Analytic marginalisation can be useful when our mean function $\boldsymbol{\mu}(\theta)$ contains linear nuisance parameters, as we can remove them from the mean function and analytically marginalise them out, reducing the number of free parameters. As a simple example, $(\boldsymbol{\beta}_t)_i = 1$ and $\mathbf{K}_\lambda = c^2\mathbf{I}_{N_\lambda}$ would allow us to marginalise over independent additive offsets to each light curve, which each have a Gaussian prior with standard deviation $c$.

\subsubsection{Blind Signal Separation (BSS)}
\label{sec:BSS}

Blind Signal Separation (BSS) refers to algorithms which distinguish different components out of an observed mixture of signals, with no \textit{a priori} information about the content of the signals, and limited assumptions about its structure. A classic example is the `cocktail party' problem. For our purposes, it could be trying to distinguish the wavelength-dependence of instrumental systematics apart from stellar variability. PCA and factor analysis may be considered examples of BSS algorithms, the goal of the method is what's different, the emphasis is on separating and distinguishing each component.

Independent Component Analysis (ICA) is strongly associated with BSS and has been used within exoplanet research before \citep{Waldmann1, Morello2015}. It works by assuming signals present in the data are non-Gaussian, optimising individual components to maximise some measure of non-Gaussianity, e.g. the kurtosis of the signal amplitude over time \citep{ICA}. In contrast, methods like PPCA, GPPCA, factor analysis, and GP factor analysis, all assume a mixture of Gaussian signals. As noted in Equation~\ref{eq:PCA_vecs}, PPCA components are degenerate up to an arbitrary orthonormal transformation labelled $\mathbf{W}$ (and factor analysis has a similar degeneracy), which limits their ability to separate components. In fact, PCA only distinguishes each component by maximising the variance of each successive component, so a mixture of signals with similar amplitudes will be poorly distinguished by PCA.

\begin{figure*}
    \centering
    \includegraphics[width=\textwidth]{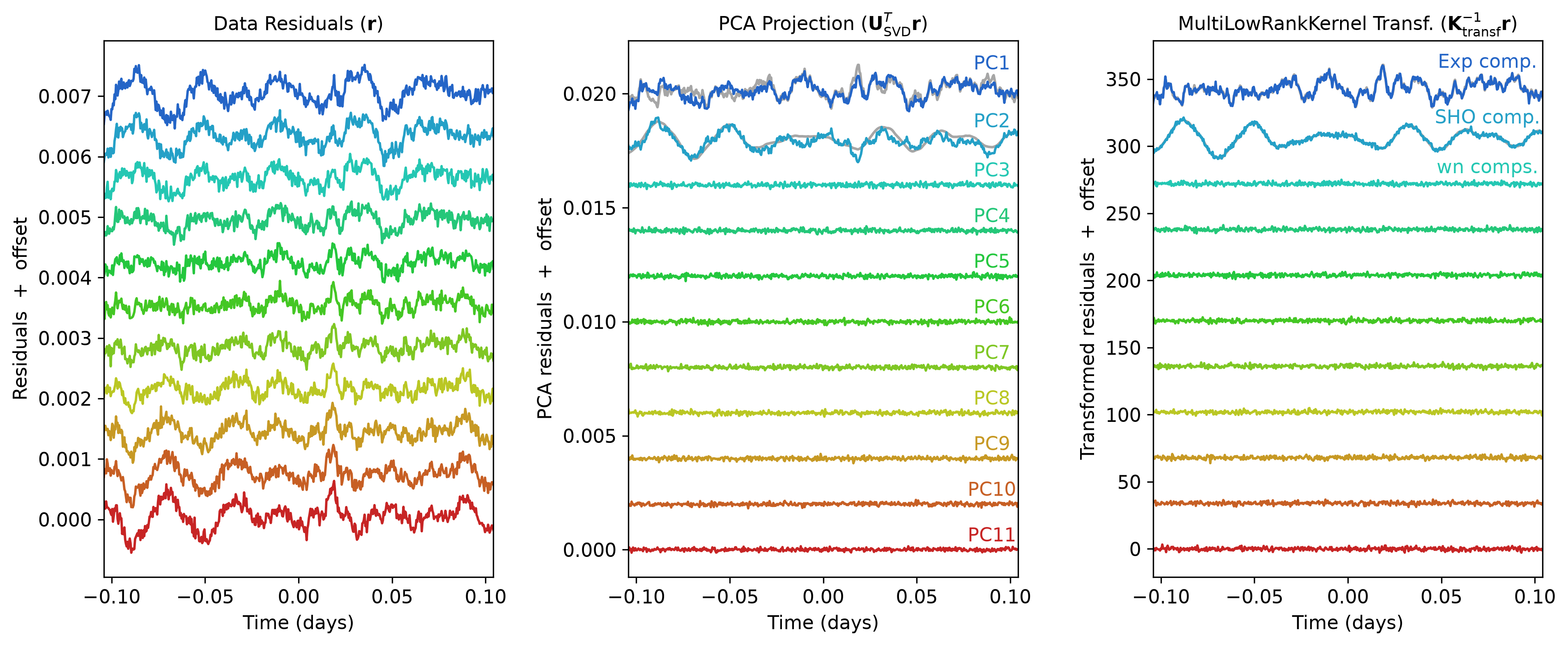}
    \caption{The \textsc{MultiLowRankKernel} can separate low-rank components based on different time correlations, while PCA can struggle to distinguish them when they have a similar amplitude. The left plot shows simulated systematics generated with a sum of $N_\alpha = 2$ low-rank kernel terms, an exponential time kernel with a constant $100$ ppm amplitude in wavelength, and a simple harmonic oscillator time kernel with wavelength amplitudes $\boldsymbol{\alpha}_\lambda$ varying linearly from 172 ppm to -172 ppm (i.e. it is anti-correlated between the top and bottom light curves, and zero in the middle light curve). The time-dependence of the injected signals is overplotted in grey in both the middle and right plots (with a rescaling applied): the two largest PCA components recover a mixture of these two signals as they have similar amplitudes (middle plot), while the best-fit \textsc{MultiLowRankKernel} model fits for the wavelength-dependence of these components and separates their time-dependence almost exactly (right plot), visualised using the transformed residuals $\tilde{\mathbf{r}} = \mathbf{K}_\mathrm{transf}^{-1} \mathbf{r}$, as this projects each systematics component onto separate light curves.} 
    \label{fig:transformation_visual}
\end{figure*}

Unlike PCA or factor analysis, GPPCA and GP factor analysis can uniquely distinguish components with similar amplitudes if they have different time covariance. We demonstrate this in Figure~\ref{fig:transformation_visual} on a simulated dataset with two low-rank kernel terms, fitting for each wavelength component initialised from random normal vectors, optimising using Limited-Memory BFGS, and correctly extracting the injected time-varying signals. Depending on the kernel function, we can distinguish signals based on whether they vary smoothly or sharply in time, whether they are periodic, or whether they correlate with some other choice of input regressor. For example, the wavelength-dependence of instrumental systematics could be distinguished from stellar variability if they vary on different timescales. In fact, this is already similar to an existing method of BSS introduced by \citet{Belouchrani1997} called Second-Order Blind Identification (SOBI). SOBI calculates autocovariance matrices with different amounts of time lag applied when cross-correlating the residuals, to distinguish components which are correlated on different timescales. They prove that components can be distinguished based on different time covariance up to a sign flip (or complex phase shift), provided each time kernel has a different length scale or shape.

Interestingly, the transformed residuals $\tilde{\mathbf{R}}$ that the \textsc{MultiLowRankKernel} calculates to simplify log-likelihood calculations can also be used to visualise each component in time, as highlighted in Figure~\ref{fig:transformation_visual}. This is a consequence of the transformed residuals being distributed according to the transformed covariance matrix $\tilde{\mathbf{R}} \sim \mathcal{N}(\mathbf{0}, \tilde{\mathbf{K}})$, provided we are modelling the overall covariance matrix correctly. For orthogonal $\boldsymbol{\alpha}_{\lambda; i}$, their transformed covariance matrix is block-diagonal, thus the transformed residuals isolate statistically independent components. The \textsc{MultiLowRankKernel} can thus be viewed as a method of BSS in a Bayesian framework, for signals which can be modelled as Gaussian processes in one dimension. We note that even when their wavelength-dependence can be distinguished, if $\boldsymbol{\alpha}_{\lambda; i}$ are non-orthogonal, the projection to calculate their time components is not unique, the QR-decomposition used by the \textsc{MultiLowRankKernel} will just be one choice of orthonormal basis for the set of vectors, and the transformed components will not be statistically independent. For Figure~\ref{fig:transformation_visual}, the components were simulated to be orthogonal so that we can visualise each component separately.

\subsubsection{Fourier-based GP Optimisations}
\label{sec:circulant}

Suppose we have data on a regularly-sampled 2D input grid, where the input grid has periodic boundary conditions (i.e. the endpoints become correlated with each other), and we have uniform white noise. This is a regime where Fourier-based GP optimisations can be performed exactly, and it has an interesting connection with the \citet{Rakitsch2013} method implemented by the \textsc{LuasKernel}.

In this case, our covariance matrix can be written as:
\begin{equation}
    \mathbf{K} = \sigma^2 \mathbf{I}_{N_\lambda} \otimes \mathbf{I}_{N_t} + \mathbf{K}_\lambda \otimes \mathbf{K}_t,
\end{equation}
One consequence of our assumptions is that both $\mathbf{K}_\lambda$ and $\mathbf{K}_t$ will be symmetric circulant matrices, i.e. they are characterised by a single row of values which get cycled to the right one row at a time, for example if $N_\lambda = 6$, then we can express $\mathbf{K}_\lambda$ as
\begin{equation}
    \mathbf{K}_\lambda = 
\begin{bmatrix}
c_0 & c_1 & c_2 & c_3 & c_2 & c_1 \\
c_1 & c_0 & c_1 & c_2 & c_3 & c_2 \\
c_2 & c_1 & c_0 & c_1 & c_2 & c_3 \\
c_3 & c_2 & c_1 & c_0 & c_1 & c_2 \\
c_2 & c_3 & c_2 & c_1 & c_0 & c_1 \\
c_1 & c_2 & c_3 & c_2 & c_1 & c_0
\end{bmatrix}
\label{eq:circulant_matrix}
\end{equation}
One property of circulant matrices is that their eigenvectors are given by Fourier modes, i.e. we can express the eigenvector matrix analytically as:
\begin{align}
(\mathbf{Q}_{K_\lambda})_{jk} &= \frac{1}{\sqrt{N_\lambda}}\omega^{(j-1)(k-1)}, \\
\omega &= \exp\left(\frac{2\pi i}{N_\lambda}\right)
\end{align}
where $\omega$ is a root of unity. The eigenvalues can also be efficiently calculated, and are related to the kernel functions power spectrum. The interesting connection here is that multiplying a vector by this eigenvector matrix is exactly equivalent to the inverse Discrete Fourier Transform (DFT), and multiplying by its conjugate transpose $\mathbf{Q}_{\mathbf{K}'_\lambda}^\dagger = \mathbf{W}_{\text{DFT; }\lambda}$ applies the DFT to the data residuals along the wavelength dimension. If both $\mathbf{K}_\lambda$ and $\mathbf{K}_t$ are circulant, then the \textsc{LuasKernel} will transform residuals as
\begin{align}
    \mathbf{K}_\text{transf}^{-1} &= \mathbf{Q}_{\mathbf{K}_\lambda}^\dagger \otimes \mathbf{Q}_{\mathbf{K}_t}^\dagger = \mathbf{W}_{\text{DFT; }\lambda} \otimes \mathbf{W}_{\text{DFT; }t}
\end{align}
where our transformation of the residuals is exactly the 2D DFT, and can be performed in $\mathcal{O}(N_\lambda N_t \log (N_\lambda N_t))$ runtime using Fast Fourier Transforms (FFTs) applied along both wavelength and time axes. Note that while we have included the eigenvector matrix in a complex form to be consistent with the FFT, all the phase information is lost when we calculate the amplitude of the residuals, and there are specific FFT optimisations for real signals.

Not only is the transformation using the 2D FFT faster to perform than for arbitrary eigenvector matrices, but we also avoid any need to perform eigendecomposition as we know it analytically, removing the main bottleneck of the \textsc{LuasKernel}. Simply performing the 2D FFT on our residuals diagonalises our log-likelihood, and our transformed log-likelihood becomes a simple weighted sum of squares in Fourier space. So the \textsc{LuasKernel} becomes equivalent to the Fourier-based approach under common assumptions made for exact Fourier-based GP optimisations, regular sampling, homogeneous noise, and periodic boundaries. The \textsc{LuasKernel} is an exact generalisation for heterogeneous noise, irregular sampling, and non-periodic bounds, while the \textsc{LuasLasrachKernel} extends further by combining with fast GP optimisations to achieve linear runtime scaling in the number of time points.

\subsection{Further Extensions}
\subsubsection{Fourier-based Optimisations in Individual Dimensions}
\label{sec:fourier_methods}

There are cases where we can mix Fourier methods with our optimisations, which can remain exact when we have circulant covariance matrices in at least one of the dimensions. For example, consider a covariance matrix of the form
\begin{align}
\mathbf{K} &= \bm{\Sigma}_\lambda \otimes \mathbf{I}_{N_t} + \mathbf{K}_\lambda \otimes \mathbf{K}_{t; \text{periodic}}\\
     &= 
    \underbrace{[\bm{\Sigma}_\lambda^{\frac{1}{2}}\mathbf{Q}_{\mathbf{K}'_\lambda} \otimes \mathbf{W}_{\text{DFT; }t}^\dagger]}_{\displaystyle\mathbf{K}_\mathrm{transf}}
    \underbrace{[\mathbf{I}_{N_\lambda} \otimes \mathbf{I}_{N_t} + \bm{\Lambda}_{K'_\lambda} \otimes \bm{\Lambda}_{K'_t}]}_{\displaystyle\tilde{\mathbf{K}}}\nonumber\\
    &\phantom{=} * \underbrace{[\mathbf{Q}_{\mathbf{K}'_\lambda}^T (\bm{\Sigma}_\lambda^{\frac{1}{2}})^T \otimes \mathbf{W}_{\text{DFT; }t}]}_{\displaystyle\mathbf{K}_\mathrm{transf}^\dagger}.
\end{align}
where $\mathbf{K}_{t; \text{periodic}}$ is built using periodic kernels (plus a constant diagonal if desired), and its eigenvector matrix exactly corresponds to a DFT matrix when we have regular time-sampling observed over an integer number of periods, since the matrix is circulant (see Section~\ref{sec:circulant}). We can transform our residuals using $\mathbf{K}_\mathrm{transf}^{-1} = \mathbf{Q}_{\mathbf{K}'_\lambda}^T \bm{\Sigma}_\lambda^{-\frac{1}{2}} \otimes \mathbf{W}_{\text{DFT; }t}$, which can be performed using a Fast Fourier Transform along the time-dimension, combined with a simultaneous diagonalisation in wavelength. The log-likelihood would have a runtime scaling of $\mathcal{O}(N_\lambda^3 + N_\lambda^2 N_t + N_\lambda N_t \log N_t)$, close to linear with the number of time points. Other Fourier-based methods could be explored in future work for the case of non-uniform time-sampling, or non-periodic time kernels, although these approaches are often approximate.

\subsubsection{Low-Rank Approximations of Covariance Matrices}
\label{sec:low_rank_approx}

\begin{figure}
    \centering
    \includegraphics[width=\columnwidth]{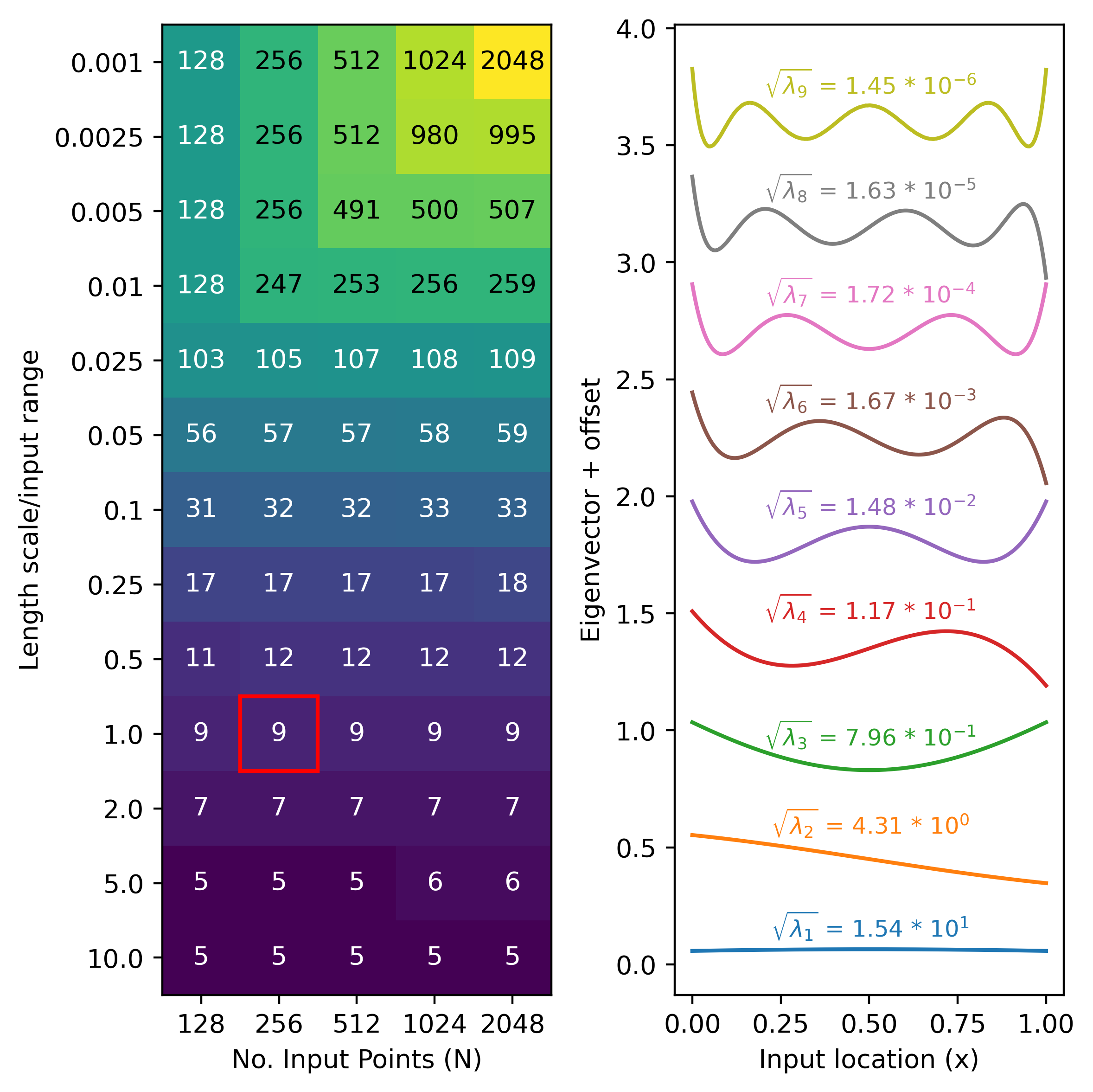}
    \caption{Left: Number of eigenvalues $\lambda >10^{-12}$ for a covariance matrix built using a squared exponential kernel on $N$ regularly sampled input locations, with $h = 1$ and length scale given relative to the full input range. Longer length scales result in very few significant eigenvectors. Right: Nine most significant eigenvectors for the case of $N = 256$ and length scale equal to the input range. The square root of the eigenvalue $\sqrt{\lambda}$ gives the standard deviation of the Gaussian prior placed on that eigenvector by the GP log-likelihood.}
    \label{fig:sq_exp_eigenvectors}
\end{figure}

We have frequently distinguished between 2D kernel terms (i.e. the product of two kernel functions) and low-rank kernel terms (see Section~\ref{sec:types_of_2D_kernels}). On a grid of inputs, 2D kernel terms produce Kronecker product covariance matrices
\begin{equation}
\mathbf{K}_\mathrm{2D} = \mathbf{K}_{\lambda; \text{2D}} \otimes \mathbf{K}_{t; \text{2D}}
\end{equation}
where both $\mathbf{K}_{\lambda; \text{2D}}$ and $\mathbf{K}_{t; \text{2D}}$ may be full-rank. In contrast, optimisations like \textsc{MultiLowRankKernel} focus on sums of low-rank kernel terms, i.e.
\begin{equation}
\mathbf{K}_\mathrm{low-rank} = \sum_i \boldsymbol{\alpha}_{\lambda; i}^{} \boldsymbol{\alpha}_{\lambda; i}^T \otimes \mathbf{K}_{t; i}.
\end{equation}
Re-writing the 2D kernel term using the eigendecomposition of $\mathbf{K}_{\lambda; \text{2D}}$ reveals the connection between these:
\begin{equation}
\mathbf{K}_\mathrm{2D} = \sum_i^\mathrm{Rank(K_\lambda)} \lambda_i \mathbf{v}_{\lambda; i}^{} \mathbf{v}_{\lambda; i}^T \otimes \mathbf{K}_{t; \text{2D}}.
\end{equation}
While the 2D kernel terms are flexible enough to handle full-rank covariance matrices, for smooth kernel functions like the squared exponential and rational quadratic kernels, the number of significant eigenvectors can often be very low. Figure~\ref{fig:sq_exp_eigenvectors} highlights this for the squared exponential kernel, measuring the number of eigenvalues $>10^{-12}$ as a function of length scale, and the number of input points. This means we can often closely approximate the squared exponential kernel using a small number of low-rank terms. For example, benchmarking shows that even $N_\alpha = 16$ can be highly scalable, even without orthogonal components (see Figures~\ref{fig:MultiLowRankKernel_Nalpha16_cost} and \ref{fig:MultiLowRankKernel_Nalpha16_ortho_cost}). This trick could be particularly important to avoid a mixture of a 2D kernel term and low-rank kernel terms as in the \textsc{LuasPlusMultiLowRankKernel}, and the \textsc{ExtendedLuasKernel}, as they scale less efficiently than the \textsc{MultiLowRankKernel}. There are also low-rank approximations to kernels which avoid eigendecomposition altogether \citep{Solin2014, hensman2018}, which may provide significant runtime improvements if the optimisation is bottlenecked by the cubic cost of eigendecomposition.

\subsubsection{Other GP Optimisations}
\label{sec:other_GP_opt}

Given the generality of many of the optimisations in this work, particularly highlighted by Equation~\ref{eq:blackbox_logL} for the \textsc{LuasLasrachKernel}, we can consider combining it with a wide range of existing 1D GP optimisations, each of which might offer computational or flexibility benefits. Our optimisations have been implemented within \textsc{luas} in a modular fashion, wrapping different 1D GP optimisations in a common \textsc{CovType} class which defines how to efficiently calculate matrix square roots, inverse square roots, inverses, and matrix multiplications. This modularity means that any existing \textsc{JAX} implementations of 1D GP optimisations can be imported and wrapped without re-writing each optimisation. 

While \textsc{tinygp} implements a flexible range of quasiseparable kernels, \citet{Rubenzahl2026} have introduced the JAX-based package \textsc{smolgp} to extend it to a more general class of state-space models using Kalman filtering. They used this to implement efficient integrals of GPs, useful for combining simultaneous observations with different exposure times. However, if this increased flexibility is not needed, benchmarking did not recover faster inference runtimes than quasiseparable kernels from \textsc{tinygp} \citep{Rubenzahl2026}. While Kalman filtering does not explicitly factorise the covariance matrix as a matrix square root, if it can be formulated as one then it could be better combined with more of our optimisations, but it can already be combined with the \textsc{LuasLasrachKernel} using Equation~\ref{eq:blackbox_logL}.

Hierarchical Off-Diagonal Low-Rank (HODLR) is a black box GP approximation which can perform almost linear time inference $\mathcal{O}(N \log N)$ with very limited restrictions on the choice of kernel functions \citep{Ambikasaran2013, Ambikasaran2015}. HODLR works by factoring a covariance matrix into an exact block diagonal component, combined with a series of low-rank updates which approximate regions away from the diagonal. The approximation accuracy can often be tuned to reach close to machine precision \citep{Ambikasaran2015}, although higher accuracy naturally comes at increased computational cost. While HODLR can be applied to multi-dimensional inputs directly, its computational performance is fastest for 1D inputs, while optimisations like the \textsc{LuasLasrachKernel} can not only reduce 2D GPs to separate 1D GP likelihoods, but can also reduce the problem size by exploiting sparsity. It could be used as an alternative to quasiseparable kernels, particularly when the quasiseparable structure is too restrictive.

There are plenty of other approximate methods, popular within machine learning, which may be complementary to our optimisations. Methods such as inducing point approximations estimate the GP log-likelihood by evaluating the GP on a reduced set of input locations, where these locations can be optimised to improve approximation accuracy \citep{Titsias2009}. KISS-GP uses inducing points on a grid to exploit Kronecker product optimisations for data which does not need to lie on a grid \citep{Wilson2015}. There are also approximations which are optimised for GPUs, which can exploit fast matrix-vector multiplications to iterate towards increasingly accurate approximations of the GP log-likelihood (e.g. \citealt{Gardner2018, Pleiss2018}). Combined with hardware acceleration, these methods present many options to be combined with our optimisations, in order to apply GP inference to very large datasets.

\subsection{Applications}

We list potential applications primarily focused on exoplanet observations. Just note, the main limitations that could affect these applications are the need for complete, rectangular datasets (i.e. no missing data), and separable noise (see Section~\ref{sec:white_noise}). These can be minor limitations for many spectroscopic time-series observations, as they possess a rectangular structure where occasional outliers can be clipped and replaced. The separable noise assumption could limit the use of rescaling noise estimates from data pipelines, but still allows us to fit white noise amplitudes in a data-driven way as separate functions of wavelength and time.

\subsubsection{Transiting Exoplanets with JWST}

The primary motivation for this work was to scale 2D GPs for the analysis of large spectroscopic time-series of transiting exoplanets. As data analyses typically assume the noise in each light curve is independent, this can result in unreliable results when wavelength-correlations are present. Unaccounted for wavelength-correlations can lead to inaccurate constraints from atmospheric retrievals, previously demonstrated both for transiting exoplanets \citep{Ih2021, Fortune2024}, and for directly-imaged planets \citep{Greco2016, Evert2023}. Rather than inferring wavelength-correlations when performing atmospheric retrievals \citep{Rotman2025}, recovering wavelength-correlations at the level of fitting the spectroscopic time-series can exploit much more information, and avoids degeneracies with inaccurate atmospheric models. Recovering wavelength-correlations from the spectroscopic time-series using 2D GPs has already been demonstrated for ground-based low-resolution transmission spectroscopy, utilising the \textsc{LuasKernel} optimisation \citep{Fortune2024}, but the computational scaling of this method limits its applications to larger JWST datasets.

All four JWST instruments have shown at least some evidence of wavelength-correlated noise, due to effects such as 1/f noise and spectral trace changes \citep{Ahrer2022, Holmberg2023, Bell2024, Gordon2026, Dyrek2026}. There is also evidence that these wavelength-correlations can be reduction-dependent \citep{Radica2026}, which implies that we cannot accurately compare the results of different reduction choices unless we can account for their differences in wavelength-correlations. JWST analyses have frequently found that different reduction and analysis choices can result in transit spectra that are offset from each other \citep{RoyPerez2026}, have different slopes \citep{Kirk2025, Claringbold2026}, or potentially even different features \citep{Kirk2024, Schmidt2025}. In future, we hope to constrain the role wavelength-covariance could play in such differences, as wavelength-covariance has already been shown to affect offsets and slopes of transmission spectra from VLT/FORS2 \citep{Fortune2024}.

One perpetual challenge within the field is the choice of wavelength-binning. This is the source of ongoing debate, since coarse wavelength bins should be more resilient to wavelength-correlations on short length scales \citep{Bell2024}, yet reducing the wavelength resolution of transit spectra can weaken constraints from atmospheric retrievals \citep{Davey2025}. This problem only arises because current analyses are treating light curves as independent, our optimisations offer a principled solution - we can fit at full wavelength (and time) resolution, while propagating the impact of wavelength-correlations to maintain robust results.

Even with perfectly stable instruments, at the photometric precision of JWST, astrophysical red noise from stellar granulation and asteroseismology can become a noise floor \citep{Waalkes2026, Weeks2026}. These signals are also chromatic \citep{Lund2019, Morris2020}, which could introduce further degeneracy with atmospheric constraints, but accounting for this wavelength-dependence can also help to improve our constraints \citep{Gordon2020}. Even our own Sun, which is relatively quiet in terms of stellar activity, still exhibits a time-correlated signal of $\sim 100$ ppm in optical light, largely due to solar granulation and oscillations \citep{Morris2020}. This signal is large enough to pose a significant barrier to many of the subtle variations in transit light curve geometry we may hope to measure, due to effects such as oblateness, limb-asymmetries, eclipse mapping, exo-moons, etc. GP models have long been one of the most frequently used tools to model stellar granulation, asteroseismology and stellar activity, in part due to their connection to physically-motivated stochastic processes (e.g. \citealt{Brewer2009, Foreman_Mackey_2017, Hara2025}). Viewed from a stellar astrophysics perspective, these optimisations could help us learn more about stellar variability, in combination with studying exoplanets.

\subsubsection{Ground-based High-Resolution Spectroscopic Time-Series}
\label{subsec:highres}

Time-series observations of transiting exoplanets at high spectroscopic resolution can be used to distinguish the Doppler-shifted atmospheric signal of an orbiting exoplanet, relative to the approximately stationary spectra from the host star's emission, and the Earth's own atmospheric absorption (known as tellurics). Typically these observations require a significant amount of data cleaning prior to analysis, which often includes subtracting a number of PCA or SysRem \citep{Tamuz2005} components from the reduced data (e.g. \citealt{dekok2013, Birkby2013}). An active area of research is whether this data cleaning can be improved so as to reduce the amount of exoplanet signal lost \citep{Meech2022, Maguire2024b}, particularly for slower moving targets when the Doppler shift is significantly reduced \citep{Cheverall2024}.

While GPs have been explored already for data cleaning, this was performed after reducing each order to 1D spectra \citep{Meech2022}, rather than fitting a 2D GP directly to each order. Combining 2D GPs with existing likelihood-based high-resolution retrieval approaches \citep{brogiline2018, Gibson2020} could be a promising method for reducing exoplanet signal loss. As highlighted in Figure~\ref{fig:model_relationships}, PCA and SysRem are closely linked to 2D GP kernels such as the \textsc{LuasLasrachKernel} and \textsc{MultiLowRankKernel}, but the GP approaches offer a number of potential advantages, where we could modify these approaches with smoothing priors on the time and/or wavelength components, and marginalise over a space of systematics models rather than subtracting off a single best-fit model. Perhaps the most important change would be the ability to efficiently joint-fit the systematics model and an exoplanet signal, applying the cleaning to the data residuals, rather than to the data which still contains the exoplanet signal.

We leave the application of this to future work, but to give a runtime estimate based on our benchmarking, Figure~\ref{fig:MultiLowRankKernel_Nalpha16_cost} tests a GP kernel using the \textsc{MultiLowRankKernel} with $N_\alpha = 16$, which has flexibility equivalent to 16 components of PCA or SysRem, and with GP priors on components, e.g. if we want to weight some components to vary over particular timescales or wavelength scales. If we have a single high-resolution spectroscopic order with $N_\lambda = 1024$ and $N_t = 128$, we achieve a 1.5 ms log-likelihood runtime on an M4 MacBook Air. This model also has linear runtime scaling with the number of datapoints, with a runtime scaling of $\sim\mathcal{O}(N_\alpha^3 N_t \bar{J}_{k_t}^2 + N_\alpha N_\lambda N_t)$, and joint-fits of multiple spectral orders could be parallelised across cores on a CPU or GPU.

\subsubsection{Extreme Precision Radial Velocities (EPRV)}
\label{sec:radial_velocity}

Extreme precision radial velocity (EPRV) analyses have been utilising 2D GPs - referred to as multidimensional GPs in the field - for many years (e.g. \citealt{Rajpaul2015, Gilbertson2020, Barragan2022}), where they have typically been used to joint-fit stellar radial velocity signals along with a number of activity indicators. Existing GP optimisations have already heavily optimised these analyses when joint-fitting a small number of activity indicators (e.g. \citealt{Gordon2020, Delisle2020, sleaf}). Since these existing analyses are normally in the regime where there are at least as many time-correlated low-rank terms as there are parallel time-series (i.e. $N_\alpha \geq N_\lambda$), the \textsc{MultiLowRankKernel} cannot provide a runtime advantage over these existing optimisations through its sparsity exploiting transformations (see Section~\ref{subsec:multilowrankkernel}).

However, our optimisations do offer some advantages if extending these models in certain ways. Accounting for measurement covariance between activity indicators has been discussed before \citep{Gilbertson2020}, and is a flexibility improvement our optimisations offer at virtually no extra cost (see Figure~\ref{fig:SingleKronTermKernel_densereadout_cost}). Measurement covariance could be beneficial to include since the activity indicators are often derived from the same spectra as the radial velocities, and thus are exposed to the same noise draw. Joint-fitting increasing numbers of parallel time-series can scale more efficiently with the \textsc{MultiLowRankKernel} for the same number of kernel terms, e.g. compare the runtimes in Figure~\ref{fig:GordonKernel_Nalpha1_cost} for the existing \citet{Gordon2020} method for just $N_\alpha = 1$ low-rank kernel term, with Figures~\ref{fig:MultiLowRankKernel_Nalpha4_cost}, \ref{fig:MultiLowRankKernel_Nalpha8_cost} and \ref{fig:MultiLowRankKernel_Nalpha16_cost} with $N_\alpha = 4, 8$ and $16$ low-rank kernel terms, respectively. This could be very useful as the identification of new activity indicators \citep{McWilliam2026, Costes2026}, or fitting RVs from multiple wavelength regions \citep{YuAigrain2024, Langford2025}, are active areas of research. Even fitting systematics directly to the high-resolution spectroscopic time-series with 2D GPs could be a possibility, as discussed in the previous section.

\subsubsection{Fitting Imaging Time-series at the Pixel-Level}

Another area of interest has been the push to make use of, or even fit, the individual pixel time-series, such as in JWST time-series \citep{Fortune2025, Connors2025, Gordon2026}, although there is a long history of this for analyses of Spitzer \citep{BLISS_mapping, demingknutson2015, Morello2015}, Kepler/K2 \citep{Wang2016, Luger2016, Pope2019}, and TESS \citep{Hattori2022}. These methods vary in their methodology, (e.g. using GPs, PCA, ICA), but as demonstrated in Section~\ref{sec:existing_methods}, these methods have links to highly-scalable GP optimisations. \citet{Fortune2025} introduced the pixel-fitting method, where pixel time-series from JWST/MIRI photometric data are joint-fit using the \textsc{LuasKernel}, but the exact kernel function used could now be implemented with \textsc{LuasLasrachKernel} to provide linear scaling with the number of time points, rather than cubic scaling before. Comparing the runtimes for a (64, 1024) dataset, similar to the dataset in \citet{Fortune2025}, using benchmarking in Figures~\ref{fig:LuasKernelND_cost} and \ref{fig:LuasLasrachKernel_cost}, this switch could reduce log-likelihood runtimes from 74 ms to 0.9 ms on an M4 MacBook Air. Alternatively, more flexible kernels such as the \textsc{ExtendedLuasKernel} could be used to simultaneously model multiple systematics sources with different timescales, as previously all systematics sources were limited to sharing the same timescale.

Extending this approach to joint-fitting pixel light curves from spectroscopic time-series would be a novel approach, replacing spectral trace extraction with a Bayesian inference approach, where we joint-model the changes in shape of the spectral trace, detector readout noise, and possible background contaminants. This could be combined with simultaneously inferring our astrophysical model parameters directly from the pixel time-series (similar to \citealt{Fortune2025}), or alternatively spectroscopic time-series could be inferred from the pixel time-series and fit separately. This approach could be very useful for NIRCam Grism time-series, where correlated readout noise is particularly challenging to separate from the spectral trace due to the fast read direction on the detector \citep{Schlawin2020}. Recently, \citet{Qiu2026} have suggested a related approach, using multidimensional GPs to account for readout noise from CCD detectors, using it to infer the covariance matrix of uncertainty in stellar spectra.

\subsubsection{Other Applications in Astronomy}

GPs are ubiquitous across astronomy, and even multidimensional GPs have already been used for a range of applications. \citet{pulsar2DGPs} apply a 2D GP approximation based on \textsc{celerite} to fit pulsar profile timing variations, useful for probing the ISM. Optimisations such as the \textsc{LuasLasrachKernel} could be used for the same kernel function and would be exact and significantly more scalable. Multiband AGN time variability has also been recently suggested to benefit from a 2D GP model which exactly corresponds to the \textsc{MultiLowRankKernel} \citep{Das2026}, provided the observations lie on a grid of datapoints. Spectroscopic time-series of brown dwarf variability present another potential application, analogous to the current use of PCA to identify spectroscopic components of variability, but with GP priors on the time components (see Section~\ref{subsec:GPPCA}).



Image reconstruction with interferometric data often uses some form of regularisation to place smoothing priors on the images being recovered, such as to calculate the first image of a black hole \citep{blackholeimagepaper4}. A 2D GP can be used as a regulariser within a principled statistical framework, weighting towards more physically plausible images with an explicit, interpretable prior. 2D GPs have already been optimised for image reconstruction by combining variational inference with Fourier-based GP optimisations \citep{MGVI, Edenhofer2024}, and are often applied within a Bayesian framework known as Information Field Theory \citep{ensslin2013}. Our optimisations provide alternative approaches to scale GPs to large images, which are exact and flexible, and have connections with existing Fourier-based optimisations (see Section~\ref{sec:fourier_methods}).





\subsubsection{Machine Learning Applications}

Related 2D GP models have been in use across fields such as geostatistics, computational biology, signal-processing, and machine learning, a summary of which is provided in \citet{Bruinsma2019}. Multi-task GPs, also known as multi-output GPs, are regression models which recover the covariance between different features/tasks over multiple samples, helping to understand features which are correlated, while also being useful for predicting unseen features \citep{multitaskGPs}. The covariance matrices used in these models can be in the same form as the \textsc{LuasKernel}:
\begin{equation}
    \mathbf{K} = \bm{\Sigma}_\mathrm{feature} \otimes \mathbf{I}_{N_\text{sample}} + \mathbf{K}_\text{feature} \otimes \mathbf{K}_\text{sample}
\end{equation}
A simple example of this in \citet{multitaskGPs} is to predict student exam scores from features about the school and students, with observations made for multiple students each year. The correlation between each feature and the student's exam score is modelled within the feature covariance matrix $\mathbf{K}_\text{feature}$, and the correlation between different samples/years is modelled within the sample covariance matrix $\mathbf{K}_\text{sample}$. There can be additional measurement covariance between features in the same sample (e.g. correlations between features for a given student), modelled by the $\bm{\Sigma}_\mathrm{feature}$ matrix, which motivated the work in \citet{Rakitsch2013}. Within the machine learning literature, our \textsc{LuasLasrachKernel} could be viewed as combining the work of \citet{Rakitsch2013} to handle measurement covariance between features, while still utilising fast 1D GP optimisations for correlations between samples, similar to \citet{Bruinsma2019}.

Linear Mixing Models are another form of 2D GP popular within machine learning, which are essentially the GP factor analysis/\textsc{MultiLowRankKernel} model focused on in Section~\ref{subsec:GPFA}, which is particularly useful as a method of dimensionality reduction (e.g. \citealt{GPFA, Wu2022}). There are many papers which focus on optimising them, although these have so far required approximations \citep{Teh2005, LMC_without_noise_or_wgibbssampling}, scale less efficiently \citep{Gordon2020, spatiotemporal_sarkka}, or have more stringent restrictions on the choice of covariance matrices to remain exact \citep{Bruinsma2019, diag_noise_GP}. Our \textsc{MultiLowRankKernel} is both exact and particularly fast at dimensionality reduction, as it scales very efficiently when the number of components to extract is much smaller than the dimensions of the dataset.

\subsubsection{Geostatistics}

Geostatistics often makes use of multidimensional GPs to model random fields over the Earth's surface. This draws from the original application of a GP, to predict the concentrations of metal ores over a given region, using measurements taken over a limited set of locations \citep{krige1951}. It has since been applied to many remote sensing problems, such as interpolating over gaps in satellite imagery \citep{gap_fill_GP}, and modelling environmental variables such as groundwater salinity \citep{groundwater_GP_prediction}. One popular GP model in use is called the Linear Model of Coregionalisation, which can be expressed as
\begin{equation}
    \mathbf{K} = \sum_{i=1}^L \mathbf{B}_{q; i} \otimes \mathbf{K}_\text{location; i}
\end{equation}
where $\mathbf{B}_{q; i}$ are covariance matrices between $N_\nu$ variables measured at every location, and $\mathbf{K}_\text{location; i}$ model correlations between $N_\mathrm{loc}$ different locations; while the sum over $L$ can handle different correlation structures over multiple distance scales. We note that this simplifies to the \textsc{SeparableKernel} when $L = 1$ (although existing optimisations in the field exist for this scenario), and the \textsc{LuasKernel} or \textsc{LuasLasrachKernel} when $L = 2$. Due to the typically non-grid 2D locations of each measurement, \textsc{celerite}-based optimisations may not easily be applied since they require sortable inputs, but combining the \textsc{LuasLasrachKernel} with approximate methods such as HODLR \citep{Ambikasaran2015} or inducing points could efficiently handle popular kernel choices such as the exponential and squared exponential kernel on 2D non-grid inputs.

Spatiotemporal GPs, as the name suggests, can be used to model data with dimensions of space and time, and are closely linked to methods in geostatistics. Their applications include epidemiology \citep{Diggle2013}, geophysical modelling \citep{remote_sensing_deep_GPs}, and even weather forecasting (e.g. \citealt{GP_climate_modelling, Kisi2025}). Optimisations such as the \textsc{LuasLasrachKernel} and \textsc{MultiLowRankKernel} could be useful for optimising these approaches and scaling to large datasets, while maintaining exact inference with highly flexible kernel functions. Compared with popular machine learning approaches like deep neural networks, GPs can be more interpretable, and produce predictions with more principled uncertainty quantification, crucial for many of these domains where risk mitigation is important. 

\section{Summary and Conclusions}

This work has introduced a selection of new, exact, highly scalable GP optimisations focused on two-dimensional rectangular datasets. These optimisations combine Kronecker product algebra with fast GP optimisations, including \textsc{celerite}, to permit inference on millions of datapoints across a flexible range of kernel choices and models. We've demonstrated links between these methods and popular algorithms like PCA, SysRem, and 2D Fast Fourier Transforms. These optimisations have an exciting range of applications which we will explore in future work, some of which include:
\begin{itemize}
  \item Efficiently accounting for wavelength- and time-correlated systematics in exoplanet transit spectroscopy
  \item Data-cleaning of ground-based high-resolution observations within a Bayesian framework
  \item Scaling Extreme Precision Radial Velocity techniques to larger datasets
  \item Inference applied directly to pixel time-series to accurately model detector systematics
  \item Gaussian process PCA to identify and distinguish smooth time components in datasets
  \item Fast analytical marginalisation over linear parameters with Gaussian priors which are separable in each dimension
\end{itemize}


\section*{Data Availability}

Our algorithms are implemented within the open-source Python package \textsc{luas}, which is available \href{https://github.com/markfortune/luas}{here}. Documentation and tutorials are located on \href{https://luas.readthedocs.io/en/latest/}{readthedocs}.


\section*{Acknowledgements}

The authors wish to acknowledge the Irish Centre for High-End Computing (ICHEC) for the provision of computational facilities and support. We acknowledge and pay respect to the traditional owners of the land on which Macquarie University is situated, upon whose unceded, sovereign, ancestral lands we work. We pay respects to their Ancestors and descendants, who continue cultural and spiritual connections to Country. 

Large Language Models (LLMs) were used to search for relevant literature, particularly to study related work outside the authors' domains of expertise, such as within machine learning, geostatistics, and signal processing. LLMs also identified useful literature in order to sample the GP predictive covariance, and to improve numerical stability. They aided in formatting some figures and LaTeX equations. While coding assistants were helpful for some testing and documentation of code, their use beyond this was limited to experimenting with algorithms and answering questions about implementations. LLMs did not develop the GP optimisations introduced in this work, nor write the text in this work, and the implementations of these algorithms were written from scratch with minimal LLM input beyond basic autocomplete coding assistance.

We are grateful to the developers of \textsc{JAX}, \textsc{NumPy}, \textsc{SciPy}, \textsc{Matplotlib}, \textsc{tinygp}, \textsc{pandas}, \textsc{iPython}, and \textsc{jaxoplanet} \citep{JAX, numpy, scipy, matplotlib, Foreman-Mackey2024, pandas, ipython,  jaxoplanet}, as these packages were used extensively in this work.
\bibliographystyle{paslike-pasa}
\bibliography{references}

\appendix
\counterwithin{figure}{section}                      
\renewcommand{\thefigure}{\thesection\arabic{figure}} 

\section{The \textsc{DoubleMultiLowRankKernel} Decomposition}
\label{app:doublemultilowrankdecomp}

\subsection{Transformations of Semiseparable Matrices}
\label{app:householder_quasisep}

If a matrix such as $\mathbf{K}_{\lambda; i}$ is semiseparable or quasiseparable, we would like to maintain this property for $\mathbf{K}'_{\lambda; i}$, defined as
\begin{equation}
    \mathbf{K}'_{\lambda; i} = \mathbf{Q}_{\mathbf{A}'}^T\mathbf{D}_\lambda^{-\frac{1}{2}}\mathbf{K}_{\lambda; i}\mathbf{D}_\lambda^{-\frac{1}{2}}\mathbf{Q}_{\mathbf{A}'}^{},
\end{equation}
where $\mathbf{D}_\lambda$ is diagonal and $\mathbf{Q}_{\mathbf{A}'}$ consists of a product of Householder matrices. We show this for semiseparable matrices, but this naturally extends to the quasiseparable matrices which \textsc{tinygp} implements without modification, since they are a more flexible class of matrices which include semiseparable matrices as a special case.

Semiseparable matrices can be represented as:
\begin{equation}
    \mathbf{K} = \mathbf{D} + \text{tril}(\mathbf{U}\mathbf{V}^T) + \text{triu}(\mathbf{V}\mathbf{U}^T)
    \label{eq:semiseparable}
\end{equation}
where $\mathbf{D}$ is diagonal and $\mathbf{U}$, $\mathbf{V}$ are both of shape $(N, J)$, where $J$ is the semiseparable rank. $\text{tril}()$ and $\text{triu}()$ are operators which zero out any elements which do not lie in the strictly lower/upper triangular part of the matrix.

Transformations with diagonal matrices, such as $\mathbf{D}_\lambda^{-\frac{1}{2}}\mathbf{K}_{\lambda; i}\mathbf{D}_\lambda^{-\frac{1}{2}}$, can be easily implemented while maintaining semiseparability, by transforming as:
\begin{align}
    \mathbf{D}' &= \mathbf{D}_\lambda^{-\frac{1}{2}}\mathbf{D}\mathbf{D}_\lambda^{-\frac{1}{2}}\\
    \mathbf{U}' &= \mathbf{D}_\lambda^{-\frac{1}{2}}\mathbf{U}\\
    \mathbf{V}' &= \mathbf{D}_\lambda^{-\frac{1}{2}}\mathbf{V},
\end{align}
which has no effect on the semiseparable rank $J$, i.e. we don't need to add any columns to $\mathbf{U}$ or $\mathbf{V}$. We will show that the transformations using $\mathbf{Q}_{\mathbf{A}'}$ and $\mathbf{Q}_{\mathbf{B}'}$, which equal a product of $N_\alpha$ and $N_\beta$ Householder matrices respectively, will increase the semiseparable ranks of each matrix by $2N_\alpha$ and $2N_\beta$ respectively.

\subsubsection{Householder Transformations as Low-Rank Updates}

Multiplication of a matrix on the left and right by a Householder matrix is equivalent to adding/subtracting three vector outer products
\begin{align}
\mathbf{H}\mathbf{K}\mathbf{H} &= (\mathbf{I} - 2 \mathbf{u} \mathbf{u}^T)\mathbf{K}(\mathbf{I} - 2 \mathbf{u} \mathbf{u}^T)
\end{align}
However, we can express our three outer products as just two symmetric outer products by defining:
 \begin{align}
 \mathbf{p}_1 &= - \mathbf{K}\mathbf{u} + 2(\mathbf{u} \cdot \mathbf{K} \mathbf{u})\mathbf{u} ,\\
 \mathbf{p}_2 &= \mathbf{K}\mathbf{u},
 \end{align}
where we see that
  \begin{align}
 \frac{\mathbf{p}_1 \mathbf{p}_1^T}{\mathbf{u} \cdot \mathbf{K} \mathbf{u}} & = +\frac{\mathbf{K}\mathbf{u} (\mathbf{K}\mathbf{u})^T}{\mathbf{u} \cdot \mathbf{K} \mathbf{u}} - 2 \mathbf{u} (\mathbf{K}\mathbf{u})^T - 2 (\mathbf{K}\mathbf{u})\mathbf{u}^T + 4 (\mathbf{u} \cdot \mathbf{K} \mathbf{u}) \mathbf{u} \mathbf{u}^T,\\
 -\frac{\mathbf{p}_2 \mathbf{p}_2^T}{\mathbf{u} \cdot \mathbf{K} \mathbf{u}} & = -\frac{\mathbf{K}\mathbf{u}  (\mathbf{K}\mathbf{u})^T}{\mathbf{u} \cdot \mathbf{K} \mathbf{u}},\\
 \implies  \mathbf{H}\mathbf{K}\mathbf{H} &= \mathbf{K} + \frac{\mathbf{p}_1 \mathbf{p}_1^T}{\mathbf{u} \cdot \mathbf{K} \mathbf{u}} -\frac{\mathbf{p}_2 \mathbf{p}_2^T}{\mathbf{u} \cdot \mathbf{K} \mathbf{u}}.
 \end{align}
 To maintain semiseparability, we now must show that these low-rank updates can be written in semiseparable form\footnote{We could alternatively handle low-rank updates using the Sherman-Morrison-Woodbury Formula (e.g. \citealt{blockinverseformula}), but expressing these transformed matrices within the common framework of semiseparable and quasiseparable matrices simplifies derivations and implementation}.

 \subsubsection{Low-Rank Updates to Semiseparable Matrices}

If we have a symmetric, low-rank update $\mathbf{K} \to \mathbf{K} + c \mathbf{p} \mathbf{p}^T$, this can be added to a semiseparable matrix in the form of Equation~\ref{eq:semiseparable}, e.g. by appending $\mathbf{p}$ as an extra column of $\mathbf{U}$, and $c \mathbf{p}$ as an extra column of $\mathbf{V}$. We must also add the diagonal of $c \mathbf{p} \mathbf{p}^T$ to the diagonal matrix $\mathbf{D}$. For both semiseparable and quasiseparable cases, each rank-1 update increases the rank of the matrices from $J \rightarrow J + 1$. Since each Householder transformation $\mathbf{H}\mathbf{K}\mathbf{H}$ requires a rank-2 update, if we have to apply this transformation successively for a stack of $N_\alpha$ Householder matrices, this will increase the rank of any quasiseparable matrix $\mathbf{K}$ from $J \rightarrow J + 2 N_\alpha$.


\subsection{Decomposing $\tilde{\mathbf{K}}$ for the \textsc{DoubleMultiLowRankKernel}}
\label{app:K_tilde_doublemultilowrankkernel}
Our transformed covariance matrix is in the form of a 2x2 block matrix:
\begin{align}
\label{eq:K_tilde_double}
\tilde{\mathbf{K}} = \begin{bmatrix}
\left(\begin{aligned}
\mathbf{I}_{N_\alpha} &\otimes \mathbf{I}_{N_t} +\\
\sum_{i=1}^{N_\alpha} \mathbf{a}_i\mathbf{a}_i^T &\otimes \mathbf{K}'_{t;i} +\\
\sum_{j=1}^{N_\beta} [\mathbf{K}'_{\lambda;j}]_{A} &\otimes \left[\begin{smallmatrix}
\mathbf{b}_j\\
\mathbf{0}
\end{smallmatrix}\right]\left[\begin{smallmatrix}
\mathbf{b}_j\\
\mathbf{0}
\end{smallmatrix}\right]^T
\end{aligned}
\right)
& \sum_{j=1}^{N_\beta} [\mathbf{K}'_{\lambda;j}]_{B} \otimes \left[\begin{smallmatrix}
\mathbf{b}_j\\
\mathbf{0}
\end{smallmatrix}\right]\left[\begin{smallmatrix}
\mathbf{b}_j\\
\mathbf{0}
\end{smallmatrix}\right]^T\\
\rule{0pt}{4.5ex} \sum_{j=1}^{N_\beta} [\mathbf{K}'_{\lambda;j}]_{C} \otimes \left[\begin{smallmatrix}
\mathbf{b}_j\\
\mathbf{0}
\end{smallmatrix}\right]\left[\begin{smallmatrix}
\mathbf{b}_j\\
\mathbf{0}
\end{smallmatrix}\right]^T 
& \left(\begin{aligned}\mathbf{I}_{N_\lambda - N_\alpha} &\otimes \mathbf{I}_{N_t} +\\
\sum_{j=1}^{N_\beta} [\mathbf{K}'_{\lambda;j}]_{D} &\otimes \left[\begin{smallmatrix}
\mathbf{b}_j\\
\mathbf{0}
\end{smallmatrix}\right]\left[\begin{smallmatrix}
\mathbf{b}_j\\
\mathbf{0}
\end{smallmatrix}\right]^T\end{aligned}
\right)
\end{bmatrix}
\end{align}
For the \textsc{MultiLowRankKernel}, $\tilde{\mathbf{K}}$ could be expressed as a simpler block diagonal matrix (Equation~\ref{eq:multilowrank_ktilde}), while $\tilde{\mathbf{K}}$ in Equation~\ref{eq:K_tilde_double} has non-zero $\mathbf{B}$ and $\mathbf{C}$ blocks. We can use the following formula for a general symmetric 2x2 block matrix
\begin{equation}
\tilde{\mathbf{K}}^{-\frac{1}{2}} =
\begin{bmatrix}
\mathbf{A}^{-\frac{1}{2}} & \mathbf{0}\\
-(\mathbf{D} - \mathbf{C}\mathbf{A}^{-1}\mathbf{B})^{-\frac{1}{2}}\mathbf{C}\mathbf{A}^{-1} & (\mathbf{D} - \mathbf{C}\mathbf{A}^{-1}\mathbf{B})^{-\frac{1}{2}}
\end{bmatrix},
\label{eq:block_inv_sqrt}
\end{equation}
known as the block Cholesky factorisation \citep{blockinverseformula}. Splitting our transformed residuals between the first $N_\alpha$ light curves $\tilde{\mathbf{r}}_A$ and the remaining light curves $\tilde{\mathbf{r}}_D$, we can calculate:
\begin{align}
\tilde{\mathbf{K}}^{-\frac{1}{2}}\tilde{\mathbf{r}} &=
\begin{bmatrix}
\mathbf{A}^{-\frac{1}{2}} & \mathbf{0}\\
-(\mathbf{D} - \mathbf{C}\mathbf{A}^{-1}\mathbf{B})^{-\frac{1}{2}}\mathbf{C}\mathbf{A}^{-1} & (\mathbf{D} - \mathbf{C}\mathbf{A}^{-1}\mathbf{B})^{-\frac{1}{2}}
\end{bmatrix}
\begin{bmatrix}
\tilde{\mathbf{r}}_A\\
\tilde{\mathbf{r}}_D
\end{bmatrix},\nonumber\\
&= \begin{bmatrix}
\mathbf{A}^{-\frac{1}{2}}\tilde{\mathbf{r}}_A\\
(\mathbf{D} - \mathbf{C}\mathbf{A}^{-1}\mathbf{B})^{-\frac{1}{2}}(-\mathbf{C}\mathbf{A}^{-1}\tilde{\mathbf{r}}_A + \tilde{\mathbf{r}}_D)
\end{bmatrix}.
\label{eq:block_mat_inv_sqrt}
\end{align}
Once we have $\tilde{\mathbf{K}}^{-\frac{1}{2}} \tilde{\mathbf{r}}$, we simply square and sum it to get $\tilde{\mathbf{r}}^T \bm{\tilde{K}}^{-1} \tilde{\mathbf{r}}$ for our transformed log-likelihood. The multiplication by $\mathbf{C}$ is cheap as it is very sparse, we can simply multiply by all $N_\alpha N_\beta^2 (N_\lambda - N_\alpha)$ non-zero values in $\mathcal{O}(N_\alpha N_\beta^2 (N_\lambda - N_\alpha))$ time. The only matrices we need to decompose are $\mathbf{A}$ and $\mathbf{D} - \mathbf{C}\mathbf{A}^{-1}\mathbf{B}$. The log-determinant of a 2x2 block matrix is given by:
\begin{equation}
\log|\tilde{\mathbf{K}}| = \log|\mathbf{A}| + \log|\mathbf{D} - \mathbf{C}\mathbf{A}^{-1}\mathbf{B}|,
\label{eq:block_determinant}
\end{equation}
so we can also use our decompositions of $\mathbf{A}$ and $\mathbf{\mathbf{D} - \mathbf{C}\mathbf{A}^{-1}\mathbf{B}}$ to calculate the log-determinant of $\tilde{\mathbf{K}}$ in our transformed log-likelihood.

\subsubsection{Decomposing the $\mathbf{A}$ block}

Starting with the $(N_\alpha N_t, N_\alpha N_t)$ $\mathbf{A}$ block
\begin{equation}
[\tilde{\mathbf{K}}]_A = \mathbf{I}_{N_\alpha} \otimes \mathbf{I}_{N_t} + \sum_{i=1}^{N_\alpha} \mathbf{a}_i\mathbf{a}_i^T \otimes \mathbf{K}'_{t;i} + \sum_{j=1}^{N_\beta} [\mathbf{K}'_{\lambda;j}]_{A} \otimes \left[\begin{smallmatrix}
\mathbf{b}_j\\
\mathbf{0}
\end{smallmatrix}\right]\left[\begin{smallmatrix}
\mathbf{b}_j\\
\mathbf{0}
\end{smallmatrix}\right]^T,
\end{equation}
note that the first two terms are in the form of the \textsc{MultiLowRankKernel}, where $\mathbf{K}'_{t;i}$ can be transformed following Appendix~\ref{app:householder_quasisep} to remain quasiseparable. However, there is the addition of the $(N_\alpha, N_\alpha)$ $\mathbf{A}$ block of the transformed wavelength covariance matrices $[\mathbf{K}'_{\lambda;j}]_A$ in a Kronecker product with sparse vectors which have at most $N_\beta$ non-zero values within $\mathbf{b}_j$. If we flip the order of the Kronecker product (which is actually the form that the \citealt{Gordon2020} method will decompose it in anyway), then these terms are entirely concentrated in the top left $(N_\alpha N_\beta, N_\alpha N_\beta)$ of the $\mathbf{A}$ block of $\tilde{\mathbf{K}}$. This flip in Kronecker product is not an issue for the calculations, provided we are careful to unravel our residuals array $\mathbf{R}$ Fortran-style instead of C-style for matrix-vector products.

This small update to the top left corner of the $\mathbf{A}$ block can be performed using a $(N_\alpha N_\beta)$-wide banded matrix, which is useful since banded matrices are valid quasiseparable matrices already implemented in \textsc{tinygp}, and all quasiseparable matrices are closed under addition \citep{Eidelman1999}. This update simply has the cost of increasing the quasiseparable rank of $\mathbf{A}$ by another $N_\alpha N_\beta$, for a total rank of $J = N_\alpha(\bar{J}_{K_t} + 3N_\beta)$. This allows us to decompose $\mathbf{A}$ as a fully quasiseparable matrix, scaling as $\mathcal{O}(N_\alpha^3 N_t (\bar{J}_{K_t} + 3 N_\beta)^2)$ in runtime.

\subsubsection{Calculating $\mathbf{C} \mathbf{A}^{-1} \mathbf{B}$}

With $\mathbf{A}$ already decomposed, we must calculate $\mathbf{C} \mathbf{A}^{-1} \mathbf{B}$, where each matrix-vector product $\mathbf{A}^{-\frac{1}{2}}\mathbf{v}$ scales as $\mathcal{O}(N_\alpha^2 N_t (\bar{J}_{K_t} + 3 N_\beta))$. Naturally, we can ignore columns in $\mathbf{B}$ which consist entirely of zeros, which already limits us to just $(N_\lambda - N_\alpha)N_\beta$ non-zero column vectors. However, we can reduce the number of matrix-vector products even further by noticing that $\mathbf{B}$ consists of only $N_\alpha N_\beta$ non-zero rows. If we swap the order of the Kronecker products for $\mathbf{B}$, labelling it as $\bar{\mathbf{B}}$, then these non-zero rows are restricted to the first $N_\alpha N_\beta$ rows of $\bar{\mathbf{B}}$. We see that rather than needing all of $\mathbf{A}^{-1}$, we only need to work with its own $(N_\alpha N_\beta, N_\alpha N_\beta)$ $\mathbf{A}$ block
\begin{equation}
\begin{bmatrix}
\bar{\mathbf{B}}^T & \mathbf{0}
\end{bmatrix}\begin{bmatrix}
[\mathbf{A}^{-1}]_A & [\mathbf{A}^{-1}]_B\\
[\mathbf{A}^{-1}]_C & [\mathbf{A}^{-1}]_D
\end{bmatrix}\begin{bmatrix}
\bar{\mathbf{B}}\\
\mathbf{0}
\end{bmatrix} = \begin{bmatrix}
\bar{\mathbf{B}}^T[\mathbf{A}^{-1}]_A\bar{\mathbf{B}} & \mathbf{0}\\
\mathbf{0} & \mathbf{0}
\end{bmatrix}
\end{equation}
We want to avoid explicitly computing all of $\mathbf{A}^{-1}$, and instead just calculate $[\mathbf{A}^{-1}]_A$. To do this, we can multiply our representation of $\mathbf{A}^{-\frac{1}{2}}$ by the first $N_\alpha N_\beta$ basis vectors, and evaluate each dot product
\begin{align}
    (\mathbf{A}^{-1})_{(i, j)} &= (\mathbf{A}^{-\frac{1}{2}}\hat{e}_i)^T(\mathbf{A}^{-\frac{1}{2}}\hat{e}_j).
\end{align}
We can then calculate a matrix square root of $[\mathbf{A}^{-1}]_A$, so that we can calculate $\bar{\mathbf{B}}' = ([\mathbf{A}^{-1}]_A^{\frac{1}{2}})^T\bar{\mathbf{B}}$, where $\bar{\mathbf{B}}'^T \bar{\mathbf{B}}'$ includes all the non-zero values of $\mathbf{C}\mathbf{A}^{-1}\mathbf{B}$. Since the non-zero values of $\bar{\mathbf{B}}'$ lie in $N_\alpha N_\beta$ rows, we only need to make a rank-$(N_\alpha N_\beta)$ update to the matrix $\mathbf{D}$, which can maintain its quasiseparability. The most significant runtime cost from this is the $N_\alpha N_\beta$ matrix-vector products of each basis vector, scaling as $\mathcal{O}(N_\alpha^3 N_\beta N_t (\bar{J}_{K_t} + 3 N_\beta))$ in runtime, which is completely dominated by the previous scaling for decomposing the $\mathbf{A}$ block.

\subsubsection{Decomposing $\mathbf{D} - \mathbf{C}\mathbf{A}^{-1}\mathbf{B}$}

Since $\mathbf{C}\mathbf{A}^{-1}\mathbf{B}$ is a symmetric, rank-$N_\alpha N_\beta$ matrix, we can maintain $\mathbf{D} - \mathbf{C}\mathbf{A}^{-1}\mathbf{B}$ as a semiseparable matrix provided that $\mathbf{D}$ is semiseparable (as shown in Section~\ref{app:householder_quasisep}), and similarly for quasiseparable matrices. By doing so, we increase its rank by $N_\alpha N_\beta$. The $\mathbf{D}$ block already has a quasiseparable rank of $N_\beta(\bar{J}_{K_\lambda} + 2N_\alpha)$ due to the sum of $N_\beta$ terms which have each been transformed by $N_\alpha$ Householder transformations, so we get a total rank of $N_\beta(\bar{J}_{K_\lambda} + 3N_\alpha)$. Note that unlike the $\mathbf{A}$ block, \textsc{tinygp} will actually decompose the $\mathbf{D}$ block in the Kronecker product order it is written in from Equation~\ref{eq:K_tilde_double}, as it must be sorted into the order of its wavelength kernel inputs, so we must be careful to re-order our $\bar{\mathbf{B}}'^T \bar{\mathbf{B}}' \to \mathbf{\tilde{B}}^T \tilde{\mathbf{B}}$, so that the non-zero elements of $\mathbf{C}\mathbf{A}^{-1}\mathbf{B}$ match this sort order. 

Note that due to the sparsity of both the $\mathbf{D}$ block and $\tilde{\mathbf{B}}$, we can split $\mathbf{D} - \mathbf{C}\mathbf{A}^{-1}\mathbf{B}$ into its own 2x2 block matrix, as it can be expressed as:
\begin{align}
\mathbf{D} - \mathbf{C}\mathbf{A}^{-1}\mathbf{B} = 
\begin{bmatrix}
    \left(\begin{aligned}\mathbf{I}_{N_\lambda - N_\alpha} &\otimes \mathbf{I}_{N_\beta} +\\
\sum_{j=1}^{N_\beta} [\mathbf{K}'_{\lambda;j}]_{D} &\otimes \mathbf{b}_j \mathbf{b}_j^T - \tilde{\mathbf{B}}^T \tilde{\mathbf{B}}\end{aligned}
\right) & \mathbf{0}\\
    \mathbf{0} &  \mathbf{I}_{N_\lambda - N_\alpha} \otimes \mathbf{I}_{N_t - N_\beta}\\
    \end{bmatrix}
\end{align}
We only need to decompose the quasiseparable $((N_\lambda - N_\alpha)N_\beta, (N_\lambda - N_\alpha)N_\beta)$ $\mathbf{A}$ block of this, which scales as $\mathcal{O}(N_\beta^3 N_\lambda (\bar{J}_{K_\lambda} + 3N_\alpha)^2)$.

\subsubsection{Total Computational Cost}

We now have everything we need to multiply our transformed residuals by the matrix inverse square root of $\tilde{\mathbf{K}}$ using Equation~\ref{eq:block_mat_inv_sqrt}, and our decompositions of $\mathbf{A}$ and $\mathbf{D} - \mathbf{C}\mathbf{A}^{-1}\mathbf{B}$ allow us to cheaply calculate the block matrix determinant from Equation~\ref{eq:block_determinant}. Our total runtime scaling to decompose $\tilde{\mathbf{K}}$ works out to $\sim\mathcal{O}(N_\alpha^3 N_t (\bar{J}_{K_t} + 3 N_\beta)^2 + N_\beta^3 N_\lambda (\bar{J}_{K_\lambda} + 3N_\alpha)^2)$, as the decomposition of the $\mathbf{A}$ and $\mathbf{D} - \mathbf{C}\mathbf{A}^{-1}\mathbf{B}$ blocks dominate. There is an additional cost of $\mathcal{O}(N_\lambda N_t (N_\alpha + N_\beta))$ to transform the residuals with $N_\alpha + N_\beta$ Householder transformations. Note that the scaling is completely linear in $N_\lambda$ and $N_t$, although it can scale quickly if the product $N_\alpha N_\beta$ is large.

\section{Decomposing $\tilde{\mathbf{K}}$ for the \textsc{LuasPlusMultiLowRankKernel}}
\label{app:luasplusmultilowrankkernel}

We have reduced $\tilde{\mathbf{K}}$ to the block matrix
\begin{align}
\tilde{\mathbf{K}} = 
\begin{bmatrix}
    \left(\begin{aligned}[\bm{\Sigma}'_{\lambda}]_A &\otimes \mathbf{I}_{N_t} +\\ [\mathbf{K}'_{\lambda; \text{2D}}]_A &\otimes \bm{\Lambda}_{K'_{t; \text{2D}}}\end{aligned}\right) & \left(\begin{aligned}
    [\bm{\Sigma}'_{\lambda}]_B &\otimes \mathbf{I}_{N_t} +\\ [\mathbf{K}'_{\lambda; \text{2D}}]_B &\otimes \bm{\Lambda}_{K'_{t; \text{2D}}}\end{aligned}
\right)\\
    \left(\begin{aligned}[\bm{\Sigma}'_{\lambda}]_C &\otimes \mathbf{I}_{N_t} +\\
    [\mathbf{K}'_{\lambda; \text{2D}}]_C &\otimes \bm{\Lambda}_{K'_{t; \text{2D}}}\end{aligned}\right) & \left(
\begin{aligned}
[\bm{\Sigma}'_{\lambda}]_D &\otimes \mathbf{I}_{N_t} +\\
[\mathbf{K}'_{\lambda; \text{2D}}]_D &\otimes \bm{\Lambda}_{K'_{t; \text{2D}}} +\\
\sum_{i=1}^{N_\alpha} \mathbf{a}_i \mathbf{a}_i^T &\otimes \mathbf{K}'_{t; i}
\end{aligned}
\right)
    \end{bmatrix}
\end{align}
with the individual covariance matrices transformed as described in Section~\ref{sec:luasplusmultilowrank}, where now the $\mathbf{D}$ block is $(N_\alpha N_t, N_\alpha N_t)$, instead of the $\mathbf{A}$ block as in Appendix~\ref{app:K_tilde_doublemultilowrankkernel}. We have two options to decompose the $\mathbf{A}$ block, either with the \textsc{LuasLasrachKernel}, or the \textsc{LuasKernel}. The \textsc{LuasLasrachKernel} approach can permit linear runtime scaling along one of the dimensions when combined with quasiseparable kernels, while the \textsc{LuasKernel} is more general and is not limited to quasiseparable kernels, but typically cannot achieve linear scaling in either dimension.

\subsection{Decomposing $\mathbf{A}$ with the \textsc{LuasKernel}}
\label{app:luasplusmultilowrankkernel_luaskernel}

In this case, we fully simultaneously diagonalise the $\mathbf{A}$ block, following the \citet{Rakitsch2013} trick implemented as the \textsc{LuasKernel}. This allows us to write the matrix inverse square root of $\mathbf{A}$ as:
\begin{align}
    \mathbf{A}^{-\frac{1}{2}} &= [\mathbf{I}_{N_\lambda - N_\alpha} \otimes \mathbf{I}_{N_t} + \bm{\Lambda}_{[K'_{\lambda; \text{2D}}]'_A} \otimes \bm{\Lambda}_{K'_t}]^{-\frac{1}{2}}[\mathbf{Q}_{[K'_{\lambda; \text{2D}}]'_A}^T[\bm{\Sigma}'_{\lambda}]_A^{-\frac{1}{2}} \otimes \mathbf{I}_{N_t}]\\
    &= \mathbf{D}_A^{-\frac{1}{2}}[\mathbf{W}^T \otimes \mathbf{I}_{N_t}]
\end{align}
where we've relabelled using $\mathbf{D}_A$ and $\mathbf{W}$ to simplify calculations.

We can multiply by one of the two terms in the $\mathbf{B}$ block as
\begin{align}
    \mathbf{A}^{-\frac{1}{2}}[[\bm{\Sigma}'_{\lambda}]_B \otimes \mathbf{I}_{N_t}] &= \mathbf{D}_A^{-\frac{1}{2}}[\mathbf{W}^T[\bm{\Sigma}'_{\lambda}]_B \otimes \mathbf{I}_{N_t}].
\end{align}
which can be performed efficiently since $\mathbf{W}^T[\bm{\Sigma}'_{\lambda}]_B$ is a $(N_\lambda - N_\alpha, N_\lambda - N_\alpha)$ matrix multiplied by a $(N_\lambda - N_\alpha, N_\alpha)$ matrix, scaling as $\mathcal{O}(N_\alpha (N_\lambda - N_\alpha)^2)$ in runtime cost. Since both terms are diagonal in the time dimension, we can treat this as $N_t$ independent calculations, and store our result as a $(N_\lambda - N_\alpha, N_\alpha, N_t)$ tensor. We can add to it the same calculation for the second term in the $\mathbf{B}$ matrix $[\mathbf{K}'_{\lambda; \text{2D}}]_B \otimes \bm{\Lambda}_{K'_{t; \text{2D}}}$. We can then contract the tensor with itself along the first axis, scaling as $\mathcal{O}(N_\alpha^2 (N_\lambda - N_\alpha) N_t)$ in runtime, to yield a $(N_\alpha, N_\alpha, N_t)$ tensor which includes all non-zero elements of $\mathbf{C}\mathbf{A}^{-1}\mathbf{B}$.

\subsection{Decomposing $\mathbf{A}$ with the \textsc{LuasLasrachKernel}}

Note that $\mathbf{A}$, $\mathbf{B}$ and $\mathbf{C}$ are already block diagonal in the time dimension, and so the calculation of $\mathbf{C}\mathbf{A}^{-1}\mathbf{B}$ can be split up as follows
\begin{align}
    \mathbf{C}\mathbf{A}^{-1}\mathbf{B} = \sum_i^{N_t}&\left([\bm{\Sigma}'_{\lambda}]_B + \lambda_{K'_{t; \text{2D}}; i} [\mathbf{K}'_{\lambda; \text{2D}}]_B\right)^T \left([\bm{\Sigma}'_{\lambda}]_A + \lambda_{K'_{t; \text{2D}}; i}[\mathbf{K}'_{\lambda; \text{2D}}]_A\right)^{-1} \nonumber\\
    &\left([\bm{\Sigma}'_{\lambda}]_B + \lambda_{K'_{t; \text{2D}}; i} [\mathbf{K}'_{\lambda; \text{2D}}]_B\right) \otimes \hat{\mathbf{e}}_i\hat{\mathbf{e}}_i^T
\end{align}
where the $(N_\alpha N_t, N_\alpha N_t)$ matrix $\mathbf{C}\mathbf{A}^{-1}\mathbf{B}$ only contains $N_\alpha^2 N_t$ non-zero elements since $[\bm{\Sigma}'_{\lambda}]_B$ and $[\mathbf{K}'_{\lambda; \text{2D}}]_B$ only have $N_\alpha$ columns.

Similar to the \textsc{LuasLasrachKernel}, if $\bm{\Sigma}_{\lambda}$ and $\mathbf{K}_{\lambda; \text{2D}}$ are quasiseparable, then it can be much faster to decompose them than eigendecomposition. Each block has a quasiseparable rank of $J = J_{\Sigma_\lambda} + J_{K_{\lambda; \text{2D}}} + 4 N_\alpha$ after applying $N_\alpha$ Householder transformations to both $\bm{\Sigma}_{\lambda}$ and $\mathbf{K}_{\lambda; \text{2D}}$. Decomposing $N_t$ separate blocks, where each matrix is $(N_\lambda - N_\alpha, N_\lambda - N_\alpha)$, scales as $\mathcal{O}((N_\lambda - N_\alpha) N_t (J_{\Sigma_\lambda} + J_{K_\lambda} + 4 N_\alpha)^2)$ in runtime.

Afterwards, the $\mathbf{C}\mathbf{A}^{-1}\mathbf{B}$ calculation requires $N_\alpha N_t$ matrix-vector products of each these blocks, scaling as $\mathcal{O}(N_\alpha (N_\lambda - N_\alpha) N_t (J_{\Sigma_\lambda} + J_{K_\lambda} + 4N_\alpha))$, which is dominated by the previous scaling to decompose the blocks. Similar to Appendix~\ref{app:luasplusmultilowrankkernel_luaskernel}, the result of this can be stored as a $(N_\lambda - N_\alpha, N_\alpha, N_t)$ tensor, which we contract along the first axis to recover a $(N_\alpha, N_\alpha, N_t)$ tensor which contains all non-zero values in $\mathbf{C}\mathbf{A}^{-1}\mathbf{B}$.

\subsection{Decomposing $\mathbf{D} - \mathbf{C}\mathbf{A}^{-1}\mathbf{B}$}

The remaining calculation is the same regardless of the approach used to calculate decompose $\mathbf{A}$ and calculate $\mathbf{C}\mathbf{A}^{-1}\mathbf{B}$. We simply build the $(N_\alpha N_t, N_\alpha N_t)$ $\mathbf{D}$ block and subtract $\mathbf{C}\mathbf{A}^{-1}\mathbf{B}$. We can then decompose $\mathbf{D} - \mathbf{C}\mathbf{A}^{-1}\mathbf{B}$ using general Cholesky decomposition because we cannot exploit any quasiseparability, which scales as $\mathcal{O}(N_\alpha^3 N_t^3)$ in runtime, and costs $\mathcal{O}(N_\alpha^2 N_t^2)$ in memory.

\subsection{Total Computational Cost}

We recover an overall runtime scaling dominated by $\sim\mathcal{O}((1 + N_\alpha^3) N_t^3 + N_\lambda N_t^2 + N_\lambda N_t (J_{\Sigma_\lambda} + J_{K_\lambda} + 4 N_\alpha)^2)$ if we exploit quasiseparability in the wavelength dimension, linear in $N_\lambda$. If we do not use any quasiseparable kernels, we still recover a scaling of $\sim\mathcal{O}((1 + N_\alpha^3) N_t^3 + N_\lambda^3 + N_\lambda N_t(N_t + N_\lambda + N_\alpha^2))$, for which benchmarking is presented in Figures~\ref{fig:LuasPlusMultiLowRankKernel_Nalpha1_cost} and \ref{fig:LuasPlusMultiLowRankKernel_Nalpha1_A100_cost} on CPU and GPU respectively. 

The memory scaling for the quasiseparable case is dominated by building the $\mathbf{D} - \mathbf{C}\mathbf{A}^{-1}\mathbf{B}$ matrix, and the decomposition of the $N_t$ quasiseparable blocks of $[\tilde{\mathbf{K}}]_A$, working out to $\sim\mathcal{O}(N_\alpha^2 N_t^2 + N_\lambda N_t (J_{\Sigma_\lambda} + J_{K_\lambda} + 4 N_\alpha))$ which is also linear in $N_\lambda$. The more general case using the \textsc{LuasKernel} builds full wavelength covariance matrices, and has a memory scaling of $\mathcal{O}(N_\lambda^2 + N_\alpha^2 N_t^2 + N_\alpha N_\lambda N_t)$, including the tensor stored in Appendix~\ref{app:luasplusmultilowrankkernel_luaskernel}.

Alternatively, we can flip the kernel dimensions around and recover a runtime scaling linear in the number of time points, provided we exploit quasiseparability in the time dimension, which Figures~\ref{fig:LuasPlusMultiLowRankKernel_Nbeta1_cost} and ~\ref{fig:LuasPlusMultiLowRankKernel_Nbeta4_cost} demonstrate.

\section{Decomposing $\tilde{\mathbf{K}}$ for the \textsc{ExtendedLuasKernel}}
\label{app:extendedluaskernel}

We can write $\tilde{\mathbf{K}}$ as the block matrix
\begin{align}
\tilde{\mathbf{K}} = \begin{bmatrix}
\left(\begin{aligned}
    \mathbf{I}_{N_\lambda - N_\alpha} &\otimes \mathbf{I}_{N_t} + \\
    [\mathbf{K}'_{\lambda; \text{2D}}]_A &\otimes \mathbf{K}'_{t; \text{2D}} +\\
    \sum_{j=1}^{N_\beta} [\mathbf{K}'_{\lambda; j}]_A &\otimes \left[\begin{smallmatrix}
\mathbf{0}\\
\mathbf{b}_{j}\\
\end{smallmatrix}\right]\left[\begin{smallmatrix}
\mathbf{0}\\
\mathbf{b}_{j}
\end{smallmatrix}\right]^T \end{aligned}\right)
& \left(\begin{aligned}
[\mathbf{K}'_{\lambda; \text{2D}}]_B &\otimes \mathbf{K}'_{t; \text{2D}} +\\
\sum_{j=1}^{N_\beta} [\mathbf{K}'_{\lambda; j}]_B &\otimes \left[\begin{smallmatrix}
\mathbf{0}\\
\mathbf{b}_{j}\\
\end{smallmatrix}\right]\left[\begin{smallmatrix}
\mathbf{0}\\
\mathbf{b}_{j}
\end{smallmatrix}\right]^T
\end{aligned}\right)\\
 \left(\begin{aligned}
[\mathbf{K}'_{\lambda; \text{2D}}]_C &\otimes \mathbf{K}'_{t; \text{2D}} +\\
\sum_{j=1}^{N_\beta} [\mathbf{K}'_{\lambda; j}]_C &\otimes \left[\begin{smallmatrix}
\mathbf{0}\\
\mathbf{b}_{j}\\
\end{smallmatrix}\right]\left[\begin{smallmatrix}
\mathbf{0}\\
\mathbf{b}_{j}
\end{smallmatrix}\right]^T
\end{aligned}\right) & \left(
\begin{aligned}
\mathbf{I}_{N_\alpha} &\otimes \mathbf{I}_{N_t} +\\
[\mathbf{K}'_{\lambda; \text{2D}}]_D &\otimes \mathbf{K}'_{t; \text{2D}} +\\
\sum_{i=1}^{N_\alpha} \mathbf{a}_{i}\mathbf{a}_{i}^T &\otimes \mathbf{K}'_{t; i} +\\
\sum_{j=1}^{N_\beta} [\mathbf{K}'_{\lambda; j}]_D &\otimes \left[\begin{smallmatrix}
\mathbf{0}\\
\mathbf{b}_{j}\\
\end{smallmatrix}\right]\left[\begin{smallmatrix}
\mathbf{0}\\
\mathbf{b}_{j}
\end{smallmatrix}\right]^T
\end{aligned}
\right)
    \end{bmatrix}
\end{align}
where the $\mathbf{A}$ block is simply in the form of the \textsc{LuasPlusMultiLowRankKernel} and can be decomposed readily. We make use of the \textsc{LuasKernel} method of decomposing it, as it simplifies the $\mathbf{C}\mathbf{A}^{-1}\mathbf{B}$ calculation, which gives a runtime scaling of $\sim\mathcal{O}((1 + N_\beta^3) N_\lambda^3 + N_t^3 + N_\lambda N_t(N_\lambda + N_t + N_\beta^2))$, while the $\mathbf{D} - \mathbf{C}\mathbf{A}^{-1}\mathbf{B}$ matrix will be decomposed using Cholesky factorisation with a runtime scaling of $\mathcal{O}(N_\alpha^3 N_t^3)$.

\subsection{Calculation of $\mathbf{C}\mathbf{A}^{-1}\mathbf{B}$}

We further subdivide the $\mathbf{A}$, $\mathbf{B}$ and $
\mathbf{C}$ blocks into their own 2x2 block matrices, this time splitting the $(N_t, N_t)$ time covariance matrices as $[N_t - N_\beta, N_\beta]$:
\begin{align}
[\tilde{\mathbf{K}}]_A = 
\begin{bmatrix}
    \left(\begin{aligned}\mathbf{I}_{N_\lambda - N_\alpha} &\otimes \mathbf{I}_{N_t} +\\ [\mathbf{K}'_{\lambda; \text{2D}}]_A &\otimes [\mathbf{K}'_{t; \text{2D}}]_A\end{aligned}\right) & \begin{aligned}
    [\mathbf{K}'_{\lambda; \text{2D}}]_A &\otimes [\mathbf{K}'_{t; \text{2D}}]_B\end{aligned}
\\
    \begin{aligned}[\mathbf{K}'_{\lambda; \text{2D}}]_A &\otimes [\mathbf{K}'_{t; \text{2D}}]_C\end{aligned} & \left(
\begin{aligned}
\mathbf{I}_{N_\lambda - N_\alpha} &\otimes \mathbf{I}_{N_\beta} +\\
[\mathbf{K}'_{\lambda; \text{2D}}]_A &\otimes [\mathbf{K}'_{t; \text{2D}}]_D +\\
\sum_{j=1}^{N_\beta} [\mathbf{K}'_{\lambda; j}]_A &\otimes \mathbf{b}_j \mathbf{b}_j^T
\end{aligned}
\right)
    \end{bmatrix}.
    \label{eq:K_A_tilde}
\end{align}
Note that since we are splitting the second term in the Kronecker product, the $\mathbf{A}$, $\mathbf{B}$ and $\mathbf{C}$ blocks will actually be mixed between each other rather than fully separated, but the block matrix inversion formulae are still valid.

Then write the $\mathbf{B}$ block, let us first label the following matrices as
\begin{align}
\mathbf{B}_{D; 1} &= [\mathbf{K}'_{\lambda; \text{2D}}]_B \otimes \mathbf{I}_{N_\beta}\\
\mathbf{B}_{D; 2} &= \sum_{j=1}^{N_\beta}[\mathbf{K}'_{\lambda; j}]_B \otimes \mathbf{b}_j \mathbf{b}_j^T
\end{align}
and we then write the $\mathbf{B}$ block as the sum of 2x2 block matrices
\begin{equation}
[\tilde{\mathbf{K}}]_B = \begin{bmatrix}
    [\mathbf{K}'_{\lambda; \text{2D}}]_B \otimes \mathbf{I}_{N_t - N_\beta} & \mathbf{0}\\
    \mathbf{0} & \mathbf{B}_{D; 1}
    \end{bmatrix}[\mathbf{I}_{N_\alpha} \otimes \mathbf{K}'_{t; \text{2D}}] + \begin{bmatrix}
    \mathbf{0} & \mathbf{0}\\
    \mathbf{0} & \mathbf{B}_{D; 2}
    \end{bmatrix}
\end{equation}
Using this form for the $\mathbf{B}$ block, we split up the $\mathbf{C}\mathbf{A}^{-1}\mathbf{B}$ calculation into separate terms of the form
\begin{equation}
\begin{bmatrix}
\mathbf{C}_A & \mathbf{0}\\
\mathbf{0} & \mathbf{C}_D
\end{bmatrix}\begin{bmatrix}
[\mathbf{A}^{-1}]_A & [\mathbf{A}^{-1}]_B\\
[\mathbf{A}^{-1}]_C & [\mathbf{A}^{-1}]_D
\end{bmatrix}
\begin{bmatrix}
\mathbf{B}_A & \mathbf{0}\\
\mathbf{0} & \mathbf{B}_D
\end{bmatrix} = \begin{bmatrix}
\mathbf{C}_A[\mathbf{A}^{-1}]_A\mathbf{B}_A & \mathbf{C}_A[\mathbf{A}^{-1}]_B\mathbf{B}_D\\
\mathbf{C}_D[\mathbf{A}^{-1}]_C\mathbf{B}_A & \mathbf{C}_D[\mathbf{A}^{-1}]_D\mathbf{B}_D
\end{bmatrix}
\end{equation}
and add everything together as
\begin{align}
\mathbf{C}\mathbf{A}^{-1}\mathbf{B} = [\mathbf{I}_{N_\alpha} \otimes \mathbf{K}'_{t; \text{2D}}]&\begin{bmatrix}
\mathbf{C}_A[\mathbf{A}^{-1}]_A\mathbf{B}_A & \mathbf{C}_A[\mathbf{A}^{-1}]_B\mathbf{B}_{D; 1}\\
\mathbf{C}_{D; 1}[\mathbf{A}^{-1}]_C \mathbf{B}_A & \mathbf{C}_{D; 1}[\mathbf{A}^{-1}]_D\mathbf{B}_{D; 1}
\end{bmatrix}[\mathbf{I}_{N_\alpha} \otimes \mathbf{K}'_{t; \text{2D}}]\nonumber\\
+[\mathbf{I}_{N_\alpha} \otimes \mathbf{K}'_{t; \text{2D}}]&\begin{bmatrix}
\mathbf{0} & \mathbf{C}_A[\mathbf{A}^{-1}]_B\mathbf{B}_{D; 2}\\
\mathbf{0} & \mathbf{C}_{D; 1}[\mathbf{A}^{-1}]_D\mathbf{B}_{D; 2}
\end{bmatrix}\nonumber\\
+&\begin{bmatrix}
\mathbf{0} & \mathbf{0}\\
\mathbf{C}_{D; 2}[\mathbf{A}^{-1}]_C\mathbf{B}_A & \mathbf{C}_{D; 2}[\mathbf{A}^{-1}]_D\mathbf{B}_{D; 1}
\end{bmatrix}[\mathbf{I}_{N_\alpha} \otimes \mathbf{K}'_{t; \text{2D}}]\nonumber\\
+&\begin{bmatrix}
\mathbf{0} & \mathbf{0}\\
\mathbf{0} & \mathbf{C}_{D; 2}[\mathbf{A}^{-1}]_D\mathbf{B}_{D; 2}
\end{bmatrix}
\label{eq:CA_inv_B_extended}
\end{align}
where we can build the $(N_\alpha N_t, N_\alpha N_t)$ inner block matrices in memory, and after this we can perform the required matrix multiplications by the $[\mathbf{I}_{N_\alpha} \otimes \mathbf{K}'_{t; \text{2D}}]$ terms in $\mathcal{O}(N_\alpha^3 N_t^3)$ time. Therefore, we just need to be able to build these inner block matrix terms to calculate $\mathbf{C}\mathbf{A}^{-1}\mathbf{B}$. Using the 2x2 block matrix inverse formula for $\mathbf{A} = [\tilde{\mathbf{K}}]_A$, we can calculate each of these terms as
\begin{equation}
\begin{bmatrix}
\mathbf{C}_A\mathbf{A}_A^{-1}\mathbf{B}_A + (\mathbf{C}_A\mathbf{A}_A^{-1}\mathbf{A}_B)[\mathbf{A}^{-1}]_D(\mathbf{A}_B^T\mathbf{A}_A^{-1}\mathbf{B}_A) & -(\mathbf{C}_A\mathbf{A}_A^{-1}\mathbf{A}_B)[\mathbf{A}^{-1}]_D\mathbf{B}_{D; i}\\
-\mathbf{C}_{D; i}[\mathbf{A}^{-1}]_D(\mathbf{A}_B^T\mathbf{A}_A^{-1}\mathbf{B}_A) & \mathbf{C}_{D; i}[\mathbf{A}^{-1}]_D\mathbf{B}_{D; i}
\end{bmatrix}
\end{equation}
where we have from Equation~\ref{eq:K_A_tilde}:
\begin{align}
    \mathbf{A}_A &= \mathbf{I}_{N_\lambda - N_\alpha} \otimes \mathbf{I}_{N_t} + [\mathbf{K}'_{\lambda; \text{2D}}]_A \otimes [\mathbf{K}'_{t; \text{2D}}]_A,\\
    \mathbf{A}_B &= [\mathbf{K}'_{\lambda; \text{2D}}]_A \otimes [\mathbf{K}'_{t; \text{2D}}]_B,\\
    \mathbf{A}_C &= [\mathbf{K}'_{\lambda; \text{2D}}]_A \otimes [\mathbf{K}'_{t; \text{2D}}]_C,\\
    \mathbf{A}_D &= \mathbf{I}_{N_\lambda - N_\alpha} \otimes \mathbf{I}_{N_\beta} + [\mathbf{K}'_{\lambda; \text{2D}}]_A \otimes [\mathbf{K}'_{t; \text{2D}}]_D + \sum_{j=1}^{N_\beta} [\mathbf{K}'_{\lambda; j}]_A \otimes \mathbf{b}_j \mathbf{b}_j^T\\
    [\mathbf{A}^{-1}]_D &= (\mathbf{A}_D - \mathbf{A}_C \mathbf{A}_A^{-1} \mathbf{A}_B)^{-1}
\end{align}
where $[\mathbf{A}^{-1}]_D$ is calculated following Appendix~\ref{app:luasplusmultilowrankkernel}, as it is also needed for the \textsc{LuasPlusMultiLowRankKernel} calculation. 



\subsubsection{The $\mathbf{C}_A\mathbf{A}_A^{-1}\mathbf{B}_A$ Term}

The $\mathbf{C}_A\mathbf{A}_A^{-1}\mathbf{B}_A$ term in Equation~\ref{eq:CA_inv_B_extended} can be calculated by simultaneously diagonalising $\mathbf{A}_A$ in both dimensions, similar to the \textsc{LuasKernel}. To simplify the notation, we will write its diagonal matrix of eigenvalues as
\begin{align}
\bm{\Lambda}_{A_A} = \mathbf{I}_{N_\lambda - N_\alpha} \otimes \mathbf{I}_{N_t - N_\beta} + \bm{\Lambda}_{[\mathbf{K}'_{\lambda; \text{2D}}]_A} \otimes \bm{\Lambda}_{[\mathbf{K}'_{t; \text{2D}}]_A}
\end{align}
and we define the $(N_\lambda - N_\alpha, N_\alpha)$ matrix:
\begin{equation}
    \mathbf{V}_B = [\mathbf{K}'_{\lambda; \text{2D}}]_B.
\end{equation}
We can then efficiently compute $\mathbf{C}_A\mathbf{A}_A^{-1}\mathbf{B}_A$ as
\begin{align}
    &\mathbf{C}_A\mathbf{A}_A^{-1}\mathbf{B}_A = [\mathbf{V}_B^T \otimes \mathbf{I}_{N_t - N_\beta}][\mathbf{Q}_{[\mathbf{K}'_{\lambda; \text{2D}}]_A} \otimes \mathbf{Q}_{[\mathbf{K}'_{t; \text{2D}}]_A}]\bm{\Lambda}_{A_A}^{-1}\nonumber\\
    &\phantom{\mathbf{C}_A\mathbf{A}_A^{-1}\mathbf{B}_A = }[\mathbf{Q}_{[\mathbf{K}'_{\lambda; \text{2D}}]_A}^T \otimes \mathbf{Q}_{[\mathbf{K}'_{t; \text{2D}}]_A}^T][\mathbf{V}_B \otimes \mathbf{I}_{N_t - N_\beta}]\\
     &= [\mathbf{I}_{N_\alpha} \otimes \mathbf{Q}_{[\mathbf{K}'_{t; \text{2D}}]_A}][\mathbf{V}_B'^T \otimes \mathbf{I}_{N_t - N_\beta}]\bm{\Lambda}_{A_A}^{-1}[\mathbf{V}_B' \otimes \mathbf{I}_{N_t - N_\beta}][\mathbf{I}_{N_\alpha} \otimes \mathbf{Q}_{[\mathbf{K}'_{t; \text{2D}}]_A}^T]\\
     &= \left[\mathbf{I}_{N_\alpha} \otimes \mathbf{Q}_{[\mathbf{K}'_{t; \text{2D}}]_A}\right]\underbrace{\left[\sum_{i=1}^{{N_t - N_\beta}}\mathbf{V}_B'^T \bm{\Lambda}_{A_A; t_i}^{-1}\mathbf{V}_B' \otimes \hat{\mathbf{e}}_i^{} \hat{\mathbf{e}}_i^T\right]}_{\mathbf{M}_1}\left[\mathbf{I}_{N_\alpha} \otimes \mathbf{Q}_{[\mathbf{K}'_{t; \text{2D}}]_A}^T\right]
     \label{eq:M1}
\end{align}
Although $\mathbf{M}_1$ is block diagonal, the simplest approach for performing the outer matrix multiplications is simply to build it as a full $(N_\alpha (N_t - N_\beta), N_\alpha (N_t - N_\beta))$ matrix, costing $\sim\mathcal{O}(N_\alpha^2 N_t^2)$ in memory. The dominant runtime cost of this calculation is the simultaneous eigendecomposition of both matrices, scaling as $
\sim\mathcal{O}(N_\lambda^3 + N_t^3 + N_\lambda N_t (N_\lambda + N_t))$, with the calculation of $\mathbf{M}_1$ scaling as $\mathcal{O}(N_\alpha^2 N_\lambda N_t)$ in runtime, and the $\mathcal{O}(N_\alpha^3 N_t^3)$ matrix multiplications of the outer terms will be later factored out by the $\mathbf{B}_Q$ term in Equation~\ref{eq:B_Q}.

\subsubsection{The $\mathbf{A}_B^T\mathbf{A}_A^{-1}\mathbf{B}_A$ and $\mathbf{C}_A\mathbf{A}_A^{-1}\mathbf{A}_B$ Terms}

Factoring using the eigendecompositions of $[K'_{\lambda; \text{2D}}]_A$ and $[\mathbf{K}'_{t; \text{2D}}]_B$ we have already calculated:
\begin{align}
    &\mathbf{A}_B^T\mathbf{A}_A^{-1}\mathbf{B}_A = [[\mathbf{K}'_{\lambda; \text{2D}}]_A \otimes [\mathbf{K}'_{t; \text{2D}}]_B^T]\left(\mathbf{I}_{N_\lambda - N_\alpha} \otimes \mathbf{I}_{N_t} + [\mathbf{K}'_{\lambda; \text{2D}}]_A \otimes [\mathbf{K}'_{t; \text{2D}}]_A\right)^{-1}\nonumber\\
    &\phantom{\mathbf{A}_B^T\mathbf{A}_A^{-1}\mathbf{B}_A = }* [\mathbf{V}_B \otimes \mathbf{I}_{N_t - N_\beta}]\\
    &= [\mathbf{Q}_{[K'_{\lambda; \text{2D}}]_A}\bm{\Lambda}_{[K'_{\lambda; \text{2D}}]_A} \otimes [\mathbf{K}'_{t; \text{2D}}]_B^T \mathbf{Q}_{[\mathbf{K}'_{t; \text{2D}}]_A}]\bm{\Lambda}_{A_A}^{-1}[\mathbf{V}_B' \otimes \mathbf{I}_{N_t - N_\beta}][\mathbf{I}_{N_\alpha} \otimes \mathbf{Q}_{[\mathbf{K}'_{t; \text{2D}}]_A}^T]\\
    &= [\mathbf{Q}_{[K'_{\lambda; \text{2D}}]_A}\bm{\Lambda}_{[K'_{\lambda; \text{2D}}]_A} \otimes [\mathbf{K}'_{t; \text{2D}}]_B^T \mathbf{Q}_{[\mathbf{K}'_{t; \text{2D}}]_A}]\left[\sum_{i=1}^{N_t - N_\beta}\bm{\Lambda}_{A_A; i}^{-1}\mathbf{V}_B' \otimes \hat{\mathbf{e}}_i\hat{\mathbf{e}}_i^T\right] \nonumber\\
    &\phantom{= }[\mathbf{I} \otimes \mathbf{Q}_{[\mathbf{K}'_{t; \text{2D}}]_A}^T]\\
    &= \underbrace{\left[\sum_{i=1}^{N_t - N_\beta}\mathbf{Q}_{[K'_{\lambda; \text{2D}}]_A}\bm{\Lambda}_{[K'_{\lambda; \text{2D}}]_A}\bm{\Lambda}_{A_A; i}^{-1}\mathbf{V}_B' \otimes [\mathbf{K}'_{t; \text{2D}}]_B^T \mathbf{Q}_{[\mathbf{K}'_{t; \text{2D}}]_A}\hat{\mathbf{e}}_i^{}\hat{\mathbf{e}}_i^T\right]}_{\mathbf{M}_2}[\mathbf{I}_{N_\alpha} \otimes \mathbf{Q}_{[\mathbf{K}'_{t; \text{2D}}]_A}^T]
    \label{eq:M2}
\end{align}
where $\mathbf{M}_2$ is $((N_\lambda - N_\alpha) N_\beta, N_\alpha(N_t - N_\beta))$, which contributes a memory scaling of $\sim\mathcal{O}(N_\alpha N_\beta N_\lambda N_t)$.

\subsubsection{Combining Terms to Calculate $\mathbf{C}\mathbf{A}^{-1}\mathbf{B}$}

For consistency, we will label:
\begin{align}
    \mathbf{M}_3 = \mathbf{B}_{D; 1} &= [\mathbf{K}'_{\lambda; \text{2D}}]_B \otimes \mathbf{I}_{N_\beta}\\
\mathbf{M}_4 = \mathbf{B}_{D; 2} &= \sum_{j=1}^{N_\beta}[\mathbf{K}'_{\lambda; j}]_B \otimes \mathbf{b}_j \mathbf{b}_j^T
\end{align}
where $\mathbf{M}_3$ and $\mathbf{M}_4$ are both $((N_\lambda - N_\alpha)N_\beta, N_\alpha N_\beta)$ matrices and can be directly calculated in $\sim\mathcal{O}(N_\alpha N_\beta^2 N_\lambda)$ runtime and memory.

We will also define 
\begin{equation}
\label{eq:B_Q}
    \mathbf{B}_{Q} = \begin{bmatrix}\mathbf{Q}_{[\mathbf{K}'_{t; \text{2D}}]_A} & \mathbf{0}\\
                                    \mathbf{0} & \mathbf{I}_{N_\beta}\end{bmatrix}
\end{equation}
which we use to deal with the extra factor which was factored out of our expressions for $\mathbf{M}_1$ and $\mathbf{M}_2$ in Equations~\ref{eq:M1} and \ref{eq:M2}, but not for $\mathbf{M}_3$ or $\mathbf{M}_4$. We can now rewrite entirely in terms of precomputed terms:
\begin{align}
\mathbf{C}\mathbf{A}^{-1}\mathbf{B} =\phantom{111111}&\nonumber\\
[\mathbf{I}_{N_\alpha} \otimes \mathbf{K}'_{t; \text{2D}}\mathbf{B}_{Q}]&\begin{bmatrix}
\mathbf{M}_1 + \mathbf{M}_2^T[\mathbf{A}^{-1}]_D\mathbf{M}_2 & -\mathbf{M}_2^T[\mathbf{A}^{-1}]_D\mathbf{M}_3\\
-\mathbf{M}_3^T[\mathbf{A}^{-1}]_D\mathbf{M}_2 & \mathbf{M}_3^T[\mathbf{A}^{-1}]_D\mathbf{M}_3
\end{bmatrix}[\mathbf{I}_{N_\alpha} \otimes \mathbf{B}_{Q}^T\mathbf{K}'_{t; \text{2D}}]\nonumber\\
+[\mathbf{I}_{N_\alpha} \otimes \mathbf{K}'_{t; \text{2D}}\mathbf{B}_{Q}]&\begin{bmatrix}
\mathbf{0} & -\mathbf{M}_2^T[\mathbf{A}^{-1}]_D\mathbf{M}_4\\
\mathbf{0} & \mathbf{M}_3^T[\mathbf{A}^{-1}]_D\mathbf{M}_4
\end{bmatrix}\nonumber\\
+&\begin{bmatrix}
\mathbf{0} & \mathbf{0}\\
-\mathbf{M}_4^T[\mathbf{A}^{-1}]_D\mathbf{M}_2 & \mathbf{M}_4^T[\mathbf{A}^{-1}]_D\mathbf{M}_3
\end{bmatrix}[\mathbf{I}_{N_\alpha} \otimes \mathbf{B}_{Q}^T\mathbf{K}'_{t; \text{2D}}]\nonumber\\
+&\begin{bmatrix}
\mathbf{0} & \mathbf{0}\\
\mathbf{0} & \mathbf{M}_4^T[\mathbf{A}^{-1}]_D\mathbf{M}_4
\end{bmatrix}
\end{align}
In practice, we can first compute $\mathbf{M}_i' = [\mathbf{A}^{-\frac{1}{2}}]_D \mathbf{M}_i$ for $i = 2, 3, 4$, and then cheaply compute each term $\mathbf{M}_i^T[\mathbf{A}^{-1}]_D\mathbf{M}_j = \mathbf{M}_i'^T \mathbf{M}_j'$ afterwards. The cost of the matrix inverse square root calculations are $\mathcal{O}(N_\alpha N_\beta^2 N_\lambda^2 N_t)$ for $\mathbf{M}_2'$, which dominates over the $\mathcal{O}(N_\alpha N_\beta^3 N_\lambda^2)$ scaling for $\mathbf{M}_3'$ and $\mathbf{M}_4'$.

\subsubsection{Total Computational Cost}

None of the terms in the calculation of $\mathbf{C}\mathbf{A}^{-1}\mathbf{B}$ adds an additional runtime scaling term which dominates except for the $\mathcal{O}(N_\alpha N_\beta^2 N_\lambda^2 N_t)$ calculation of $\mathbf{M}_2'$. Adding this to the cost to decompose $\mathbf{A}$ and $\mathbf{D} - \mathbf{C}\mathbf{A}^{-1}\mathbf{B}$ yields a $\mathcal{O}((1 + N_\alpha^3)N_t^3 + (1 + N_\beta^3)N_\lambda^3 + N_\alpha N_\beta^2 N_\lambda^2 N_t + N_\lambda N_t^2)$ runtime scaling. Our total memory scaling is dominated by building $[\tilde{\mathbf{K}}]_D$, $\mathbf{A}_D$, and $\mathbf{M}_2$ in memory, working out to $\mathcal{O}(N_\alpha^2 N_t^2 + N_\beta^2 N_\lambda^2 + N_\alpha N_\beta N_\lambda N_t)$ in total.

\section{GP Predictive Mean and Covariance Calculation}
\label{sec:GP_predictions}

To calculate the distribution of models the GP is fitting to the data, we need the GP predictive mean and predictive covariance
\begin{align}
    \label{eq:predictive_mean_formula}
    \boldsymbol{\upmu}_\text{pred} &= \boldsymbol{\upmu}_* + \mathbf{K}_{*}\mathbf{K}^{-1}(\mathbf{y} - \boldsymbol{\upmu}) \\
    \mathbf{K}_\text{pred} &= \mathbf{K}_{**} - \mathbf{K}_{*}^T\mathbf{K}^{-1}\mathbf{K}_{*}.
    \label{eq:predictive_covariance_formula}
\end{align}
The GP mean is already straight-forward to calculate given that we can calculate the log-likelihood for each optimisation, as we can already calculate $\mathbf{K}^{-1}\mathbf{r} = (\mathbf{K}^{-\frac{1}{2}})^T\mathbf{K}^{-\frac{1}{2}}\mathbf{r}$ efficiently. Multiplication by $\mathbf{K}_{*}$ is simple if the predicted points also lie on a grid, in which case it equals a sum of Kronecker products, for which matrix-vector multiplications can be performed efficiently \citep{Saatchi2011}.

\begin{figure*}
    \centering
    \includegraphics[width=0.9\textwidth]{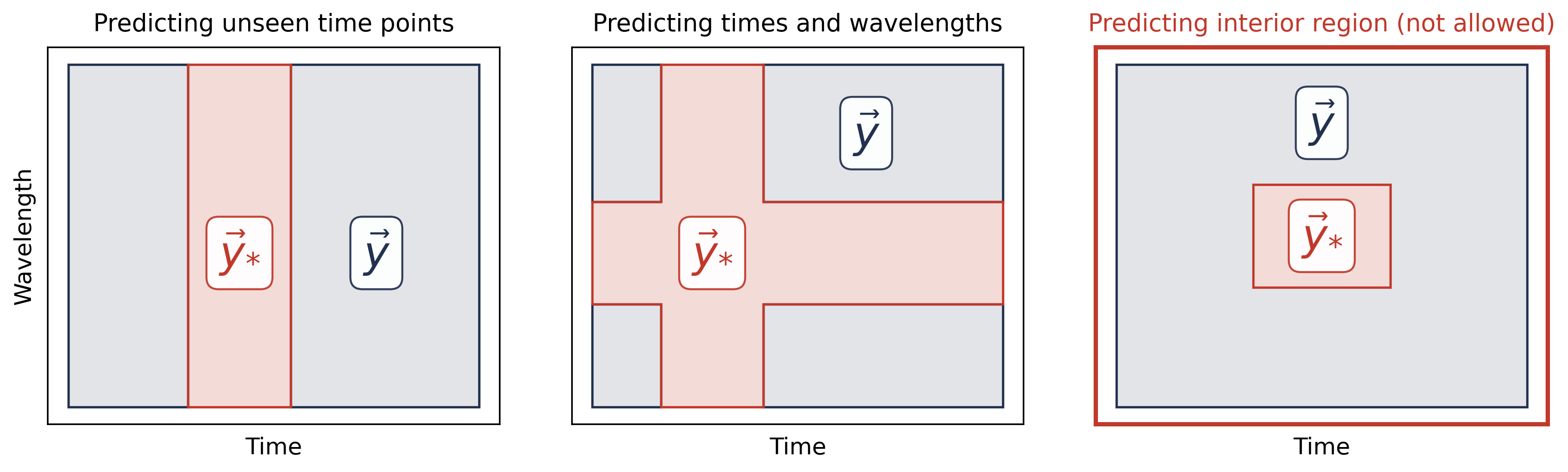}
    \caption{GP regression and prediction requires the training points $\mathbf{y}$ to lie on a complete input grid, which is satisfied in the left and middle plots, but not the rightmost plot. The prediction locations $\mathbf{y}_*$ are not required to lie on a grid, but we do require the combined training and prediction locations to lie on a grid, which all plots satisfy.}
    \label{fig:predicted_locations}
\end{figure*}

\begin{figure}
    \centering
    \includegraphics[width=0.8\columnwidth]{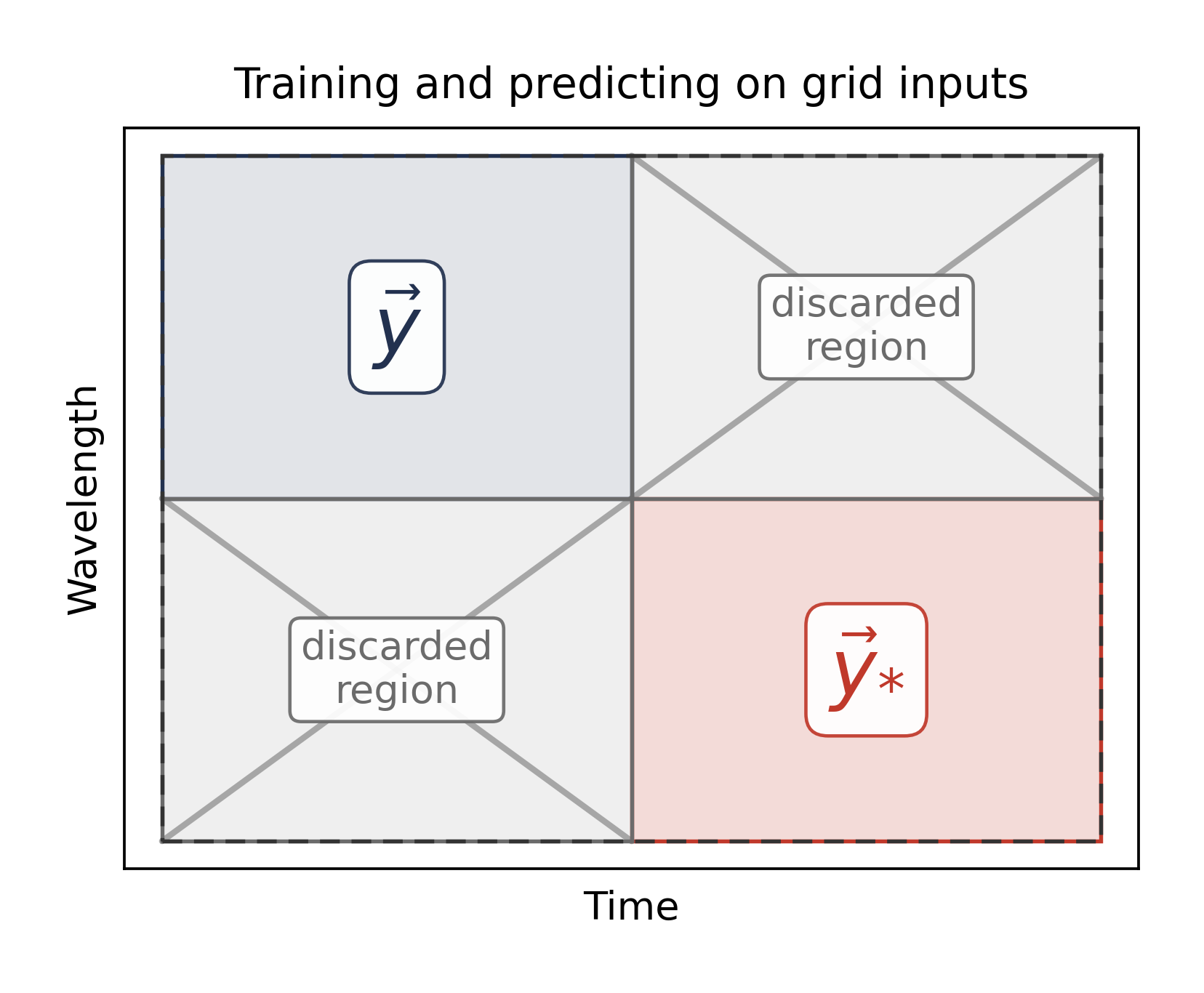}
    \caption{We can always train on a grid of locations and predict values on a different input grid, even though the union of these points doesn't form a grid directly. We do this by predicting at additional locations, so that the union of trained and predicted points forms a complete grid. We can then simply discard these additional predicted values.}
    \label{fig:predicted_locations2}
\end{figure}

The GP predictive covariance is more challenging to calculate, there's no obvious factorisation which allows us to efficiently decompose $\mathbf{K}_\text{pred}$ directly for all optimisations. However, we can exploit our existing optimisations provided that not only do the observed inputs lie on a grid, but also the union of the observed and predicted inputs lie on a grid. For example, we could predict values at unseen times or wavelengths, or even predict at both unseen times and unseen wavelengths, but the region we train on must always be a grid, and the combination of points we train and predict on must be a grid, as visualised in Figure~\ref{fig:predicted_locations}. We can also handle the case of both the training points and the predicted points being on a grid, useful for upsampling or downsampling a dataset, by utilising a trick shown in Figure~\ref{fig:predicted_locations2}.

For this method, we first concatenate our observed and predicted input locations together as $\mathbf{X}_\text{full} = ([\boldsymbol{\lambda}, \boldsymbol{\lambda}_*], [\mathbf{t}, \mathbf{t}_*])$ and construct the covariance matrix using any of the previously mentioned kernel functions, which will have the block structure
\begin{equation}
\mathbf{K}_\text{full} =
\begin{bmatrix}
\mathbf{K} & \mathbf{K}_*\\
\mathbf{K}_*^T & \mathbf{K}_{**}
\end{bmatrix}.
\end{equation}
The covariance matrix for the trained points $\mathbf{K}$, and the covariance matrix on the union of trained and predicted points $\mathbf{K}_\text{full}$, both lie on complete grids and therefore calculations with them can be optimised using any of the GP optimisations in this work. This means that we can perform cross-validation and sample from the GP predictive distribution efficiently, provided that we can perform these calculations using decompositions of the matrices $\mathbf{K}$ and $\mathbf{K}_\text{full}$.

Note that the inverse of $\mathbf{K}_\text{full}$ as given by the formula for the inverse of a 2x2 block matrix is
\begin{align}
\mathbf{K}_\text{full}^{-1} &=
\begin{bmatrix}
\mathbf{K}^{-1} + \mathbf{K}^{-1}\mathbf{K}_*\mathbf{K}_\mathrm{pred}^{-1}\mathbf{K}_*^T\mathbf{K}^{-1} & -\mathbf{K}^{-1}\mathbf{K}_*\mathbf{K}_\mathrm{pred}^{-1}\\
-\mathbf{K}_\mathrm{pred}^{-1}\mathbf{K}_*^T\mathbf{K}^{-1} & \mathbf{K}_\mathrm{pred}^{-1}
\end{bmatrix},
\label{eq:GP_predictive_cov_inv}
\end{align}
where the $\mathbf{D}$ block of $\mathbf{K}_\text{full}^{-1}$ matches the inverse of $\mathbf{K}_\mathrm{pred}$ as defined in Equation~\ref{eq:predictive_covariance_formula}. This allows us to multiply vectors by $\mathbf{K}_\text{pred}^{-1}$ by decomposing $\mathbf{K}_\text{full}$.

\subsection{Cross-Validation Calculation}
\label{eq:cross_val}

Suppose we want to perform cross-validation using our GP model, where we favour GP predictive models which maximise the probability of drawing the unseen data, i.e. maximise the log-probability:
\begin{align}
\log p(y_* | \boldsymbol{\upmu}_\text{pred}, \mathbf{K}_\text{pred}) = &-\frac{1}{2}\mathbf{r}_*^T\mathbf{K}_\text{pred}^{-1}\mathbf{r}_* - \frac{1}{2}\log|\mathbf{K}_\text{pred}| - \frac{N_*}{2}\log 2\pi
\end{align}
where the residuals in our predictions are calculated as $\mathbf{r}_* = \mathbf{y}_* - \boldsymbol{\upmu}_\text{pred}$, where we use our GP predictive mean from Equation~\ref{eq:predictive_mean_formula} to calculate predictions on unseen points using the training data $\mathbf{y}$.

We then zero-pad the unseen residuals and multiply by the inverse of $\mathbf{K}_\mathrm{full}$ in Equation~\ref{eq:GP_predictive_cov_inv}
\begin{align}
\begin{bmatrix}
\mathbf{0} & \mathbf{r}_*^T
\end{bmatrix}\mathbf{K}_\text{full}^{-1}\begin{bmatrix}
\mathbf{0}\\
\mathbf{r}_*
\end{bmatrix} &= \mathbf{r}_*^T\mathbf{K}_\text{pred}^{-1}\mathbf{r}_*,
\end{align}
We can express this zero-padding of the unseen residuals as a matrix multiplication using the non-square $(N + N_*, N_*)$ matrix
\begin{equation}
    \mathbf{P}_* = \begin{bmatrix}
\mathbf{0}\\
\mathbf{I}_{N_*}
\end{bmatrix}
\end{equation}
which allows us to define a (non-square) matrix inverse square root of $\mathbf{K}_\mathrm{pred}$ as $\mathbf{K}_\text{pred}^{-\frac{1}{2}} = \mathbf{K}_\text{full}^{-\frac{1}{2}}\mathbf{P}_*$, which satisfies $\mathbf{K}_\text{pred}^{-1} = (\mathbf{K}_\text{pred}^{-\frac{1}{2}})^T\mathbf{K}_\text{pred}^{-\frac{1}{2}}$.

For the determinant of the GP predictive covariance $|\mathbf{K}_\text{pred}|$, smaller values of which correspond to more confident predictions on the unseen data, we note our formula for the determinant of a block matrix from Equation~\ref{eq:block_determinant}, which for $\mathbf{K}_\text{full}$ yields:
\begin{align}
\log|\mathbf{K}_\text{full}| &= \log|\mathbf{K}| + \log|\mathbf{K}_{**} - \mathbf{K}_*^T\mathbf{K}^{-1}\mathbf{K}_*|\\
\log|\mathbf{K}_\text{full}| &= \log|\mathbf{K}| + \log|\mathbf{K}_\text{pred}|
\end{align}
So using any of the GP optimisations presented, we can calculate the log determinant $|\mathbf{K}|$ using the trained inputs $\mathbf{X}$, calculate the log determinant $|\mathbf{K}_\text{full}|$ by using all inputs $\mathbf{X}_\text{full}$, and then the difference gives the GP predictive covariance log determinant $\log|\mathbf{K}_\text{pred}|$.

Our cross-validation likelihood can thus be written as:
\begin{align}
\log p(\mathbf{y}_* | \boldsymbol{\upmu}_\text{pred}, \mathbf{K}_\text{pred}) = -&\frac{1}{2}\left[\begin{smallmatrix}
\mathbf{0} & 
\mathbf{r}_*^T
\end{smallmatrix}\right]\mathbf{K}_\text{full}^{-1}\left[\begin{smallmatrix}
\mathbf{0}\\ 
\mathbf{r}_*
\end{smallmatrix}\right] - \frac{1}{2}\log|\mathbf{K}_\text{full}|\nonumber\\
- &\frac{N_*}{2}\log 2\pi + \frac{1}{2}\log|\mathbf{K}|
\end{align}

\subsection{Sampling from the Predictive Covariance}
\label{eq:GP_sampling}

If we want to take samples from $\mathbf{K}_\mathrm{pred}$, we can use a corresponding non-square matrix square root for $\mathbf{K}_\mathrm{pred}$, equivalent to a sampling method known as Matheron's rule (e.g. \citealt{Wilson2020}). The matrix square root can be expressed as:
\begin{equation}
\mathbf{K}_\mathrm{pred}^{\frac{1}{2}} = 
\begin{bmatrix}
    -\mathbf{K}_*^T\mathbf{K}^{-1} & \mathbf{I}_{N_*}
\end{bmatrix}\mathbf{K}_\mathrm{full}^{\frac{1}{2}}
\label{eq:conditional_matrix_sqrt}
\end{equation}
which we can validate as a matrix square root by multiplying it out:
\begin{align}
    \mathbf{K}_\mathrm{pred}^{\frac{1}{2}} (\mathbf{K}_\mathrm{pred}^{\frac{1}{2}})^T &= \begin{bmatrix}
    -\mathbf{K}_*^T\mathbf{K}^{-1} & \mathbf{I}_{N_*}
\end{bmatrix}\begin{bmatrix}
\mathbf{K} & \mathbf{K}_*\\
\mathbf{K}_*^T & \mathbf{K}_{**}
\end{bmatrix} \begin{bmatrix}
    -\mathbf{K}^{-1}\mathbf{K}_*\\
    \mathbf{I}_{N_*}
\end{bmatrix}\\
&= \mathbf{K}_{**} - \mathbf{K}_{*}^T\mathbf{K}^{-1}\mathbf{K}_{*}
\end{align}
Therefore, we can draw samples from $\mathbf{K}_\mathrm{pred}$ by taking a random normal vector matching the shape of the combined training and prediction locations $(N_\lambda + N_{\lambda; *}, N_t + N_{t; *})$, and multiplying by $\mathbf{K}_\mathrm{pred}^{\frac{1}{2}}$ in Equation~\ref{eq:conditional_matrix_sqrt}. This procedure still produces valid draws from $\mathbf{K}_\mathrm{pred}$ despite the non-square matrix square root, as can be verified using the definition of the sample covariance:
\begin{align}
    \mathbb{E}[(\mathbf{r} - \bar{\mathbf{r}})(\mathbf{r} - \bar{\mathbf{r}})^T] &= \mathbf{K}_\mathrm{pred}^{\frac{1}{2}}\mathbb{E}[(\mathbf{z} - \bar{\mathbf{z}})(\mathbf{z} - \bar{\mathbf{z}})^T](\mathbf{K}_\mathrm{pred}^{\frac{1}{2}})^T\\
    &= \mathbf{K}_\mathrm{pred}^{\frac{1}{2}}\mathbf{I}_{N + N_*}(\mathbf{K}_\mathrm{pred}^{\frac{1}{2}})^T\\
    &= \mathbf{K}_\mathrm{pred}
\end{align}

\section{Extending to Higher Dimensions}
\label{app:higher_dim}

We can implement the \textsc{SeparableKernel}, \textsc{LuasKernel}, and \textsc{LuasLasrachKernel} for arbitrary numbers of dimensions $d$ with minimal modification. The \textsc{SeparableKernel} simply follows the algorithms in \citet{Saatchi2011}, decomposing each of the $d$ covariance matrices in the Kronecker product one-by-one. We simply extend this by allowing each matrix to be decomposed by any matrix square root, rather than specifically Cholesky decomposition. The data residuals are now a $d$-dim tensor, and can be transformed using algorithms in \citet{Saatchi2011}. The \textsc{LuasKernel} is extended similarly, we just perform the same simultaneous diagonalisation trick looping over all dimensions, applying eigendecompositions in each dimension. The transformation of the residuals then follows the same algorithm. For 3D datacubes, both methods yield a scaling of $\mathcal{O}(N_m^3 + N_\lambda^3 + N_t^3 + N_m N_\lambda N_t (N_m + N_\lambda + N_t))$ for some third dimension of size $N_m$. See Figures~\ref{fig:LuasKernelND_3D_m64_cost} and \ref{fig:LuasKernelND_3D_m64_A100_cost} for benchmarking examples on both CPU and GPU for the \textsc{LuasKernel} with $N_m = 64$.

We can also extend the \textsc{LuasLasrachKernel} to arbitrary dimensions with a modification to the \textsc{LuasKernel} approach. We perform simultaneous diagonalisation on all dimensions except for one of them, which still creates a block diagonal matrix, although potentially a very large one. For example, suppose we take a set of $N_m$ spectroscopic time-series of the same shape (for example $N_m$ different spectral orders), of shape $(N_m, N_\lambda, N_t)$, and we have some 3D kernel function consisting of two Kronecker products:
\begin{equation}
    \mathbf{K} = \bm{\Sigma}_m \otimes \bm{\Sigma}_\lambda \otimes \bm{\Sigma}_t + \mathbf{K}_m \otimes \mathbf{K}_\lambda \otimes \mathbf{K}_t
    \label{eq:3D_rakitsch}
\end{equation}
We decompose the matrix using the transformation:
\begin{align}
    \mathbf{K} &= \underbrace{[\bm{\Sigma}_m^{\frac{1}{2}} \mathbf{Q}_{K'_m} \otimes \bm{\Sigma}_\lambda^{\frac{1}{2}} \mathbf{Q}_{K'_\lambda} \otimes \mathbf{I}_{N_t}]}_{\mathbf{K}_\mathrm{transf}} \underbrace{[\mathbf{I}_{N_m} \otimes \mathbf{I}_{N_\lambda} \otimes \bm{\Sigma}_t + \bm{\Lambda}_{K'_m} \otimes \bm{\Lambda}_{K'_\lambda} \otimes K_t]}_{\tilde{\mathbf{K}}}\nonumber\\
    &\phantom{=} \underbrace{[\mathbf{Q}_{K'_m}^T (\bm{\Sigma}_m^{\frac{1}{2}})^T \otimes \mathbf{Q}_{K'_\lambda}^T (\bm{\Sigma}_\lambda^{\frac{1}{2}})^T \otimes \mathbf{I}_{N_t}]}_{\mathbf{K}_\mathrm{transf}^T}
    \label{eq:3D_luaslasrach}
\end{align}
The block diagonal transformed covariance matrix $\tilde{\mathbf{K}}$ is now similar to the result in 2D, but with the eigenvalues of all other dimensions stacked together:
\begin{align}
 \tilde{\mathbf{K}} &= \mathbf{I}_{N_m N_\lambda} \otimes \bm{\Sigma}_t + (\bm{\Lambda}_{K'_m} \otimes \bm{\Lambda}_{K'_\lambda}) \otimes \mathbf{K}_t
 \end{align}
The number of dense blocks to decompose in $\tilde{\mathbf{K}}$ will be equal to the product of the number of non-zero eigenvalues for each dimension we eigendecompose. Our example yields a scaling of $\mathcal{O}(N_m^3 + N_\lambda^3 + N_m N_\lambda N_t (N_m + N_\lambda + (J_{K_t} + J_{\Sigma_t})^2))$ when $\mathbf{K}_t$ and $\bm{\Sigma}_t$ are quasiseparable/diagonal. Since this scaling is very similar to the \textsc{LuasKernel}, but with a linear scaling with the number of time points, the \textsc{LuasLasrachKernel} optimisation only scales more efficiently if the other dimensions $N_m$ and $N_\lambda$ are relatively small, or if the kernels in those other dimensions are low-rank. Figures~\ref{fig:LuasLasrachKernel_3D_m64_cost} and \ref{fig:LuasLasrachKernel_3D_m64_A100_cost} benchmark the \textsc{LuasLasrachKernel} on 3D datacubes for $N_m = 64$ on both CPU and GPU, which gives comparable performance to the \textsc{LuasKernel} for the same dataset shapes, but which does perform faster for higher numbers of time points.

\section{Numerical Stability}
\label{app:stability}

The numerical stability of these algorithms was very high from our testing, although we still recommend double precision for any of these optimisations. We compared the calculations from each of our algorithms benchmarked in Appendix~\ref{app:benchmarking} to Cholesky decomposition of the full constructed covariance matrix for $N_\lambda = 16$ and $N_t = 128$. We computed both the log-likelihood and the gradients of the log-likelihood with respect to the hyperparameters (assuming zero mean function). We kept the kernel height scales fixed to unity, and randomly draw white noise amplitudes in wavelength with unit mean and a standard deviation of 0.1. We found the relative difference in the log-likelihood values were $<10\times$ machine precision ($\epsilon_\mathrm{double}\sim 2.2 * 10^{-16}$) compared to building the full covariance matrix and decomposing using Cholesky decomposition. The gradients automatically calculated by \textsc{JAX} autodiff were also typically very accurate, with relative errors on the order of $10^{-12}$. The only case where we needed to specify a more stable way for JAX to perform the gradient calculation was when taking the gradient with respect to eigendecomposition, because this gradient calculation depends on a matrix of the form
\begin{equation}
    \mathbf{F}_{ij} = \frac{1}{\lambda_i - \lambda_j}, i \neq j
\end{equation}
which becomes unstable if any eigenvalues are close or approximately equal \citep{lorentz_broadening}. Such a situation occurs frequently, so we use a solution described in \citet{lorentz_broadening} and approximate the gradient calculation by approximating $\mathbf{F}$ as:
\begin{equation}
    \mathbf{F}_{ij} \approx \frac{\lambda_i - \lambda_j}{(\lambda_i - \lambda_j)^2 + \epsilon}, i \neq j
\end{equation}
where $\epsilon$ is a small constant, which we set to $\epsilon = 10^{-16}$ during our testing. Note that $\epsilon$ will be negligible except in the scenario when $\lambda_i - \lambda_j \approx 0$, where it is needed for stability. This approximation only slightly affects the gradient calculation, with the largest relative errors we recovered on the order of $10^{-8}$, while the log-likelihood value is completely unaffected and remains exact. Note that many applications of gradients, such as gradient-based optimisers, and MCMC samplers like Hamiltonian Monte Carlo and No U-Turn Sampling, do not require exact gradients in order for inference to remain exact \citep{Hoffman2011}. 


\section{Benchmarking}
\label{app:benchmarking}

\begin{figure*}
\centering

\begin{tikzpicture}[
  bubA/.style={ellipse, draw, thick, align=center, font=\small,
               inner sep=1pt, minimum width=28mm, minimum height=20mm},
  bubB/.style={bubA, minimum width=32mm, minimum height=18mm},
  bubC/.style={bubA, minimum width=26mm, minimum height=16mm},
  lbl/.style={rectangle, draw, thick, rounded corners=2pt, align=center,
              font=\small\bfseries, inner sep=3pt,
              minimum width=32mm, minimum height=14mm},
  hdr/.style={lbl, font=\small\bfseries, fill=black!10},
  divider/.style={red, thick, dash pattern=on 3pt off 3pt},
  divlbl/.style={font=\small\itshape, red!70!black, inner sep=3pt},
  ar/.style={-{Stealth[length=3mm,width=2mm]}, thick},
  skip/.style={ar, rounded corners=6pt},
  edgelbl/.style={font=\scriptsize, inner sep=1.5pt, fill=white}
]

\pgfmathsetmacro{\RCC}{\RE}
\pgfmathsetmacro{\REE}{\RE + (\RE - \RD)}

\node[hdr] (H1) at (\CA,\RH) {No Low-Rank Kernels};
\node[hdr] (H2) at (\CB,\RH) {+ Low-Rank Time Kernels\\$\sum_i^{N_\alpha} \alpha_i(\lambda)\alpha_i(\lambda')k_{t; i}(t, t')$};
\node[hdr] (H3) at (\CC,\RH) {+ Low-Rank\\Wavelength Kernels\\$\sum_j^{N_\beta} k_{\lambda; j}(\lambda, \lambda')\beta_j(t)\beta_j(t')$};

\node[hdr] (L1) at (\CL,\RA) {Separable white noise\\$\sigma(\lambda)\delta_{\lambda\lambda'}\sigma(t)\delta_{tt'}$};
\node[hdr] (L2) at (\CL,\RB) {Quasisep. Measurement Cov.\\$s_{\lambda; quasi.}(\lambda, \lambda')\sigma(t)\delta_{tt'}$};
\node[hdr] (L3) at (\CL,\RC) {Arbitrary Measurement Cov.\\$s_{\lambda; arb.}(\lambda, \lambda')\sigma(t)\delta_{tt'}$};
\node[hdr] (L4) at (\CL,\RD) {Arbitrary Measurement Cov.\\+ 2D Kernel Term\\w/ Quasisep. Time Kernel\\$s_{\lambda; arb.}(\lambda, \lambda')\sigma(t)\delta_{tt'}$ +\\$k_{\lambda; arb.}(\lambda, \lambda')k_{t; quasi}(t, t')$};
\node[hdr] (L6) at (\CL,\RCC) {Arbitrary Measurement Cov.\\$s_{\lambda; arb.}(\lambda, \lambda')\sigma(t)\delta_{tt'}$};
\node[hdr] (L5) at (\CL,\REE) {Arbitrary Measurement Cov.\\+ 2D Kernel Term\\w/ Arbitrary Kernels\\$s_{\lambda; arb.}(\lambda, \lambda')\sigma(t)\delta_{tt'}$ +\\$k_{\lambda; arb.}(\lambda, \lambda')k_{t; arb.}(t, t')$};

\node[lbl] (A1) at (\CA,\RA) {Diagonal Covariance Matrix\\\\$\mathcal{O}(N_\lambda N_t)$};
\node[lbl] (A2) at (\CB,\RA) {\textsc{Multi}-\\\textsc{LowRankKernel}\\Ortho. Components:\\$+\mathcal{O}(N_\alpha N_t \bar{J}_{K_t}^2 + N_\alpha N_\lambda N_t)$\\\\Non-Ortho. Components:\\$+\mathcal{O}(N_\alpha^3 N_t \bar{J}_{K_t}^2 + N_\alpha N_\lambda N_t)$};
\node[lbl] (A3) at (\CC,\RA) {\textsc{DoubleMulti}-\\\textsc{LowRankKernel}\\\textmd{Fig.}~\ref{fig:DoubleMultiLowRankKernel_Nalpha1_Nbeta1_cost}, \ref{fig:DoubleMultiLowRankKernel_Nalpha1_Nbeta4_cost}, \ref{fig:DoubleMultiLowRankKernel_Nalpha4_Nbeta1_cost}, \ref{fig:DoubleMultiLowRankKernel_Nalpha4_Nbeta4_cost}\\\\$+\mathcal{O}(N_\alpha^3 N_t (\bar{J}_{K_t} + 3N_\beta)^2$\\$\phantom{\mathcal{O}(}+N_\beta^3 N_\lambda (\bar{J}_{K_\lambda} + 3N_\alpha)^2$\\$ + N_\beta N_\lambda N_t)$};

\node[lbl] (B1) at (\CA,\RB) {\textsc{SeparableKernel}\\\textmd{Fig.}~\ref{fig:SingleKronTermKernel_quasireadout_cost}\\\\$\mathcal{O}(N_\lambda J_{\Sigma_\lambda}^2 + N_\lambda N_t \bar{J}_{\Sigma_\lambda})$};
\node[lbl] (B2) at (\CB,\RB) {\textsc{Multi}-\\\textsc{LowRankKernel}\\Ortho. Components:\\\textmd{Fig.}~\ref{fig:MultiLowRankKernel_Nalpha16_ortho_cost}\\$+\mathcal{O}(N_\alpha N_t \bar{J}_{K_t}^2 + N_\alpha N_\lambda N_t)$\\\\Non-Ortho. Components:\\\textmd{Fig.}~\ref{fig:MultiLowRankKernel_Nalpha1_cost}, \ref{fig:MultiLowRankKernel_Nalpha4_cost}, \ref{fig:MultiLowRankKernel_Nalpha8_cost}, \ref{fig:MultiLowRankKernel_Nalpha16_cost}\\$+\mathcal{O}(N_\alpha^3 N_t \bar{J}_{K_t}^2 + N_\alpha N_\lambda N_t)$};

\node[lbl] (C1) at (\CA,\RC) {\textsc{SeparableKernel}\\\textmd{Fig.}~\ref{fig:SingleKronTermKernel_densereadout_cost}\\\\$\mathcal{O}(N_\lambda^3 + N_\lambda^2N_t)$};
\node[lbl] (C2) at (\CB,\RC) {\textsc{Multi}-\\\textsc{LowRankKernel}\\Ortho. Components:\\$+\mathcal{O}(N_\alpha N_t \bar{J}_{K_t}^2 + N_\alpha N_\lambda N_t)$\\\\Non-Ortho. Components:\\$+\mathcal{O}(N_\alpha^3 N_t \bar{J}_{K_t}^2 + N_\alpha N_\lambda N_t)$};

\node[lbl] (D1) at (\CA,\RD) {\textsc{LuasLasrachKernel}\\\textmd{Fig.}~\ref{fig:LuasLasrachKernel_cost}, \ref{fig:LuasLasrachKernel_A100_cost}\\\\$\mathcal{O}(N_\lambda^3 + N_\lambda^2 N_t + N_\lambda N_tJ_{K_t}^2)$};
\node[lbl] (D3) at (\CC,\RD) {\textsc{LuasPlusMulti}-\\\textsc{LowRankKernel}\\\textmd{Fig.}~\ref{fig:LuasPlusMultiLowRankKernel_Nbeta1_cost}, \ref{fig:LuasPlusMultiLowRankKernel_Nbeta4_cost}\\\\$+\mathcal{O}(N_\beta^3 N_\lambda^3 + N_\lambda N_t (J_{K_t} + 2N_\beta)^2)$};

\node[lbl] (F1) at (\CA,\RCC) {\textsc{SeparableKernel}\\\textmd{Fig.}~\ref{fig:SingleKronTermKernel_densereadout_cost}\\\\$\mathcal{O}(N_\lambda^3 + N_\lambda^2N_t)$};
\node[lbl] (F2) at (\CB,\RCC) {\textsc{Multi}-\\\textsc{LowRankKernel}\\Ortho. Components:\\\textmd{Fig.}~\ref{fig:MultiLowRankKernel_Nalpha1_poweredexp_cost}, \ref{fig:MultiLowRankKernel_Nalpha1_poweredexp_densereadout_A100_cost}\\$+\mathcal{O}(N_\alpha N_t^3 + N_\alpha N_\lambda N_t)$\\\\Non-Ortho. Components:\\$+\mathcal{O}(N_\alpha^3 N_t^3 + N_\alpha N_\lambda N_t)$};

\node[lbl] (E1) at (\CA,\REE) {\textsc{LuasKernel}\\\textmd{Fig.}~\ref{fig:LuasKernelND_cost}, ~\ref{fig:LuasKernelND_A100_cost}\\\\$\mathcal{O}(N_\lambda^3 + N_t^3 + N_\lambda^2 N_t + N_\lambda N_t^2)$};
\node[lbl] (E2) at (\CB,\REE) {\textsc{LuasPlusMulti}-\\\textsc{LowRankKernel}\\\textmd{Fig.}~\ref{fig:LuasPlusMultiLowRankKernel_Nalpha1_cost}, \ref{fig:LuasPlusMultiLowRankKernel_Nalpha1_A100_cost}\\\\$+\mathcal{O}(N_\alpha^3 N_t^3 + N_\alpha^2 N_\lambda N_t)$};
\node[lbl] (E3) at (\CC,\REE) {\textsc{Extended}-\\\textsc{LuasKernel}\\\textmd{Fig.}~\ref{fig:ExtendedLuasKernel_Nalpha1_Nbeta1_cost}, \ref{fig:ExtendedLuasKernel_Nalpha1_Nbeta1_A100_cost}, 
\ref{fig:ExtendedLuasKernel_Nalpha1_Nbeta4_A100_cost}, \ref{fig:ExtendedLuasKernel_Nalpha4_Nbeta1_A100_cost}\\\\$+\mathcal{O}(N_\beta^3 N_\lambda^3 + N_\alpha N_\beta^2 N_\lambda^2 N_t)$};

\draw[ar] (A1) -- (A2);
\draw[ar] (A2) -- (A3);
\draw[ar] (B1) -- (B2);
\draw[ar] (C1) -- (C2);
\draw[ar] (F1) -- (F2);
\draw[ar] (E1) -- (E2);
\draw[ar] (E2) -- (E3);
\draw[ar] (D1) -- (D3);       

\coordinate (WL) at ($(current bounding box.west)+(-0.35,0)$);
\coordinate (ER) at ($(current bounding box.east)+(0.35,0)$);

\pgfmathsetmacro{\yI}{(\RB*0.45+\RC*0.55}
\pgfmathsetmacro{\yII}{(\RD+\RCC)/2}
\coordinate (YI)  at (0,\yI);
\coordinate (YII) at (0,\yII);

\draw[divider] (WL |- YI) -- (ER |- YI)
   node[divlbl, above right, pos=0] {linear in $N_\lambda$; linear in $N_t$ - quasiseparable kernels in wavelength and time}
   node[divlbl, below right, pos=0] {cubic in $N_\lambda$; linear in $N_t$ - quasiseparable kernels in time};

\draw[divider] (WL |- YII) -- (ER |- YII)
   node[divlbl, above right, pos=0] {cubic in $N_\lambda$; linear in $N_t$ - quasiseparable kernels in time}
   node[divlbl, below right, pos=0] {cubic in $N_\lambda$; cubic in $N_t$ - no restrictions on any kernels};

\end{tikzpicture}
\caption{Summary of the runtime scaling for different GP optimisations introduced in this work. The particular kernel optimised is the sum of the kernel term of that row (e.g. arbitrary measurement covariance), plus some number of low-rank terms based on the column. The total runtime scaling is found by adding together all terms which have arrows pointing toward it, where the scaling for the \textsc{DoubleMultiLowRankKernel} assumes arbitrary components for the \textsc{MultiLowRankKernel}. Note that terms in the right-most column either have both low-rank time kernels and low-rank wavelength kernels (the \textsc{DoubleMultiLowRankKernel} and \textsc{ExtendedLuasKernel}), or just low-rank wavelength kernels added to it (the \textsc{LuasPlusMultiLowRankKernel}), depending on how many arrows connect toward it. Different benchmarking plots are linked to for most optimisations.}
\label{fig:scaling_flowchart}
\end{figure*}
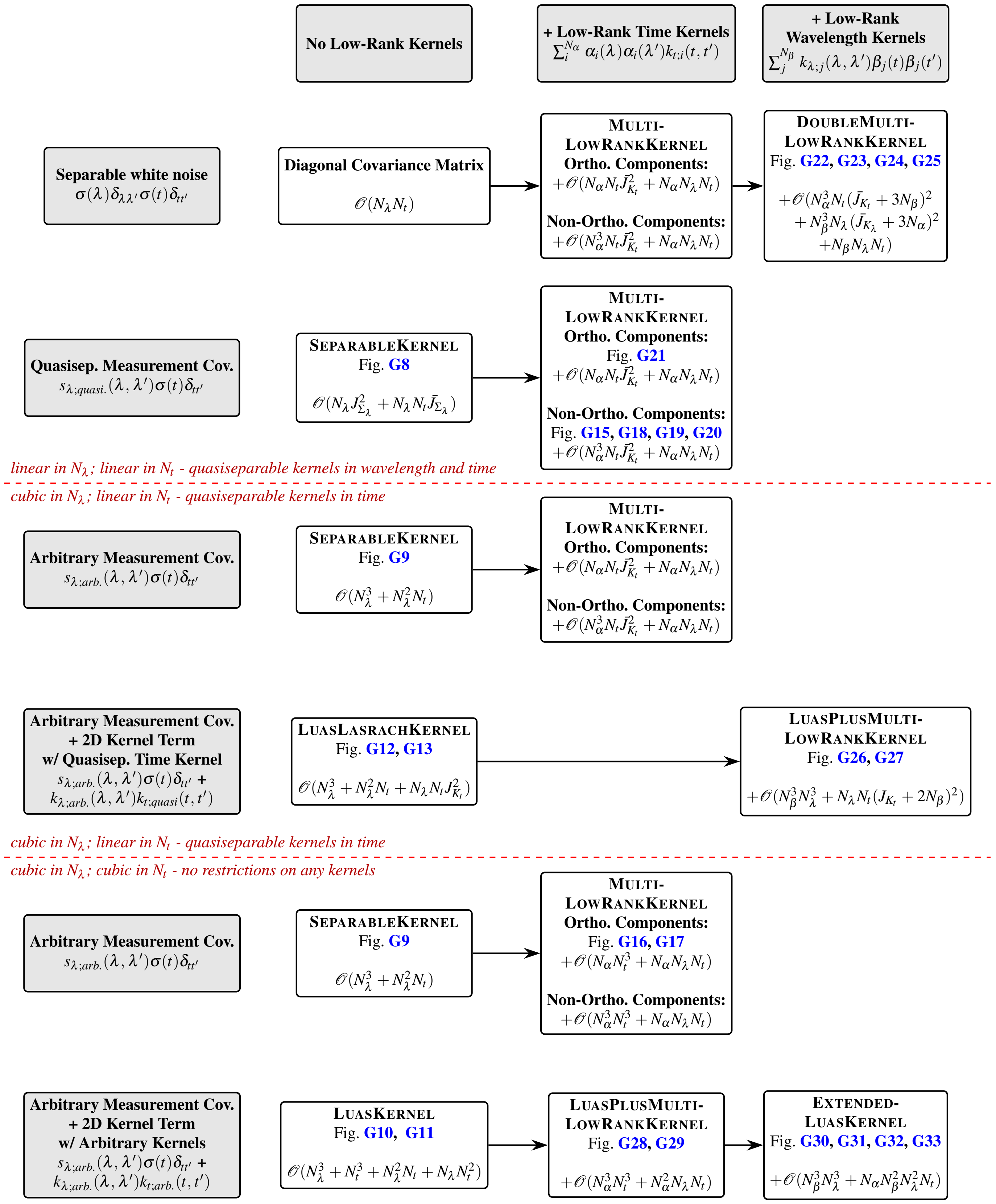

In this section, we've provided log-likelihood runtimes of the \textsc{luas} implementations of these new GP optimisations at the time of writing. Calculations were performed on both an M4 MacBook Air (10 cores), as well as on an NVIDIA A100-SXM4-40GB GPU. All calculations were performed at double precision, as calculations may not be numerically stable at single precision. Therefore, the use of a GPU with a high rate of floating point operations per second at double precision (FP64 FLOPS) is crucial, as it varies dramatically between GPUs, and is often not supported at all. We calculate these runtimes for a range of 2D dataset shapes, reporting both the log-likelihood runtime, as well as the runtime per number of datapoints (which should be constant for fully linear scaling). Calculations were performed on the CPU using \textsc{jax} version 0.10.1, \textsc{SciPy} version 1.17.1 and a recent commit of \textsc{tinygp} (Commit f3466f0; see \href{https://github.com/dfm/tinygp/commit/f3466f05f007d907a439282d4148564d65f778b9}{pull request \#269} for details). The GPU environment had \textsc{jax} version 0.9.0 for CUDA 12, \textsc{SciPy} version 1.17.0 and the same commit of \textsc{tinygp} (which features GPU optimisation for quasiseparable kernels). We exclude the one-time compile cost of JIT-compilation, typically on the order of 0.1 - 10 seconds on the MacBook, and use the method \textsc{block\_until\_ready} to avoid issues with asynchronous dispatch in JAX.

Runtimes were taken as the median over at least 10 runs. If 10 runs completed in less than 10 seconds, the number of runs was multiplied by 10 and run again, with this process terminating after greater than 10 seconds were exceeded, or at 10,000 runs. For the MacBook, background processes would occasionally cause significantly longer runtimes over durations of minutes, so a best of three separate runs was used to avoid these outliers. The GPU runtimes were significantly more consistent, so only a single run was performed for each.

\begin{figure*}[p]
    \begin{minipage}[b]{\columnwidth}
        \includegraphics[width=\linewidth]{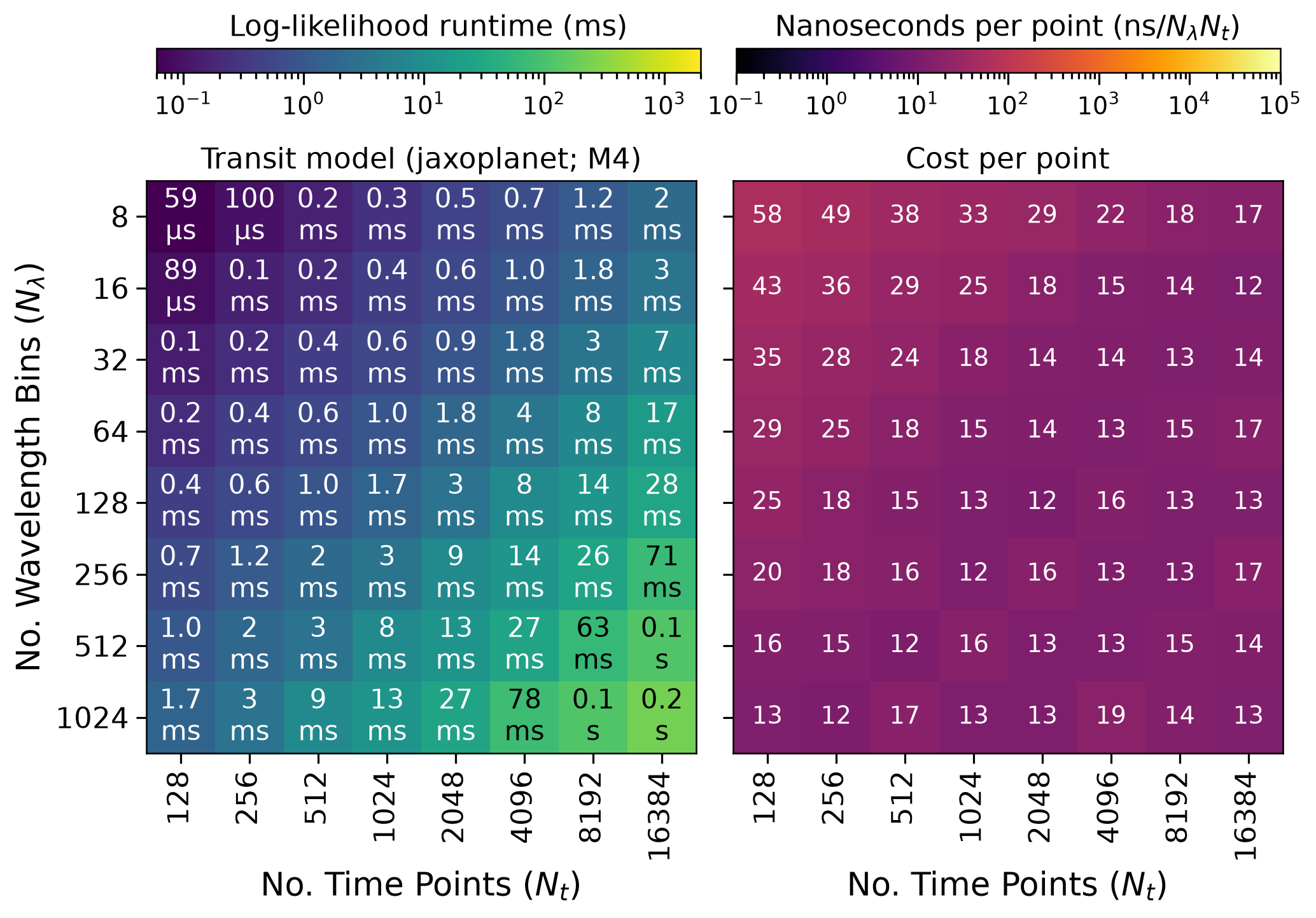}
    \end{minipage}\hfill%
    \begin{minipage}[b]{\columnwidth}
        \includegraphics[width=\linewidth]{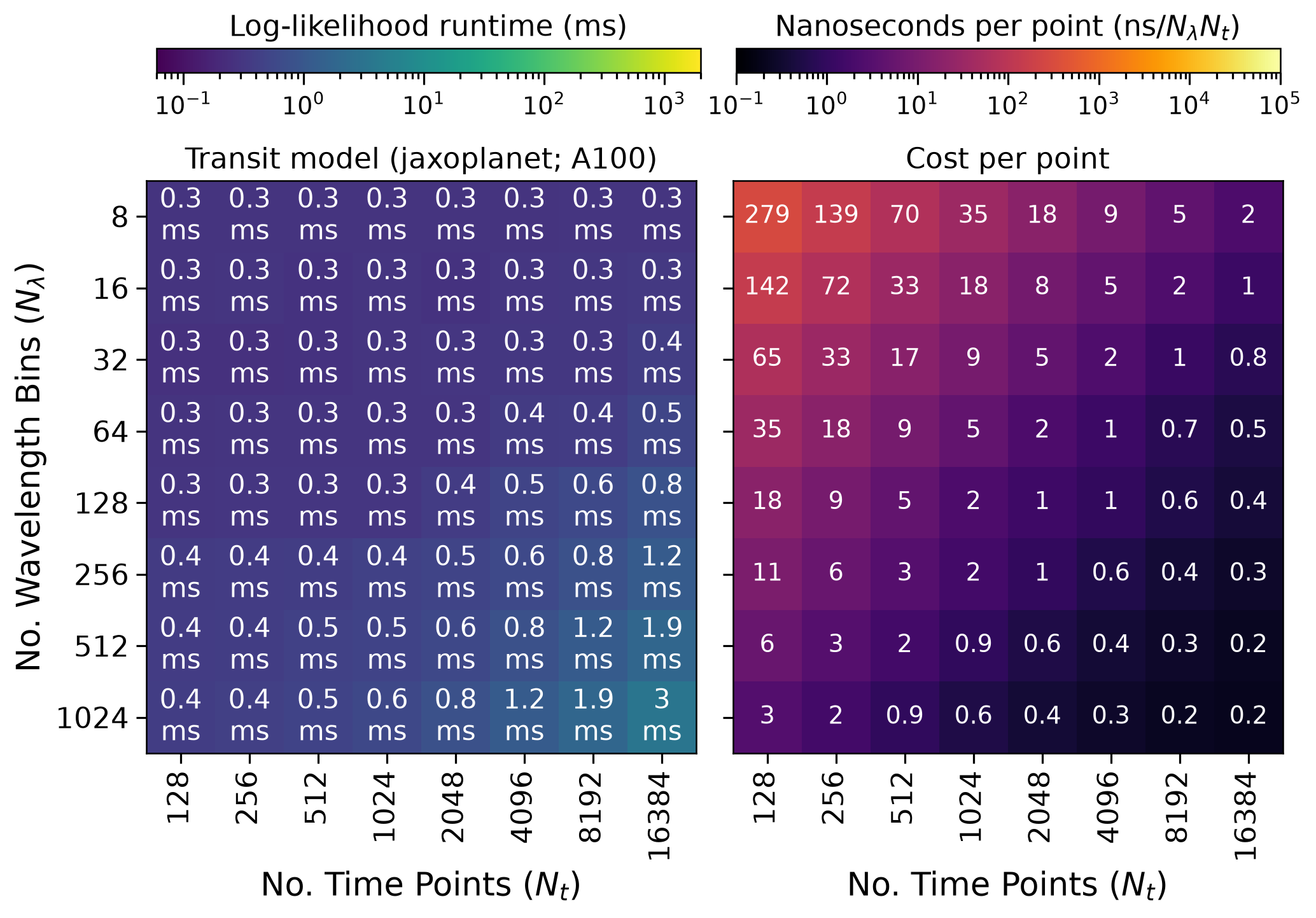}
    \end{minipage}

    \begin{minipage}[t]{\columnwidth}
        \caption{Transit light curve runtime cost using \textsc{jaxoplanet} on an M4 MacBook Air, which approaches approximately 10-20 ns per datapoint for larger datasets, more expensive than many of the GP optimisations presented in this work.}
    \label{fig:jaxoplanet_transit_cost}
    \end{minipage}\hfill%
    \begin{minipage}[t]{\columnwidth}
        \caption{Compared to Figure~\ref{fig:jaxoplanet_transit_cost}, transit modelling with \textsc{jaxoplanet} can scale significantly better for larger datasets on the A100 GPU, at which point it outperforms any of the GP optimisations in this work, even when the GP calculations are also run on a GPU.}
    \label{fig:jaxoplanet_transit_A100_cost}
    \end{minipage}

    \vspace{2\baselineskip}

    \begin{minipage}[b]{\columnwidth}
        \includegraphics[width=\linewidth]{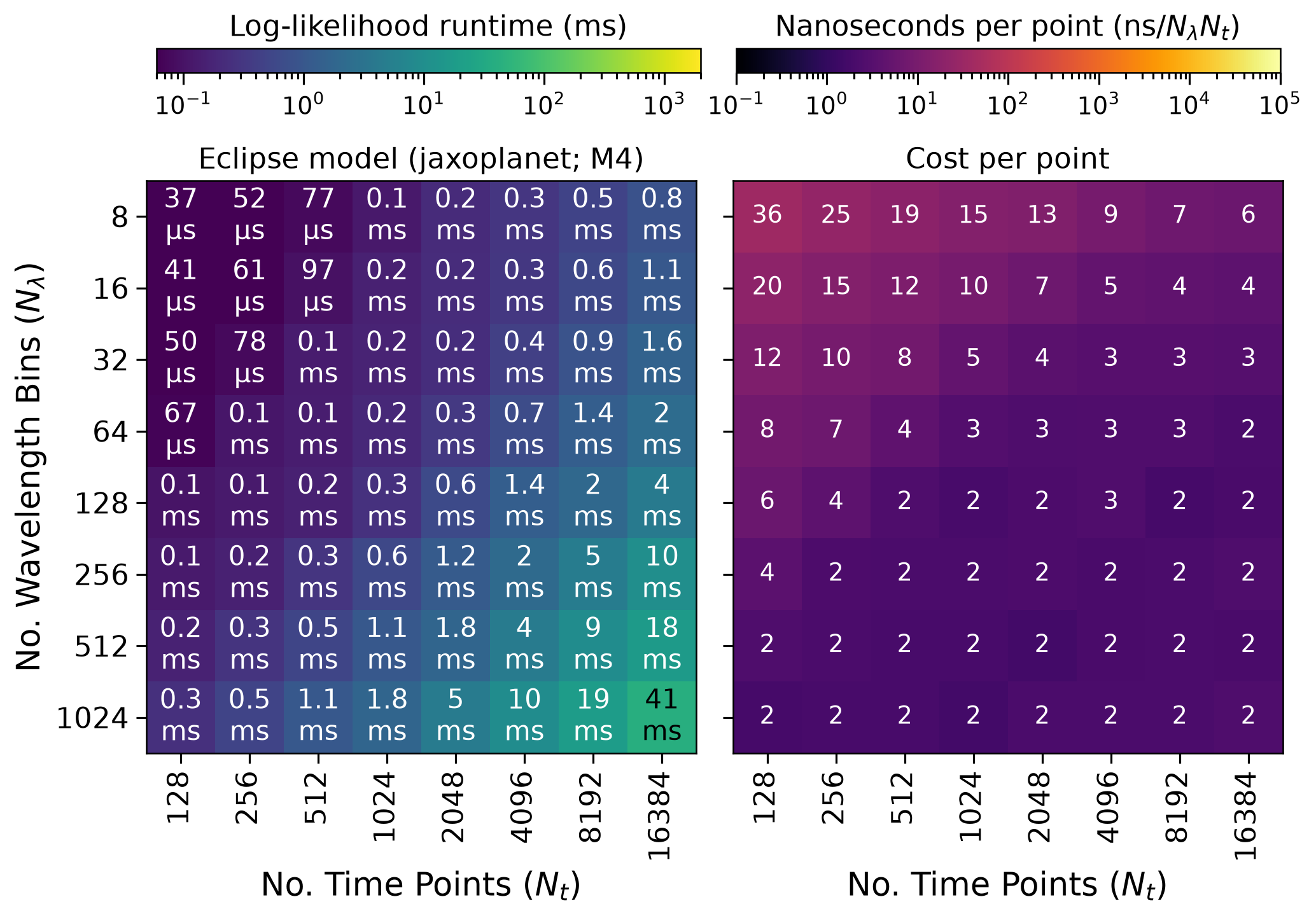}
    \end{minipage}\hfill%
    \begin{minipage}[b]{\columnwidth}
        \includegraphics[width=\linewidth]{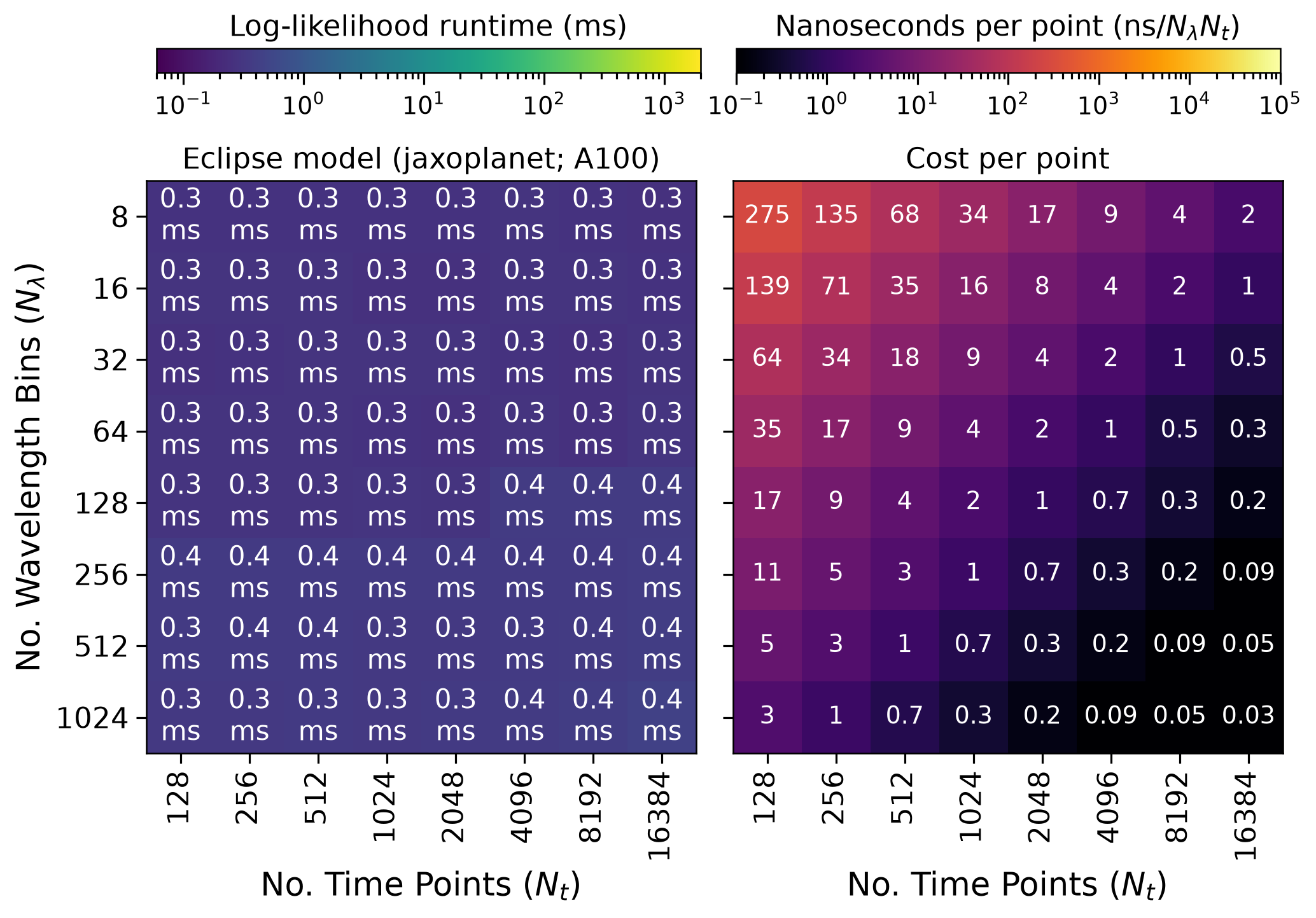}
    \end{minipage}

    \begin{minipage}[t]{\columnwidth}
        \caption{Eclipse light curve modelling can be performed significantly faster by generating a single transit light curve (also with \textsc{jaxoplanet}), and scaling it by $f_p/f_*$ for each light curve. The runtime cost approaches approximately 2 nanoseconds per datapoint for larger datasets on the MacBook, which has a similar runtime cost to some of the fastest GP optimisations implemented for this work.}
    \label{fig:jaxoplanet_eclipse_cost}
    \end{minipage}\hfill%
    \begin{minipage}[t]{\columnwidth}
        \caption{Compared to Figure~\ref{fig:jaxoplanet_eclipse_cost}, eclipse modelling with this optimised approach is trivial on the GPU and is dominated by a fixed bottleneck for these dataset sizes.}
    \label{fig:jaxoplanet_eclipse_A100_cost}
    \end{minipage}
\end{figure*}

These calculations were performed assuming zero mean, but for many applications there is a mean function we also calculate for each likelihood calculation, such as building a model of transit light curves. For many of these new GP optimisations, transit light curve modelling would be the bottleneck if running on CPU, so the runtimes for transit light curve generation using \textsc{jaxoplanet} \citep{jaxoplanet} with quadratic limb darkening are included in Figure~\ref{fig:jaxoplanet_transit_cost} for reference. We also tested the popular transit modelling package \textsc{batman} \citep{batman}, which yielded similar runtimes, typically within a factor of two of these values. Eclipse light curve modelling can be performed significantly faster in either case, as spectroscopic eclipse light curves are typically assumed to share the same planetary parameters (such as radius ratio $R_p/R_*$) without any limb darkening, which means that they all have the same shape but scaled by the planet-to-star flux ratio $f_p/f_*$. This means that we can simply generate a single transit light curve and scale it for each light curve, resulting in significant computational savings if the number of wavelengths $N_\lambda$ is high, as shown in Figure~\ref{fig:jaxoplanet_eclipse_cost} using \textsc{jaxoplanet}. Since both problems are easily parallelised, these calculations can be performed dramatically faster on GPU for large datasets, as shown in Figures~\ref{fig:jaxoplanet_transit_A100_cost} and ~\ref{fig:jaxoplanet_eclipse_A100_cost}. The cost of a general covariance matrix decomposition using Cholesky factorisation is shown in Figures~\ref{fig:GeneralKernel} and ~\ref{fig:GeneralKernel_A100} on CPU and GPU respectively, but quickly becomes intractable for larger datasets. These are compared to a wide range of new GP optimisations we introduce, on both CPU and GPU, which are arranged and linked to in Figure~\ref{fig:scaling_flowchart}, which summarises each methods runtime scaling. 

\subsection{Gradient Calculations}
\label{sec:gradients}

\begin{figure*}[p]
    \begin{minipage}[b]{\columnwidth}
        \includegraphics[width=\linewidth]{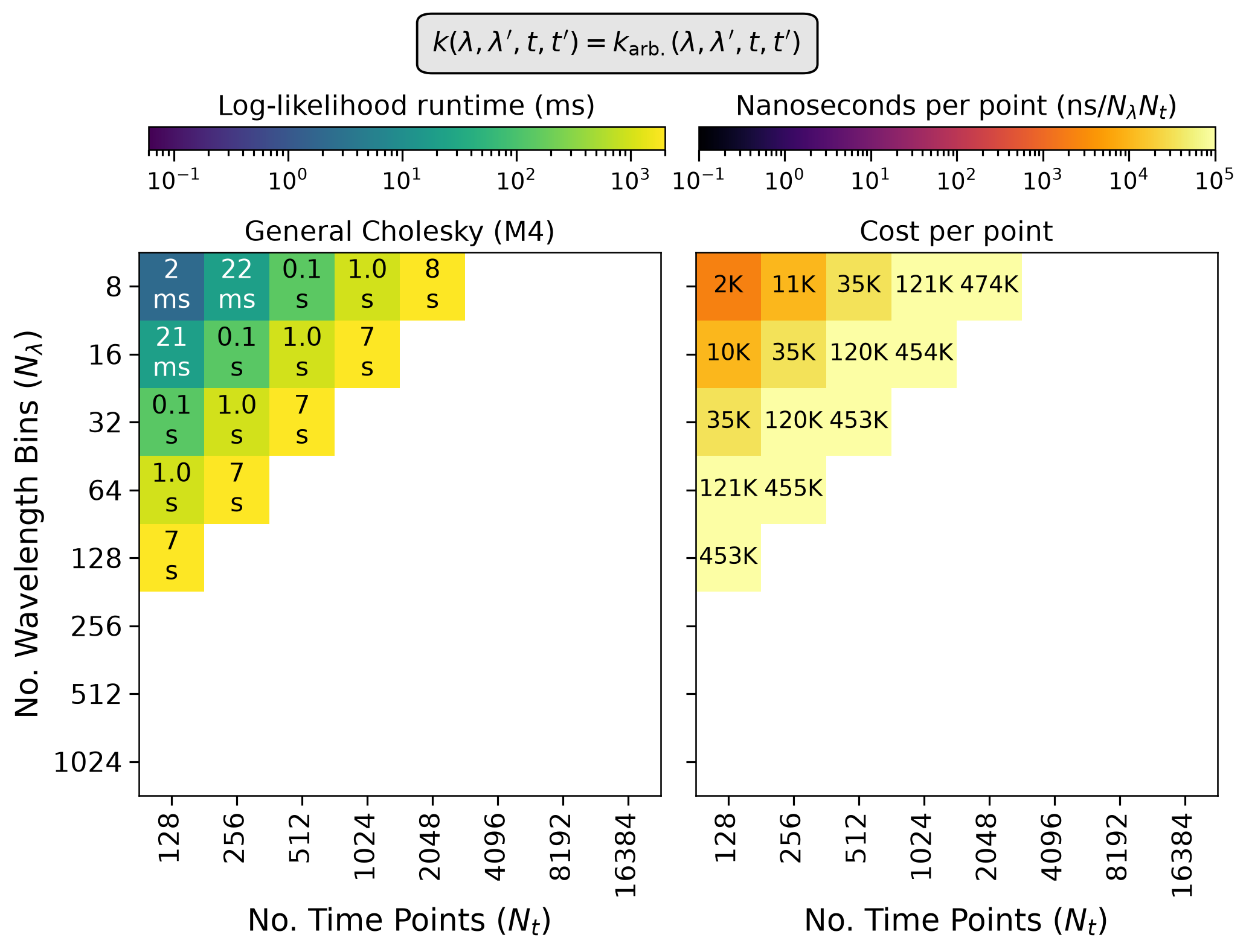}
    \end{minipage}\hfill%
    \begin{minipage}[b]{\columnwidth}
        \includegraphics[width=\linewidth]{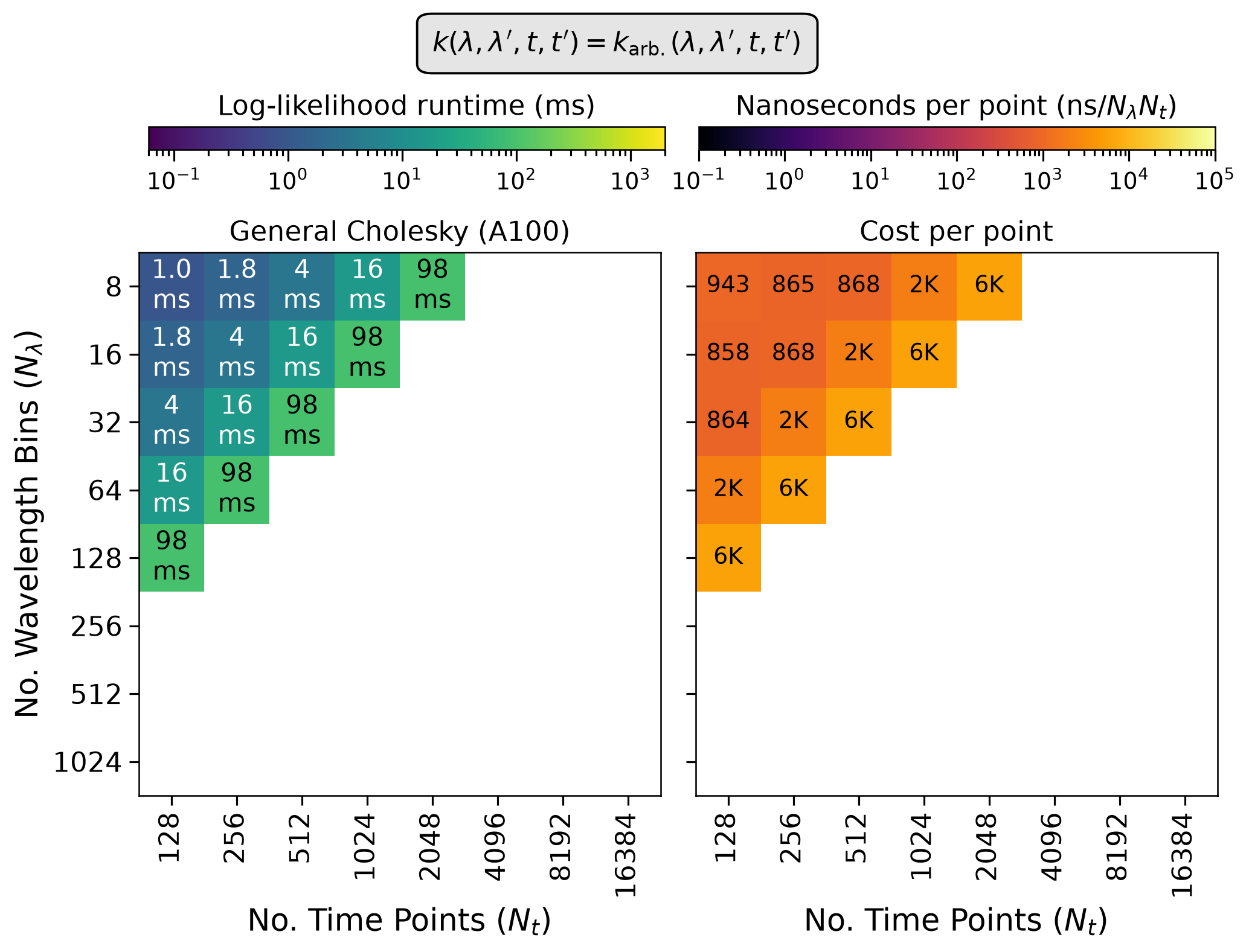}
    \end{minipage}

    \begin{minipage}[t]{\columnwidth}
        \caption{Decomposition of any general covariance matrix using Cholesky decomposition scales as $\mathcal{O}(N_\lambda^3 N_t^3)$ in runtime, which becomes prohibitively expensive even for relatively small datasets, particularly on CPU.}
    \label{fig:GeneralKernel}
    \end{minipage}\hfill%
    \begin{minipage}[t]{\columnwidth}
        \caption{Compared to Figure~\ref{fig:GeneralKernel}, running on GPU allows us to scale to slightly larger datasets for any general covariance matrix on 2D data, but due to the expensive memory scaling of $\mathcal{O}(N_\lambda^2N_t^2)$ required to store the covariance matrix, we were unable to decompose matrices with a total size larger than $(16384, 16384)$.}
    \label{fig:GeneralKernel_A100}
    \end{minipage}

    \vspace{2\baselineskip}

    \begin{minipage}[b]{\columnwidth}
        \includegraphics[width=\linewidth]{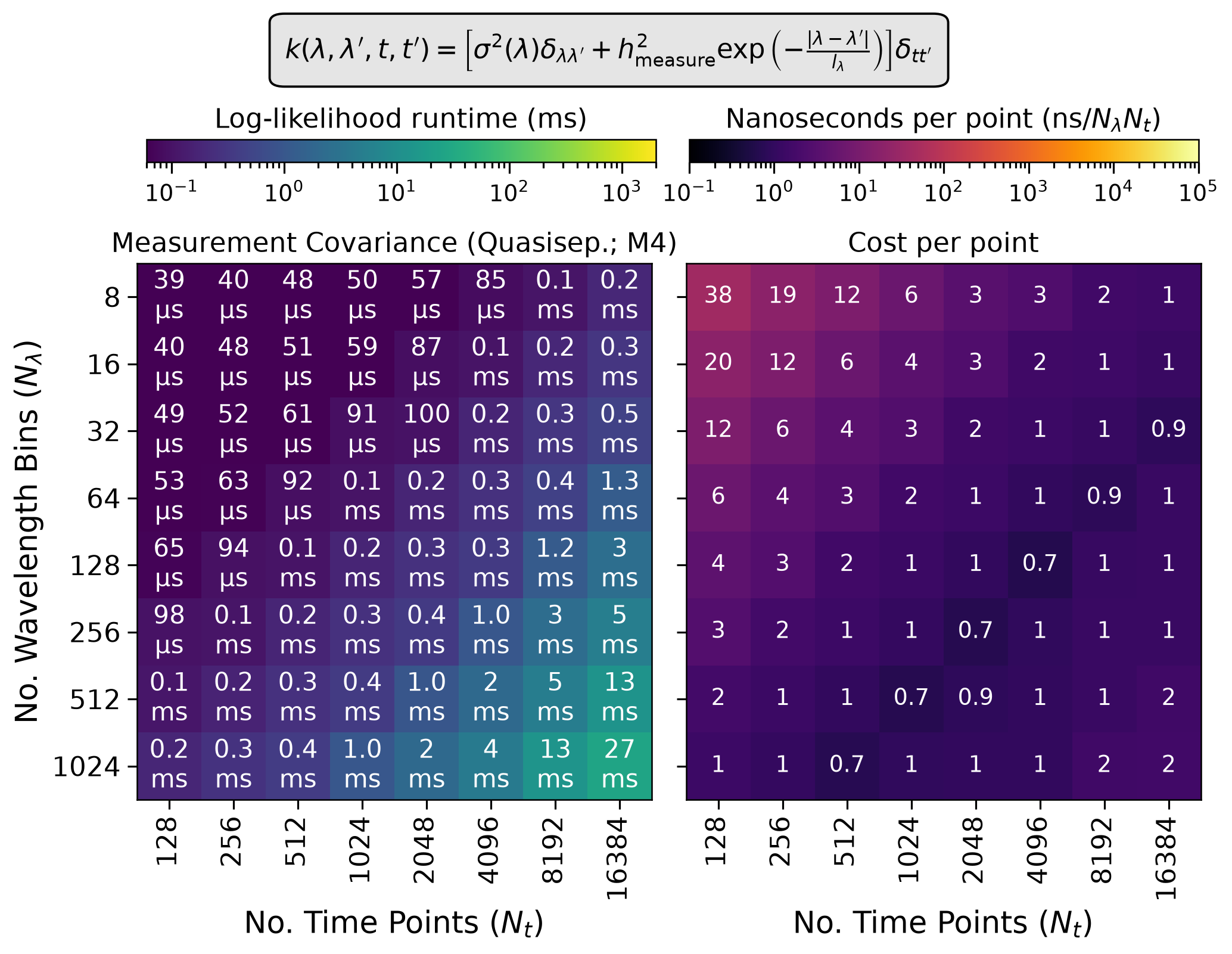}
    \end{minipage}\hfill%
    \begin{minipage}[b]{\columnwidth}
        \includegraphics[width=\linewidth]{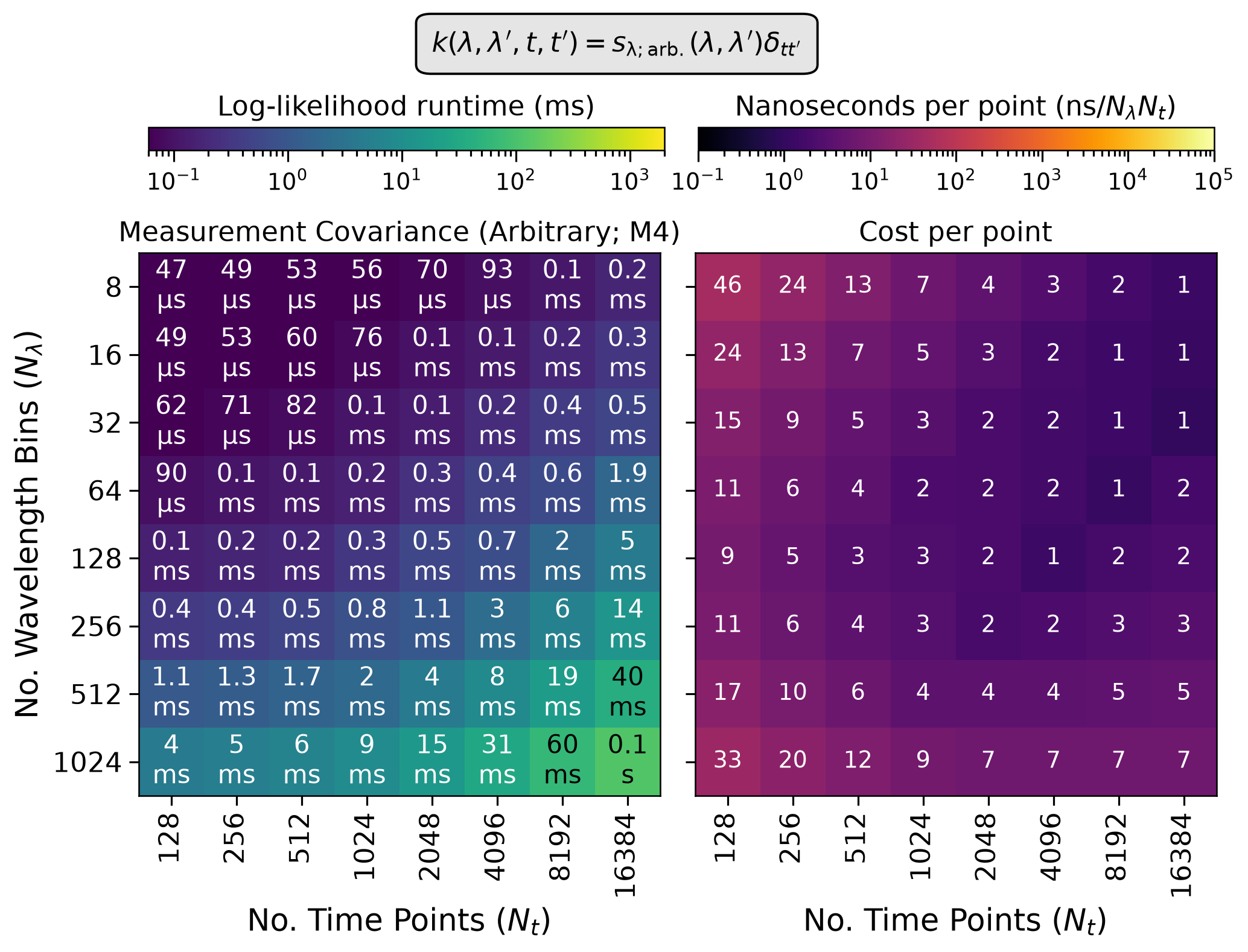}
    \end{minipage}

    \begin{minipage}[t]{\columnwidth}
        \caption{Measurement covariance which is in the form of a rank-1 quasiseparable matrix can be efficiently scaled to very large datasets even on a MacBook, as it has a linear scaling with the number of datapoints $\sim\mathcal{O}(N_\lambda J_{\Sigma_\lambda}^2 + N_\lambda N_t J_{\Sigma_\lambda})$, where $J_{\Sigma_\lambda}$ is the quasiseparable rank of the measurement covariance matrix.}
    \label{fig:SingleKronTermKernel_quasireadout_cost}
    \end{minipage}\hfill%
    \begin{minipage}[t]{\columnwidth}
        \caption{For an arbitrary measurement covariance matrix, the runtime scaling is $\mathcal{O}(N_\lambda^3 + N_\lambda^2 N_t)$ with the \textsc{SeparableKernel}, so it runs slower for a large number of wavelengths relative to Figure~\ref{fig:SingleKronTermKernel_quasireadout_cost} with quasiseparable measurement covariance.}
    \label{fig:SingleKronTermKernel_densereadout_cost}
    \end{minipage}
\end{figure*}

\begin{figure*}[p]
    \begin{minipage}[b]{\columnwidth}
        \includegraphics[width=\linewidth]{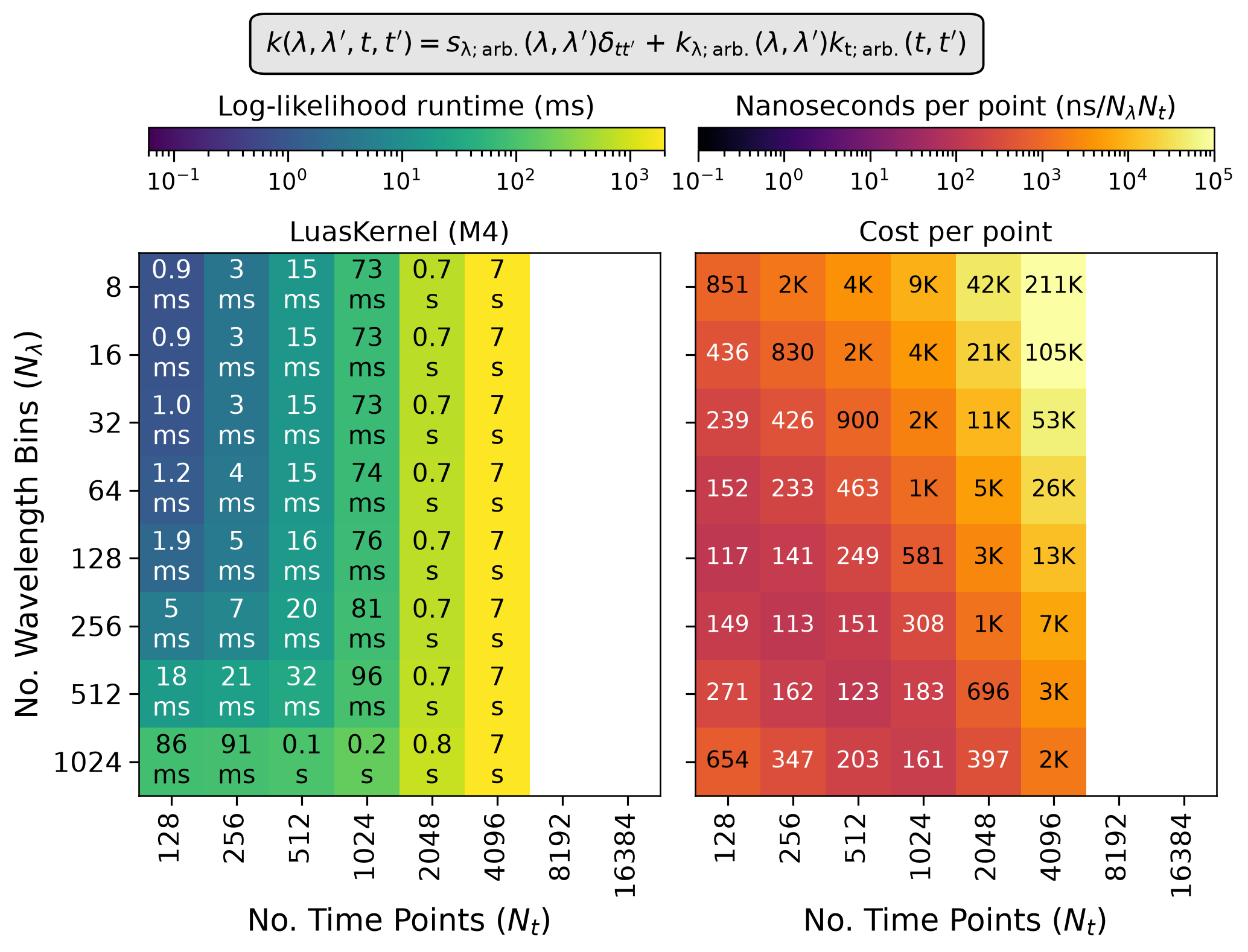}
    \end{minipage}\hfill%
    \begin{minipage}[b]{\columnwidth}
        \includegraphics[width=\linewidth]{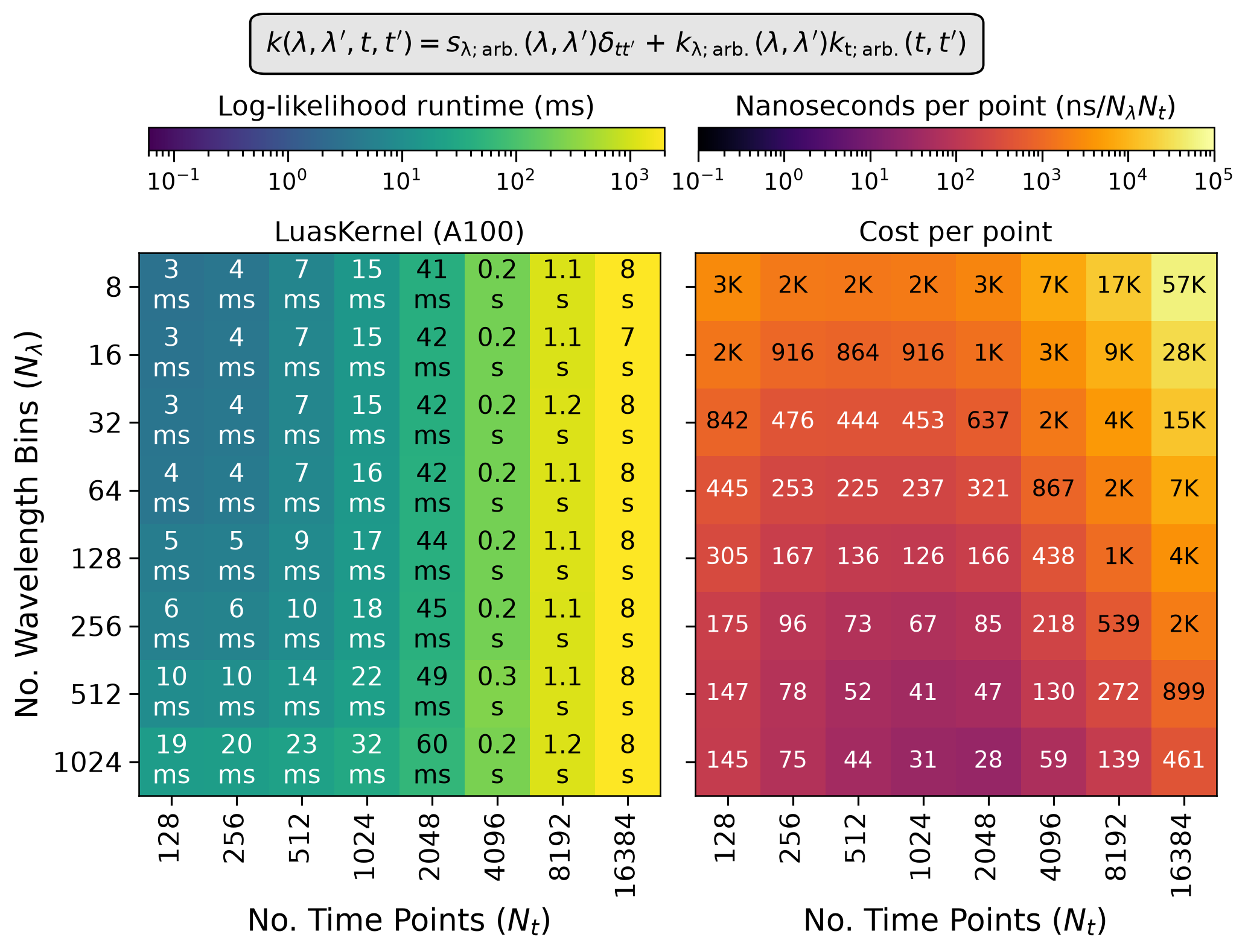}
    \end{minipage}

    \begin{minipage}[t]{\columnwidth}
        \caption{With a runtime scaling of $\mathcal{O}(N_\lambda^3 + N_t^3 + N_\lambda N_t (N_\lambda + N_t))$, the runtime cost of the \textsc{LuasKernel} is dominated by the eigendecomposition of covariance matrices in both dimensions. By comparing with Figure~\ref{fig:matrix_decomp_cost}, we can note that the top row of runtimes is only marginally larger than the cost of eigendecomposition of the time covariance matrix alone, and this remains the bottleneck even for hundreds of wavelengths.}
    \label{fig:LuasKernelND_cost}
    \end{minipage}\hfill%
    \begin{minipage}[t]{\columnwidth}
        \caption{Since eigendecomposition can be performed faster on GPU (see Figure~\ref{fig:matrix_decomp_cost}), the \textsc{LuasKernel} can be scaled to larger datasets than on CPU (shown in Figure~\ref{fig:LuasKernelND_cost}). We see that the GPU runtimes are often significantly faster, excluding the smallest datasets tested.}
    \label{fig:LuasKernelND_A100_cost}
    \end{minipage}

    \vspace{2\baselineskip}

    \begin{minipage}[b]{\columnwidth}
        \includegraphics[width=\linewidth]{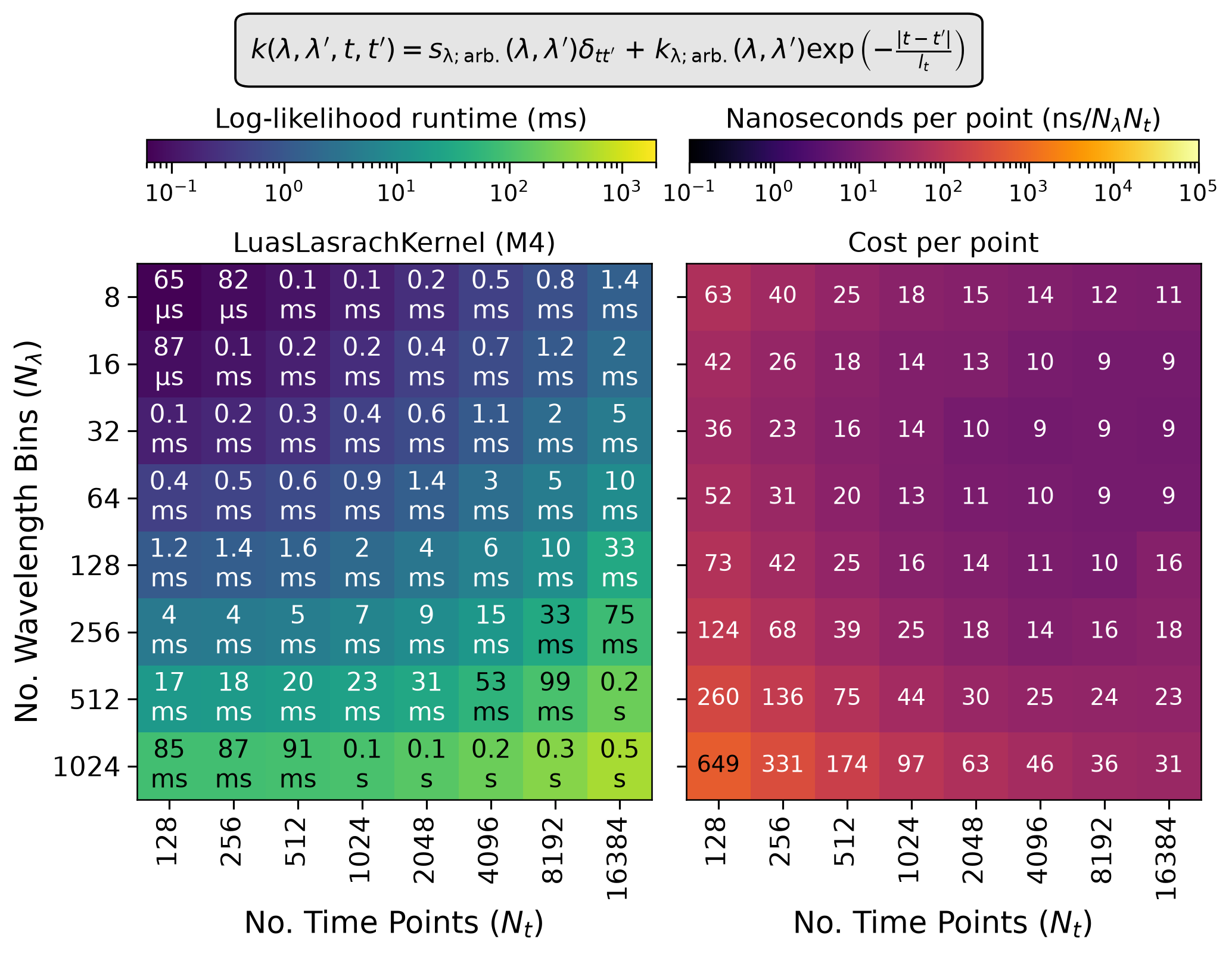}
    \end{minipage}\hfill%
    \begin{minipage}[b]{\columnwidth}
        \includegraphics[width=\linewidth]{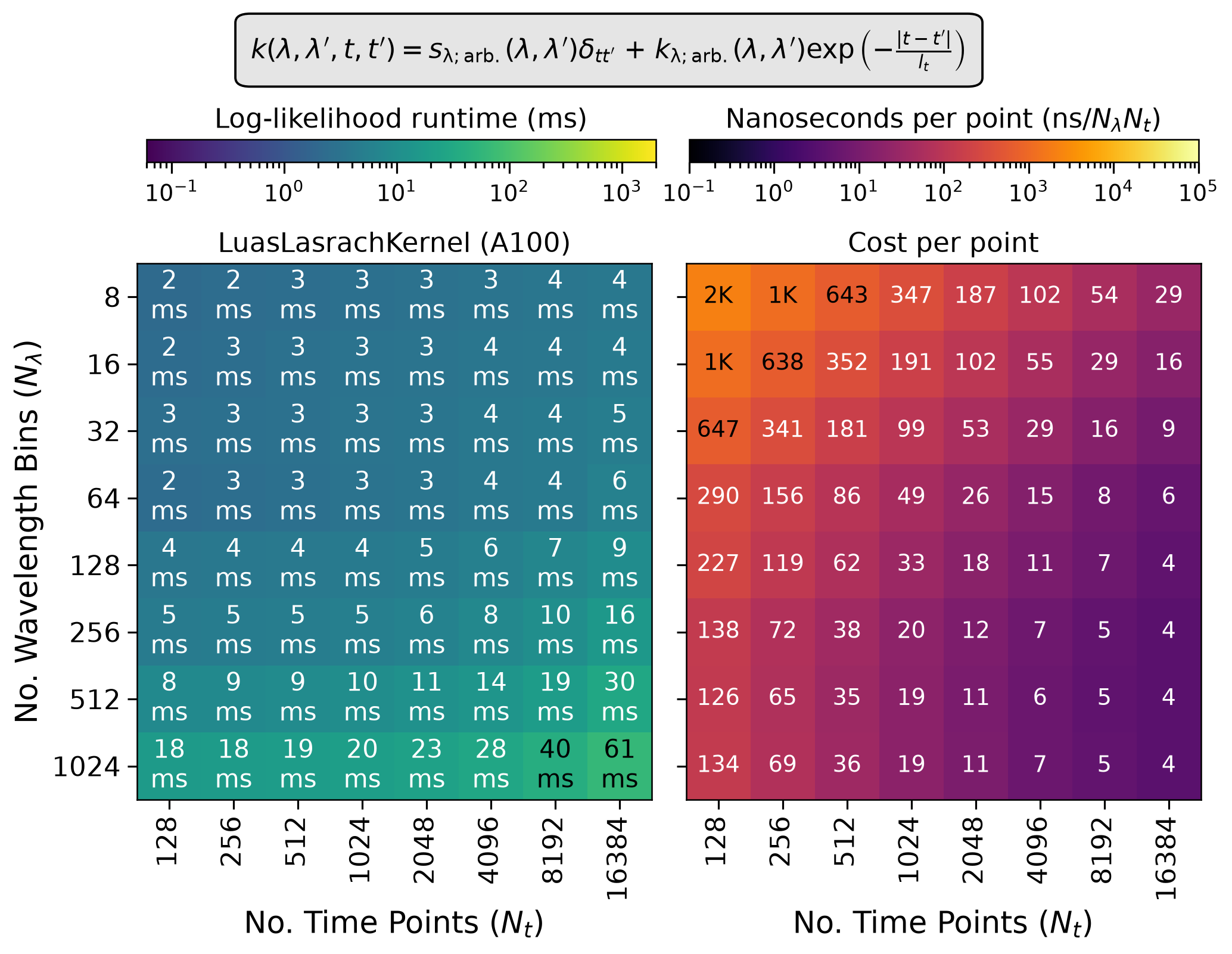}
    \end{minipage}

    \begin{minipage}[t]{\columnwidth}
        \caption{In comparison to the \textsc{LuasKernel} in Figure~\ref{fig:LuasKernelND_cost}, the \textsc{LuasLasrachKernel} can achieve dramatically better scaling in $N_t$ when the time kernel is quasiseparable, with a runtime scaling of $\mathcal{O}(N_\lambda^3 + N_\lambda^2 N_t/T + N_\lambda N_t J_{K_t}^2/T)$ for $T$ parallel workers. As we still must perform an eigendecomposition of a wavelength covariance matrix, runtimes for high $N_\lambda$ and low $N_t$ are similar to the \textsc{LuasKernel}.}
    \label{fig:LuasLasrachKernel_cost}
    \end{minipage}\hfill%
    \begin{minipage}[t]{\columnwidth}
        \caption{Compared to Figure~\ref{fig:LuasLasrachKernel_cost}, running the \textsc{LuasLasrachKernel} on GPU can also perform significantly faster on large datasets, as GPUs optimise the eigendecomposition of the wavelength covariance matrix, as well as parallelising the decomposition with many parallel workers $T$.}
    \label{fig:LuasLasrachKernel_A100_cost}
    \end{minipage}
\end{figure*}

\begin{figure*}[p]
    \begin{minipage}[b]{\columnwidth}
        \includegraphics[width=\linewidth]{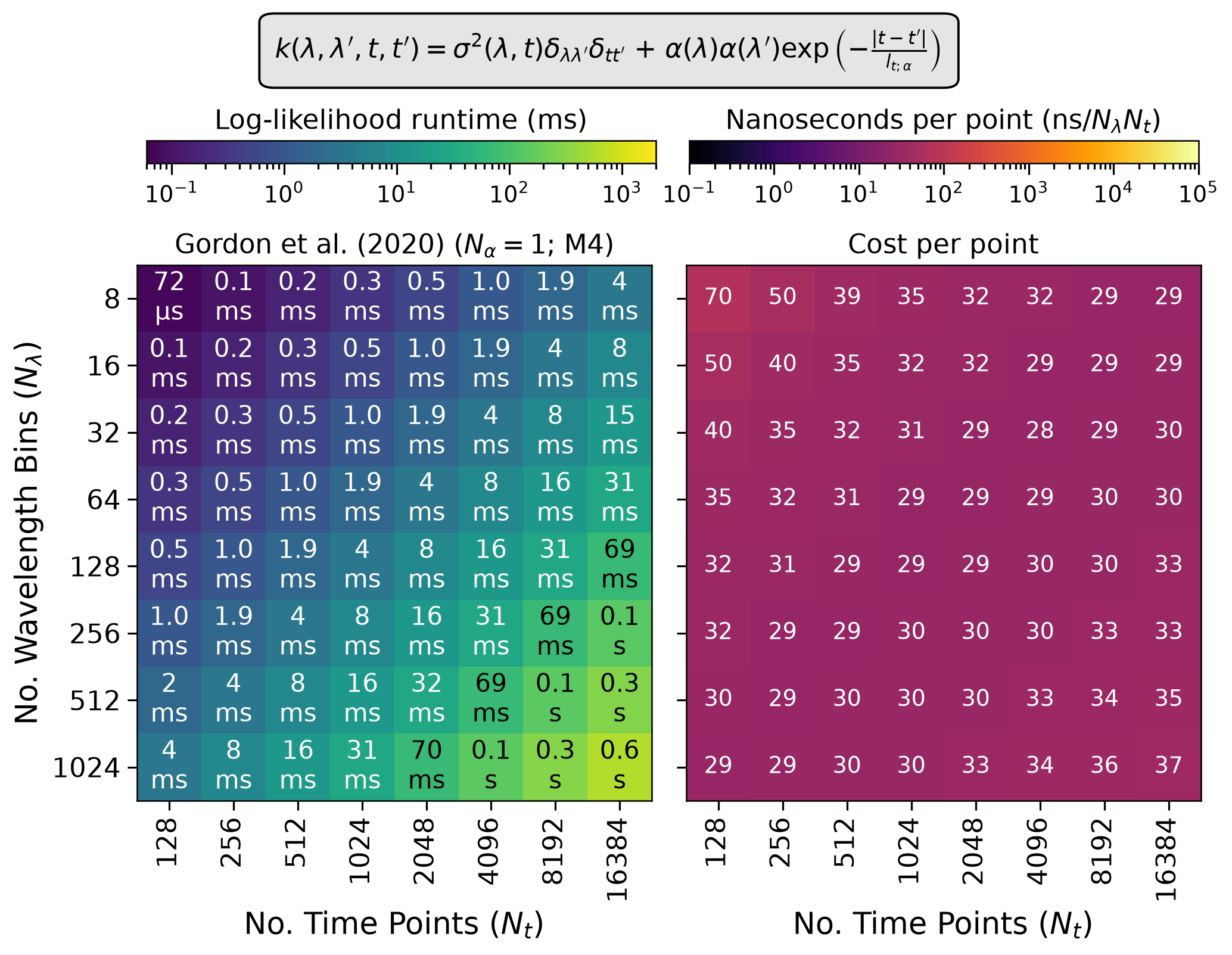}
    \end{minipage}\hfill%
    \begin{minipage}[b]{\columnwidth}
        \includegraphics[width=\linewidth]{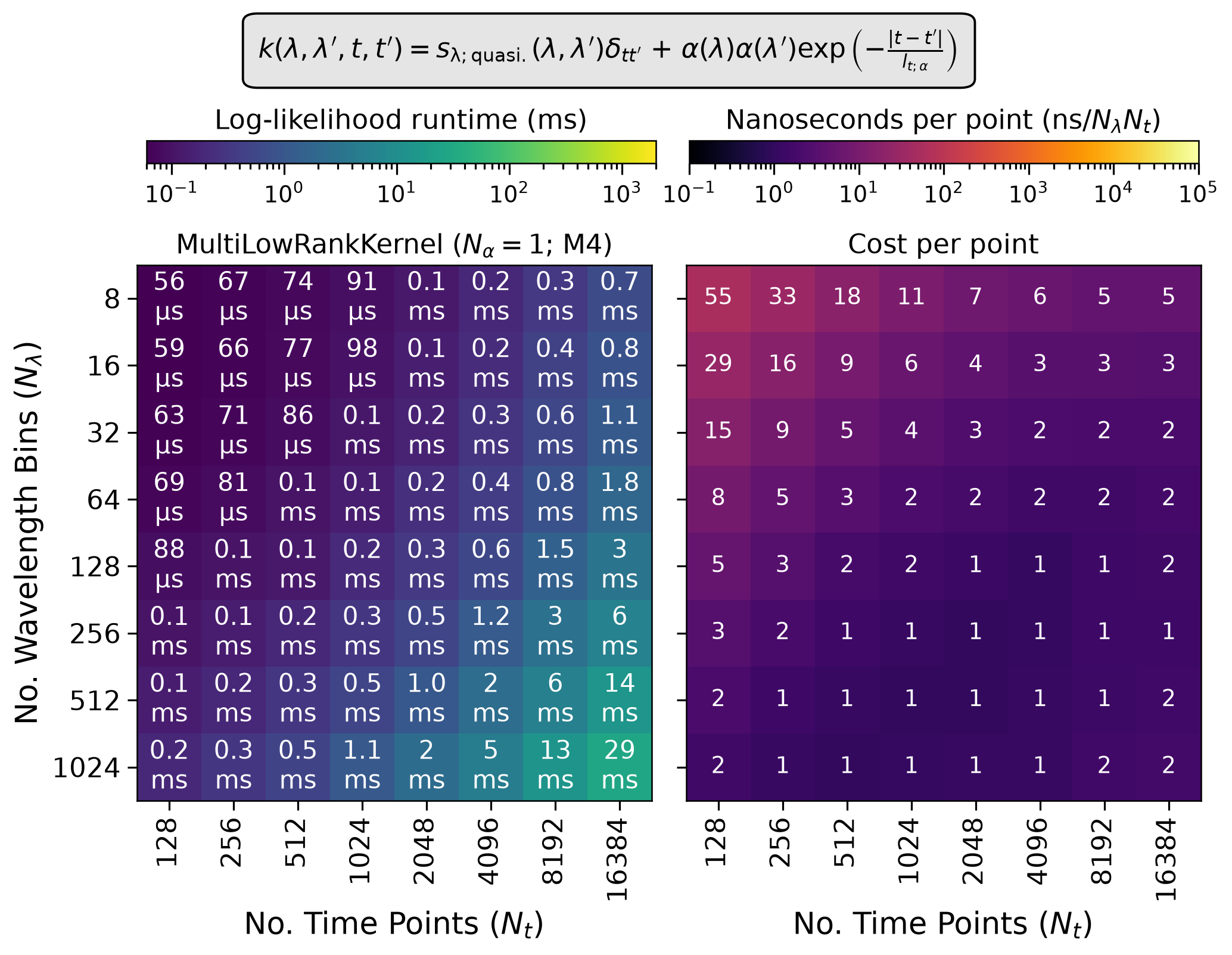}
    \end{minipage}

    \begin{minipage}[t]{\columnwidth}
         \caption{For comparison, we benchmarked the \citet{Gordon2020} optimisation for a low-rank kernel term implemented within \textsc{tinygp}. The right plot clearly shows the linear scaling of this algorithm $\mathcal{O}(N_\lambda N_t J_{K_t}^2)$, with a cost of about 30 ns/point for likelihood calculations on CPU. Unlike our optimisations, this model does not account for measurement covariance, however, it can account for white noise amplitudes which are not separable in each dimension (see Section~\ref{sec:white_noise}).}
    \label{fig:GordonKernel_Nalpha1_cost}
    \end{minipage}\hfill%
    \begin{minipage}[t]{\columnwidth}
        \caption{Compared to the \citet{Gordon2020} optimisation in Figure~\ref{fig:GordonKernel_Nalpha1_cost}, our \textsc{MultiLowRankKernel} optimisation achieves significantly faster runtimes for a similar kernel function. While both algorithms achieve a linear runtime scaling, our algorithm only requires a couple floating point operations per datapoint, and therefore costs just 1-2 ns per datapoint for larger datasets. We also include quasiseparable measurement covariance with this kernel, flexibility not introduced in \citet{Gordon2020}, but we cannot handle non-separable white noise amplitudes across the dataset, which the \citet{Gordon2020} method can handle.}
    \label{fig:MultiLowRankKernel_Nalpha1_cost}
    \end{minipage}
    
    \vspace{2\baselineskip}

    \begin{minipage}[b]{\columnwidth}
        \includegraphics[width=\linewidth]{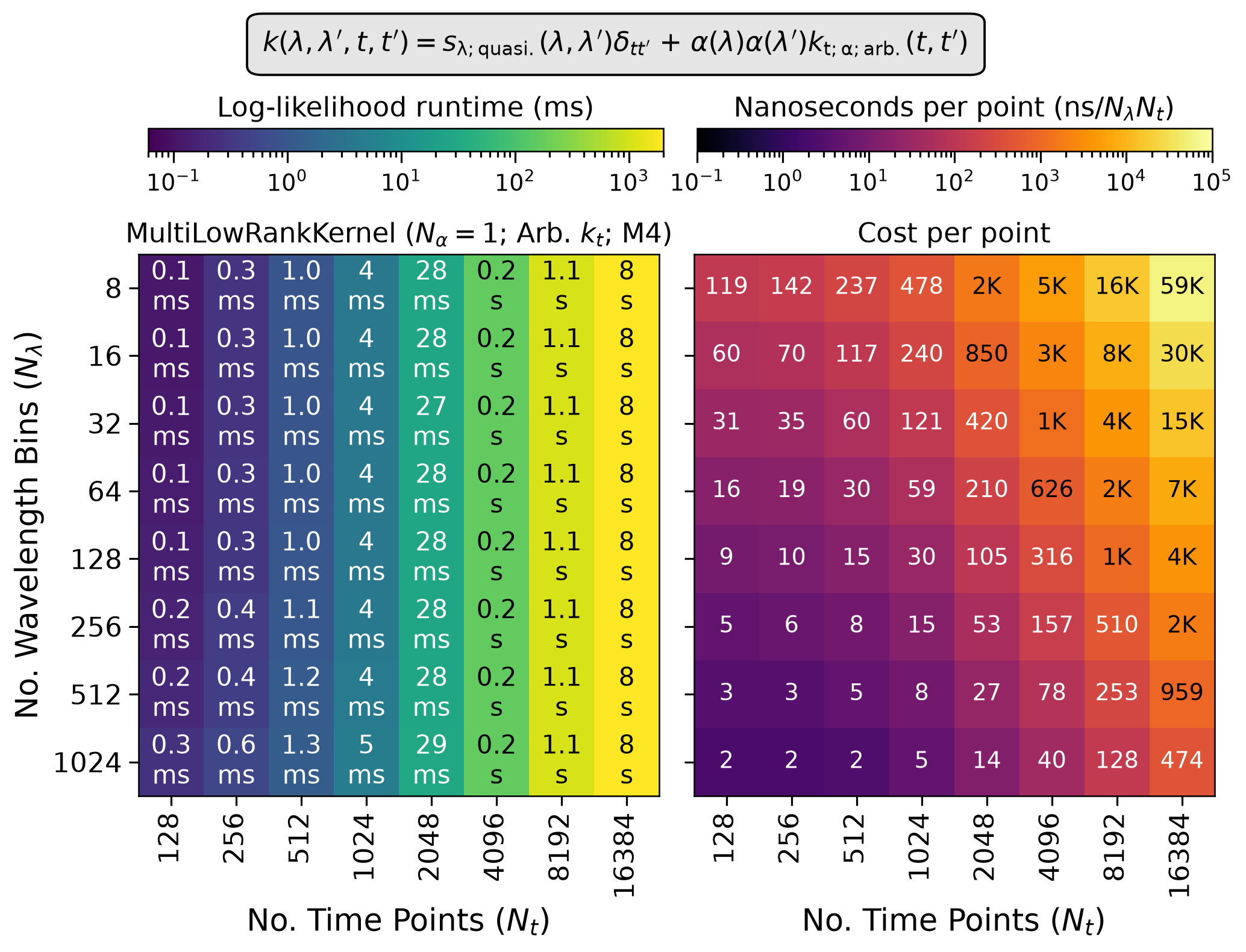}
    \end{minipage}\hfill%
    \begin{minipage}[b]{\columnwidth}
        \includegraphics[width=\linewidth]{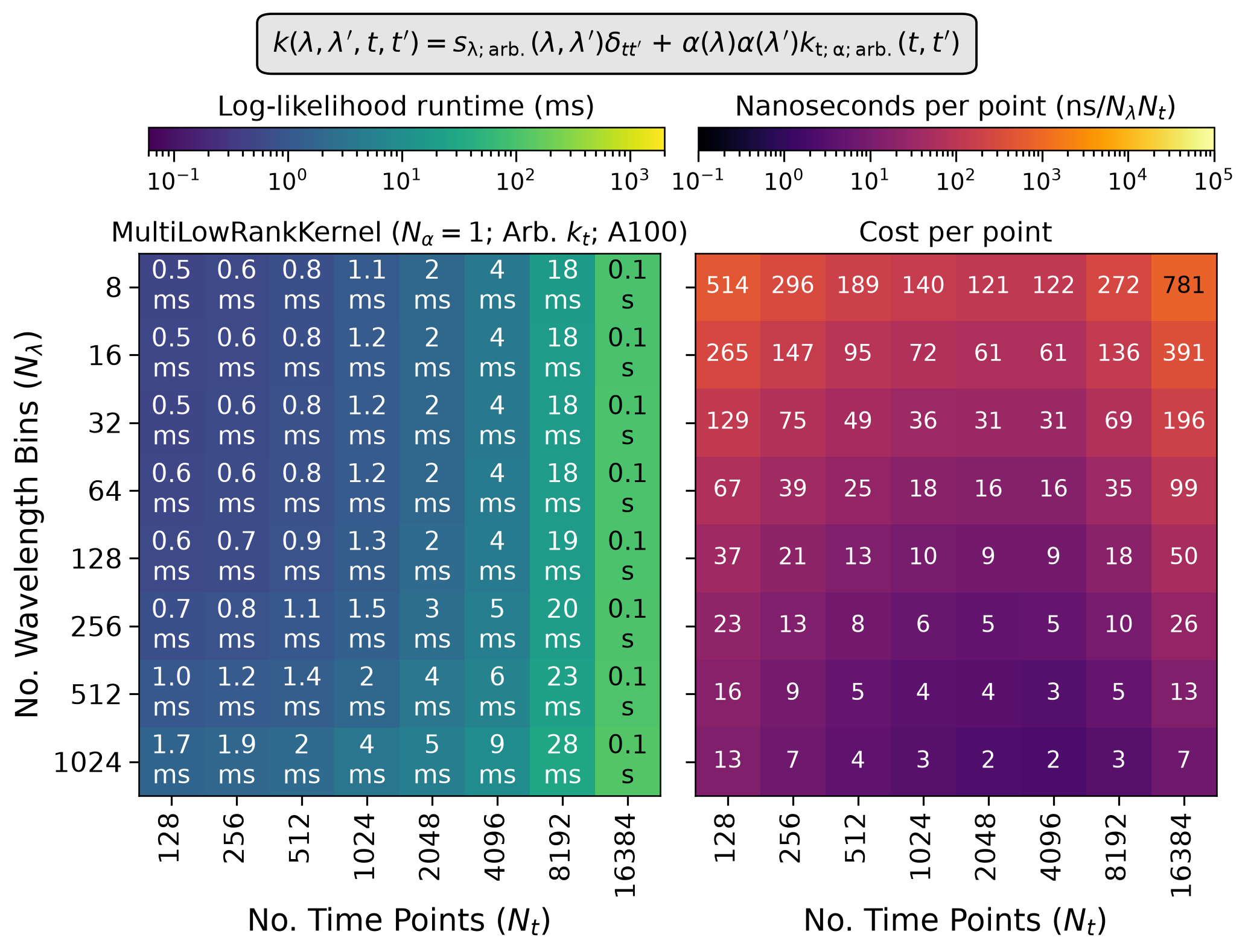}
    \end{minipage}

    \begin{minipage}[t]{\columnwidth}
    \caption{We can also extend this kernel to include any time kernel, not just quasiseparable kernels. We also include measurement covariance, but assume it is quasiseparable. With a scaling of $\mathcal{O}(N_\alpha^3 N_t^3 + N_\alpha N_\lambda N_t)$, the runtime cost is dominated by the Cholesky factorisation of a $(N_\alpha N_t, N_\alpha N_t)$ matrix on CPU, which can be seen by comparing these values with Figure~\ref{fig:matrix_decomp_cost}.}
    \label{fig:MultiLowRankKernel_Nalpha1_poweredexp_cost}
    \end{minipage}\hfill%
    \begin{minipage}[t]{\columnwidth}
        \caption{Compared to the CPU calculations in Figure~\ref{fig:MultiLowRankKernel_Nalpha1_poweredexp_cost}, the GPU can more efficiently model an arbitrary low-rank kernel term even with a more general arbitrary measurement covariance. The runtimes are similarly dominated by the $(N_\alpha N_t, N_\alpha N_t)$ matrix Cholesky decomposition, but this is much faster on GPU (Figure~\ref{fig:matrix_decomp_cost}).}
    \label{fig:MultiLowRankKernel_Nalpha1_poweredexp_densereadout_A100_cost}
    \end{minipage}
\end{figure*}

\begin{figure*}[p]
    \begin{minipage}[b]{\columnwidth}
        \includegraphics[width=\linewidth]{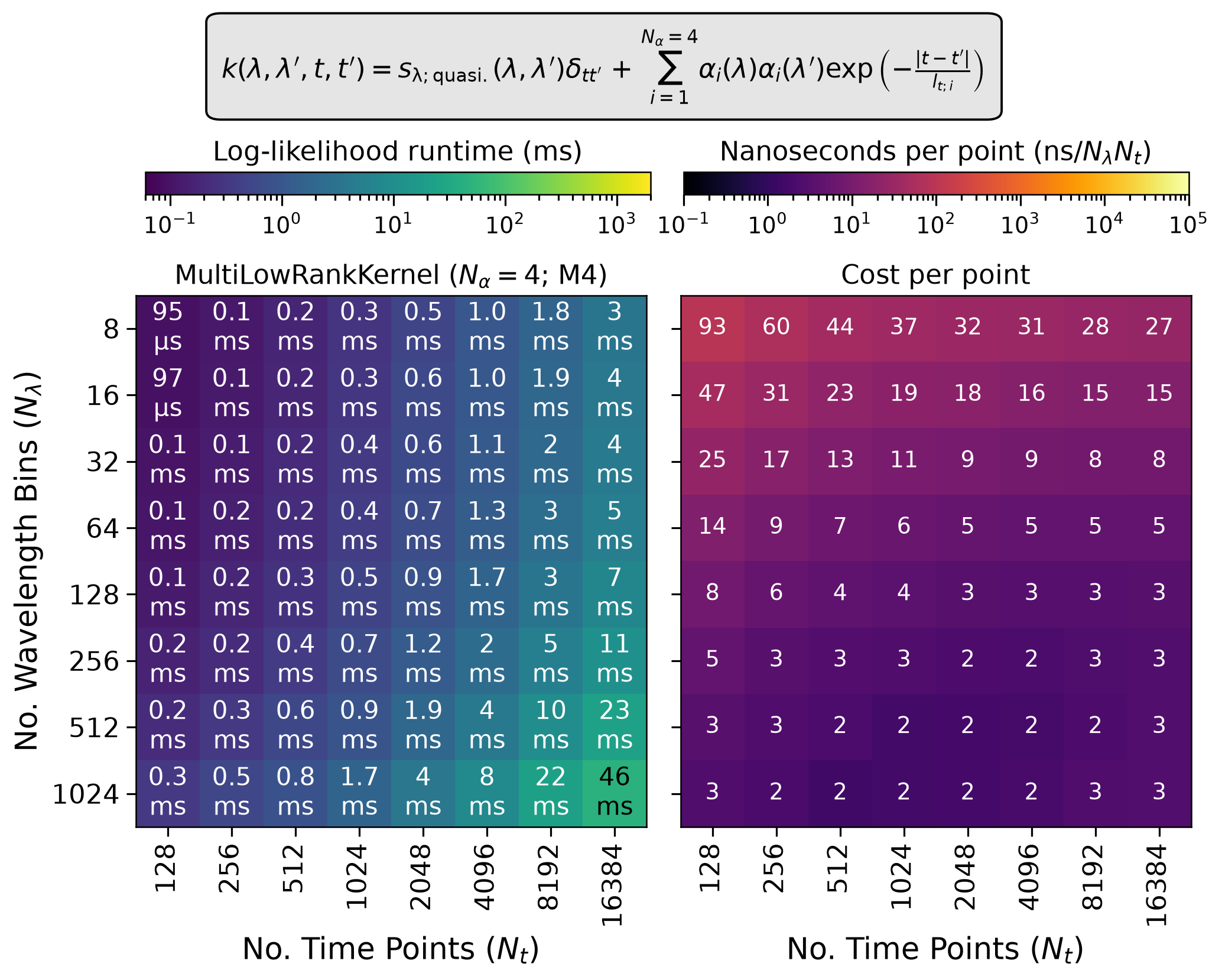}
    \end{minipage}\hfill%
    \begin{minipage}[b]{\columnwidth}
        \includegraphics[width=\linewidth]{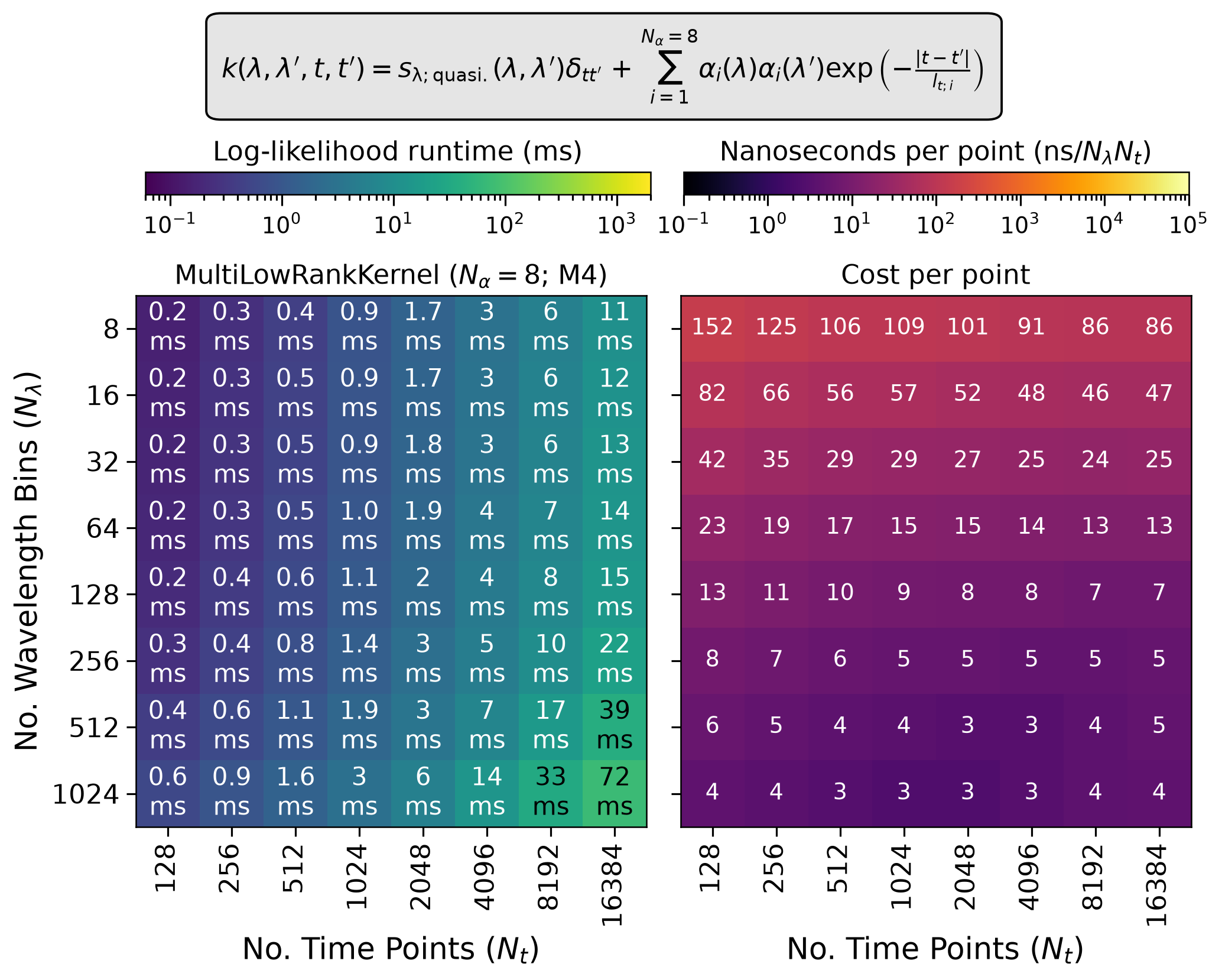}
    \end{minipage}

    \begin{minipage}[t]{\columnwidth}
         \caption{Comparing with Figure~\ref{fig:MultiLowRankKernel_Nalpha1_cost} where $N_\alpha = 1$, increasing to $N_\alpha = 4$ results in a relatively minor runtime increase, particularly for large $N_\lambda$, consistent with its runtime scaling as $\mathcal{O}(N_\alpha^3 N_t \bar{J}_{K_t}^2 + N_\alpha N_\lambda N_t)$, where the fact the quasiseparable decomposition does not scale with $N_\lambda$ mitigates the cubic scaling with $N_\alpha$.}
    \label{fig:MultiLowRankKernel_Nalpha4_cost}
    \end{minipage}\hfill%
    \begin{minipage}[t]{\columnwidth}
        \caption{Increasing further to $N_\alpha = 8$ gives a further increase in runtimes relative to Figures~\ref{fig:MultiLowRankKernel_Nalpha1_cost} and \ref{fig:MultiLowRankKernel_Nalpha4_cost} for $N_\alpha = 1, 4$ respectively, most significant when $N_\lambda$ is small and the $\mathcal{O}(N_\alpha^3 N_t \bar{J}_{K_t}^2)$ term dominates.}
    \label{fig:MultiLowRankKernel_Nalpha8_cost}
    \end{minipage}

    \vspace{2\baselineskip}

    \begin{minipage}[b]{\columnwidth}
        \includegraphics[width=\linewidth]{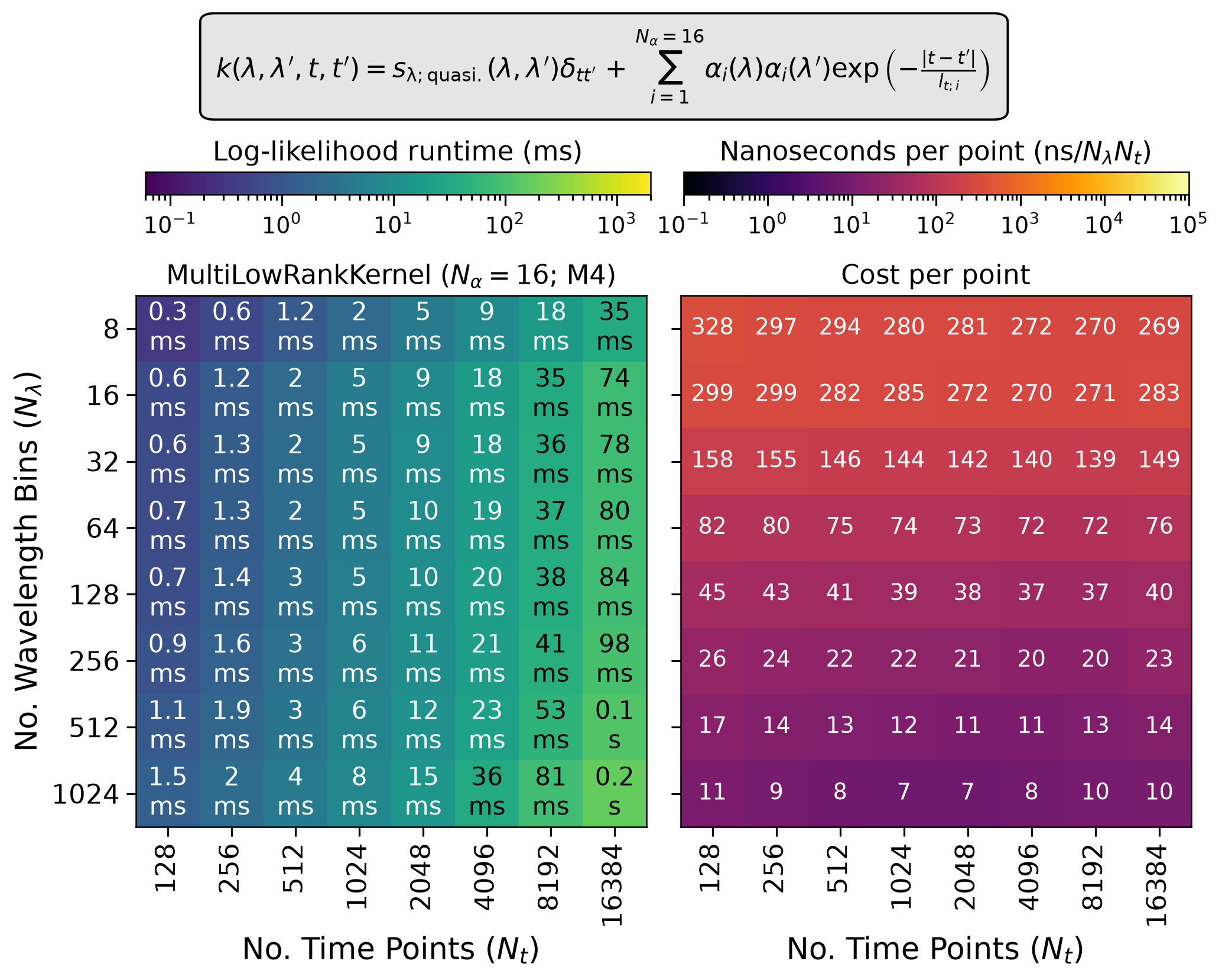}
    \end{minipage}\hfill%
    \begin{minipage}[b]{\columnwidth}
        \includegraphics[width=\linewidth]{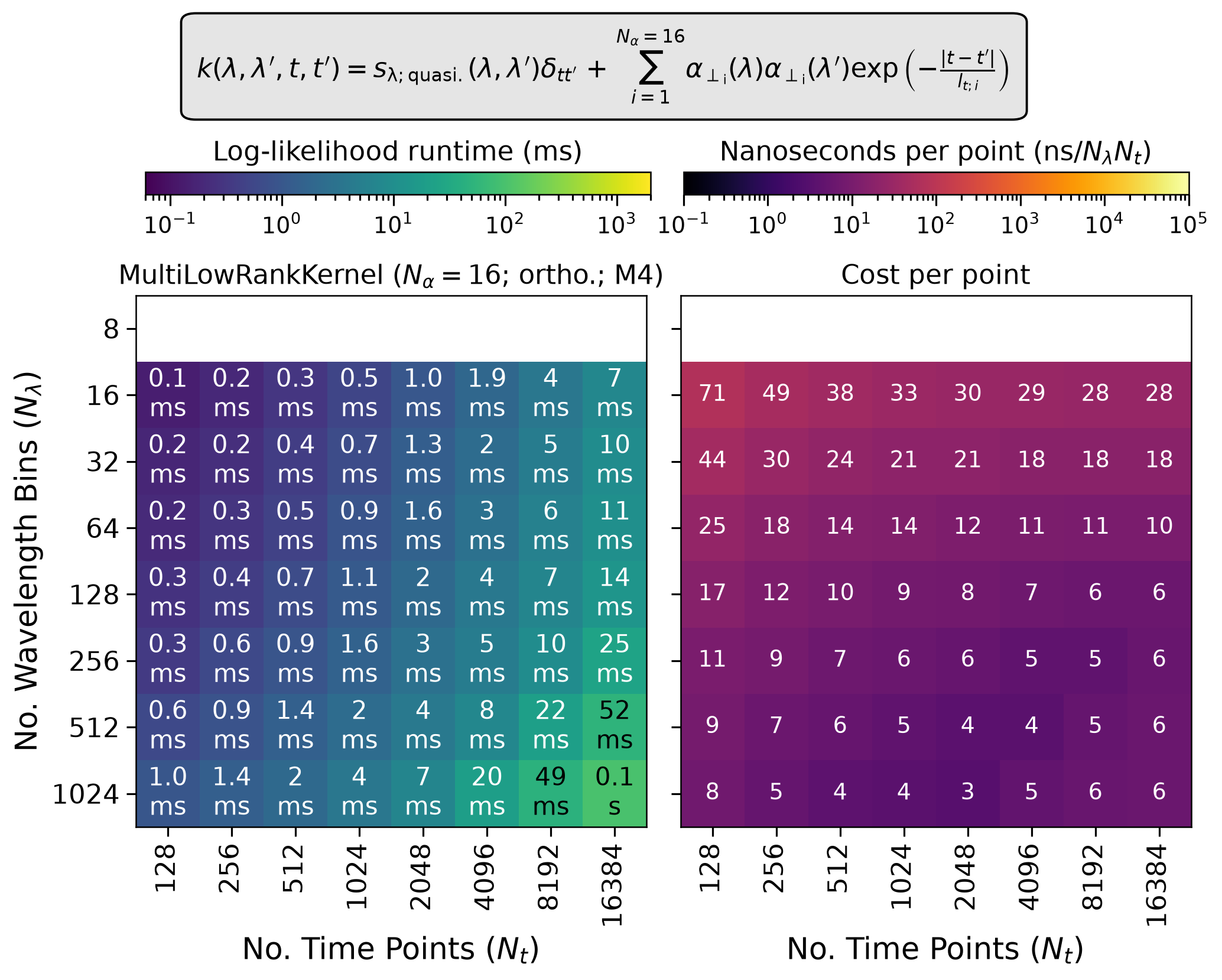}
    \end{minipage}

    \begin{minipage}[t]{\columnwidth}
    \caption{Increasing to $N_\alpha = 16$, we can see the cost of the $\mathcal{O}(N_\alpha^3 N_t \bar{J}_{K_t}^2)$ term dominates, as the runtime cost is almost constant between $N_\lambda = 16 \to 256$. Note that when $N_\lambda = 8, 16$, our method reduces to the \citet{Gordon2020} method with the addition of measurement covariance, as the transformations to exploit sparsity only work when $N_\lambda > N_\alpha$.}
    \label{fig:MultiLowRankKernel_Nalpha16_cost}
    \end{minipage}\hfill%
    \begin{minipage}[t]{\columnwidth}
        \caption{If the $\boldsymbol{\alpha}_{\lambda; i}$ vectors are orthogonal (after a whitening transformation), then the runtime scaling is instead $\mathcal{O}(N_\alpha N_t \bar{J}_{K_t}^2 + N_\alpha N_\lambda N_t)$, which performs significantly faster when $N_\lambda$ is small and $N_\alpha$ large compared with Figure~\ref{fig:MultiLowRankKernel_Nalpha16_cost}. When $N_\alpha = N_\lambda$, this is equivalent to the \textsc{LuasLasrachKernel} with different time kernel functions for each eigenvector of the wavelength covariance matrix. Note we must exclude $N_\lambda = 8$ as we cannot have 16 orthogonal vectors in this scenario.}
    \label{fig:MultiLowRankKernel_Nalpha16_ortho_cost}
    \end{minipage}
\end{figure*}

\begin{figure*}[p]
    \begin{minipage}[b]{\columnwidth}
        \includegraphics[width=\linewidth]{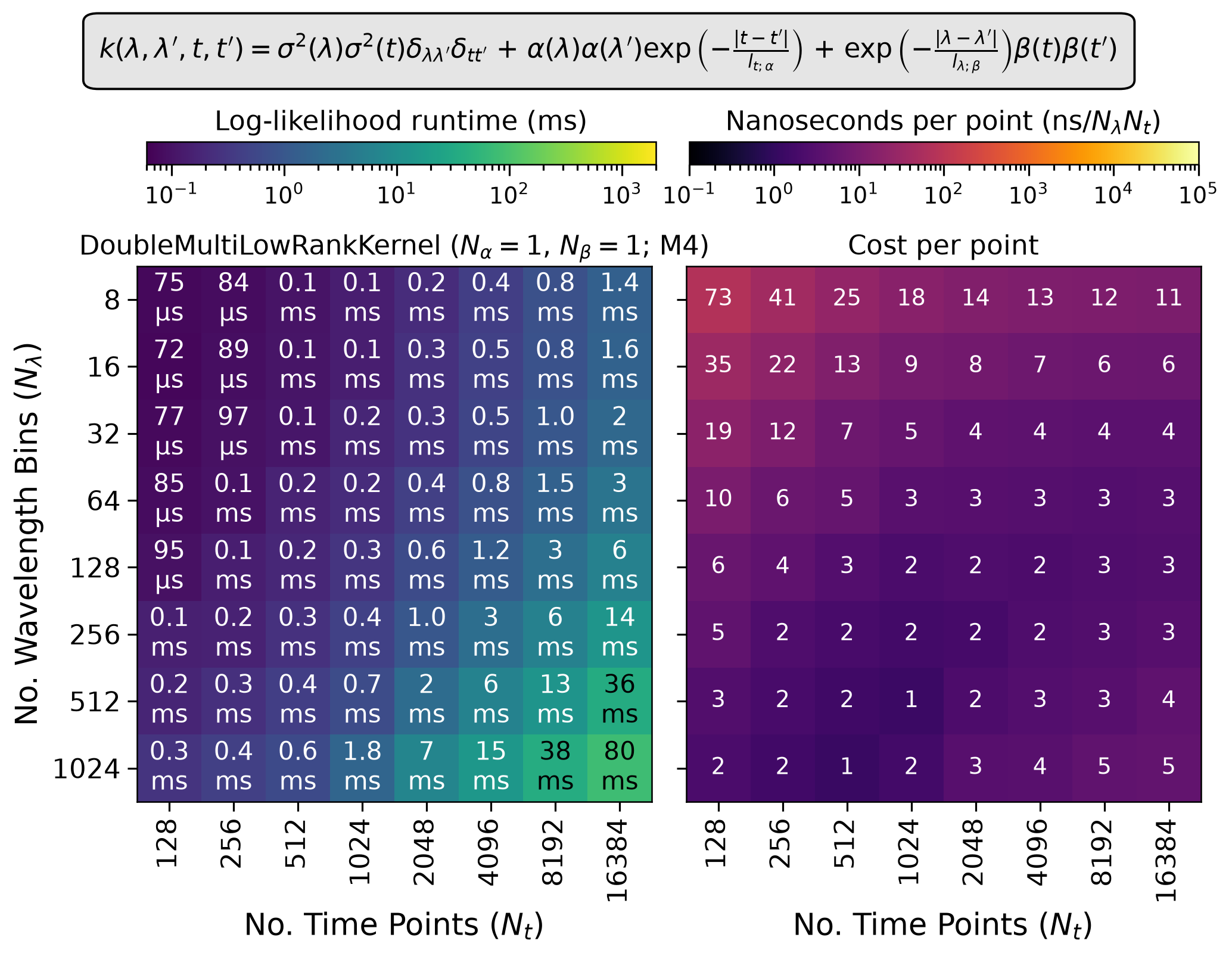}
    \end{minipage}\hfill%
    \begin{minipage}[b]{\columnwidth}
        \includegraphics[width=\linewidth]{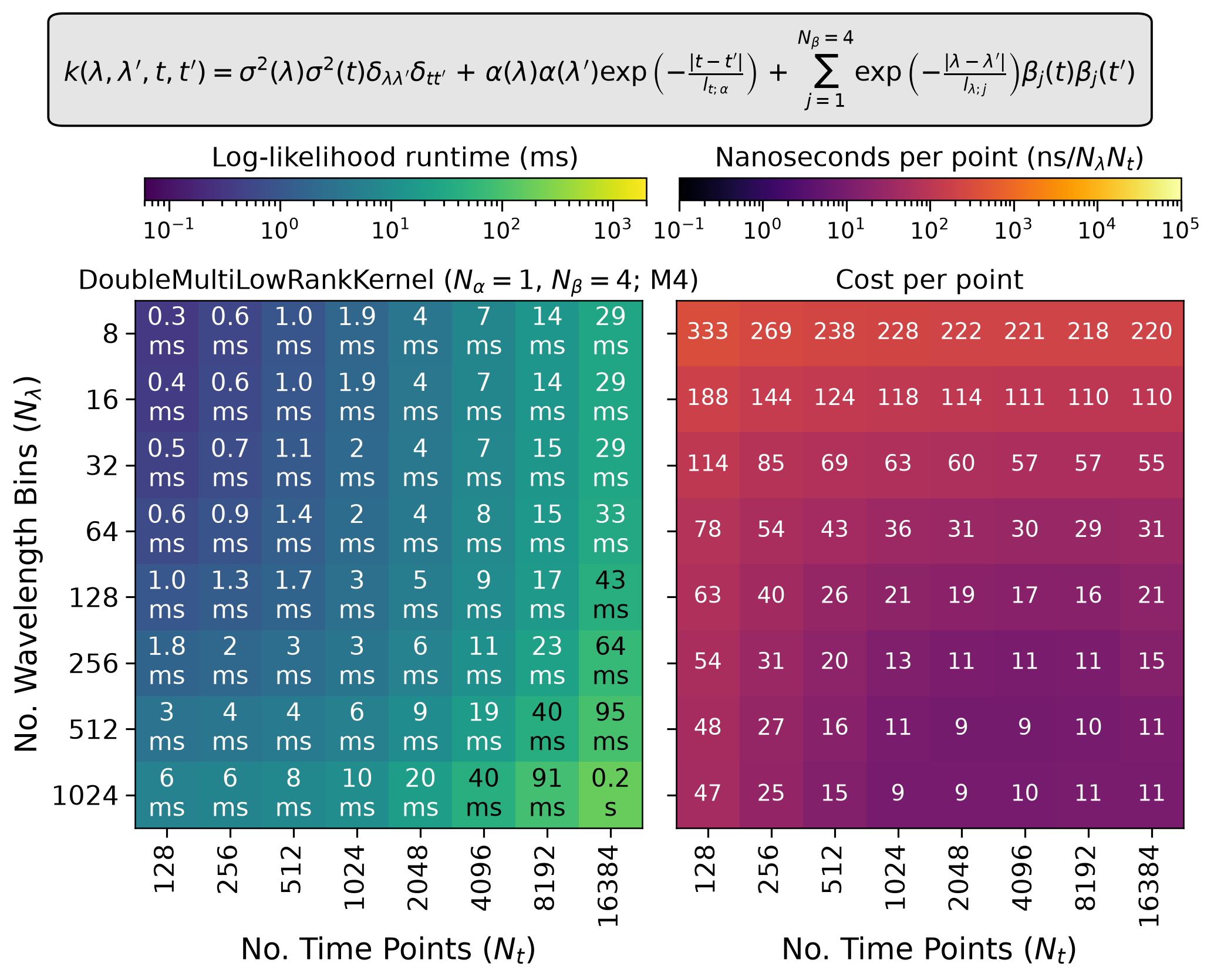}
    \end{minipage}

    \begin{minipage}[t]{\columnwidth}
         \caption{With low-rank quasiseparable kernel terms in both dimensions, we can still achieve a linear runtime scaling with the number of datapoints using the \textsc{DoubleMultiLowRankKernel}, with a runtime scaling of $\sim\mathcal{O}(N_\alpha^3 N_t (\bar{J}_{K_t} + 3 N_\beta)^2 + N_\beta^3 N_\lambda (\bar{J}_{K_\lambda} + 3N_\alpha)^2 + N_\lambda N_t (N_\alpha + N_\beta))$. Currently we can only maintain quasiseparability in both dimensions if we exclude measurement covariance, so heterogeneous white noise is modelled for this optimisation.}
    \label{fig:DoubleMultiLowRankKernel_Nalpha1_Nbeta1_cost}
    \end{minipage}\hfill%
    \begin{minipage}[t]{\columnwidth}
        \caption{Increasing from $N_\beta = 1$ in Figure~\ref{fig:DoubleMultiLowRankKernel_Nalpha1_Nbeta1_cost} to $N_\beta = 4$ is quite cheap when $N_\alpha$ is small, where $N_\beta$ corresponds to the kernel term which has its vector outer product in the longer axis for most of these data shapes.}
    \label{fig:DoubleMultiLowRankKernel_Nalpha1_Nbeta4_cost}
    \end{minipage}

    \vspace{2\baselineskip}

    \begin{minipage}[b]{\columnwidth}
        \includegraphics[width=\linewidth]{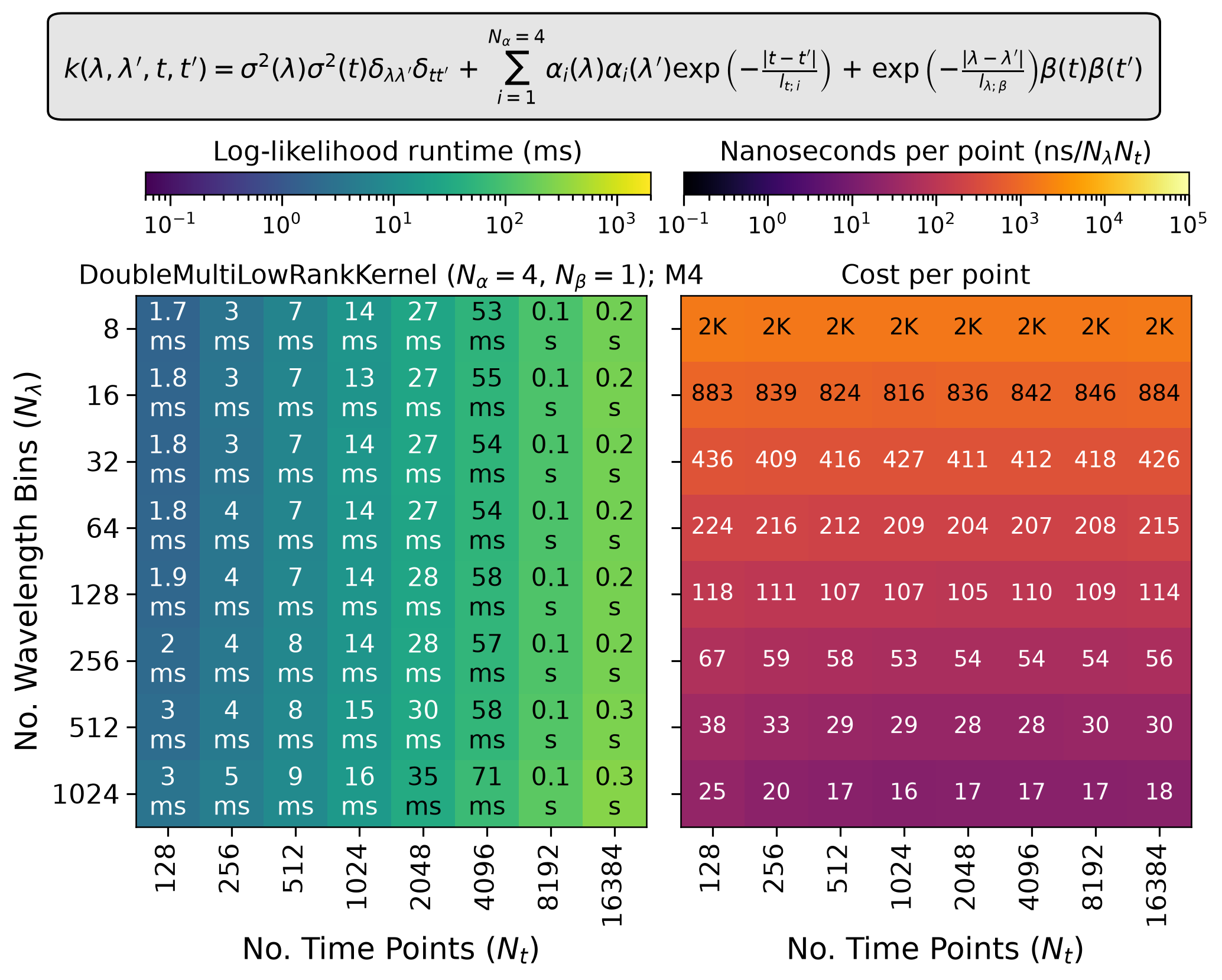}
    \end{minipage}\hfill%
    \begin{minipage}[b]{\columnwidth}
        \includegraphics[width=\linewidth]{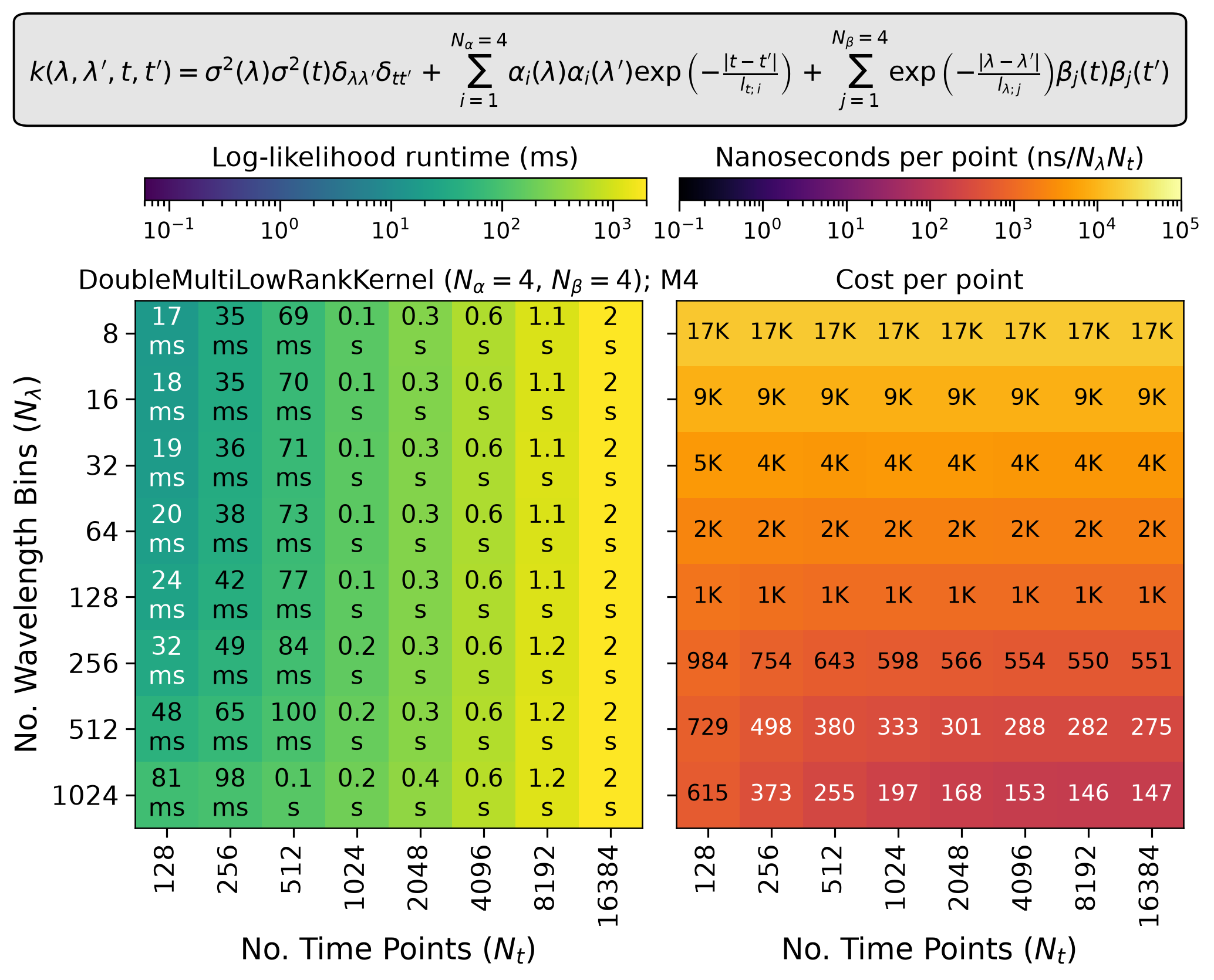}
    \end{minipage}

    \begin{minipage}[t]{\columnwidth}
    \caption{In contrast to Figure~\ref{fig:DoubleMultiLowRankKernel_Nalpha1_Nbeta4_cost}, having more kernel terms which have their vector outer product in the shorter axis scales less efficiently, particularly for a small number of wavelengths.}
    \label{fig:DoubleMultiLowRankKernel_Nalpha4_Nbeta1_cost}
    \end{minipage}\hfill%
    \begin{minipage}[t]{\columnwidth}
        \caption{The \textsc{DoubleMultiLowRankKernel} scales poorly with the product of $N_\alpha N_\beta$, we see that log-likelihood calculations are expensive even for small datasets when $N_\alpha = N_\beta = 4$, and our GPU runtimes were typically slower or similar for this kernel.}
    \label{fig:DoubleMultiLowRankKernel_Nalpha4_Nbeta4_cost}

    \end{minipage}
\end{figure*}

\begin{figure*}[p]
    \begin{minipage}[b]{\columnwidth}
        \includegraphics[width=\linewidth]{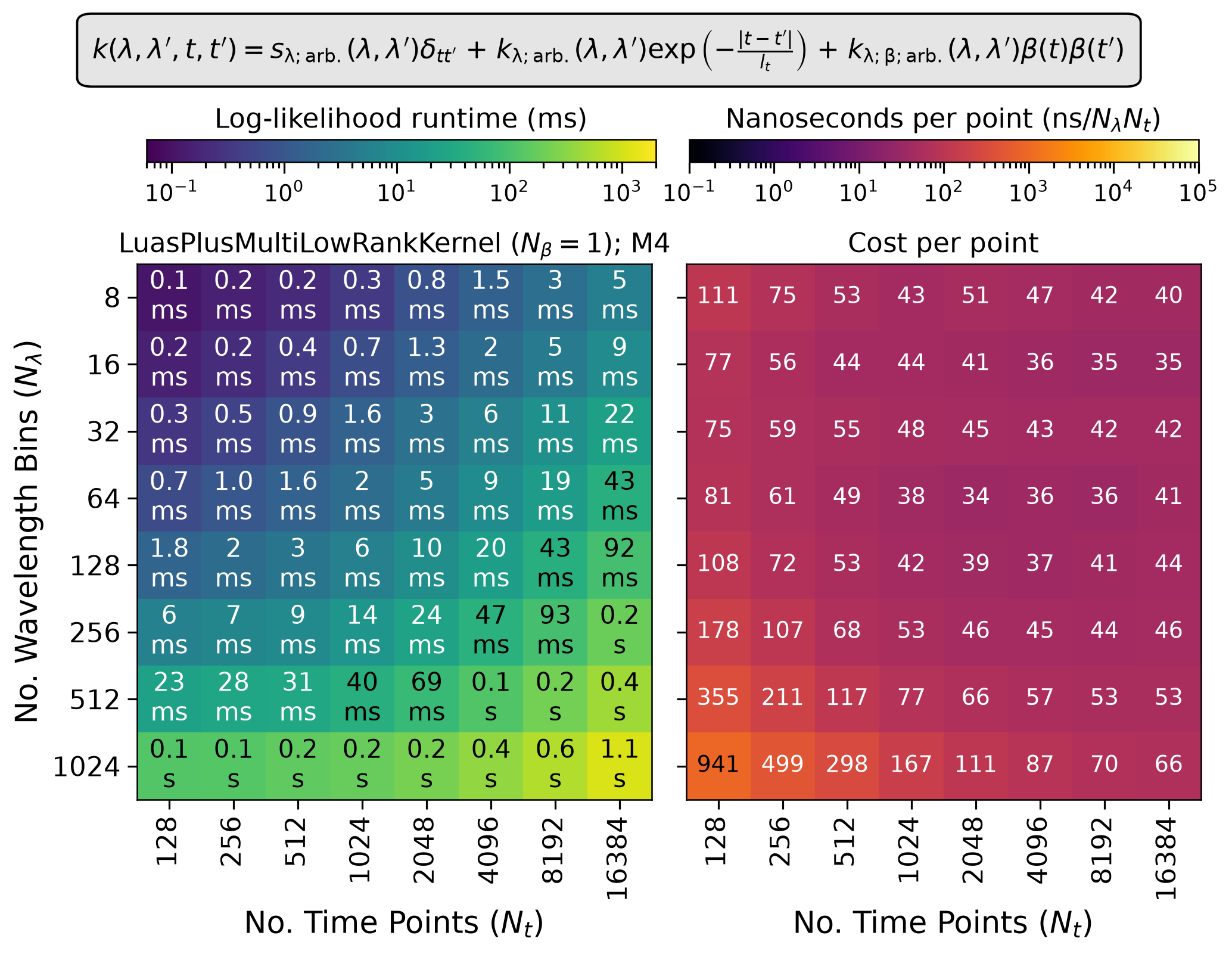}
    \end{minipage}\hfill%
    \begin{minipage}[b]{\columnwidth}
        \includegraphics[width=\linewidth]{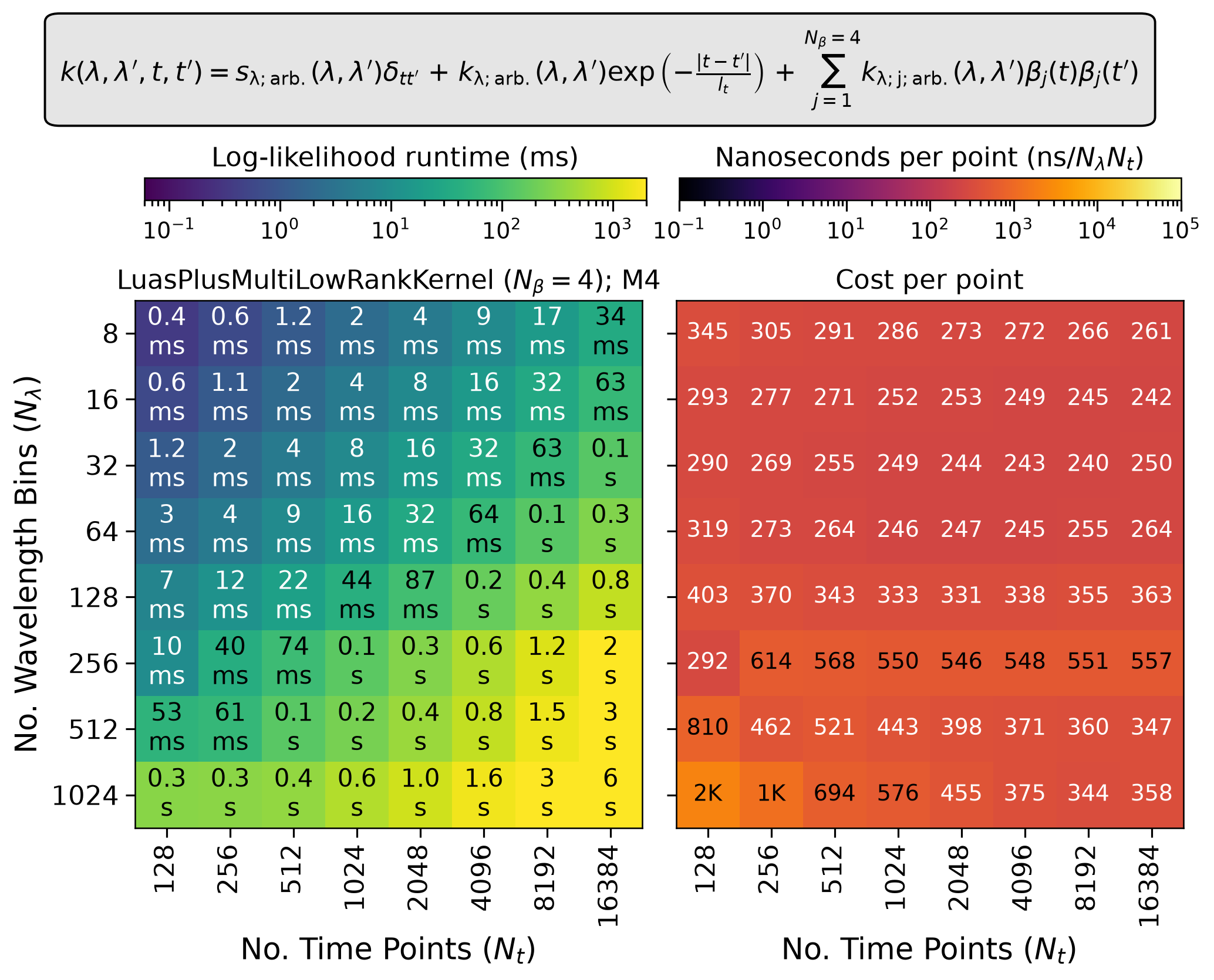}
    \end{minipage}

    \begin{minipage}[t]{\columnwidth}
         \caption{Even for the more complex scenario of arbitrary measurement covariance, a 2D kernel term with a quasiseparable time kernel, and a low-rank term, we can use the \textsc{LuasPlusMultiLowRankKernel} to scale linearly with the number of time points, provided the low-rank term has its vector outer product in time. We achieve a runtime scaling of $\mathcal{O}((1 + N_\beta^3) N_\lambda^3 + N_\lambda^2 N_t  + N_\lambda N_t (J_{K_t} + 2 N_\beta)^2)$ in this case.}
    \label{fig:LuasPlusMultiLowRankKernel_Nbeta1_cost}
    \end{minipage}\hfill%
    \begin{minipage}[t]{\columnwidth}
        \caption{Compared to Figure~\ref{fig:LuasPlusMultiLowRankKernel_Nbeta1_cost}, increasing from $N_\beta = 1$ to $N_\beta = 4$ produces a further increase in runtime, but the linear scaling with the number of time points remains.}
    \label{fig:LuasPlusMultiLowRankKernel_Nbeta4_cost}
    \end{minipage}

    \vspace{2\baselineskip}

    \begin{minipage}[b]{\columnwidth}
        \includegraphics[width=\linewidth]{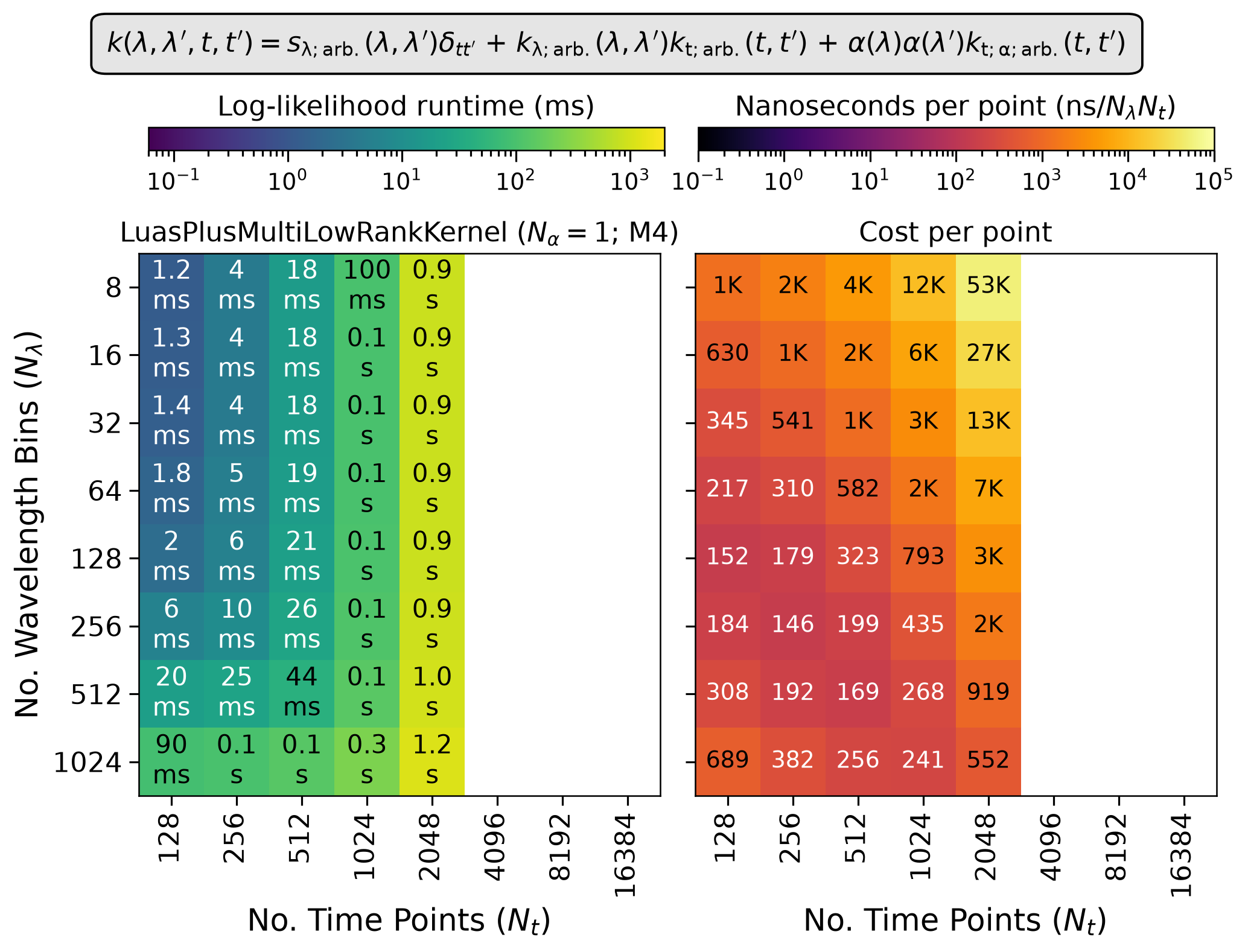}
    \end{minipage}\hfill%
    \begin{minipage}[b]{\columnwidth}
        \includegraphics[width=\linewidth]{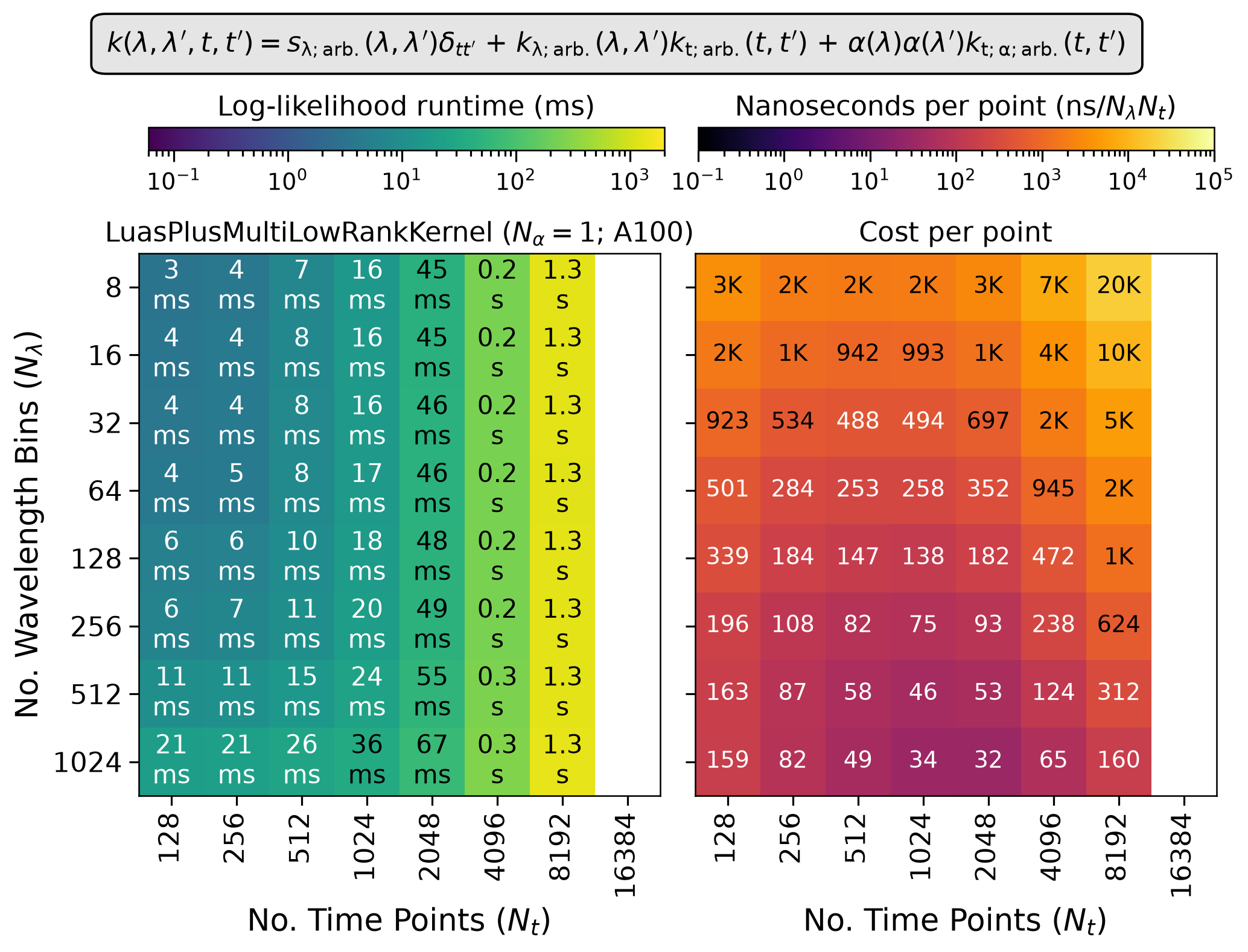}
    \end{minipage}

    \begin{minipage}[t]{\columnwidth}
    \caption{Compared to Figure~\ref{fig:LuasPlusMultiLowRankKernel_Nbeta1_cost}, we see that the runtimes are significantly more expensive when we set $N_\alpha = 1$ instead of $N_\beta = 1$. This is to be expected, because the algorithm requires an eigendecomposition in the time dimension in this case. We have used the \textsc{LuasKernel} method to decompose in this scenario (see Appendix~\ref{app:luasplusmultilowrankkernel_luaskernel}), which results in a runtime scaling of $\sim\mathcal{O}((1 + N_\alpha^3) N_t^3 + N_\lambda^3 + N_\lambda N_t(N_t + N_\lambda + N_\alpha^2))$, but doesn't require any quasiseparable kernels in either dimension. It has a similar runtime to the \textsc{LuasKernel} in Figure~\ref{fig:LuasKernelND_cost}, both primarily bottlenecked by the time covariance matrix eigendecomposition.}
    \label{fig:LuasPlusMultiLowRankKernel_Nalpha1_cost}
    \end{minipage}\hfill%
    \begin{minipage}[t]{\columnwidth}
        \caption{Compared to Figure~\ref{fig:LuasPlusMultiLowRankKernel_Nalpha1_cost}, running this kernel on GPU scales more efficiently to larger datasets, similar to the \textsc{LuasKernel} on GPU (Figure~\ref{fig:LuasKernelND_A100_cost}).}
    \label{fig:LuasPlusMultiLowRankKernel_Nalpha1_A100_cost}
    \end{minipage}
\end{figure*}

\begin{figure*}[p]
    \begin{minipage}[b]{\columnwidth}
        \includegraphics[width=\linewidth]{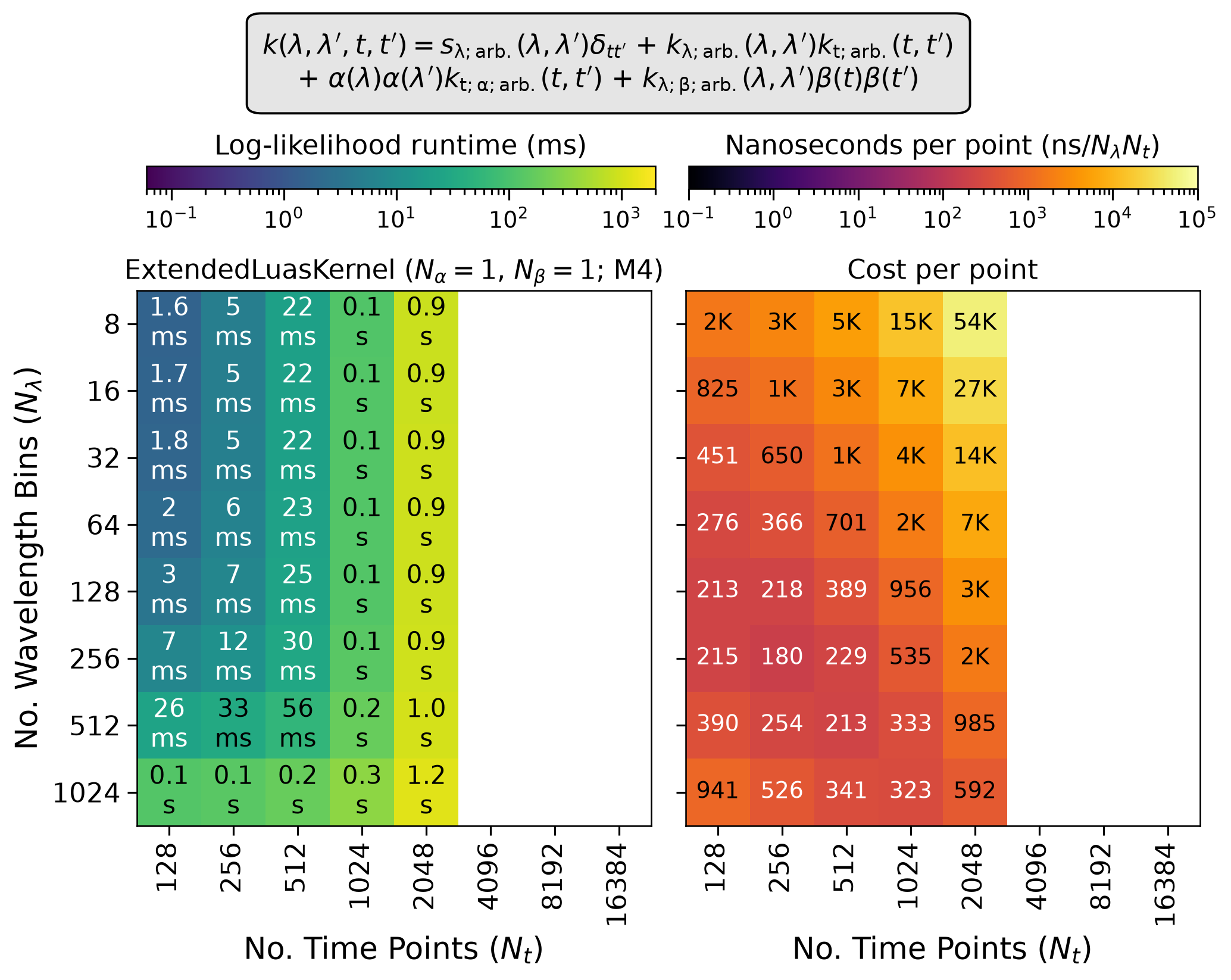}
    \end{minipage}\hfill%
    \begin{minipage}[b]{\columnwidth}
        \includegraphics[width=\linewidth]{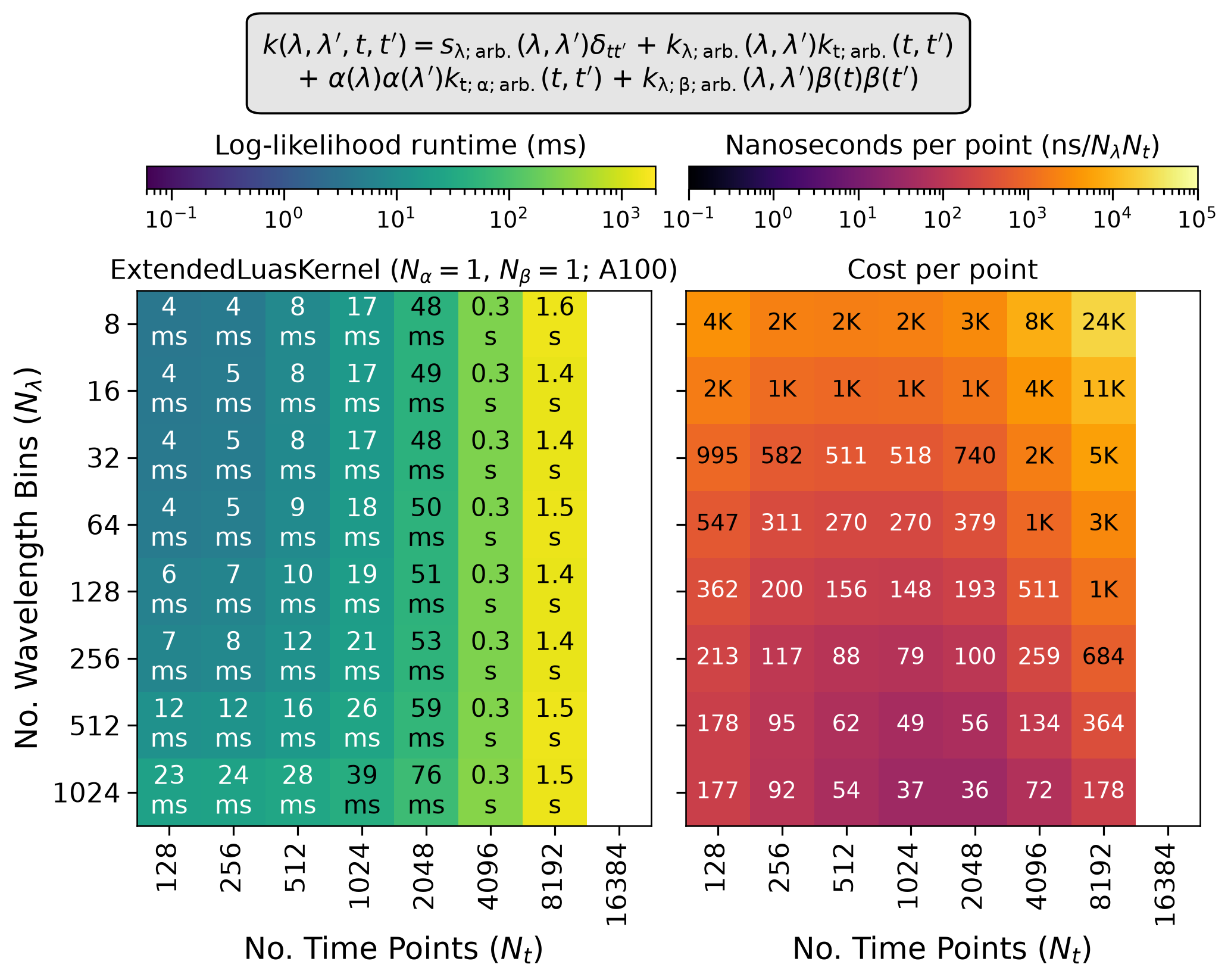}
    \end{minipage}

    \begin{minipage}[t]{\columnwidth}
         \caption{The \textsc{ExtendedLuasKernel} scales less efficiently to larger datasets, particularly because it cannot exploit quasiseparable kernels in either dimension, with a runtime scaling of $\sim\mathcal{O}((1 + N_\alpha^3)N_t^3 + (1 + N_\beta^3)N_\lambda^3 + N_\alpha N_\beta^2 N_\lambda^2 N_t + N_\lambda N_t^2)$. It is worth noting however that for the case $N_\alpha = N_\beta = 1$, we still achieve runtimes similar to the \textsc{LuasKernel} (Figure~\ref{fig:LuasKernelND_cost}) as well as the \textsc{LuasPlusMultiLowRankKernel} with $N_\alpha = 1$ (Figure~\ref{fig:LuasPlusMultiLowRankKernel_Nalpha1_cost}), so we can add these extra low-rank terms with minimal extra computational expense.}
    \label{fig:ExtendedLuasKernel_Nalpha1_Nbeta1_cost}
    \end{minipage}\hfill%
    \begin{minipage}[t]{\columnwidth}
        \caption{Re-running the same kernel in Figure~\ref{fig:ExtendedLuasKernel_Nalpha1_Nbeta1_cost} on GPU, the calculation still appears to be bottlenecked by the eigendecomposition of a time covariance matrix (comparing with Figure~\ref{fig:matrix_decomp_cost}), but this is much faster on the GPU.}
\label{fig:ExtendedLuasKernel_Nalpha1_Nbeta1_A100_cost}
    \end{minipage}

    \vspace{2\baselineskip}

    \begin{minipage}[b]{\columnwidth}
        \includegraphics[width=\linewidth]{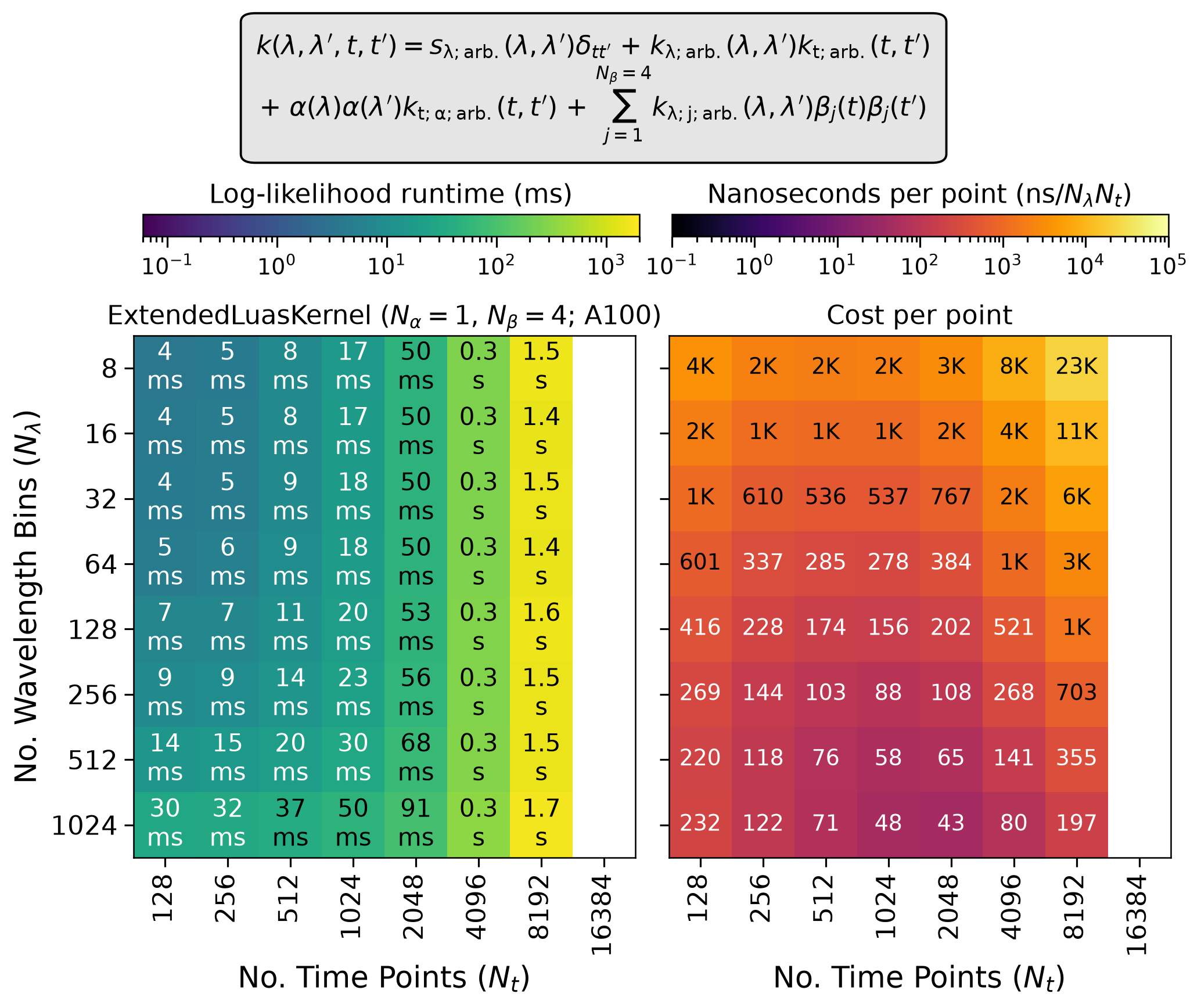}
    \end{minipage}\hfill%
    \begin{minipage}[b]{\columnwidth}
        \includegraphics[width=\linewidth]{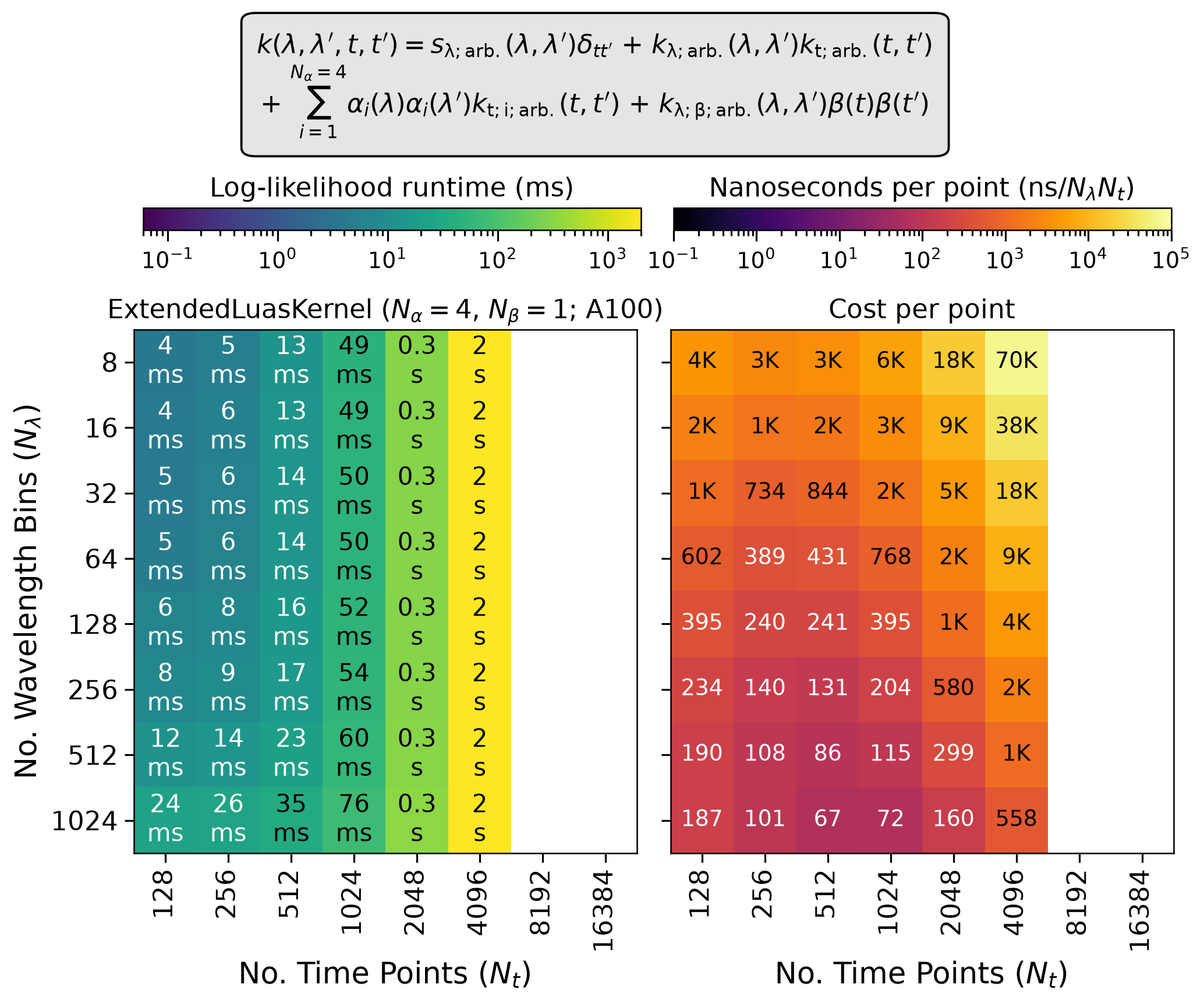}
    \end{minipage}

    \begin{minipage}[t]{\columnwidth}
        \caption{Remaining on GPU, we see that we can increase to $N_\beta = 4$ while keeping $N_\alpha = 1$ with limited change in runtime from Figure~\ref{fig:ExtendedLuasKernel_Nalpha1_Nbeta1_A100_cost}, likely because we are still bottlenecked by the same time covariance matrix eigendecomposition.}
        \label{fig:ExtendedLuasKernel_Nalpha1_Nbeta4_A100_cost}
    \end{minipage}\hfill%
    \begin{minipage}[t]{\columnwidth}
        \caption{However, increasing to $N_\alpha = 4$ while keeping $N_\beta = 1$ is much more expensive, even on the GPU. In fact, for small $N_\lambda$ this can sometimes be outperformed by a general Cholesky decomposition of the full covariance matrix (Figure~\ref{fig:GeneralKernel_A100}), although it is much more efficient for larger $N_\lambda$. This expensive calculation is mostly due to the $(N_\alpha N_t, N_\alpha N_t)$ Cholesky decomposition, combined with the additional costs due to the complex algorithm (as shown in Appendix~\ref{app:extendedluaskernel}).}
        \label{fig:ExtendedLuasKernel_Nalpha4_Nbeta1_A100_cost}
    \end{minipage}
\end{figure*}

\begin{figure*}[p]
    \begin{minipage}[b]{\columnwidth}
        \includegraphics[width=\linewidth]{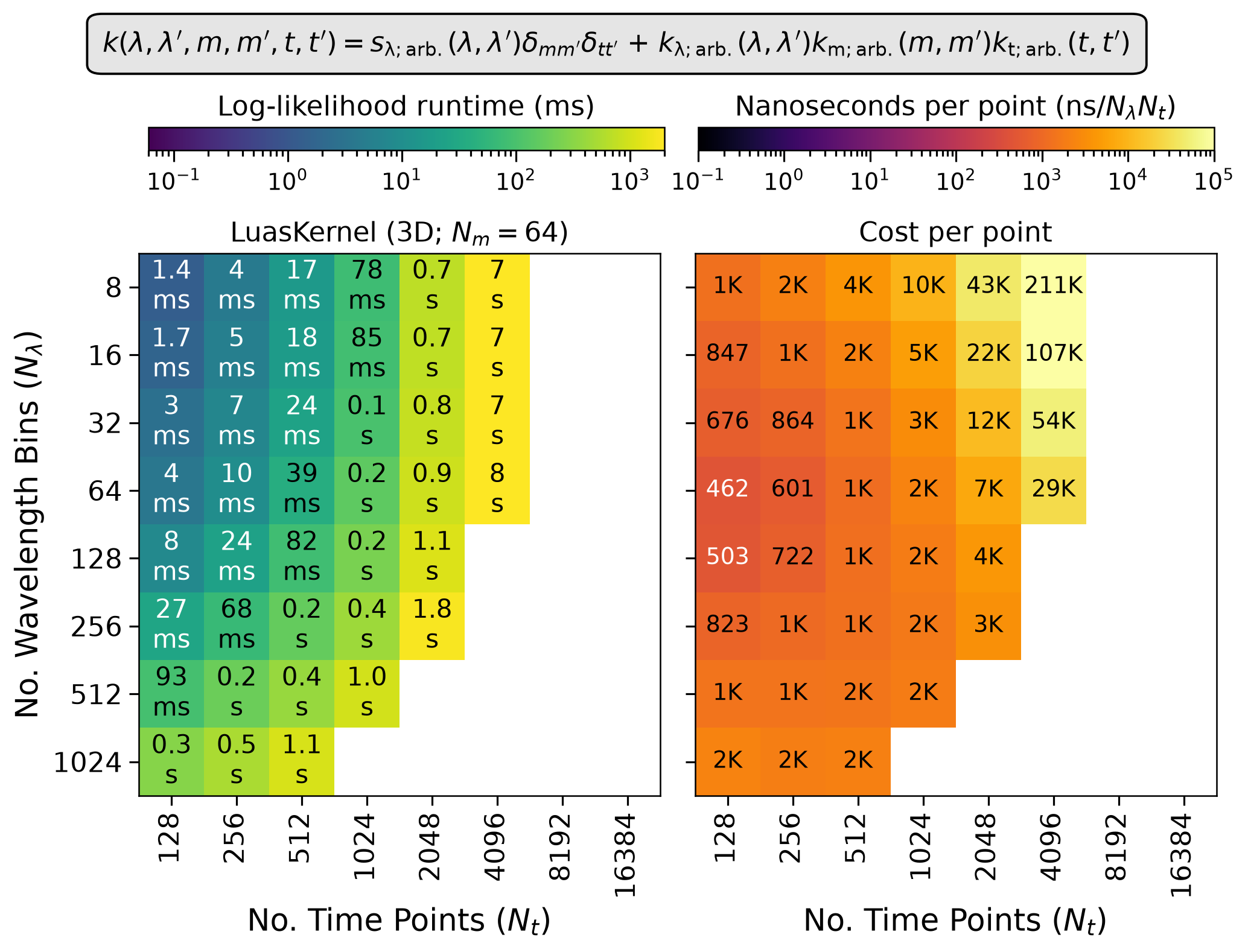}
    \end{minipage}\hfill%
    \begin{minipage}[b]{\columnwidth}
        \includegraphics[width=\linewidth]{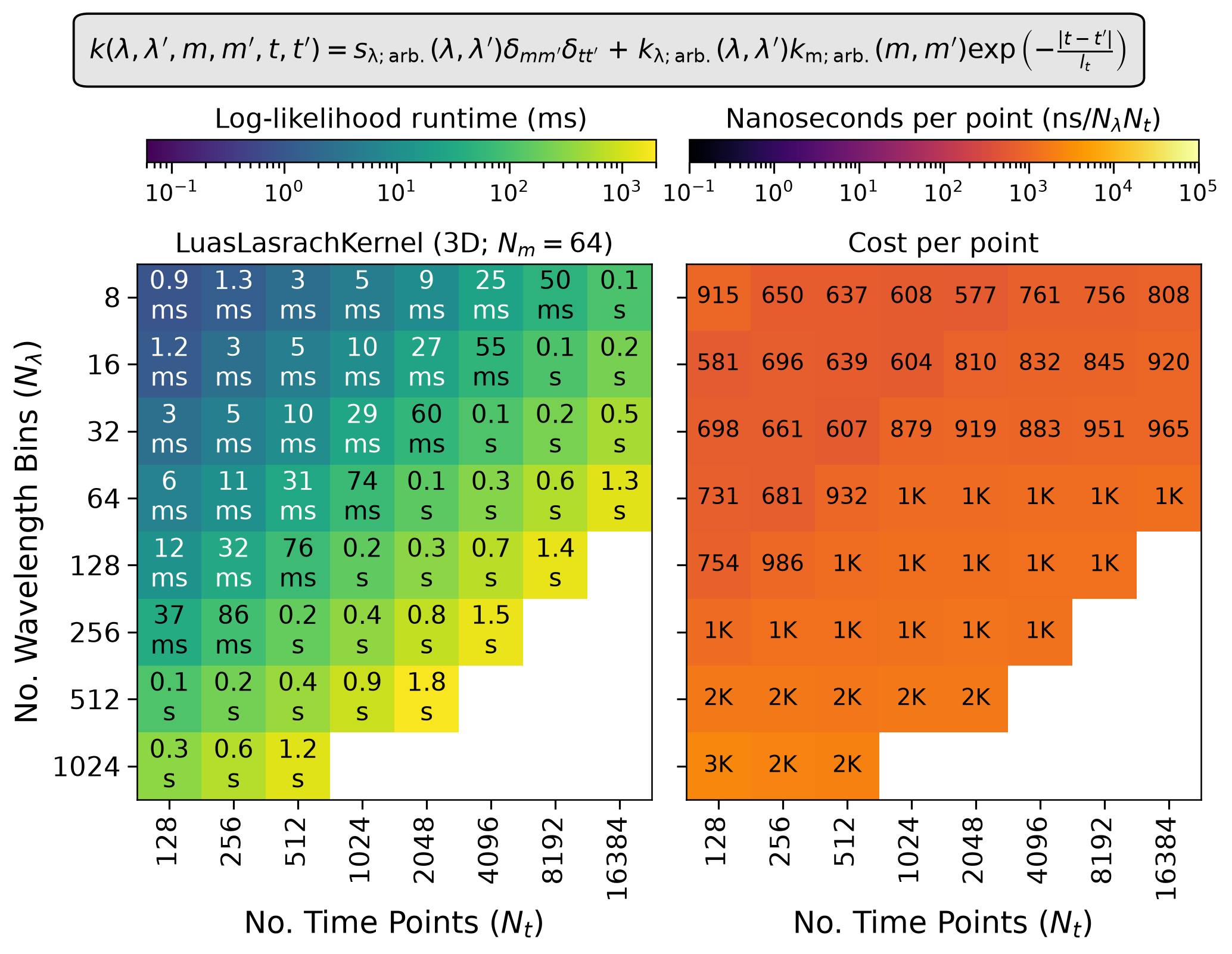}
    \end{minipage}

    \begin{minipage}[t]{\columnwidth}
         \caption{Arbitrary, separable 3D kernel function implemented using the \textsc{LuasKernel} for a third data dimension with a fixed size of $N_m = 64$. Runtimes are relatively slow on CPU, with a runtime scaling of $\mathcal{O}(N_m^3 + N_\lambda^3 + N_t^3 + N_m N_\lambda N_t (N_m + N_\lambda + N_t))$, limited both by the cost of eigendecomposition and the large matrix multiplications, compared with the same model in Figure~\ref{fig:LuasKernelND_3D_m64_A100_cost}.}
    \label{fig:LuasKernelND_3D_m64_cost}
    \end{minipage}\hfill%
    \begin{minipage}[t]{\columnwidth}
        \caption{The same as Figure~\ref{fig:LuasKernelND_3D_m64_cost} but using the \textsc{LuasLasrachKernel} with an exponential kernel in time. The $\mathcal{O}(N_m^3 + N_\lambda^3 + N_m N_\lambda N_t (N_m + N_\lambda + (J_{K_t} + J_{\Sigma_t})^2))$ runtime scaling is linear in the number of time points, but this runtime advantage is limited to quasiseparable kernels in the longest dimension, while the \textsc{LuasKernel} can use any time kernel.}
    \label{fig:LuasLasrachKernel_3D_m64_cost}
    \end{minipage}

    \vspace{2\baselineskip}

    \begin{minipage}[b]{\columnwidth}
        \includegraphics[width=\linewidth]{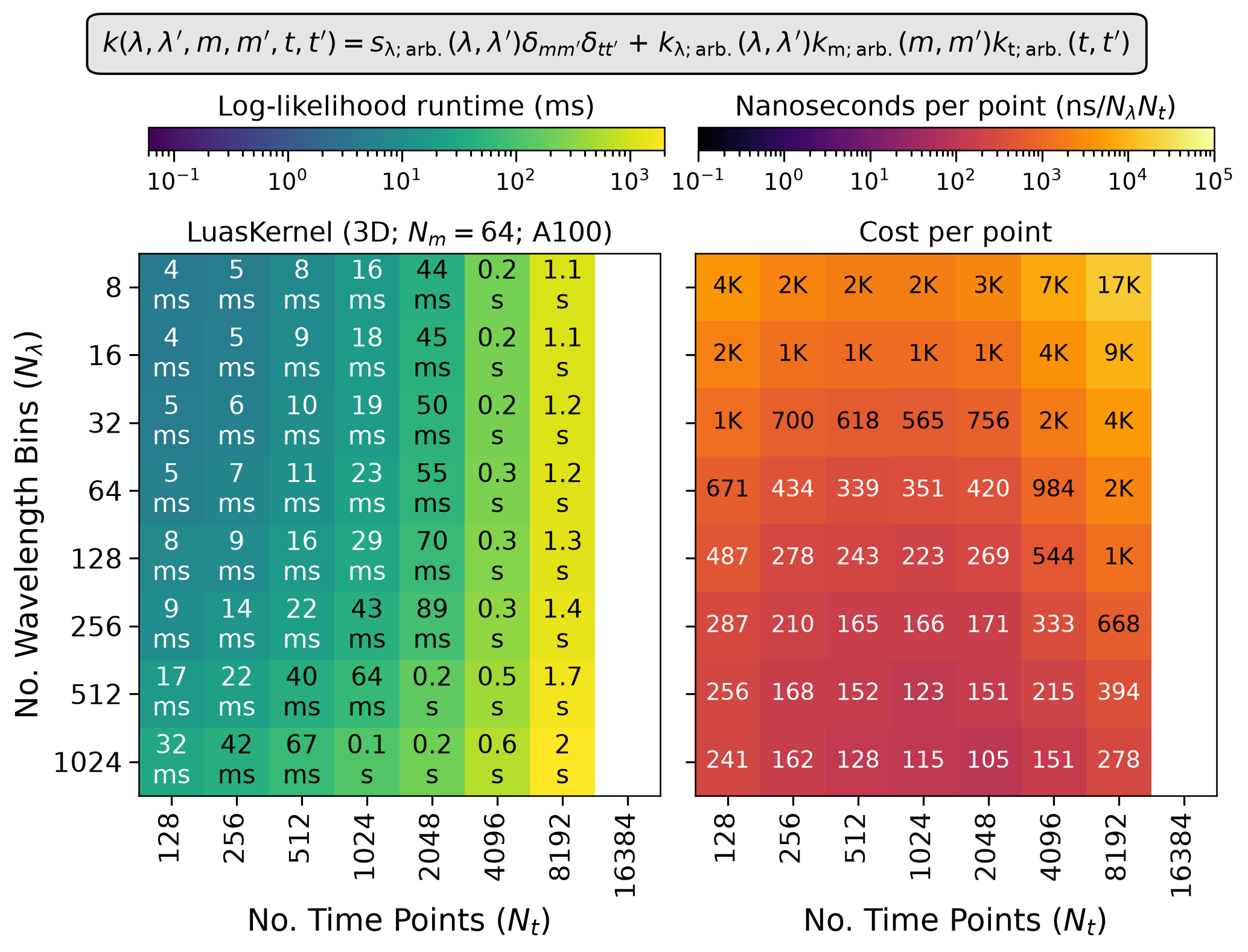}
    \end{minipage}\hfill%
    \begin{minipage}[b]{\columnwidth}
        \includegraphics[width=\linewidth]{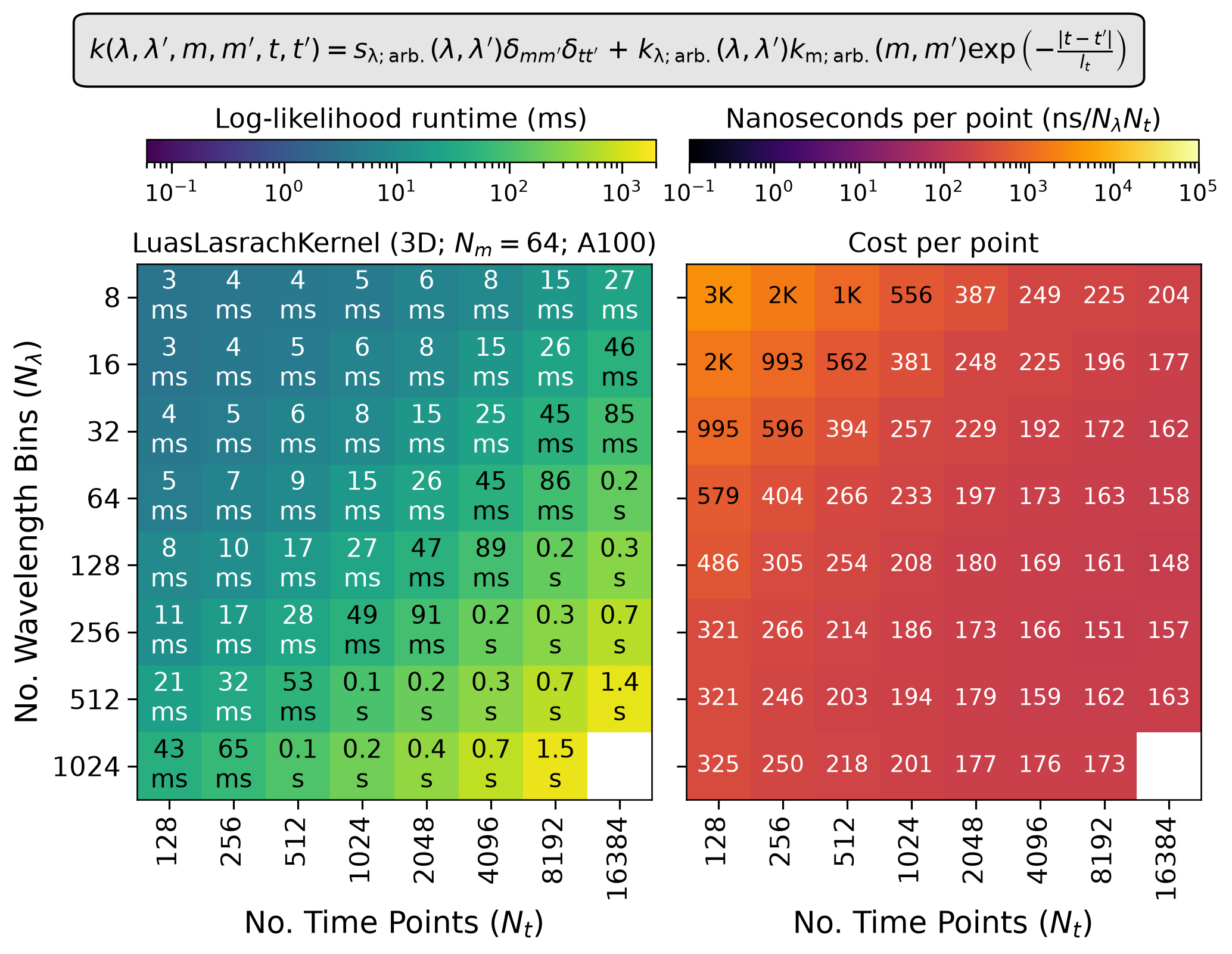}
    \end{minipage}

    \begin{minipage}[t]{\columnwidth}
    \caption{The same 3D kernel function as Figure~\ref{fig:LuasKernelND_3D_m64_cost} but implemented on GPU. The runtime cost is comparable in runtime to the 2D case in Figure~\ref{fig:LuasKernelND_A100_cost} despite having 64x as many datapoints to fit in the additional dimension. However, for the largest datacubes benchmarked the large matrix multiplications required to transform the residuals dominate the calculations.}
    \label{fig:LuasKernelND_3D_m64_A100_cost}
    \end{minipage}\hfill%
    \begin{minipage}[t]{\columnwidth}
        \caption{The same as Figure~\ref{fig:LuasKernelND_3D_m64_A100_cost} but using the \textsc{LuasLasrachKernel} with an exponential kernel in time, also on GPU. The runtimes are largely comparable although the scaling with the number of time points is slightly better. We reach datasets with $2^{29}$ datapoints (i.e. $4$ GB) for the largest datacubes benchmarked.}
    \label{fig:LuasLasrachKernel_3D_m64_A100_cost}
    \end{minipage}
\end{figure*}

We use automatic differentiation in \textsc{JAX} for gradient calculations, where the relative increase in computational cost when including gradient calculations was typically found to vary between $\sim1$-$10\times$ for all the log-likelihood runtimes. Specifically for the benchmarking calculations, if the log-likelihood runtime was $<1$ second for a given dataset shape between Figures~\ref{fig:SingleKronTermKernel_quasireadout_cost}-\ref{fig:ExtendedLuasKernel_Nalpha1_Nbeta4_A100_cost}, we also benchmarked calculating both the log-likelihood and its gradients for all hyperparameters. Averaging across all dataset shapes tested for each optimisation (excluding Figure~\ref{fig:GordonKernel_Nalpha1_cost} which implements the existing \citealt{Gordon2020} method in \textsc{tinygp}), the mean relative cost of including gradients was 4.5$\times$, while the median relative cost was 3.6$\times$. Both the mean and median relative cost were always $<10\times$ the log-likelihood cost, except for the \textsc{MultiLowRankKernel} with $N_\alpha = 16$ in Figure~\ref{fig:MultiLowRankKernel_Nalpha16_cost}, which averaged between $15$-$16\times$ more expensive for both mean and median averages. This is likely due to the large number of parameters to calculate the gradients with respect to (the vectors $\boldsymbol{\alpha}_{\lambda; i}$ alone contribute $N_\alpha N_\lambda$ free parameters). Interestingly, the \textsc{MultiLowRankKernel} with $N_\alpha = 16$ \textit{orthogonal} components in Figure~\ref{fig:MultiLowRankKernel_Nalpha16_ortho_cost} performs significantly better, with a $<5\times$ performance hit relative to its log-likelihood calculation. Regardless, no individual dataset shape tested ever cost $>20\times$ the computational cost across all figures, even for cases where there are $>10,000$ hyperparameters to calculate gradients for. 


\end{document}